\documentclass[11pt,a4paper,fleqn,twoside]{report}

\usepackage{amsmath}
\usepackage{amssymb}

\usepackage{appendix}
\usepackage{booktabs}
\usepackage{array}
\usepackage{mdwlist}

\usepackage{graphicx}
\usepackage{wrapfig}
\usepackage{subfig}

\usepackage[latin1]{inputenc} 
\usepackage[T1]{fontenc} 
\usepackage[german,english]{babel}

\usepackage{hyperref}
\usepackage{emptypage}

\graphicspath{{img/}}

\showhyphens{temperature}

\newcommand{\lbb}{ L_{\scriptstyle{bb}} }

\newcommand{\lbk}{ L_{\mathbf k \; \scriptstyle{bb}} }
\newcommand{\lbw}{ L_{\omega \; \scriptstyle{bb}} }
\newcommand{\lbv}{ L_{\nu \; \scriptstyle{bb}} }
\newcommand{\lbvs}[1]{ L^{\scriptscriptstyle{ #1 }}_{\nu \; \scriptstyle{bb}} }
\newcommand{\tss}[1]{ { \scriptscriptstyle{ #1 } } }
\newcommand{\srm}[1]{ { \scriptscriptstyle{ \mathrm{#1} } } }
\newcommand{\sucmb}{\mathrm{SU(2)}_{\mathrm{\scriptscriptstyle{CMB}}}}
\newcommand{\tcmb}{T_{\mathrm{\scriptscriptstyle{CMB}}}}
\newcommand{\Lcmb}{\Lambda_{\mathrm{\scriptscriptstyle{CMB}}}}

\title{Detection of Thermal Ground State Effects in SU(2) Yang-Mills Thermodynamics}
\author{Carlos Falquez}

\begin{document}

\pagenumbering{roman}

\setlength{\oddsidemargin}{0cm} 
\setlength{\evensidemargin}{1cm} 
\pagestyle{headings}

\begin{titlepage}
  \centering \renewcommand{\baselinestretch}{1.5} \sc
  \huge Fakult\"at f\"ur Physik \\
  \LARGE Karlsruher Institut f\"ur Technologie (KIT)\\
  \vfill \rm \large {\bf
    Diploma thesis \\
    in Physics \\
    submitted by \\
    Carlos Falquez \\
    born in Guayaquil, Ecuador}
\end{titlepage}

\thispagestyle{empty}
\cleardoublepage

\begin{titlepage}
  \centering \renewcommand{\baselinestretch}{1.5} \vspace*{3cm} \sc
  \huge
  Investigation of Thermal\\
  Ground-State Effects in\\
  SU(2) Yang-Mills Thermodynamics \vfill \rm \large {\bf
    This diploma thesis has been carried out by \\
    Carlos Falquez \\
    at the \\
    Laboratorium f\"ur Applikationen der Synchrotronstrahlung (LAS)\\
    under the supervision of \\
    Priv.-Doz.~Dr.~Ralf Hofmann}
\end{titlepage}

\thispagestyle{empty}
\cleardoublepage

\setlength{\oddsidemargin}{0cm} 
\setlength{\evensidemargin}{0.85cm}

\selectlanguage{english}

\begin{abstract}
  
  \noindent We investigate experimental consequences of the postulate
  that fundamentally photon propagation is governed by an SU(2) rather
  than a U(1) gauge principle. In the context of thermodynamics
  the SU(2) Yang-Mills theory is assumed to be in its deconfining
  phase. In addition to the results already available, we compute
  the dispersion law for the longitudinal mode, which does not
  thermalize with matter, at one-loop accuracy.
  We analyze radiometric and bolometric methods to detect
  a gap in the black-body radiation spectra for low frequencies
  and at low temperatures. We simulate the expected experimental results
  for total-power and Dicke-switch radiometry. These simulations
  indicate that the predicted SU(2) signature is measurable with presently
  available detector technology appealing to both radiometric methods.
  We also investigate non-homogeneous thermodynamics (thermomagnetic
  effect) in the adiabatic approximation.
  \vspace{12mm} 
\end{abstract}

\cleardoublepage

\setcounter{tocdepth}{2} 
\tableofcontents
\cleardoublepage
\thispagestyle{empty}

\pagenumbering{arabic}

\chapter{Introduction}
\label{chap:intro}

The discovery of the Cosmic Microwave Background (CMB) by Penzias and Wilson \cite{PenziasWilson}
is easily one of the pivotal moments in the history of cosmology
and the development of the Hot Big-Bang model.
COBE further confirmed that the CMB power spectrum follows to a high accuracy $\sim 10^{-2}$
a Planck distribution in the frequency region of $\nu$=60\,--\,600\,GHz,
\cite{COBE1994,COBE1990,weinberg2008cosmology}.

Recently, however, measurements of CMB line temperatures $T$ performed by the \\
ARCADE 2 instrument \cite{Arcade2}
at very low-frequencies ($\nu=3\,\mbox{GHz}\, \dots\, 90\,\mbox{GHz}$)
show a statistically significant ($5\,\sigma$) unexplained power excess.
Combining the ARCADE 2 data with data from the relevant literature
in the frequency region from 22 MHz to 10 GHz
results in a background power law spectrum of the form \cite{Arcade2a}
\begin{align}
  T(\nu) \, &= \, T_0+T_R\,\left(\frac{\nu}{\nu_0}\right)^\beta\,,
\label{lineT}
\end{align}
with $T_0=2.725$\,K, $\nu_0=1$\,GHz, $T_R=1.19\pm 0.14$\,K, and a spectral index of $\beta=-2.62\pm 0.04$.
It is further argued in \cite{Arcade2a} that this deviation
from the perfect black-body relation $T(\nu)=\mathrm{const}$
is neither an artifact of galactic foreground subtraction,
nor can it be attributed to distant point sources.

Remarkably, a thermodynamical field theory developed in \cite{Hofmann2005,Hofmann2007fd}
could be able to explain the power excess in \eqref{lineT}.
Namely, it is argued that the gauge group $\mathrm{U(1)_Y}$ of the photon field
in the Standard Model (SM) is not fundamental but merely emerges from a 
SU(2) Yang-Mills theory in a specific thermodynamical phase: The \emph{deconfining phase}.
This is called the $\sucmb$ hypothesis.
The theory is determined by a single parameter,
and can be fixed by specifying the \emph{critical temperature} $T_c$  
below which the thermal SU(2) gas undergoes a \emph{phase transition}
to the \emph{preconfining phase}.

Assuming the $\sucmb$ hypothesis, the excess power at low frequencies measured by the ARCADE2 radiometers
can be explained as a result of evanescent photon fields thermalized near the critical
temperature $T_c\overset{!}{=}\tcmb$ of the deconfining phase \cite{RH2009}.

That such an explanation for the excess power of the CMB is possible 
based solely on gauge symmetry considerations and finite temperature effects
motivates the search for further experimental evidence in favor of the $\sucmb$ hypothesis.
The ARCADE2 observation serves to fix $T_c$, which we use
to predict further observable consequences that can then be tested by experiment. 

The purpose of this thesis is the investigation of possible realizations of such an
experimental test.
We discuss the characteristics of the thermal radiation spectrum at low temperature and frequency
under the $\sucmb$ hypothesis. It will turn out that SU(2) gauge dynamics manifests
itself through the appearance of a temperature-dependent \emph{frequency cutoff} $\nu^*(T)$ below which
photon propagation is forbidden.
The cutoff frequency $\nu^*$ thus defines a \emph{screening region}.
At low frequencies and temperatures comparable to that of $T_c$,
this effect results in a \emph{black-body anomaly}:
The vanishing of the power output of a black-body cavity within the screening region.
This will turn out to be the key experimental effect to
search for when probing for SU(2) physics.

This thesis is organized as follows. Chapter \ref{chap:GaugeSym} reviews
elementary properties of gauge theory and corresponding symmetry groups.
A short review of Maxwell electrodynamics introduces the required mathematical
formalism which is then generalized to nonabelian gauge symmetries.
Basic concepts of topological configurations in SU(2) Yang-Mills theory are introduced.
In the final section, thermal field theory is shortly summarized.
Chapter \ref{chap:YM} serves as an introduction to our main subject: The thermodynamics
of SU(2) Yang-Mills theory, in the context of the nonperturbative approach developed
in \cite{Hofmann2005,Hofmann2007fd}.
After a review of the relevant theory,
we proceed to calculate the dispersion law of the transverse and longitudinal polarization
modes of the SU(2) photon. We also investigate the thermodynamical consistency
of one-loop resummed energy density and pressure.
In chapter \ref{chap:inhtd} we investigate the possibility of generalizing
an important assumption made in chapter \ref{chap:YM} while deriving
the thermal ground state, namely thermodynamical homogeneity (and isotropy).
It will turn out that, within the restrictions imposed by an adiabatic approximation,
certain deviations from homogeneity are indeed allowed.
Chapter \ref{chap:radiometry} is the most important section of this work.
After a quick introduction to the basics of radiation transfer, antenna theory and radiometry,
we investigate the implications of the $\sucmb$ hypothesis
on the radiometric measurement of thermal photon radiation.
We show how the existence of a screening region
can be established through purely radiometric means.
In the final section, we comment on the possibility of a bolometric experimental setup.
The appendices contain the details of certain derivations.
Essential for chapter \ref{chap:inhtd} is the result derived
in appendix \ref{app:ymeom}, where a particular
solution to the SU(2) Yang-Mills equations
in Minkowskian signature is derived.
We have attached a CD with the Mathematica source code used in this work.

Unless stated otherwise, we always work in natural units $\hbar=c=k_B=1$.


\chapter{Gauge Fields and Symmetries}
\label{chap:GaugeSym}

In this chapter, we give a very summarized survey of gauge theory
and related concepts, in order to provide the reader with the minimal
background needed for chapter \ref{chap:YM}.
We first review the basics of gauge theory and symmetry groups,
then we briefly show how Yang-Mills theory emerges by extending
the U(1) symmetry to the nonabelian gauge group SU(N).
Next, we discuss topological field configurations in the context of Euclidean Yang-Mills theory.
Finally, a brief summary of thermal field theory is given for completeness.

\section{Gauge theories and symmetry groups}

\subsection{Maxwell electrodynamics}
\label{seq:abelGT}

In the Lagrangian formalism, the dynamics of the photon field $A^\mu$ in the presence
of an external charged field of current density $j^\mu$ are described by the
\emph{Maxwell Lagrangian density} $\mathcal L_{\tss{M}}$
\begin{align}
  \mathcal L_{\tss{M}} \, \equiv\, -\frac{1}{4} F^{\mu\nu} F_{\mu\nu} - e j^\mu A_\mu\,.
  \label{EMLag}
\end{align}
where the \emph{coupling constant} $e$ characterizes the strength of the
electromagnetic interaction. It is identified with the \emph{electric charge}
of the current $j^\mu$ inducing external field.

The \emph{field strength tensor} $F_{\mu\nu}$ can be defined by the action
of the commutator $[D_\mu,D_\nu]$ on an arbitrary Lorentz scalar $\Phi$
\begin{align}
  F_{\mu\nu} \Phi \, \equiv \, \frac{i}{e}[D_\mu,D_\nu]\Phi \,,
  \label{defFmn}
\end{align}
where the \emph{covariant derivative} $D_\mu$ is defined
in terms of the \emph{gauge potential} $A_\mu$
\begin{align}
  D_\mu \,\equiv\, \partial_\mu - i e A_\mu \,.
  \label{DefCovDer1}
\end{align}
The familiar electromagnetic vector fields $\mathbf E$ and $\mathbf B$
are recovered from the field strength tensor by the identities
\begin{align}
  E^i \, &\equiv \, - F^{0i}\,,
  \label{defEfromF} \\
  B^i \, &\equiv \, - \frac12 \epsilon^{ijk} F_{jk} \,.
  \label{defBfromF}
\end{align}
It can be shown (Bohm-Aharonov effect) that the gauge potential $A_\mu$,
not the field strength tensor $F_{\mu\nu}$,
is the fundamental physical field of the theory \cite{ryder1996quantum}.

The Maxwell field strength tensor can in turn be written in terms of the gauge potential $A_\mu$
by inserting equation \eqref{DefCovDer1} in \eqref{defFmn}  
\begin{align}
  F_{\mu\nu}\, &=\, \partial_\mu A_\nu - \partial_\nu A_\mu \,.
  \label{DefF}
\end{align}
Note that the electromagnetic coupling $e$ does not appear in the expression \eqref{DefF},
since the single gauge field commutes with itself and terms such as $e^2[A_\mu,A_\nu]$ vanish.
The photon cannot \emph{interact with itself}, and does not carry charge.
We will later generalize Maxwell electrodynamics to allow for a
selfinteracting (also charged) gauge potential multiplet.

The \emph{inhomogeneous} Maxwell equations are given by
the Euler-Lagrange equations of motion derived from \eqref{EMLag}
by the principle of least action
\begin{align}
  \partial_\mu F^{\mu\nu} \, = \, j^\nu \,.
  \label{ELeqMaxwell1}
\end{align}
The \emph{homogeneous} Maxwell equations follow
from the \emph{Bianchi identity} \cite{atiyah1979,frankel2004geometry,ryder1996quantum}
\begin{align}
  \partial_\mu \tilde{F}^{\mu\nu} \, &= \, 0 \,,
  \label{ELeqMaxwell2}
\end{align}
where the \emph{dual field strength tensor} $\tilde{F}^{\mu\nu}$ is defined by
\begin{align}
  \tilde{F}^{\mu\nu} \, \equiv\, \frac{1}{2}\epsilon^{\mu\nu\sigma\rho}F_{\sigma\rho} \,.
  \label{DefDualF}
\end{align}
Here $\epsilon^{\mu\nu\sigma\rho}$ is the totally antisymmetric Levi-Civita tensor
with $\epsilon^{0 1 2 3}=1$.

The dynamics of the system described by equation. \eqref{ELeqMaxwell1} and \eqref{ELeqMaxwell2}
are \emph{invariant} under local field shifts
\begin{align}
  A_\mu \, \rightarrow \, A'_\mu \,=\, A_\mu + \frac{1}{e} \partial_\mu \xi\,,
  \label{VectorU1GT1}
\end{align}
with $\xi=\xi(x)$ an arbitrary Lorentz scalar function.
Transformations of the form \eqref{VectorU1GT1} are called \emph{gauge transformations}.
They leave the field strength tensor \eqref{DefF},
and thus the equations of motion \eqref{ELeqMaxwell1} and integrability conditions \eqref{ELeqMaxwell2}, unmodified
\begin{align}
  F^{\mu\nu} \, &\rightarrow \, F'^{\mu\nu} \, = \,F^{\mu\nu}\,.
  \label{TensorU1GT}
\end{align}
This \emph{gauge invariance} is a manifestation of redundancy in the degrees of freedom
of the theory: The photon is a massless spin-1 particle with only two transverse polarizations,
yet in \eqref{EMLag} the dynamics are given by a 4-component gauge potential $A_\mu$.
The excess polarization modes are nevertheless necessary for a covariant formalism.
Local gauge freedom also implies that the gauge potential
transforms under Lorentz rotations as a 4-vector only up to gauge transformations.
A Lorentz transformation between two frames is always associated with
a gauge transformation \cite{moriyasu1983elementary}.

In the framework of differential geometry, the field $A_\mu$
with transformation law \eqref{VectorU1GT1} can be interpreted
as the \emph{connection} on a \emph{principal} U(1)-\emph{bundle}
over the space-time manifold $\mathbb R^4$ with Minkowskian signature $(+,-,-,-)$,
U(1) being the group of 1x1 unitary matrices \cite{atiyah1979,frankel2004geometry}.
The Maxwell gauge potential $A_\mu$ is therefore also called the U(1) \emph{gauge field}.
Its emergence can be understood as the result of demanding \emph{invariance}
under \emph{local} phase transformations of the quantum state vector
$|\psi \rangle$ characterizing the physical system \cite{ryder1996quantum}
\begin{align}
  |\psi \rangle \, \rightarrow \, |\psi' \rangle   %
                                        \, &= \,\Omega(x) |\psi\rangle \,,
       \qquad
  \langle \psi' | \psi' \rangle \, \overset{!}{=} \, \langle \psi | \psi \rangle \,.
  \label{PsiGT} 
\end{align}
Here, $\Omega(x)=e^{i \xi(x)}$ is a space-time function taking values in U(1),
which we write for simplicity as $\Omega(x) \in \mathrm{U(1)}$.
Theories which exhibit invariance under a continuous group of \emph{local symmetry transformations}
are called \emph{gauge theories}.
The group U(1) is the \emph{gauge group} of the Maxwell theory.
After a U(1) phase transformation,
the Maxwell gauge field $A_\mu$ transforms as in \eqref{VectorU1GT1},
which may be rewritten as
\begin{align}
  A_\mu \, \rightarrow \, A'_\mu \,=\, \Omega A_\mu \Omega^\dagger +\frac{i}{e} \Omega \partial_\mu \Omega^\dagger\,,
  \label{VectorU1GT2}
\end{align}
for any $\Omega(x) \in \mathrm{U(1)}$.

The geometric formalism just described can be generalized to larger groups G,
while respecting Lorentz symmetries and covariance \cite{atiyah1979}.
From \eqref{PsiGT} it is clear that G must be either unitary or anti-unitary.
Following Yang and Mills \cite{YangMills54}, we replace phase transformations
with non-abelian \emph{isotopic spin} rotations of SU(2) valued fields.
The theory of SU(N) groups is reviewed in the next section.

\subsection{Lie groups and Lie algebras}

\label{sec:LieGroups}

The U(1) group underlying Maxwell electromagnetism is the \emph{abelian} group of phase rotations
(two arbitrary phase rotations commute), and a generic element of U(1) may be written as $e^{i \xi}$ with real parameter $\xi$.
In general, a \emph{continuous group} G of order $n$ may be parametrized by a set of $n$ real parameters.
A \emph{Lie group} $G$ of order $n$ is a continuous group where the set of parameters live
in a differentiable manifold $M^n$.
The operations of group multiplication and inversion are \emph{smooth maps} in $M^n$.
A \emph{compact Lie group} is a Lie group with compact manifold.

In the neighbourhood near the identity element $\mathbf 1$,
$M^n$ looks locally like $\mathbb R^n$ with flat Euclidean metric.
This region can always be given a coordinate basis.
The $n$ coordinate unit vectors $t^a$, $a=1..n$ are called the \emph{generators} of the group $G$.
Every element $g$ of $G$ near the identity can thus be written as
\begin{align}
  g \, = \, \mathbf 1 + i \omega^a t^a\,,
  \label{defInfg}
\end{align}
with $\omega_a$ infinitesimal parameters.
Note that for unitary groups U(N), $t^a$ is hermitian,
and for special unitary groups SU(N), $t^a$ is also traceless.
Elements further away from the identity are generated by the \emph{exponential map}
$e^{i\omega^a t^a}$. In the case of a matrix group, the exponential map is given
by the usual series
\begin{align}
  e^{i\omega^a t^a} \, = \, \sum_{n=0}^{\infty} \frac{i^n}{n!} \, \left( \omega^a t^a \right)^n \,.
  \label{defExpMap}
\end{align}

Let $g_1$ and $g_2$ be two elements close to the identity, from \eqref{defInfg}
we write them as $g_1=\mathbf 1 + i \omega_1^a t^a$
and $g_2=\mathbf 1 + i \omega_2^a t^a$.
Their product is given, up to first order, by the expression
$g_1 g_2 = \mathbf 1 + i (\omega_1^a + \omega_2^a) t^a + O(\omega^2)$,
so up to first order they both commute, $[g_1,g_2]=O(\omega^2)$.
Up to second order, however, the expression $g_1 g_2 g^{-1}_1 g^{-1}_2$
is not the identity, but
\begin{align}
  g_1 g_2 g^{-1}_1 g^{-1}_2 \,= \mathbf 1 - \omega_1^a \omega_2^b \, [t^a, t^b] \,.
  \label{LieComm}
\end{align}
Nevertheless, \eqref{LieComm} is still close enough to the identity
such that it can be written as in \eqref{defInfg}.
This means that the commutator $[t^a, t^b]$ must be a linear combination
of the generators $t^c$
\begin{align}
  [t^a, t^b] \, = \, i f^{abc} t^c\,.
  \label{defCommLie}
\end{align}
The factors $f^{abc} \in \mathbb R$ are called \emph{structure constants},
and can always be chosen to be completely antisymmetric.
The generators $t^c$ together with the commutator relations \eqref{defCommLie}
define the \emph{Lie algebra} of the group $G$, which we denote as $\mathfrak g$.
The \emph{rank} of the
Lie algebra is defined as the number of mutually commuting and linear independent generators.
These generators form the \emph{Cartan sub-algebra} of the group.
A Lie group $G$ containing no invariant subgroup other than the identity and $G$
itself is called \emph{simple}.
A Lie group $G$ is called \emph{semi-simple} if it does not contain any abelian
invariant subgroup other than the identity.

Finally, we review some concepts from the theory of representations \cite{gilmore2008}.
A linear representation $R$ of a group $G$
is a map which associates every element $g$ of $G$ with a linear transformation
$R(g)$ acting on a vector space $V$ of dimension $m$.
For a given basis of $V$, the transformations $R(g)$ can be written as matrices,
also of dimension $m$.
The elements of a matrix group, for example, build a representation themselves,
called the \emph{fundamental representation} of the group.
For a Lie group with $n$ generators,
the structure constants also define a matrix representation, the \emph{adjoint representation},
made up of $n \times n$ matrices
\begin{align}
  (R_c)_{ab} \, = \, - i f^{abc} \,.
  \label{defAdjRep}
\end{align}
Further discussion of Lie groups, Lie algebras, and their representations can be found in \cite{gilmore2008}.
In the following, we specialize for the group SU(2).

\section{SU(2) Yang-Mills Theory}
\label{sec:YM}

Following Yang and Mills \cite{YangMills54}, we extend the gauge symmetry of the
vacuum ($j^\mu\equiv 0$) Maxwell equations to the non-abelian compact Lie group SU(2).
The group SU(2) is defined as the group of $2\times2$ unitary matrices with unit determinant.
It is a continuous Lie group of rank 1, and its the Lie algebra is defined by
three traceless hermitian generators which we may choose as $t^a=\frac{1}{2}\sigma^a$,
$a=1..3$, where $\sigma^a$ are the Pauli matrices.
Then $\mathrm{tr}\, t^a t^b=\frac{1}{2} \delta^{ab}$,
and the structure constant is the totally antisymmetric tensor $f^{abc}=\epsilon^{abc}$.
The matrices $t^a$ define the 2-dimensional fundamental representation,
while the 3-dimensional adjoint representation is given by the structure constant as in \eqref{defAdjRep}.

Define now the Yang-Mills Lie-algebra valued connection $A_\mu$
\begin{align}
  A_\mu(x)\,=\,A^a_\mu(x) t^a \,,
  \label{defSU2A}
\end{align}
where the $A^a_\mu(x)$ are (Minkowski) space-time-dependent real functions.
The gauge potentials $A^a_\mu$ now carry space-time index $\mu$
and internal symmetry index $a$.
They are the basic dynamical variables of SU(2) Yang-Mills theory.

Generalizing \eqref{VectorU1GT2}, we demand that under SU(2) gauge transformations,
the Lie-algebra valued potential transform as
\begin{align}
  A_\mu \, \rightarrow \, A'_\mu \,=\, \Omega A_\mu \Omega^\dagger +\frac{i}{g} \Omega \partial_\mu \Omega^\dagger\,,
  \label{defAGTSU2}
\end{align}
where the local transformation $\Omega(x)=e^{i\omega^a t^a}$ is now an element of SU(2), $\Omega(x) \in \mathrm{SU(2)}$, and $g$
is the Yang-Mills \emph{gauge coupling}.
Note that this implies that the individual gauge potentials $A_\mu^a=2\mathrm{tr}\,A_\mu t^a$
transform in the adjoint representation, up to an inhomogeneous term.
A field $\phi^a$ transforming homogeneously $\phi^a\rightarrow \phi'^a = \Omega^{ab}_{\mathrm{adj}}\,\phi^b$
under the adjoint representation $\Omega^{ab}_{\mathrm{adj}}\equiv 2 \mathrm{tr} \,\, t^a \Omega \, t^b \,\Omega^\dagger$
is also called an \emph{adjoint field}.

As in Maxwell electrodynamics, we introduce the SU(2) gauge covariant derivative $D_\mu$.
The operator $D_\mu$ is constructed so that a field and its covariant derivative
transform in the same way.
Acting on a field $\phi$ that transforms bi-homogeneously under the fundamental representation,
$\phi\rightarrow \Omega \phi \Omega^\dagger$, $D_\mu$ acts as a SU(2) matrix
\begin{align}
  D_\mu  \, &= \,  \mathbf 1_{2\times2} \partial_\mu -igA_\mu \,.
  \label{defCovDerSU2f}
\end{align}
This results in a bi-homogeneous transformation law for $D_\mu \phi$
\begin{align}
  D_\mu\phi  \, & \rightarrow \, \Omega D_\mu\phi \Omega^\dagger\,.
  \label{defCovDerSU2fO}
\end{align}
Similarly, on an adjoint field $\phi^a$, $D_\mu$ acts as
\begin{align}
  (D_\mu \phi)^a \, &= \, [D_\mu,\phi ]^a \, = \, \partial_\mu \phi^a + g f^{abc} A_\mu^b \phi^c\,,
  \label{defCovDerSU2c}
\end{align}
so that the transformation for the adjoint field $(D_\mu \phi)^a$ reads
\begin{align}
  (D_\mu \phi)^a\, & \rightarrow \,\Omega^{ab}_{\mathrm{adj}}\, (D_\mu \phi)^b \,.
  \label{defCovDerSU2cO}
\end{align}
A Lie-valued Yang-Mills field strength tensor $F_{\mu \nu}$
can be defined as in \eqref{defFmn}
by its action on a field $\phi$ living in the fundamental representation
\begin{align}
  F_{\mu \nu} \phi \, &= \,\frac{i}{g} [D_\mu , D_\nu] \phi \,.
  \label{defFYM1}
\end{align}
Calculating the commutator in \eqref{defFYM1} gives
\begin{align}
  F_{\mu\nu} \,&=\, \partial_\mu A_\nu - \partial_\nu A_\mu -ig\left[ A_\mu, A_\nu \right]\,,
  \label{defgF1}
\end{align}
written in components
\begin{align}
  F^a_{\mu\nu} = \partial_\mu A^a_\nu - \partial_\nu A^a_\mu +g f^{abc}  A^b_\mu A^c_\nu \,.
  \label{defgF2}
\end{align}
Observe that, compared with the strength tensor \eqref{DefF} of the U(1) theory,
expression \eqref{defgF2}  has an extra term which is antisymmetric and bilinear in the
gauge fields $A^a_\mu$, and proportional to $g$.
It describes gauge field self-interactions, and is what makes \emph{non-abelian} gauge theories
fundamentally different from their abelian Maxwell counterpart.

By construction, the Yang-Mills field strength tensor $F_{\mu \nu}$ transforms
in the fundamental representation
\begin{align}
  F_{\mu\nu} \rightarrow \Omega F_{\mu\nu} \Omega^\dagger\,.
  \label{defGTF}
\end{align}
This allows us to define a gauge invariant pure Yang-Mills SU(2) Lagrangian
$\mathcal L_{\tss{YM}}$ 
\begin{align}
  \mathcal L_{\tss{YM}} \, & \equiv \, -\frac{1}{4} F^{a\mu\nu} F^a_{\mu\nu}
  \notag \\
                           &=\,  -\frac{1}{2} \, \mathrm{tr} \, F^{\mu\nu} F_{\mu\nu}\,.
  \label{defgF4}
\end{align}
The term \emph{pure} refers to the absence in \eqref{defgF4} of any field other than the SU(2) gauge field.
From the behaviour of the field strength tensor under gauge transformations \eqref{defGTF}
it is easily seen that the Lagrangian defined in \eqref{defgF4} is SU(2) gauge invariant.
Further, Lagrangian \eqref{defgF4}
respects Lorentz invariance,
renormalizability and CP-Invariance.
A term proportional to $F^a_{\mu\nu} \tilde{F}^{a\mu\nu}$, where the SU(2) dual field strength tensor is
defined as in \eqref{DefDualF},
$\tilde{F}^{a\mu\nu} \equiv \frac{1}{2} \varepsilon^{\mu\nu\rho\sigma} F_{a\rho\sigma}$,
may be added to the SU(2) Lagrangian \eqref{defgF4} if the demand for CP-Invariance is relaxed,
but this will not concern us in this work.

The SU(2) Yang-Mills action $\mathcal S_{\tss{YM}}$ is defined as 
\begin{align}
  \mathcal S_{\tss{YM}} \, & \equiv \, \int\,\mathrm{d^4} x\,\mathcal L_{\tss{YM}}
  \notag \\
  & = \, -\frac{1}{4} \int \,\mathrm{d^4}x\,F^{a\mu\nu} F^a_{\mu\nu} \,.
  \label{defYMS}
\end{align}
The Euler-Lagrange equations of motion can be derived from \eqref{defYMS}
by the principle of least action. We obtain the \emph{SU(2) Yang-Mills equations of motion}
\begin{align}
  D_\mu F^{\mu\nu} = 0\,.
  \label{defgF5}
\end{align}
As in the Maxwell case, the Bianchi identity of differential geometry
results in the integrability conditions \cite{atiyah1979,Lenz05}
\begin{align}
  D_\mu \tilde{F}^{\mu\nu} = 0 \,.
  \label{defgF7}
\end{align}
Note that \eqref{defgF5} follows from \eqref{defgF7} if the field strength tensor
satisfies
\begin{align}
  F^a_{\mu\nu} \, = \, \tilde{F}^a_{\mu\nu} \,,
  \label{defSD}
\end{align}
in which case $F_{\mu\nu}$ is called \emph{selfdual},
or
\begin{align}
  F^a_{\mu\nu} \, = \, -\tilde{F}^a_{\mu\nu} \,.
  \label{defASD}
\end{align}
in which case $F_{\mu\nu}$ is called \emph{anti-selfdual}.
Because of their importance, such field configurations are discussed in the next section.

\subsection{Selfdual field configurations}

\label{sec:selfdual}

We have seen that for \emph{(anti) selfdual field configurations}
\begin{align}
   F^a_{\mu\nu} \, = \, \pm \, \tilde{F}^a_{\mu\nu} \,,
  \label{defSelfDual}
\end{align}
the SU(2) Yang-Mills equations \eqref{defgF5} are \emph{automatically satisfied},
since \eqref{defgF5} then follows from the geometric identity \eqref{defgF7}.
This is significant, since equation \eqref{defSelfDual} is \emph{first-order},
yet \eqref{defSelfDual} automatically implies the \emph{second-order}
Yang-Mills equations.

There are no known solutions for the selfdual equation \eqref{defSelfDual}
in 4D Minkowski SU(2) Yang-Mills theory.
Consider now Euclidean 4D Yang-Mills, which is arrived at by
letting the metric become $g_{\mu\nu}\rightarrow -\delta_{\mu\nu}$
and rotating the real time-like component of the 4-vectors
\begin{subequations}
\begin{align}
  x^0 \, &\rightarrow\,ix^4             \,, \\
  A^a_0 \, &\rightarrow\, -i A^a_4      \,.
\end{align}
\label{Etransf}
\end{subequations}
with real $x^4$, $A^a_4$.
It can be shown that the Euclidean Yang-Mills action $\mathcal S^{\tss{E}}_{\tss{YM}}$
in $\mathbb R^4$ (that is, the action \eqref{defYMS} after rotation \eqref{Etransf})
can be decomposed as
\begin{align}
  \mathcal S^{\tss{E}}_{\tss{YM}}  & = \, \frac{1}{4} \int \,\mathrm{d^4}x\, \left[ \pm F^{a\mu\nu} \tilde{F}^a_{\mu\nu}
  + \frac{1}{2} \left( F^a_{\mu\nu} \mp \tilde{F}^a_{\mu\nu} \right)^2\right] \,.
  \label{defEYMS}
\end{align}
This is the celebrated \emph{Bogomol'nyi decomposition} \cite{bogomol1976stability}.
The action \eqref{defEYMS} is invariant under space-time translations.
The canonical \emph{Euclidean energy-momentum tensor} $\theta_{\mu \nu}$
then follows from Noether's theorem \cite{greiner1996field}
as the Noether current resulting from translation invariance, and
can be calculated from the Lagrangian \eqref{defgF4}
after rotation \eqref{Etransf} (up to a 4-divergence term).
In the case of pure Yang-Mills theory it can be written as \cite{Polyakov1977}
\begin{align}
  \theta_{\mu \nu} \, &= \, \frac{1}{4} \mathrm{tr} \left[ \left( F_{\mu \lambda} - \tilde{F}_{\mu \lambda} \right)
                                               \left( F_{\nu \lambda} + \tilde{F}_{\nu \lambda} \right)
                                             + (\mu \leftrightarrow \nu)
                                       \right]
  \label{defEMTensor}
\end{align}

Note that for (anti) selfdual field configurations
the energy-momentum tensor \eqref{defEMTensor} \emph{vanishes}.
This guarantees that (anti) selfdual field configurations
have zero energy-momentum, and \emph{do not propagate}.
Further, the Euclidean action \eqref{defSelfDual}
becomes
\begin{align}
  \mathcal S^{\tss{E}}_{\tss{YM}}  & = \, \pm \frac{1}{4} \int \,\mathrm{d^4}x\, F^{a\mu\nu} \tilde{F}^a_{\mu\nu} \,,
  \label{defEYMSsd}
\end{align}
with $+$/$-$ sign for selfdual/anti-selfdual field configurations.

To further interpret this result,
rescale the field strength by absorbing the coupling constant
in the gauge field: $g A_\mu \rightarrow A_\mu$.
Then $F^a_{\mu\nu} \rightarrow \frac{1}{g} F^a_{\mu\nu}$,
and the expression \eqref{defEYMSsd} goes to
\begin{align}
  \mathcal S^{\tss{E}}_{\tss{YM}} 
  \, & \rightarrow \, \pm \frac{1}{4g^2} \int \,\mathrm{d^4}x\, F^{a\mu\nu} \tilde{F}^a_{\mu\nu} \,.
  \label{defEYMSsd2}
\end{align}
Consider now field configurations with finite action \eqref{defEYMSsd2}.
Following \cite{atiyah1979}, we assume finite action configurations
define a field strength which vanishes at infinity. That is,
$F_{a\mu\nu}(x) \to 0$ sufficiently fast as $|x|\to \infty$.
The gauge potential $A_\mu(x)$ of finite action configurations
should therefore also tend to zero. But because of  \emph{gauge freedom},
this limit is zero only up to a gauge transformation \eqref{defAGTSU2}
\begin{align}
  \lim_{|x| \to \infty} {A_\mu(x)} \, \sim \, \Omega(x) \partial_\mu \Omega^\dagger(x)\,.
  \label{defLimAInf}
\end{align}
The gauge transformation $\Omega(x)$ need only be defined at large $|x|$,
say over a sphere $S^3$ at the boundary of Euclidean 4-space-time.
In this sense, $\Omega(x)$ defines a continuous map
\begin{align}
  \Omega\,:\,S^3 \, \rightarrow \, \mathrm{SU(2)}
  \label{defContMap}
\end{align}
But such a map has a well defined \emph{integer invariant}:
The \emph{Pontryagin index} (topological charge) $k$ which counts the number of 
times the map $\Omega$ wraps around the SU(2) manifold.
Thus, in Euclidean Yang-Mills theories, asymptotically
flat potentials can be distinguished by its Pontryagin index.
It is a \emph{topological invariant} of the field configuration:
Smooth deformations of the gauge field away from the Euclidean boundary $S^3$
cannot change the value of $k$.
For the group SU(2), the index $k$ is given by the gauge invariant expression
\cite{atiyah1979}
\begin{align}
  8 \pi^2 k \, &\equiv \, \frac{1}{4} \int \,\mathrm{d^4}x\, F^{a\mu\nu} \tilde{F}^a_{\mu\nu} \,.
  \label{defPIndex}
\end{align}
For (anti)selfdual field configurations,
the topological invariant \eqref{defPIndex} coincides exactly
with the Yang-Mills action \eqref{defEYMSsd2}.
In general, the Euclidean Yang-Mills action \eqref{defEYMS}
is given by the topological term \eqref{defPIndex} plus a sum of bilinears
in the strength field.
The \emph{minimal-action} configuration within a given topological sector
of topological charge $k$ is thus given by (anti) selfdual field configurations
with action $\mathcal S = \frac{8 \pi^2}{g^2} |k|$.

\subsection{Topological classification}
\label{sec:topoconf}

Here we introduce some terminology we will make use of later while discussing
SU(2) Yang-Mills thermodynamics in chapter \ref{chap:YM}.
Following Gross, Pisarski and Yaffe \cite{GrossYaffe1981},
classical gauge field configurations in Euclidean space-time
with finite action can be classified according to
topological charge, holonomy and magnetic charge.

\subsubsection{Topological charge}

As we saw in the preceding section, a well defined index $k$,
topological in nature, can be defined as
\begin{align}
  8 \pi^2 k \, &\equiv \, \frac{1}{4} \int \,\mathrm{d^4}x\, F^{a\mu\nu} \tilde{F}^a_{\mu\nu} \,.
  \tag{\ref{defPIndex}}
  \label{defPIndex2}
\end{align}
This expression divides different field configurations into \emph{topological sectors}
classified by the integer $k$.

Let us summarize here the results of the preceding section.
In  each topological sector $k$, the configuration with minimal Euclidean action
\eqref{defEYMS} is given by the solutions to the selfdual
equations \eqref{defSelfDual}.
The Yang-Mills equations of motion are automatically satisfied
for selfdual fields, thus reducing second order field equations of the type
\eqref{defgF5}, \eqref{defgF7} to the first order selfdual equation \eqref{defSelfDual}.
Selfdual field configurations in Euclidean space-time \emph{cannot propagate},
since they have a vanishing energy-momentum tensor \eqref{defEMTensor}.

\subsubsection{Holonomy}

The \emph{holonomy} of a field configuration is given by the eigenvalues 
of the\emph{Polyakov loop} defined as a time-like \emph{Wilson loop} $L$
\begin{align}
  L(\vec x) \,& = \, \mathcal P \, \exp{i\int_{0}^{\beta}}\,\mathrm{d} \tau \, A_4(\tau,\vec x) \,,
  \label{defPolLoop}
\end{align}
where $\mathcal P$ denotes path ordering and $A_4$ is the Euclidean-rotated gauge field \eqref{Etransf}.

\subsubsection{Magnetic charge}

The magnetic charge is defined as the coefficient $m$ in the expression
for the \emph{magnetic flux}
$\mathbf B = m \frac{\mathbf r}{|\mathbf r|^3}$, evaluated at spatial infinity
\cite{Diakonov2009}.

\section{Thermal field theory}

Recall from quantum statistical mechanics that the \emph{grand canonical partition function} $Z(\beta)$
for a quantum system, which is defined by the Hamilton operator $\hat H$ and
vanishing chemical potential $\mu =0$,
can be written as an integral over a complete set of eigenfunctions $| q \rangle $ of $\hat H$
\begin{align}
  Z(\beta)\,=\, \mathrm{tr}\, e^{-\beta \hat H} \,= \, \int \mathrm{d}q\, \langle q | e^{-\beta \hat H} | q \rangle \,,
  \label{defZq}
\end{align}
where as usual $\beta \equiv 1/T$ is the inverse temperature.
It can be shown \cite{Kapusta,LeBellac} that by rotating to \emph{Euclidean space-time} \eqref{Etransf},
the partition function \eqref{defZq} for a quantum field $q(x)$
can be written as a transition amplitude by integrating
over all field configurations that periodic in Euclidean time.
In the case of pure SU(2) Yang-Mills theory, the role of $q(x)$ is played
by the gauge field $A_\mu(x)$, rotated to Euclidean time.
The \emph{pure Yang-Mills partition function} is then given by
\begin{align}\label{SU2pf}
  Z \equiv \int\limits_{A_\mu(0,\vec x)=A_\mu(\beta,\vec x)} \mathcal D A_\mu e^{- \mathcal S^{\tss{E}}_{\tss{YM}} } \,,
\end{align}
where the field configurations are Lie-Algebra valued $A_\mu\equiv A_\mu^a t^a$
with group generators $t_a$ normalized as usual, and
the Euclidean Yang-Mills action $\mathcal S^{\tss{E}}_{\tss{YM}}$ was defined in \eqref{defEYMS}
Integration \eqref{SU2pf} goes over all periodic field configurations $A_\mu(\tau=0,\vec x)=A_\mu(\tau=\beta,\vec x)$.
In this formalism, the Euclidean space-time $\mathbb{R}^4$
goes over to the Euclidean torus $\mathbb{R}^4 \rightarrow S_1 \times \mathbf R^3$.
Selfdual solutions on $\mathbb{R}^4 \rightarrow S_1 \times \mathbf R^3$ are called \emph{calorons}.

Expression \eqref{SU2pf} is used to
derive other thermodynamical quantities such as the pressure, energy, etc.
Perturbatively, the partition function $\ln{Z}$ can be expanded
in \emph{bubble diagrams}.
Further details can be found in the literature  \cite{Kapusta,LeBellac}.


\chapter[SU(2) Yang-Mills Thermodynamics]{SU(2) Yang-Mills Thermodynamics in the Deconfining Phase}
\label{chap:YM}

In this chapter, we review a nonperturbative approach to the thermodynamics of the
SU(2) Yang-Mills gas developed in \cite{Hofmann2005,Hofmann2007}.
It is shown in \cite{Hofmann2005} that such an approach results
in three distinct thermodynamical phases:
confining, preconfining and deconfining. Each phase is characterized by the emergence
of a scalar field describing the thermal ground state dynamics.
In this thesis we are interested in the physics of the deconfining phase,
and the possibility of experimental detection of ground state effects in this regime.
Experimental evidence for SU(2) dynamics hinges on the appearance of a
\emph{frequency cutoff} $\nu^*$ below which thermal propagation
of the SU(2) photon is forbidden.
This effect can be detected by measuring the \emph{antenna temperature} of an antenna immersed
into a thermalized photon gas.
The radiometry of the SU(2) photon is discussed in chapter \ref{chap:radiometry}.

This chapter is structured as follows.
In section \ref{sec:NonPapp} we briefly review the nonperturbative approach to the deconfining
phase, outlining the basic assumptions and results.
Section \ref{sec:thermalGS} introduces the resulting \emph{effective theory} and some of its
consequences, for example, the emergence of nonzero \emph{vacuum energy}.
In section \ref{sec:thermalpart} we list the basic thermal excitations of the theory
and give the \emph{Feynman rules} necessary to calculate effective radiative corrections.
The purpose of this introductory review is not to give a complete description
of the theoretical approach in \cite{Hofmann2005,Hofmann2007}, but to introduce the
concepts and mathematical results necessary for an understanding of the physics behind the $\nu^*$ cutoff.

In section \ref{sec:su2cmb} we review the hypothesis that thermalized photon propagation
is des-cribed by the tree-level massless mode of a Yang-Mills SU(2) theory in the
deconfining phase. This hypothesis fixes the critical temperature of the deconfining phase
to be slightly lower than the baseline CMB temperature $\tcmb=2.726 \mathrm K$.

Our own results are presented in the last two sections.
In section \ref{sec:polarization} we calculate radiative corrections to the \emph{propagation of massless modes},
both transverse and longitudinal. A formula for the temperature-dependent
frequency cutoff $\nu^*(T)$ of the transverse polarization is derived,
and the longitudinal polarization is shown to be nonpropagating.
Finally, section \ref{sec:resumm} is dedicated to a discussion of thermal resummation
and thermodynamical consistency.

\section[A nonperturbative approach]{A nonperturbative approach to Yang-Mills thermodynamics}

\label{sec:NonPapp}

The sector of topologically nontrivial, selfdual, nonpropagating
pure Yang-Mills configurations on $S_1 \times \mathbf R^3$
Euclidean space-time (calorons) is in general not accessible
by perturbation methods.
The quantum amplitude $\mathcal A$ of such configurations is
non-analytic in the coupling constant $g$,
$\mathcal A\sim\exp{(-\frac{8 \pi^2}{g^2})}$, and thus missed
at all orders of perturbation theory.

A thermodynamical approach to this problem was developed in \cite{Hofmann2005,Hofmann2007}
that solves this difficulty by integrating out topologically nontrivial
degrees of freedom through a spatial coarse-graining over such field configurations.
This results in an effective theory containing topologically trivial excitations
interacting with an effective \emph{macroscopic scalar field} $\phi^a$
living in the adjoint representation of the gauge group SU(2).
The field $\phi^a$ also introduces with itself a \emph{mass scale}
$\Lambda$ which is the only parameter of the theory.

The effective theory features a thermal ground state above which
topologically trivial excitations fluctuate.
In order to construct such an effective theory
two basic constraints for infinite-volume thermodynamics
should be obeyed \cite{Hofmann2005}.
First, in the absence of external perturbations, any nonzero
field resulting from the ensemble average and spatial coarse-graining
over nonpropagating field configurations must be, (modulo gauge transformations),
\emph{spatially homogeneous and isotropic}.
Furthermore, such a field can only be constant over all space, and rotationally invariant.
Similarly, the gauge invariant field resulting from such averaging and coarse-graining operations cannot explicitly depend on time.
The basic considerations leading us to the unique definition of a macroscopic
scalar field are sketched in this section.

\subsection{Basic roadmap}

\label{sec:BasicRoad}

The nonperturbative approach to SU(2) Yang-Mills Thermodynamics developed in
\cite{Hofmann2005,Hofmann2007} \emph{uniquely derives} an effective field theory
starting from the fundamental Euclidean Yang-Mills action considered in the
thermodynamical limit of infinite volume.
Here, we summarize the main line of analysis
underlying the emergence of a macroscopic scalar field \cite{HofmannLudescher2010}.

\begin{enumerate}
  \item \label{list:top} Consider the topologically nontrivial, selfdual sector of periodic Euclidean
    field configurations. Independently of their topological parameters, these field
    excitations have a vanishing (classical) energy-momentum tensor, and thus do not describe
    propagating degrees of freedom.
  \item \label{list:mult} Let a Lorentz invariant gauge multiplet $\phi^a$ be defined by coarse-graining
    over such configurations in \ref{list:top}. The multiplet $\phi^a$ is thus
    by construction \emph{nonpropagating} (recall section \ref{sec:selfdual}).
    But this means $\phi^a$ acts as a \emph{static},    \emph{spatially homogeneous} background.
  \item \label{list:phi} Spatial isotropy would be broken by such a Lorentz tensor
    unless $\phi^a$ were a \emph{scalar field}.
    Further, spatial homogeneity requires that the scalar field $\phi^a$ have a \emph{constant modulus} $|\phi|$.
    Since $\phi^a$  by construction cannot propagate and emerges by coarse-graining over
    periodic configurations in Euclidean time, its equation of motion produces
    -- in a given gauge -- an Euclidean-time periodic phase
    $\hat \phi^a(\tau + \beta) = \hat \phi^a(\tau)$ with no space dependence.
    Also, since the \emph{classical action} of selfdual configurations with topological charge $k$
    is \emph{independent of temperature} $\beta^{-1}$, $S=\frac{8 \pi^2 |k|}{g^2}$ (see \ref{sec:selfdual}),
    \emph{no explicit $\beta$ dependence} is allowed in the definition of $\hat \phi^a$.
    Finally, if the scalar gauge multiplet is thermalized with the sector of trivial topological charge $k=0$,
    it must carry charge under the gauge group SU(2) (that is, transform nontrivially under gauge rotations).
    Then the index $a$ is just a SU(2) valued \emph{color} index.
  \item It can be shown \cite{HerbstHofmann2004} that there exists a \emph{unique construction}
    for a family of phases $\{ \hat{ \phi^a}\}$
    living in the \emph{adjoint} representation satisfying the conditions described in \ref{list:phi}.
    This construction is necessarily nonlocal, and
    only admits contributions from $|k|=1$ (Harrington-Shepard) (anti)calorons of trivial holonomy.
    Non-trivial holonomy calorons can be \emph{apriori excluded} from contributing to $\phi^a$ since
    such configurations are known to be unstable under trivial-topology fluctuations \cite{Diakonov2004}
    (The average Polyakov line is zero, $\langle \mathrm{tr} L \rangle = 0$ \cite{Diakonov2009}).
    Configurations of topological charge $|k|>1$ are also excluded, by Lorentz symmetry arguments \cite{HerbstHofmann2004}
    so that $\phi^a$ emerges from spatial coarse graining over only \emph{trivial holonomy calorons of topological charge $|k|=1$}.
    It turns out that the set $\{\hat{ \phi^a}\}$ defines the kernel of - and \emph{completely determines} a -
    second order differential operator $\mathcal D$.
    The fact that the scalar field $\phi^a$ both has vanishing Euclidean energy density and belongs to the kernel
    of $\mathcal D$ can be exploited to \emph{fix the gauge invariant effective potential $V\left( \phi \right)$ uniquely}
    for $\phi^a$. This action also introduces a mass scale as an integration constant,
    the \emph{Yang-Mills scale} $\Lambda$. 
  \item Since no object other than the inert adjoint scalar field $\phi^a$ may emerge from the sector
    of topologically nontrivial fundamental configurations, the coupling of $\phi^a$ with the sector of
    zero topological charge must be minimal. That is, it should be realized via a covariant derivative
    $D_\mu = \partial_\mu + a_\mu$. Higher dimensional operators are forbidden by perturbative
    renormalizability \cite{Hofmann2005}.
\end{enumerate}

Thus, we arrive at a scalar field $\phi$ and an effective Lagrangian $\mathcal L_{\srm{eff}}$
describing interactions with the sector of topologically trivial charge $k=0$.
The \emph{effective theory} described by the scalar field $\phi$ 
and the Lagrangian $\mathcal L_{\srm{eff}}$ is discussed in some detail in the next sections.

\section{SU(2) thermal ground state in the deconfining phase}
\label{sec:thermalGS}

\begin{figure}[t]
  \centering
  \includegraphics[width=120mm]{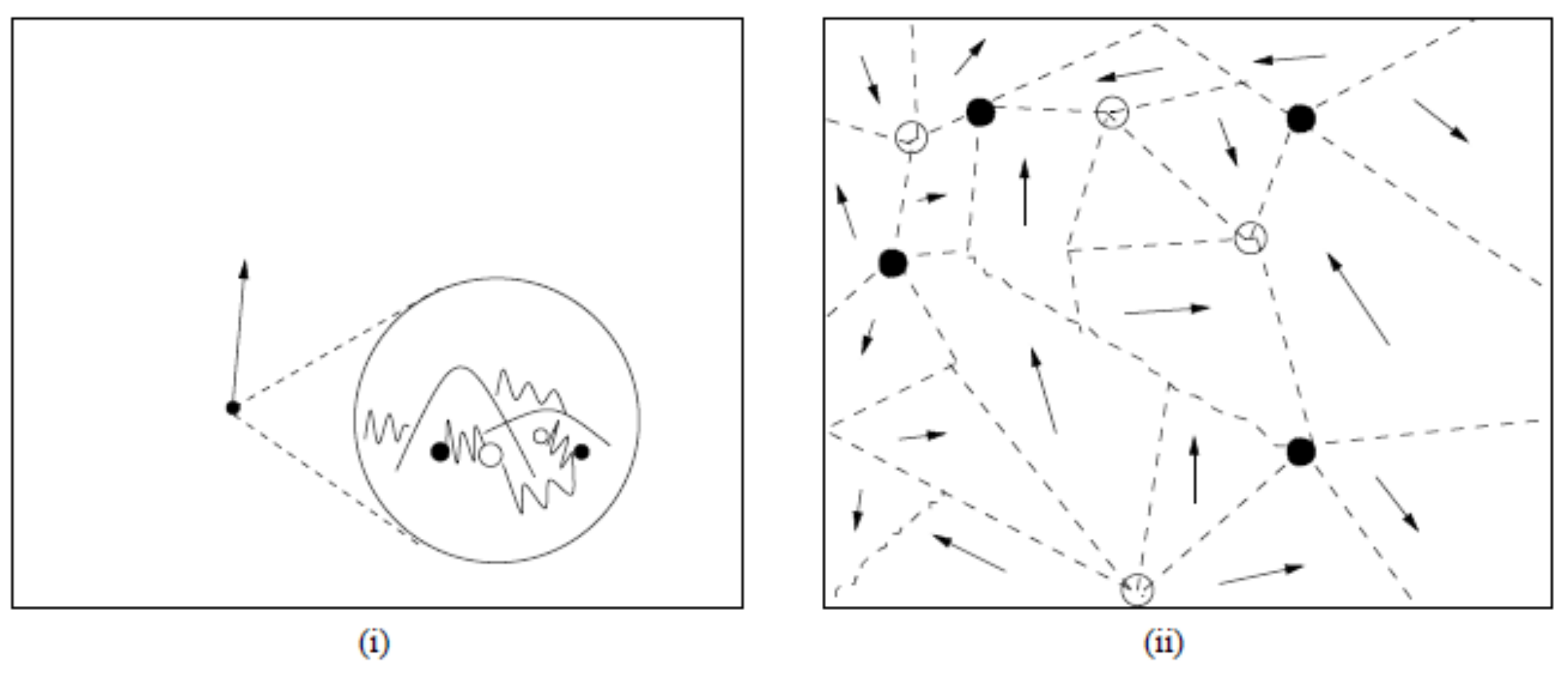}
  \caption{A spatially homogeneous and isotropic static field $\phi^a$ is
  generated by spatial coarse-graining over interacting calorons of trivial holonomy
  and topological charge $|k|=1$, shown in (i) as fluctuations inside a spherical
  integration volume. The inert field $\phi^a$ and the pure-gauge configuration
  $a_\mu^{\srm{gs}}$ which solves the effective Yang-Mills equations
  together define a \emph{thermal ground state}.
  (ii) depicts domainization occurring
  from interactions between calorons. Domain walls intersections are identified
  as \emph{magnetic monopoles}. Domainization effects can be taken into account
  by loop-expanding the quantity under investigation about the thermal ground state
  in the effective theory. Figure taken from \cite{hofmann2011thermodynamics}.
  }
  \label{fig:coarse}
\end{figure}

\subsection{Inert adjoint scalar field}

\label{sec:scalarfield}

As described in \ref{sec:BasicRoad}, upon \emph{spatial coarse graining}
over the caloron sector of the \emph{fundamental} SU(2) theory
a Lorentz invariant multiplet $\phi^a$ is generated which describes
ground state physics up to a phase transition temperature given by the
Yang-Mills scale of the theory.
The \emph{modulus} of $\phi^a$ does not depend on space or time, and is given by
\begin{align}
\left| \phi \right| \, &\equiv \, \sqrt{\frac{1}{2}\, \mathrm{tr}\, \phi^2 }
  \label{defModPhi} \\
  \, &= \, \sqrt{\frac{\Lambda^3\beta}{2\pi}}\,,
\label{PhiMod}
\end{align}
with the inverse temperature $\beta \equiv \frac{1}{T}$.
The \emph{equations of motion} ruling over the dynamics of $\phi^a$
are derived from an (effective) Euclidean Lagrangian of the form
\begin{align}
  \mathcal L_{\phi} = \mathrm{tr}\, \left[ \left( \partial_\tau \phi \right)^2 + V\left( \phi^2 \right) \right] \,,
  \label{defPhiLag}
\end{align}
the potential $V$ is given by the expression
\begin{align}
  V(\phi) \, &= \, \mathrm{tr}\, \frac{\Lambda^6}{\phi^2}\,,
      \label{PhiV} 
\end{align}
with the inverse of the scalar field defined as $\phi^{-1}\equiv\frac{\phi}{|\phi|^2}$.
The \emph{mass scale} $\Lambda$ emerges as an integration constant and serves
as the \emph{Yang-Mills scale} of the theory, defining a phase transition temperature
$T_c\equiv\lambda_c\frac {\Lambda}{2\pi}$ which bounds the deconfining phase from below.
The \emph{critical dimensionless temperature} $\lambda_c$ is defined by the evolution
of the gauge coupling $e(\lambda)$ in the \emph{effective theory} (see section \ref{sec:thermalgs}).
At $\lambda_c$, the gauge coupling exhibits a logarithmic singularity so that
$e(\lambda)=-4.59\log(\lambda-\lambda_c)+18.42$.
Numerically, $\lambda_c=13.87$ \cite{Hofmann2009}.
The transition temperature $T_c$ is the \emph{only parameter of the theory}.
To make contact with experiment, this parameter must be fixed so that quantitative
calculations may be performed. This will be done in section \ref{sec:su2cmb}.

Up to global gauge transformations, a solution to the Euler-Lagrange equations of motion determined by
\eqref{defPhiLag} and \eqref{PhiV} is given by
\begin{align}
  \phi=2\,\sqrt{\frac{\Lambda^3\beta}{2\pi}}\,t_1\,\exp(\pm\frac{4\pi i}{\beta}t_3\tau)\,.
  \label{PhiDefL}
\end{align}
The field $\phi$ represents a spatially homogeneous background for the dynamics of
coarse-grained, propagating, trivial-topology gauge fields.
It will also \emph{break the gauge symmetry} SU(2) down to its Abelian subgroup U(1).
The effect of this symmetry breaking
will be to confer a temperature-dependent mass to two of the three available vector modes.
This is discussed in section \ref{sec:su2modes}.

\subsection{Effective action and bare thermal ground state}

\label{sec:thermalgs}

In section \ref{sec:BasicRoad}, it was shown how to construct a scalar field resulting
from spatial coarse-graining over Harrington-Shepard calorons.
The procedure integrates out the topologically nontrivial sector
of the fundamental theory.
The dynamics of the topologically trivial remainder
is then determined by an effective theory valid up to a scale $|\phi|$,
characterized by  an effective Lagrangian $\mathcal{L}_{\srm{eff}}$
describing the interaction of the scalar field $\phi$
with the topologically trivial sector ($k=0$).
It has been shown \cite{Hofmann2005} that $\mathcal{L}_{\srm{eff}}$
can be uniquely defined -- up to a \emph{maximal energy resolution $|\phi|$} --
and is given by
\begin{align}
  \mathcal{L}_{\srm{eff}}[a_\mu] &=
        \mathrm{tr}\,\left(\frac{1}{2}\, G_{\mu\nu}G_{\mu\nu}+(D_\mu\phi)^2+\frac{\Lambda^6}{\phi^2}\right)\,,
  \label{SU2EffectLag}
\end{align}
where $a_\mu=a_\mu^a\,t_a$ is a propagating, trivial-topology gauge field
in the adjoint representation remaining after coarse-graining.
The \emph{effective field strength} is defined as usual
\begin{align}
  G_{\mu\nu} \equiv \partial_\mu a_\nu - \partial_\nu a_\mu -ie [a_\mu,a_\nu] = G^a_{\mu\nu}\,t_a \,,
  \label{DefEffG}
\end{align}
with $e$ the \emph{effective gauge coupling}.
The effective coupling constant $e$ is not the same as the coupling constant of the fundamental theory.
The evolution of this coupling constant as function of dimensionless temperature $e(\lambda)$
follows from demanding thermodynamical consistency \cite{Hofmann2005}.
As already mentioned,  the coupling behaves logarithmically at the \emph{ critical temperature } $\lambda_c$,
$e(\lambda)|_{\lambda_c} \sim -\log(\lambda - \lambda_c)$,
and the theory undergoes a \emph{second order phase transition} to the preconfining phase.
We remark here that, for temperatures higher than $\lambda_c$,
local gauge transformations can always be found such that
the Polyakov loop $\mathrm{Pol}[a_\mu^{\srm{bg}}]$
is transformed from $\mathrm{Pol}=-{\bf 1}_2$
to $\mathrm{Pol}={\bf 1}_2$.
This reflects the electric ${\bf Z}_2$ degeneracy of the thermal ground state estimate,
and the theory for $\lambda > \lambda_c$ is thus \emph{deconfining}.
Further details can be found in the literature \cite{Hofmann2005,Hofmann2007}.
The covariant derivative $D_\mu$ imposes \emph{minimal coupling} between
the effective $k=0$ gauge field and the $|k|=1$ coarse-grained inert scalar field,
and is written as
\begin{align}
  D_\mu \phi \, &= \, \partial_\mu\phi-ie[a_\mu,\phi] \,.
  \label{DefEffCov}
\end{align}

Note how the effective action in \eqref{SU2EffectLag} at resolution $|\phi|$
splits into a classical Yang-Mills term describing the propagation and interactions among topologically
trivial fluctuations,
an interaction term between $k=0$ and $|k|=1$ configurations
given by the square of the covariant derivative acting on $\phi$, $(D_\mu\phi)^2$
(Higgs mechanism for two out of three propagating directions in the SU(2) algebra),
and a potential $V$ for the field $\phi$ \cite{Hofmann2005}.

The Euler-Lagrange equations governing the effective theory can be read
from the Lagrangian \eqref{SU2EffectLag}
\begin{align}
  D_{\mu}G_{\mu\nu} \, &= \, ie[\phi,D_\nu\phi]\,.
  \label{EffEoM}
\end{align}
The inert scalar ground state configuration \eqref{PhiDefL} and the \emph{pure gauge configuration}
\begin{align}
  a^{\srm{gs}}_\mu \, &= \, \mp\delta_{\mu 4}\frac{2\pi}{e\beta}\,t_3\,,
  \label{DefAPureG}
\end{align}
are \emph{ground state solutions} of \eqref{EffEoM} since
$G_{\mu\nu}[a^{\srm{gs}}_\kappa]=D^\nu[a^{\srm{gs}}_\kappa]\phi= 0$.
Apart from (small) radiative corrections and up to global gauge transformations,
\eqref{PhiDefL} and \eqref{DefAPureG} constitute a good first estimate of the thermal ground state
of the effective theory in the deconfining phase.
Accordingly, a good first estimate of the \emph{energy density of the thermal ground state} is given by
\begin{align}
  \rho^{\srm{gs}} \, &\equiv \, \mathcal {L}_{\srm{eff}}[a^{\srm{gs}}_\mu]
  \,=\, \mathrm{tr}\,\frac{\Lambda^6}{\phi^2}=4\pi\Lambda^3\,T\,,
  \label{GSenergy}
\end{align}
which is interpreted as a temperature dependent cosmological constant.
Together, $\phi$ and $a^{\srm{gs}}_\mu$ represent an \emph{a priori estimate
of the bare thermal ground state} which turns out to be quite accurate.

For later use, we write \eqref{GSenergy} in SI units.
The temperature $T$ is rescaled as $T \rightarrow k_B T$,
the mass scale $\Lambda$ as $\Lambda \rightarrow \frac{\Lambda}{\hbar c}$,
and the \emph{energy density of the thermal ground in the deconfining phase}
depends on the temperature through the expression
\begin{align}
  \rho^{\srm{gs}} = 4 \pi \left( \frac{\Lambda}{\hbar c} \right)^3 k_B T\,.
  \label{GSenergySI}
\end{align}
This energy density will later serve as a criterion for thermodynamical consistency
in the presence of external fields.

\section{Thermal quasiparticles and Feynman rules} 
\label{sec:thermalpart}

\subsection{Mass spectrum}
\label{sec:su2modes}

In the effective theory of deconfining Yang-Mills thermodynamics,
each broken generator $t_a$ of the original gauge symmetry,
causes the original massless mode associated with $t_a$ to acquire a mass $m_a$
\cite{Hofmann2007}
\begin{align}
  m^2_a=-2e^2 \mathrm {tr} \, \left[ \phi,t_a \right]^2 \, .
  \label{ModeMasses}
\end{align}
To interpret \eqref{ModeMasses} physically, we impose \emph{unitary gauge} for the thermal ground state
\begin{subequations}
\begin{align}
  \phi \,&=\, 2\,|\phi|\,t_3\,, \\
  a_\mu^{\srm{gs}} \,&=\, 0\,.
\end{align}
  \label{GSgauge}
\end{subequations}
In this gauge we have
\begin{subequations}
\begin{align}
  m \, & \equiv \, m_1 \, =\, m_2\, =\, 2e(\lambda) |\phi| \,=\, 2 \Lambda  \,\frac{ e(\lambda)}{\sqrt{\lambda}} \, = \,
        2\Lambda \, e\left( \frac{2 \pi}{\Lambda} T \right)  \,\sqrt{\frac{\Lambda}{2\pi T}} \,,
    \label{TLHmass}\\
    m_3 \, &= \, 0\,.
  \label{TLMmass}
\end{align}
  \label{ModeMasses2}
\end{subequations}
In the fundamental theory, all gauge modes are massless, 
while in the effective theory two of the three gauge modes acquire mass at tree-level.
Note that the mass $m$ in \eqref{TLHmass} is temperature-dependent.
Recall that, at the critical temperature $\lambda_c \equiv \frac{2 \pi}{\Lambda} T_c$,
the effective coupling $e(\lambda)$ shows a logarithmic singularity
$e(\lambda)|_{\lambda_c} \sim -\log(\lambda - \lambda_c)$ for $\lambda \searrow \lambda_c$.
This signals a phase transition to the preconfining regime.
Near this critical temperature, the mass $m$ in \eqref{TLHmass} diverges,
and interactions between massless and massive modes decouple.
That is, radiative corrections to \eqref{TLMmass} becomes exactly zero.

The unbroken gauge subgroup U(1) is responsible for the remaining
\emph{massless} mode after coarse-graining. It is this remaining
symmetry, and the vanishing of radiative corrections to \eqref{TLMmass}
at $T_c$, that will lead us to speculate about the fundamental origin
of the Standard Model photon (see section \ref{sec:su2cmb}).

But first, we review the Feynman rules in the effective theory.

\subsection{Finite temperature Feynman rules in unitary gauge}

\label{sec:feynmanrules}

In unitary-Coulomb gauge, which is a physical, completely fixed gauge,
the real-time propagators of the fields are obtained by a Wick rotation to real
time\footnote{This method becomes problematic for self interactions
and loop calculations of higher order than two \cite{DolanJackiw}.
It has been shown that such difficulties do not apply
to our approach to SU(2) thermodynamics \cite{Hofmann2006}.
Note also that a change of signature from Euclidean to Minkowskian is trivial
as far as the physics of the ground state estimate of section \ref{sec:thermalgs}
and the associated dynamical gauge-symmetry breaking are concerned:
neither ground state pressure, energy density or quasiparticle mass
depend on Euclidean time.
}
of the usual Matsuraba imaginary time propagators \cite{Kapusta,LandsmanWeert, LeBellac}.
In the real-time formulation of finite-temperature field theory
\cite{LandsmanWeert} it possible to discern quantum from
thermal fluctuations.
To tell the former from the latter is necessary since quantum
fluctuations away from the mass-shell will be limited by the physical
resolution $|\phi|$ in the  unitary-Coulomb gauge.
Also, the fields $a_\mu^{1,2}$ may only propagate thermally \cite{Hofmann2005}.
The Feynman rules for the SU(2) vertices are taken from the literature as well,
by substituting the fundamental gauge constant $g\to e$.

To conform with the nomenclature of \cite{SHG2007} we refer to
the fluctuative SU(2) directions 1 and 2 as tree-level heavy (TLH) and
to direction 3 as tree-level massless (TLM).
In unitary-Coulomb gauge \eqref{GSgauge} the \emph{real-time bare thermal propagators}
in momentum space are then given by \cite{SHG2007}
\begin{align}
  D^{\srm{TLH},0}_{ab,\mu\nu}(p) &= -\delta_{ab} \tilde{D}_{\mu\nu}
        \left[2\pi\delta(p^2-m^2)n_B(|p_0|/T)\right]\,,
  \label{DTLH} \\
  D^{\srm{TLM},0}_{ab,\mu\nu}(p) &= -\delta_{ab}
  \left\lbrace P^T_{\mu\nu}\left[\frac{i}{p^2}+2\pi\delta(p^2)n_B(|p_0|/T)\right]
  -i\frac{u_\mu u_\nu}{\mathbf{p}^2}\right\rbrace \,,
  \label{DTLM}
\end{align}
where $n_B(x)=1/(e^x-1)$ denotes the Bose-Einstein distribution function and
\begin{align}
  \tilde{D}_{\mu\nu} &= \left( g_{\mu\nu}-\frac{p_\mu p_\nu}{m^2} \right)\,.
  \label{defDT}
\end{align}
The four-vector $u_\mu=(1,0,0,0)$ specifies the restframe of the thermal bath
and the \emph{3-space transverse} projection operator $P^{\mu\nu}_T$ is defined as
\begin{align}
  P^{00}_T & = P^{0i}_T = P^{i0}_T = 0\,,\\
  P^{ij}_T & = \delta^{ij} - p^{i}p^{j}/\textbf{p}^2\,.
\end{align}
TLM modes carry a color index 3 while TLH modes have a color index 1
and 2.
\begin{figure}[t]
  \centering
  \includegraphics[width=110mm]{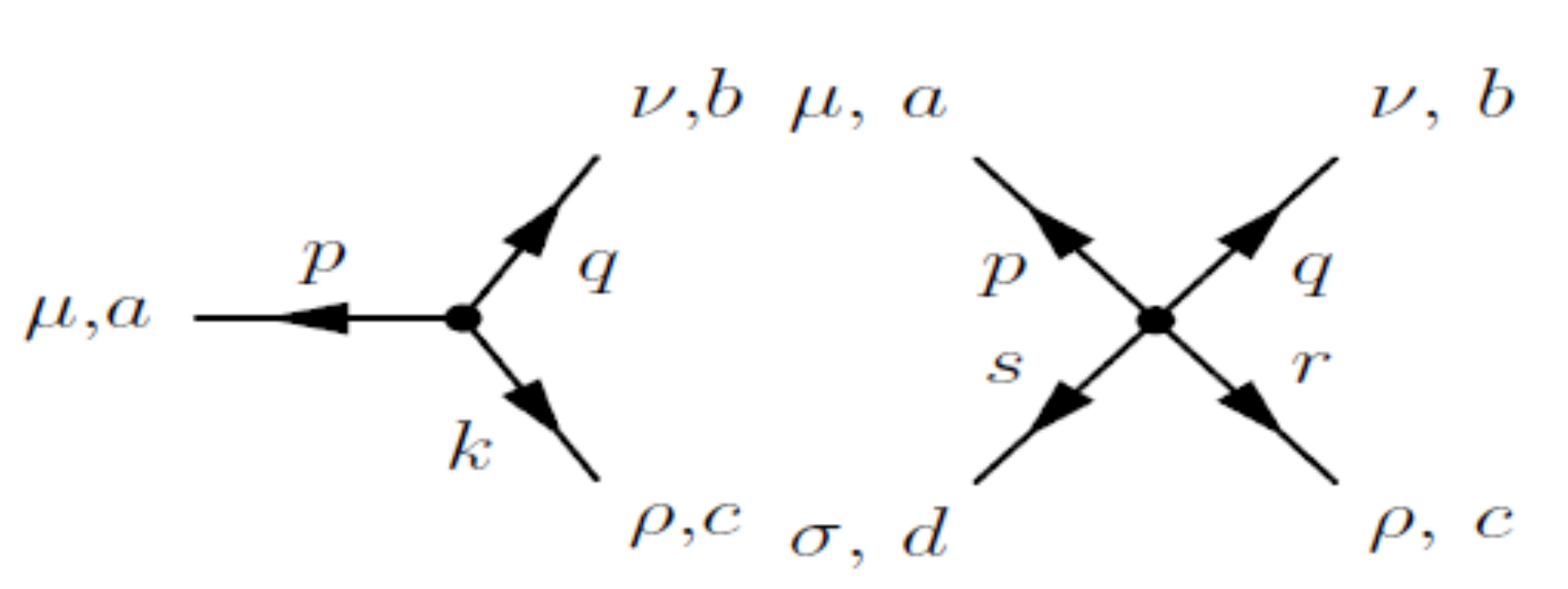}
  \caption{ 3- and 4- vertices. Taken from \cite{SHG2007}}
  \label{fig:3and4vert}
\end{figure}
The three- and four-gauge-boson vertices are given as (see figure \ref{fig:3and4vert})
\begin{align}
    \Gamma^{\mu\nu\rho}_{[3]abc}(p,q,k) \, =& \,  e (2\pi)^4\delta(p+q+k) \epsilon_{abc} \notag \\
    &\, \times\, \left[g^{\mu\nu}(q-p)^\rho + g^{\nu\rho}(k-q)^\mu + g^{\rho\mu}(p-k)^\nu\right]
    \label{threeV}\\
    \Gamma^{\mu\nu\rho\sigma}_{[4]abcd}(p,q,s,r) \, =& \,
               -ie^2(2\pi)^4\delta(p+q+s+r) [ \notag \\
    & \quad \epsilon_{abe}\epsilon_{cde}(g^{\mu\rho}g^{\nu\sigma}-g^{\mu\sigma}g^{\nu\rho}) \notag\\
    & +\epsilon_{ace}\epsilon_{bde}(g^{\mu\nu}g^{\rho\sigma}-g^{\mu\sigma}g^{\nu\rho}) \notag\\
    & +\epsilon_{ade}\epsilon_{bce}(g^{\mu\nu}g^{\rho\sigma}-g^{\mu\rho}g^{\nu\sigma})]\,.
    \label{fourV}
\end{align}
Finally, recall that, before calculation,
each loop-diagram has to be divided by $i$
and the number of its vertices (symmetry factor) \cite{LandsmanWeert}.

The maximal off-shellness of effective gauge modes
as well as the momentum transfer in four-vertices
are constrained by coarse-graining down to a resolution $|\phi|$ \cite{Hofmann2005,Hofmann2007fd}.
TLM modes \eqref{DTLM} may never be off-shell beyond 
\begin{align}
  |p^2| \le |\phi|^2\,.
  \label{constraint1}
\end{align}
TLH modes on the other hand may only fluctuate thermally, that is, on-shell.
Condition \eqref{constraint1} \emph{also fixes the momentum transfer} in a three-vertex by momentum conservation.
Momentum transfers in an effective four vertex cannot be larger than
$|\phi|^2$ \cite{Hofmann2005}.

We now have all the information on hand that we need to calculate in the effective thermal theory.
The only step left is to \emph{fix the scale} $\Lambda$ so that we may
get quantitative results. This is achieved by identifying the Standard Model
U(1) propagating photon with the massless SU(2) mode in the effective theory.
This is discussed in section \ref{sec:su2cmb}.

Afterwards, we introduce the
polarization tensor for the TLM mode,
and proceed to calculate its influence on photon propagation.
This is the subject of section \ref{sec:polarization}.

\section{The $\mathrm{SU(2)}_{\mathrm{\scriptscriptstyle{CMB}}}$ hypothesis and some implications}
\label{sec:su2cmb}

The nonperturbative approach to the thermodynamics of the SU(2)
Yang-Mills gas reviewed in the preceding sections can now be summarized
as follows: An adjoint scalar field $\phi^a$ is defined by coarse-graining
over trivial-holonomy field configurations with topological charge $|k|=1$.
A thermal ground state composed of $\phi^a$ and
the pure-gauge field configuration $a_\mu^{\srm{gs}}$,
considered together with topologically trivial ($|k|=0$) gauge field fluctuations,
is seen to  \emph{effectively} describe the fundamental SU(2) Yang-Mills theory at finite-temperature.
In the effective theory, two of the three fundamental gauge modes of the local SU(2) symmetry
become massive due to $\phi^a$ by the Higgs mechanismus.
The remaining U(1) symmetry is responsible for the masslessnes of the third gauge mode.

At this point, it is worthwhile to consider whether this symmetry is realized in nature.
An intriguing possibility is the identification with the $\mathrm{U(1)_{Y}}$ gauge group
of the Standard Model.  As argued in \cite{PSA2005,hofmann20062,RHLeip2007,RH2009,JHEP2007},
this would immediatly fix the critical temperature $T_c$ with the cosmic microwave background
temperature $\tcmb= \, 2.725\,\mathrm K$. It follows for the SU(2) Yang-Mills scale 
\begin{align}
  \Lambda \, \equiv \, \Lcmb \, &= \, 2\pi\,\frac{k_B T_c}{\lambda_c} 
  \,=\, 2\pi\,\frac{k_B \tcmb}{\lambda_c}
 \label{defSU2const} \\
  \,&=\, 1.70433\,\cdot\,10^{-23}\,\mathrm{J}  \notag \\
  \,&=\, 1.06376\,\cdot\,10^{-4}\,\mathrm{eV}  \notag  \,,
\end{align}
with $\lambda_c =  13.87$.
Once we have fixed the Yang-Mills scale $\Lambda\equiv \, \Lcmb$
we can proceed to calculate the polarization tensor of the TLM mode.
This is done in the next section.

\section{Polarization tensor and photon propagation}

\label{sec:polarization}

Having identified the massless mode of the coarse-grained effective theory
with the familiar U(1) photon, we now proceed to investigate
what effect the existence of a thermal ground state may have in the
physics of photon propagation.
We expect the thermal ground state to interact with the massless
mode via the massive vectors. This interaction is encoded in the
\emph{polarization tensor} $\Pi^{\mu\nu}(p_0,\vec{p})$, which we
compute perturbatively at one-loop level.
The rest of the chapter is dedicated to the main results of this computation.

\subsection{Prerequisites}

\label{sec:polarizationIntro}

The dispersion relations for the  massless gauge modes are given by
the poles of the full thermal propagator $D^{TLM}_{ab,\mu\nu}(k)$.
In general, any \emph{dressed} thermal propagator $\mathcal D_{\mu\nu}$
can be written as the geometric series
\begin{align}
  \mathcal D_{\mu\nu} = \mathcal D^0_{\mu\nu} +  \mathcal D^0_{\mu\sigma} \Pi^{\sigma\rho} \mathcal D^0_{\rho\nu} + \ldots\,,
  \label{fullPropSeries}
\end{align}
where $ \mathcal D^0_{\mu\nu}$ is the bare thermal propagator
and the \emph{polarization tensor} $\Pi^{\mu\nu}$ is defined as the sum of
\emph{one-particle irreducible} (1PI) diagrams with 2 external lines \cite{peskin1995introduction}.
The series \eqref{fullPropSeries} can be written as
\begin{align}
    \mathcal D_{\mu\nu} = \mathcal D^0_{\mu\nu} +  \mathcal D^0_{\mu\sigma} \Pi^{\sigma\rho} \mathcal D_{\rho\nu}\,.
     \label{fullDSeries}
\end{align}
Defining the inverse propagator as $ \mathcal D^{-1}_{\mu\sigma} \mathcal D^{\sigma\nu} = g_\mu{}^\nu$,
we get an equation for the polarization tensor
\begin{align}
  \Pi_{\mu\nu} = \mathcal D^{-1}_{\mu\nu} - {\mathcal D^0}^{-1}_{\mu\nu}\,.
  \label{pieq}
\end{align}
Further, $\Pi^{\mu\nu}$ is transverse \cite{LandsmanWeert}
\begin{align}
  p^\mu \Pi_{\mu\nu} = 0\,.
\end{align}
Here, $p^\mu$ is the 4-momentum vector of the thermal excitation.
By separating the space-time transverse projector operator
\begin{align}
  P^{\mu\nu} \equiv g^{\mu\nu} - p^\mu p^\nu / p^2\,,
\end{align}
in the space transverse projector $P_T$ and space longitudinal projector $P_L$
\begin{align}
  \begin{split}
    P_T^{00} &= P_T^{0i}=P_T^{i0}=0 \,, \\
    P_T^{ij} &= \delta^{ij} - p^i p^j / \mathbf{p}^2 \,, \\
    P_L^{\mu\nu}& = -P^{\mu\nu} - P_T^{\mu\nu}\,,
  \end{split}
\end{align}
we can expand the space-time transverse polarization tensor $\Pi_{\mu\nu}$ as
\begin{align}
  \Pi_{\mu\nu}(p_0,\mathbf p) \, &= \, G( p_0,\mathbf  p) {P_T}_{\mu\nu} + F(p_0,\mathbf p ) {P_L}_{\mu\nu}\,,
  \label{defPolTensor}
\end{align}
with scalar functions G and F.
Before analytic continuation from imaginary time, the full interacting propagator $D^{TLM}_{ab,\mu\nu}(k)$
for the massless mode of the effective theory can be written in the \emph{Coulomb gauge}
in the Euclidean formulation as the linear combination \cite{Kapusta,SHG2007}
\begin{align}
D^{TLM}_{ab,\mu\nu}(p) \, &=\, -\delta_{ab}
\left\{ P^T_{\mu\nu}\frac{1}{G-p^2}+\frac{p^2}{\mathbf p^2}\frac{1}{F-p^2}\,u_\mu u_\nu \right\}\,,
\label{defFullD}
\end{align}
where $u^\mu=(1,0,0,0)$ specifies the \emph{restframe} of the many-body system. Note that,
as expected, while the first term in the right hand side of expression \eqref{defFullD}
describes \emph{transverse propagation},
$P_T^{\mu\nu} D^{TLM}_{\mu\nu} \sim (G-p^2)^{-1}$,
the second term describes \emph{longitudinal propagation}
$P_L^{\mu\nu} D^{TLM}_{\mu\nu} \sim (F-p^2)^{-1}$.
The corresponding \emph{dispersion relations} are then determined for all modes
by the poles of \eqref{defFullD}.
For the \emph{transverse modes}, this results in
\begin{align}
  p_0^2 \, &= \, \mathbf p^2 +  G(p_0,\mathbf p)\,.
  \label{dispRelT1}
\end{align}
Separate $p_0\,=\, \omega -i\gamma$, and assume \emph{weak damping} $\gamma / \omega \ll 1$.
The \emph{transverse dispersion relation} is then given by
\begin{align}
  \omega^2(\mathbf p) \, &= \, \mathrm{Re}\, p_0^2 \, =\, \mathbf p^2 +   \mathrm{Re}\,G(\omega(\mathbf p),\mathbf p)\,,
  \label{dispRelT}\\
  \gamma(\mathbf p)   \, &=\,  \mathrm{Im}\, p_0^2   \, =\,-\frac{1}{2\omega(\mathbf p)}\mathrm{Im}\,G(\omega(\mathbf p),\mathbf p)\,.
  \label{dispRel2T}
\end{align}
Similarly, for the longitudinal mode
\begin{align}
 \omega^2_L(\mathbf p) \, &=\, \mathbf p^2 +   \mathrm{Re}\,F(\omega_L(\mathbf p),\mathbf p)\,,
\label{dispRelL}\\
  \gamma_L(\mathbf p)\, &=\, -\frac{1}{2\omega_L(\mathbf p)}\mathrm{Im}\,F(\omega_L(\mathbf p),\mathbf p)\,,
\label{dispRel2L}
\end{align}
where we added the subscript $L$ to differentiate from the transverse relations.
Taking $\mathbf{p}$ parallel to the $z$-axis, we can extract from \eqref{defPolTensor}
expressions for the scalar functions $G$, $F$ as components of $\Pi_{\mu\nu}$
\begin{align}
  G(\omega(\mathbf p),\mathbf p)\, &=\, \Pi_{11}(\omega(\mathbf p),\mathbf p) \,=\, \Pi_{22}(\omega(\mathbf p),\mathbf p) \,,
  \label{GofPi} \\
  F(\omega_L(\mathbf p),\mathbf p)\, &=\, \frac{p^2}{\mathbf p^2} \Pi_{00} \,.
  \label{FofPi}
\end{align}

\subsection{Polarization tensor at the one-loop level}

\begin{figure}[t]
  \centering
  \includegraphics[width=140mm]{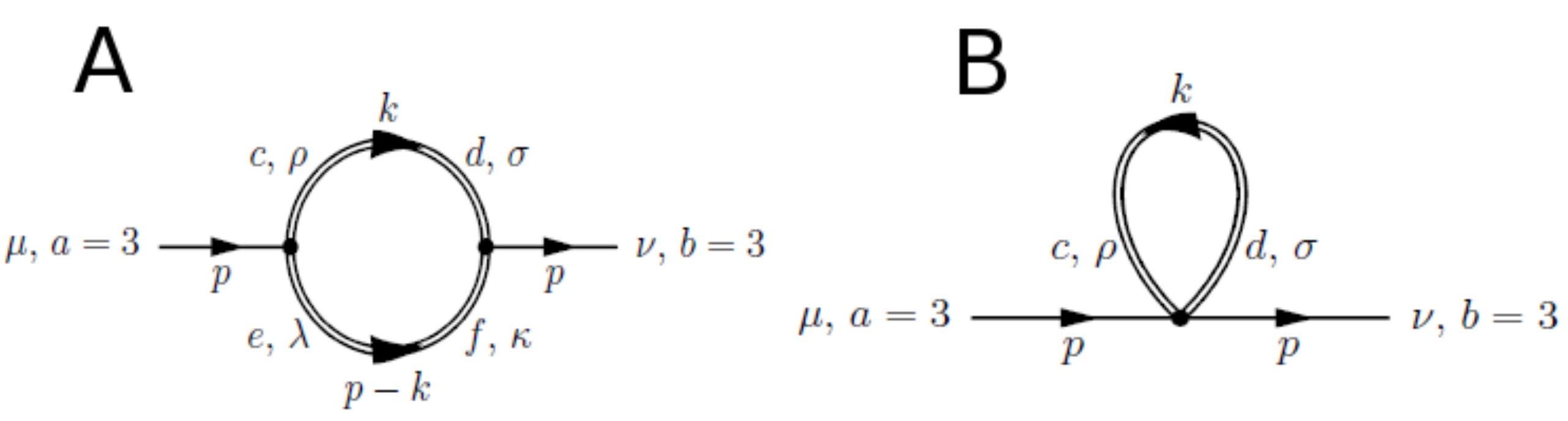}
  \caption{The photon is represented by single-line propagators, double-line propagators correspond to massive modes.
  Taken from \cite{SHG2007}.}
  \label{fig:feynman}
\end{figure}
In order to calculate the polarization tensor one has to sum all 1PI diagrams
to the required precision.
Figure \ref{fig:feynman} shows the diagrams contributing to the
photon polarization tensor $\Pi^{\mu\nu}$ at the one-loop level
as a function of incoming momentum $p$.
The total polarization tensor is just the sum of the two diagrams
\begin{align}
  \Pi^{\mu\nu}(p)\,=\, \Pi^{\mu\nu}_{A}(p) + \Pi^{\mu\nu}_{B}(p)\,.
  \label{TotalPolTensor}
\end{align}
Diagram A can be written by following the Feynman rules described in section \ref{sec:feynmanrules},
which results in 
\begin{align}
\Pi^{\mu\nu}_{A}(p)\,=\,
            \frac{1}{2i}\int\frac{d^4k}{(2\pi)^4} e^2 &
            \; \epsilon_{ace}[g^{\mu\rho}(-p-k)^\lambda+g^{\rho\lambda}(k-p+k)^\mu+g^{\lambda\mu}(p-k+p)^\rho] \notag\\
            &\times (-\delta_{cd})\left(g_{\rho\sigma}-\frac{k_\rho k_\sigma}{m^2}\right) 
            \left[\frac{i}{k^2-m^2}+2\pi\delta(k^2-m^2)\,n_B(|k_0|/T) \right] \notag\\
            &\times \epsilon_{dbf}[g^{\sigma\nu}(-k-p)^\kappa+g^{\nu\kappa}(p+p-k)^\sigma+g^{\kappa\sigma}(-p+k+k)^\nu] \notag\\
            &\times (-\delta_{ef})\left(g_{\lambda\kappa}-\frac{(p-k)_\lambda(p-k)_\kappa}{(p-k)^2}\right) \notag\\
            &\times \left[\frac{i}{(p-k)^2-m^2}+2\pi\delta( (p-k)^2-m^2)\,n_B(|p_0-k_0|/T) \right]\,.
  \label{1LdiagramA}
\end{align}
Here $m=2e|\phi|=2e\sqrt{\Lambda^3/2\pi T}$ is the induced mass of the
two other gauge modes in the effective theory
(see \eqref{ModeMasses2}), and $n_B(x)$ is the Bose-Einstein function.

Diagram B is associated with the following expression
\begin{align}
   \Pi^{\mu\nu}_{B}(p)\,=\,
        \frac{1}{i} \int \frac{d^4k}{(2\pi)^4} &
        (-\delta_{ab}) \left( g_{\rho\sigma}-\frac{k_\rho k_\sigma}{m^2} \right)
        \left[ 2 \pi \delta (k^2-m^2) n_B(|k_0|/T) \right]  \notag \\
        & \times (-ie^2)
         [ \epsilon_{abe}\epsilon_{cde}(g^{\mu\rho}g^{\nu\sigma}-g^{\mu\sigma}g^{\nu\rho}) \notag\\
        &\qquad\quad
        +\epsilon_{ace}\epsilon_{bde}(g^{\mu\nu}g^{\rho\sigma}-g^{\mu\sigma}g^{\nu\rho}) 
        + \epsilon_{ade}\epsilon_{bce}(g^{\mu\nu}g^{\rho\sigma}-g^{\mu\rho}g^{\nu\sigma})]\,.
  \label{1LdiagramB}
\end{align}
As shown in \cite{LH2008}, $G$ and $F$ may be calculated selfconsistently by numerical methods
from this expressions for the polarization tensor. This is described in the following sections.

\subsection{Transversal modes}
\label{sec:transmodes}

Transversal propagation is determined by the scalar function $G(\omega,\mathbf p)$,
which can be calculated from the polarization tensor by use of equation \eqref{GofPi}.
In both \eqref{1LdiagramA} and \eqref{1LdiagramB}, the \emph{off-shellness} of the $k$ integration
is restricted in the effective theory by the maximal resolution $|\phi|$ after coarse-graining, see \eqref{constraint1}.
This results in the \emph{support condition}
\begin{align}
  |\left( p+k \right)^2| \,&=\,|G + 2 p\cdot k + 4 e^2 |\phi|^2 | \, \leq \, |\phi|^2 \, = \, \frac{\Lambda^3}{2 \pi T} \,,
  \label{supportcond}
\end{align}
where we have used $p^2=G$ and $k^2=4e^2 |\phi|^2$.
The $k$ integration in \eqref{1LdiagramA} and \eqref{1LdiagramB}
is to be taken over all values of $k$ such that \eqref{supportcond} is fullfilled.
It can be shown that diagram A, which is given by equation \eqref{1LdiagramA}, is identically zero
over the support \eqref{supportcond} \cite{LH2008}.
Since \eqref{1LdiagramA} is the only imaginary contribution
at the one-loop level, there is no finite transverse photon width \eqref{dispRel2T}.

The scalar function $G(\omega,\mathbf p)$ can then be calculated
up to one-loop from diagram B, that is equation \eqref{1LdiagramB},
by integrating over the support determined by the inequality \eqref{supportcond}.
To evaluate \eqref{1LdiagramB}, transform the integral to dimensionless variables
\begin{align}
  \xi \, &\equiv \, \frac{ k_3}{|\phi|}\,,\quad\quad
  \rho\, \equiv \,  \frac{1}{|\phi|} \sqrt{ k_1^2+ k_2^2}\,, \quad\quad
  \varphi \, \equiv \,  \arctan{\frac{k_2}{k_1}} \,.
  \label{defXiRho}
\end{align}
Equation \eqref{GofPi} gives $G$ when $\mathbf p$ is taken parallel to the $z$-axis.
Recall the definition of \emph{dimensionless temperature}
\begin{align}
  \lambda \,&\equiv \, \frac{2\pi T}{\Lambda}\,,
  \label{defTempDim}
\end{align}
and obtain an integral expression for the radiatively induced
\emph{screening function} $G$, normalized to $G\rightarrow G/T^2$
\begin{align}
  \frac{G}{T^2}\,&=\,\frac{\Pi^{11}}{T^2} \, = \, \frac{\Pi^{22}}{T^2}  \notag\\
                 &=\,       \int \mathrm d \xi\,\int \mathrm d\rho\,\,
                        e(\lambda)^2\lambda^{-3}\left(-4+\frac{\rho^2}{4e(\lambda)^2}\right)\,\rho\,
                                \frac{n_B\left(2\pi \lambda^{-3/2}\sqrt{\rho^2+\xi^2+4e(\lambda)^2}\right)}{\sqrt{\rho^2+\xi^2+4e(\lambda)^2}}
   \label{defG2}\\
                 &\equiv \, \int \mathrm d \xi\,\int \mathrm d\rho\,\, h_G \left( \xi,\rho, \lambda \right)\,,
\label{defG}
\end{align}
where we have noted the $\lambda$-dependence of the coupling $e$ by writing it explicitly.
The integration must be done over the support \eqref{supportcond}, which we also write in dimensionless variables.
Define the \emph{dimensionless momentum} $X$
\begin{align}
  X \, &\equiv \, \frac{|\mathbf p|}{T} \, = \, \frac{|p_3|}{T} \,,
  \label{defX}
\end{align}
and the \emph{support function} $s_{\pm}$
\begin{align}
  s_{\pm}\left(  \xi,\rho,\lambda, X, g \right) &\equiv
  \left| \frac{g \lambda^{3}}{(2\pi)^2}
        + \frac{\lambda^{3/2}}{\pi}\left(\pm\sqrt{X^2 + g}\sqrt{\rho^2+\xi^2+4e^2}- \xi X\right)
        + 4e^2\right| \,.
  \label{supportDimless}
\end{align}
Then the support condition \eqref{supportcond} can be written in dimensionless form
as \cite{LH2008}
\begin{align}
  s_{\pm}\left(  \xi,\rho,\lambda, X, g \right) \, \le \, 1 \,,
  \label{supportDimlessCond}
\end{align}
with $g\equiv G/T^2$.
The doubling of inequality \eqref{supportcond} comes from writing the delta function in \eqref{1LdiagramB}
in terms of $k^0$ \cite{SHG2007}
\begin{align}
  \delta \left( k^2 - m^2 \right) \, = \, \delta \left( k^0 + \sqrt{|\mathbf k|^2 + m^2} \right)
                        + \delta \left( k^0 - \sqrt{|\mathbf k|^2 + m^2} \right) \,.
  \label{doubDelta}
\end{align}
Define the function $H_G\left( \lambda, X,g \right)$
as the integral of $h_G\left( \xi,\rho, \lambda \right)$
summed over the support defined by $s_{\pm}\left( \xi,\rho,\lambda, X, g \right) \le 1$
\begin{align}
  H_G\left( \lambda, X,g \right) \,  \equiv \, \sum_{\sigma=+,-} \int_{-\infty}^{\infty}
        \mathrm d \xi\,\int_{0}^{\infty} \mathrm d\rho\,\,
        \theta\left(1-s_\sigma\left(  \xi,\rho,\lambda, X, g \right)\right) \,
        h_G\left( \xi,\rho, \lambda \right)\,.
  \label{defNumIntH}
\end{align}
where $\theta(x)$ is the Heaviside step function.
Equation \eqref{defG2} then becomes
\begin{align}
  g - \, H_G\left( \lambda, X, g\right) = 0 \,.
  \label{NumAlgoG}
\end{align}
It turns out that \eqref{NumAlgoG} is enough to uniquely determine $G$ \cite{LH2008}.
For a given temperature $T$ and momentum
$\mathbf{p} = |\mathbf p |\, \mathbf{ \hat z} $,
the integral \eqref{defNumIntH} can be calculated numerically as a function of
$\lambda$, $X$ and $g$.
The value of $g$ consistent with its defining equation \eqref{defG}
can then be determined by searching for the root of expression \eqref{NumAlgoG}
using standard numerical methods.
This root $g$ of \eqref{NumAlgoG} happens to be unique for each momentum and temperature.
This is all done using the Mathematica numerical software.
Our algorithmic procedure is then as follows:
For a given fixed temperature $T$ and various values of $X$ within
an interval $[0,X_{max}]$, the value of $g$ consistent with \eqref{defG}
is calulated by finding the root of \eqref{NumAlgoG}.
Each point is then used to interpolate $g(X,T)$ in $X$ over $[0,X_{max}]$.

\begin{figure}[p]
  \centering
  \subfloat[The dimensionless screening function $G/T^2$ calculated for a given temperature $T$ and
  momentum $X=p_3 / T$ by solving for the unique root of equation \eqref{NumAlgoG}.
  To obtain a complete plot in a given interval of X, the values of $G$ for a given $T$
  were successively calculated for different values of X partitioning the interval.
  The curves shown correspond to the temperatures
  $T=5\mathrm K$ (red), $T=8\mathrm K$ (blue) and $T=10\mathrm K$ (green).
  ]{\label{fig:G_at_5K_8K_10K}\includegraphics[width=130mm]{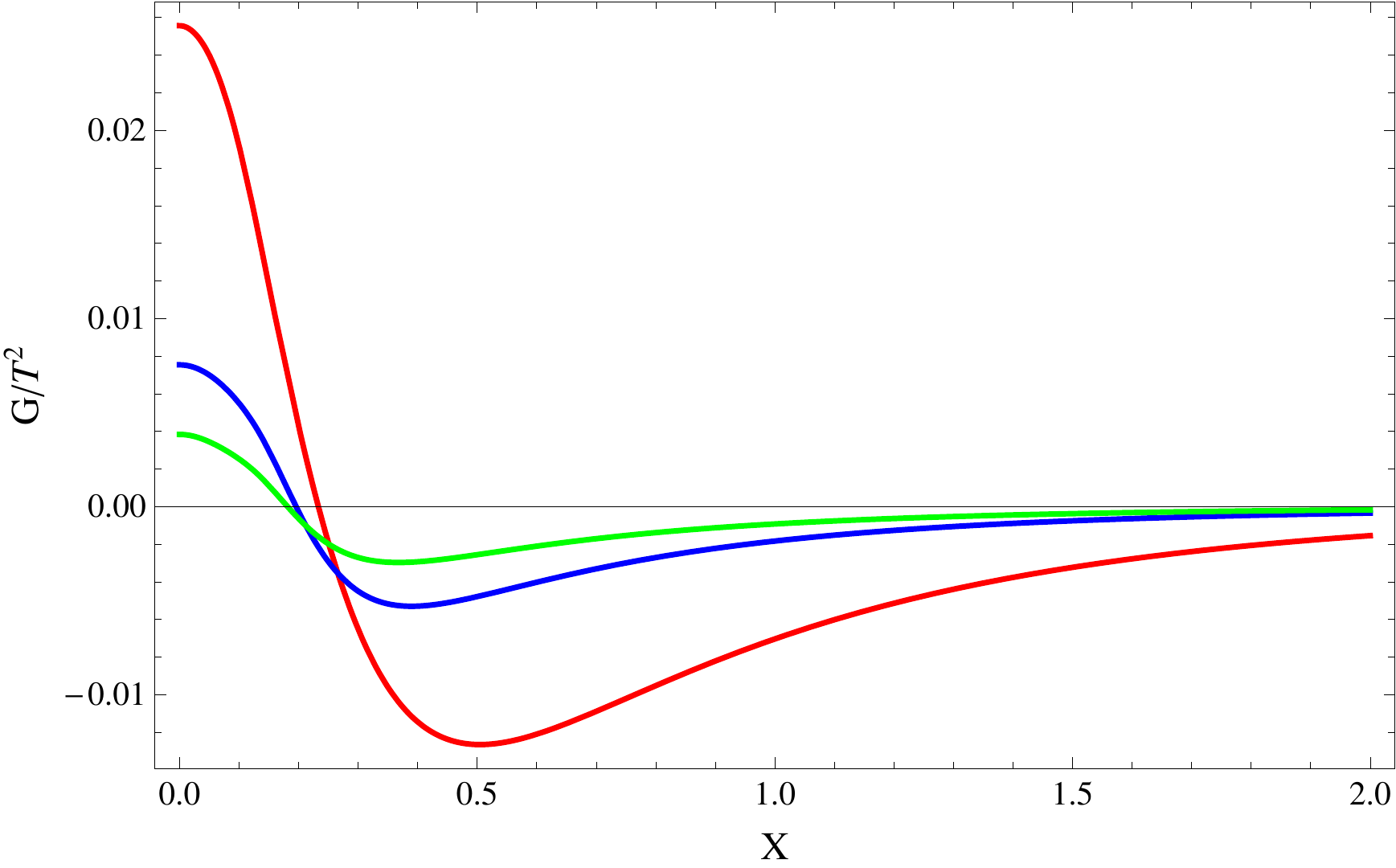}} \\
  \subfloat[The dimensionless dispersion law $Y(X)=\sqrt{X^2+G/T^2}$ calculated from $G$.
  Note that the value of the cutoff frequency $Y^* \equiv Y(X=0)$ for each temperature falls with $T$.
  The plotted temperatures are $T=5\mathrm K$ (red), $T=8\mathrm K$ (blue) and $T=10\mathrm K$ (green).
  The U(1) dispersion law $Y=X$ is shown as a black line.
  ]{\label{fig:disp_at_5K_8K_10K}\includegraphics[width=130mm]{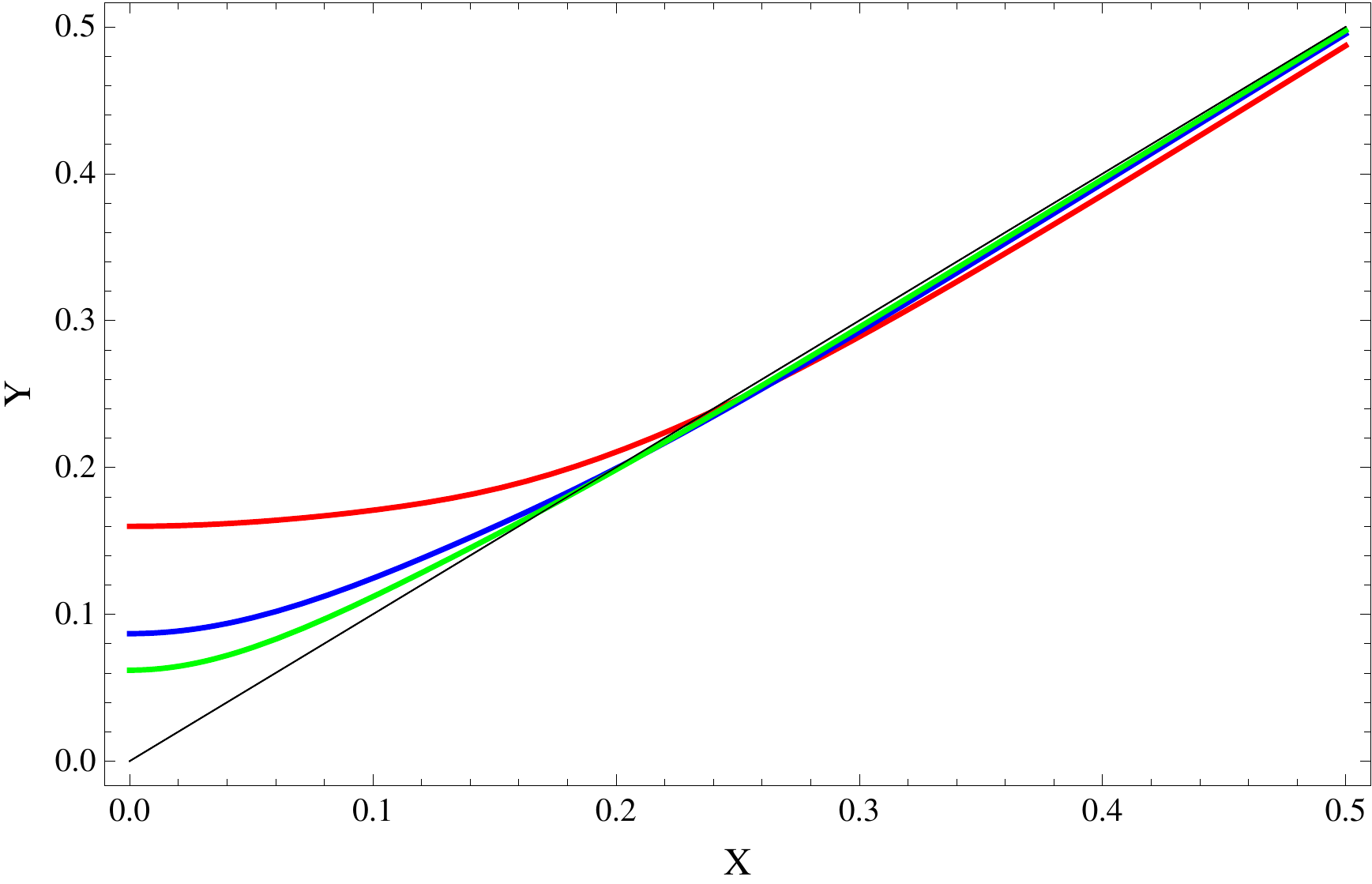}}
  \caption{
   Transverse screening mass $G/T^2$ (a) and dispersion relations (b).
  }\label{fig:Gplots}
\end{figure}

In figure \ref{fig:G_at_5K_8K_10K} we have plotted the resulting function $G(X,T)/T^2$
for different temperatures.
Note the existence of both an \emph{screening} region
where $G(X,T)>0$ and an \emph{antiscreening} region where $G(X,T)<0$.
Asymptotically, $G$ goes to zero, its maxima and minima also 
tend to zero with increasing $T$.

With $G$ nonzero we expect the dispersion law \eqref{dispRelT}
to deviate from the massless case $G=0$.
Defining the \emph{dimensionless frequency}
\begin{align}
  Y\, &\equiv \, \frac{\omega}{T} \,,
  \label{defY}
\end{align}
we write the transverse dispersion law \eqref{dispRelT} as
\begin{align}
  Y \, &= \, \sqrt{X^2 + \frac{G(X,T)}{T^2}} \,.
  \label{dispLawY}
\end{align}
This dispersion relation for the transverse modes is plotted
in figure \ref{fig:disp_at_5K_8K_10K} for different temperatures.
The \emph{cutoff frequency} $\omega^*$ is defined as
\begin{align}
  \omega^*(T) \, \equiv \, \sqrt{G(X=0,T)} \, = \, Y(X=0,T) T \, = \, Y^*(T) T\,.
  \label{defYstern}
\end{align}
As seen in figure \ref{fig:disp_at_5K_8K_10K}, the cutoff frequency
is the \emph{minimal} frequency available for transverse propagation at a given temperature. 
From the same figure is also clear that $Y^*$ decreases with increasing temperature.
The functional relation $Y^*(T)$ is very important for our experimental
considerations, since it defines the upper boundary of the \emph{screening region} $Y \leq Y^*$
where photon propagation is forbidden.
The experimental consequences of such screening effects will be investigated in chapter \ref{chap:radiometry}.
In the next section, an expression for $Y^*(T)$ will be derived from curve fitting.

\subsection{Some properties of the photon screening function G}

\label{sec:Gprop}

In this section we give numerical fits for three characteristic points of the
screening function. We work in SI units, so as to build an intuition on experimental
orders of magnitude.

\begin{figure}[p]
  \centering
  \subfloat[The set of cutoff frequencies $\{\nu^*_i\}$ calculated from a discrete set of temperatures
  $\{T_i\}$ between 2 and 165 Kelvin, shown here as black dots, overlapped with the curve
  $\nu^*(T) \,=\, 0.22+42.71\cdot T^{-0.53}$ fitted from the dataset $\{\nu^*_i,T_i\}$.
  ]{\label{fig:cutoff_fit}\includegraphics[width=150mm]{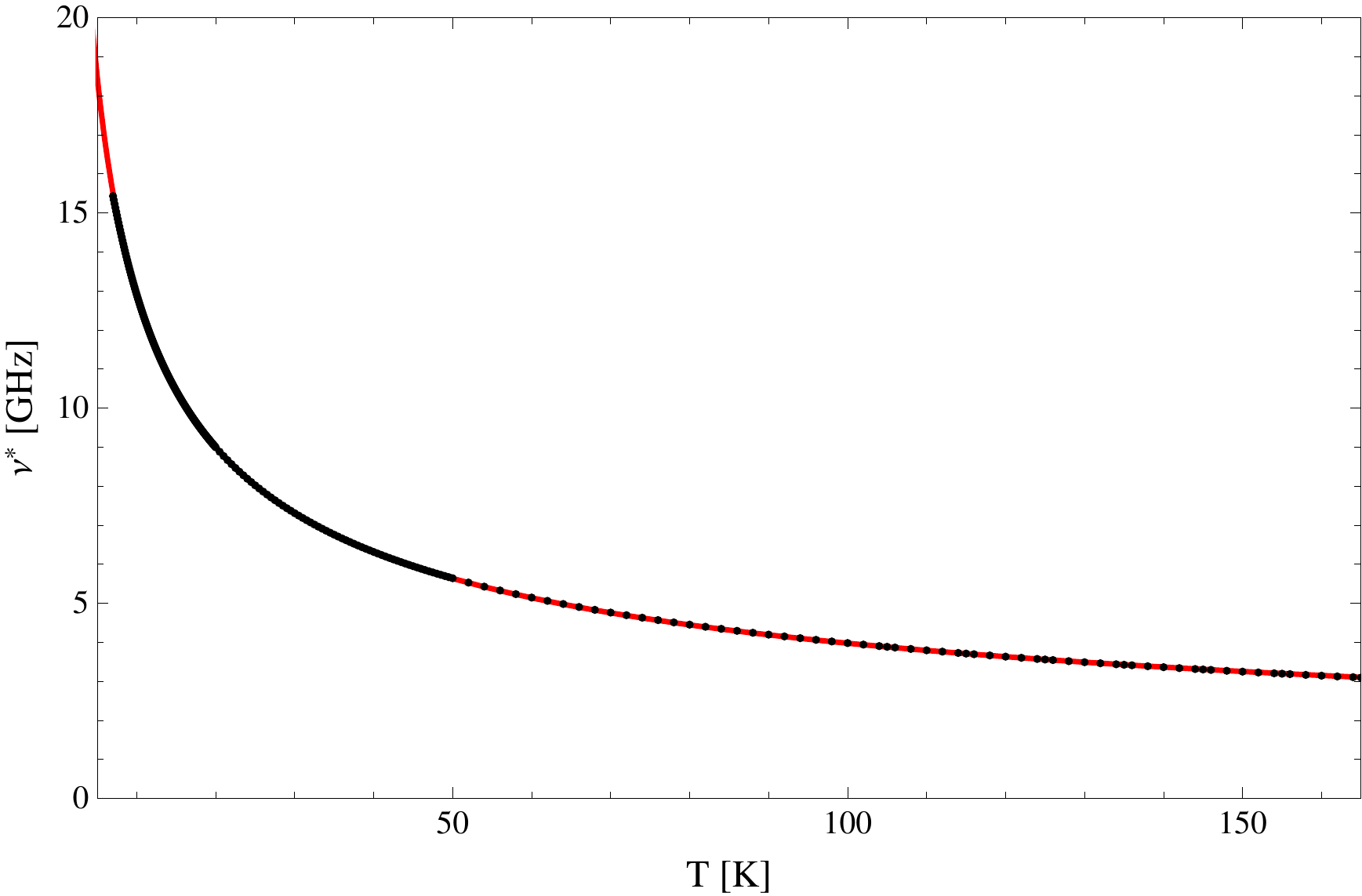}} \\
  \subfloat[Closeup of the dataset $\{\nu^*_i,T_i\}$ near the temperature $T=4.5$K for which
  the cutoff frequency $\nu^*$ rapidly falls to zero.
  The temperature region between 2K and 7K was not included in the calculation of best-fit parameters.
  ]{\label{fig:cutoff_data-close}\includegraphics[width=70mm]{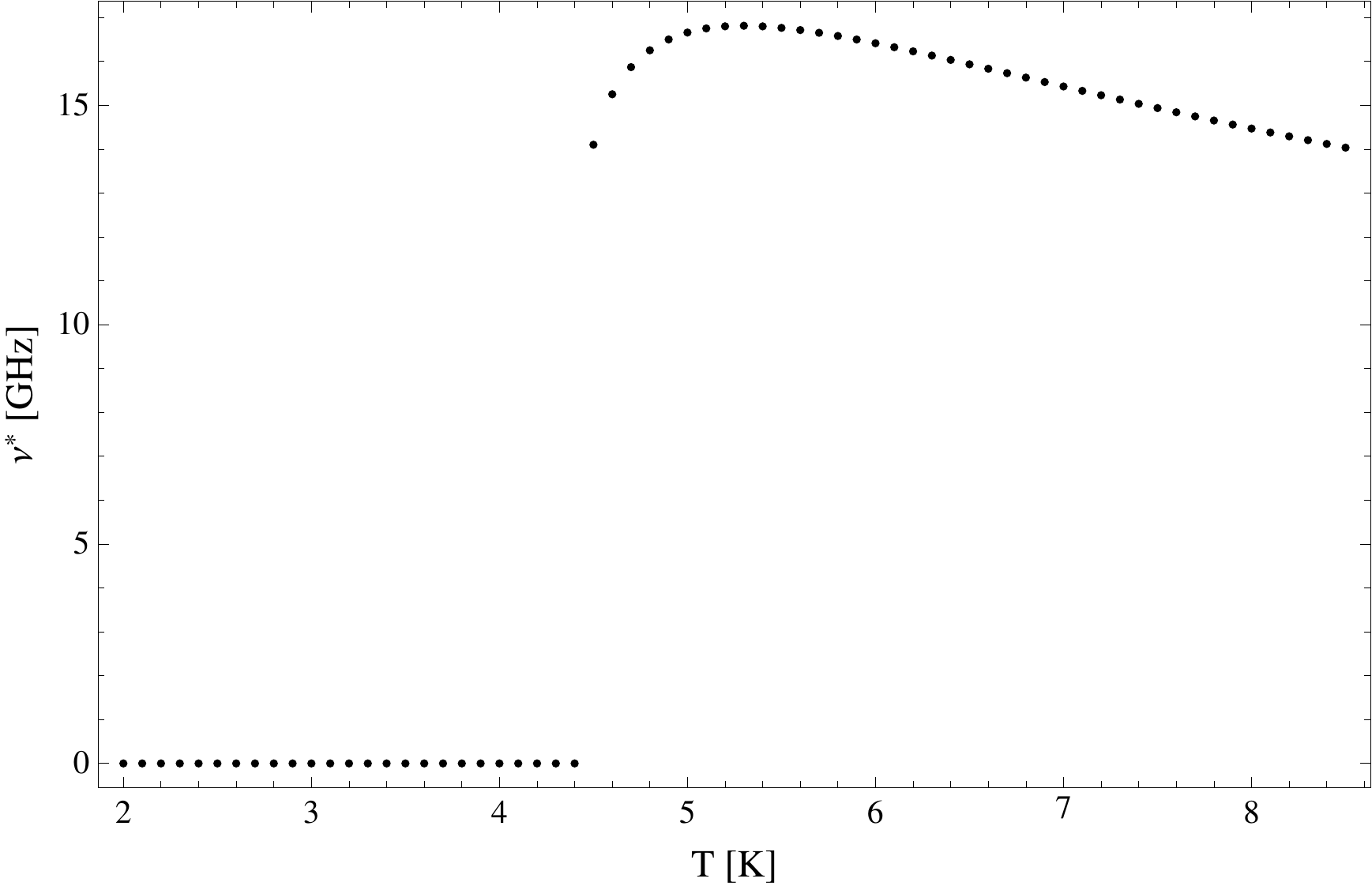}}
  \hspace{5mm}
  \subfloat[Closeup of the dataset $\{\nu^*_i,T_i\}$ near $T=7$K
  which was used as the lower bound of the curve fit.
  ]{\label{fig:cutoff_fit-close}\includegraphics[width=70mm]{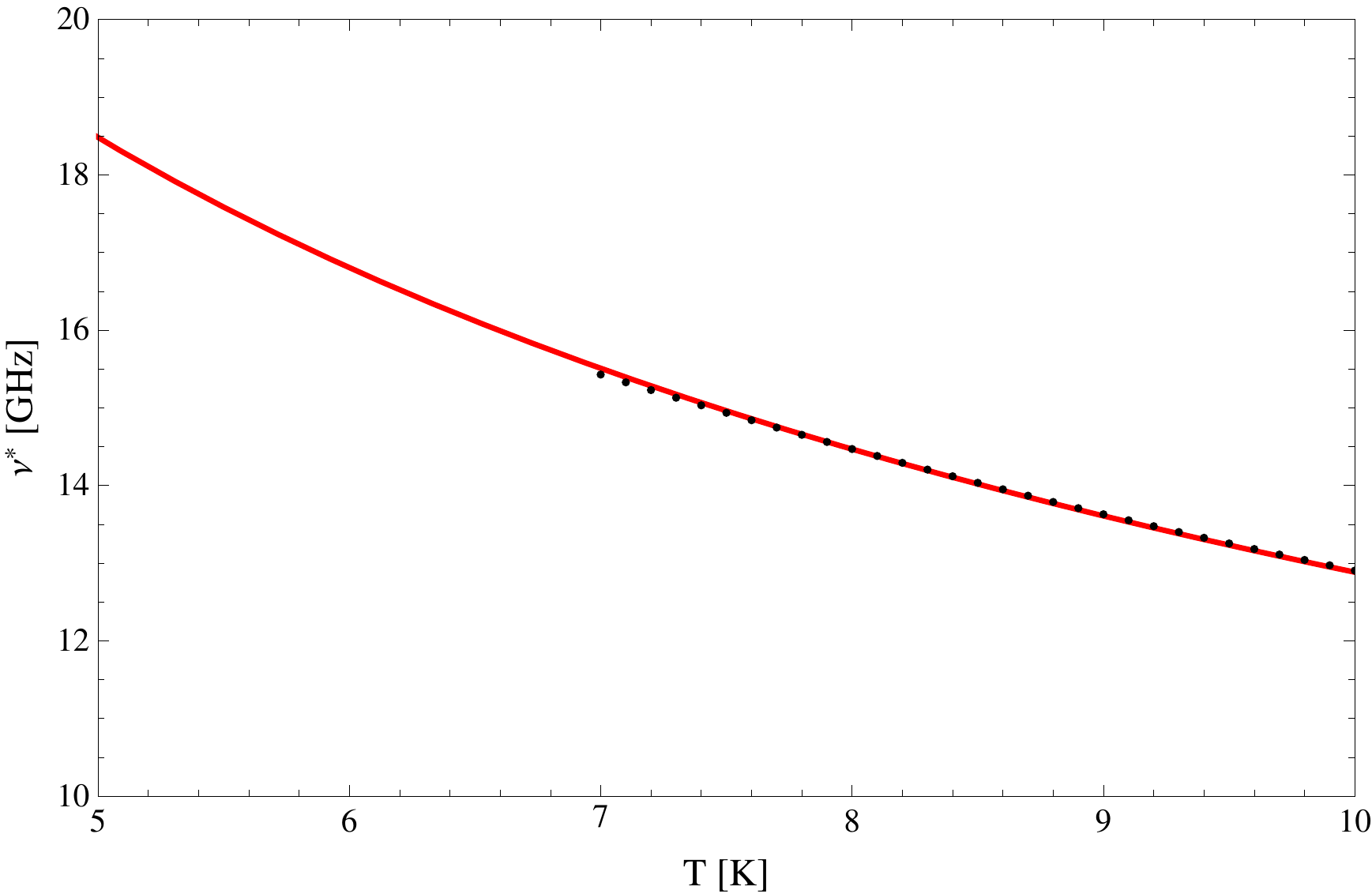}}
  \caption{The cutoff frequency $\nu^*$ was fitted to a power function in $T$,
  using a dataset $\{\nu^*_i,T_i\}$ calculated from discrete set of temperatures $\{T_i\}$ between 2 and 165 Kelvin.
  The original set had $i=1,..,306$ data-points.
  The first 50 of those, corresponding to temperatures between 2K and 7K where ignored
  in the calculation of best-fit parameters.
  Taking $\nu^*$ in GHz and $T$ in Kelvin, the best fit for the remaining 256 points in the dataset
  is given by the power law
  $\nu^*(T)\,=\,0.22+42.7 T^{-0.53}$ \,.
  }\label{fig:cutoff_freq}
\end{figure}

In section \ref{sec:transmodes} we defined the \emph{cutoff frequency} as the
minimal available frequency for any given temperature, $h\nu^*(T) = c\sqrt{G(|\mathbf p|=0,T)}$.
We calculated $\nu^*$ for temperatures between 2 and 165 Kelvin.
The results are depicted in figure \ref{fig:cutoff_freq}.
Figure \ref{fig:cutoff_data-close} shows how $\nu^*$ grows with decreasing temperature,
reaching a maximum at about 5.4 Kelvin of about $\nu^*=17GHz$, then decreases and suddenly
drops to zero below 4.5K.
To calculate the fit, we have assumed a \emph{power law} functional relation of the form
\begin{align}
\nu^*(T)\,=\, c + b T^\gamma\,.
  \label{nuFitExp}
\end{align}
We excluded from the fit data points below 7K, as seen in figure \ref{fig:cutoff_data-close}.
The Mathematica software was used to fit the data points to \eqref{nuFitExp}, resulting in the expression
\begin{align} 
    \frac{\nu^*(T)}{\text{GHz}} \, &= \, 42.71 \left( \frac{T}{\text{K}} \right)^{-0.53} + 0.22\,.
  \label{nuSternFit}
\end{align}
How well this fit matches the data is shown in figure \ref{fig:cutoff_freq}.

\begin{figure}[t]
  \centering
  \includegraphics[width=140mm]{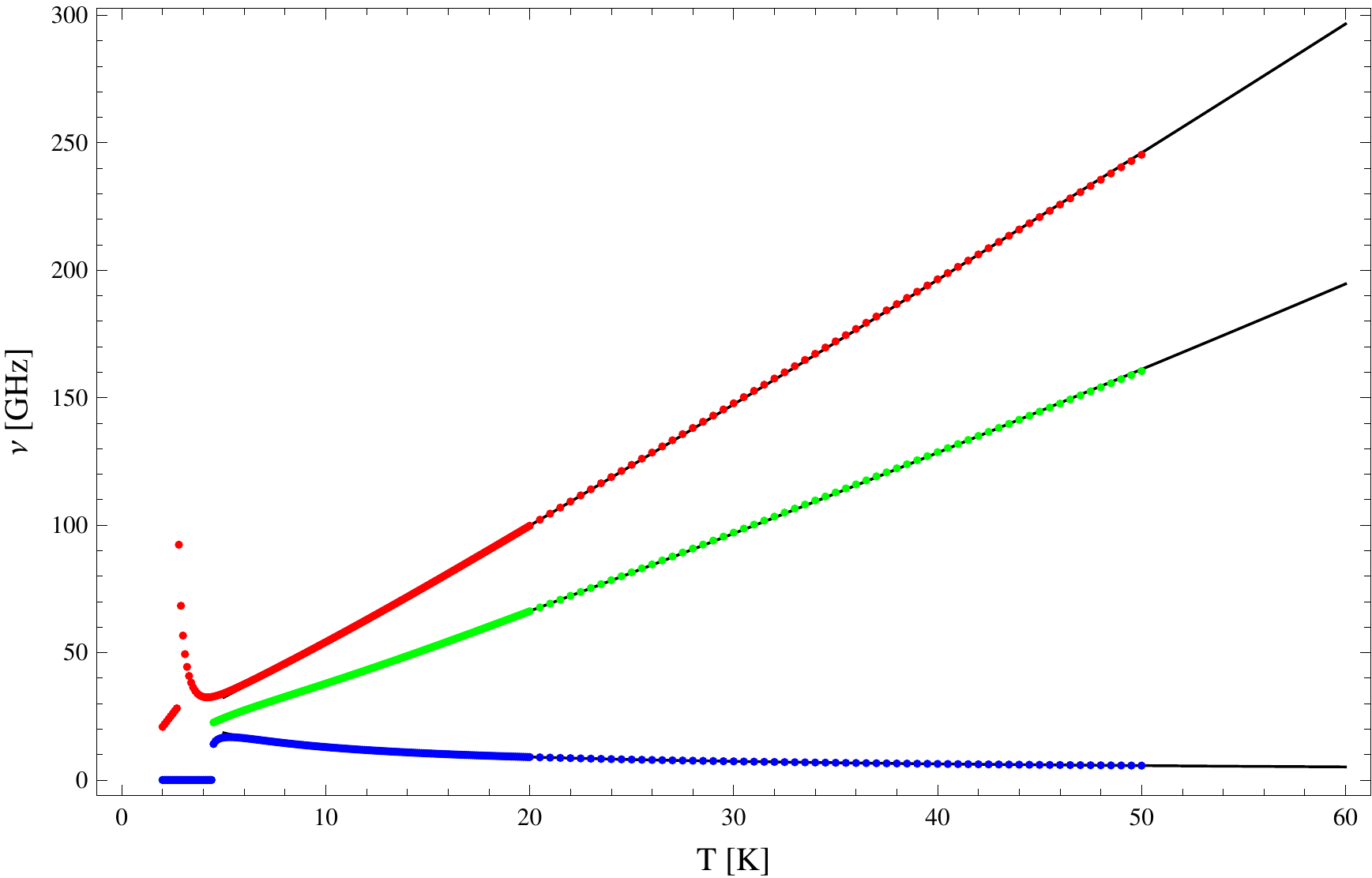}
  \caption{The $T$-dependence of the characteristic points of $G$:
           $\nu^*(T)$ (blue), $\nu_c(T)$ (green), and $\nu_M(T)$ (red).
           Taken from \cite{falquez2010modification}.
           }
  \label{fig:Ypoints}
\end{figure}

Another interesting characteristic frequency is the \emph{crossover frequency}
$\nu_c$, defined as
\begin{align}
    G(\nu_c,T) \, &= \, 0\,.
   \label{nuCdef}
\end{align}
This is the frequency at which the screening function $G$ goes from screening to antiscreening.
Just as $\nu^*$, $\nu_c$ drops to zero below 4.5 Kelvin.
A fit of the data points to a power law of the form \eqref{nuFitExp} gives as result
\begin{align}
   \frac{\nu_c(T)}{\text{GHz}} &= 1.83 \left( \frac{T}{\text{K}} \right)^{1.12} + 13.48 \,.
  \label{nuCFit}
\end{align}
This is plotted in figure \ref{fig:Ypoints}.

In chapter \ref{chap:radiometry} we introduce the \emph{radiance} function of a black emitter.
It will be shown that the ratio between an SU(2) and an U(1) emitter is given by the term
\begin{align}
\left(1 - \frac{c^2 G}{(h\nu)^2}\right)\theta \left( \nu - \nu^* \right) \,.
 \label{U1SU2ratio}
\end{align}
The maximal difference between U(1) and SU(2) radiance in the antiscreening regime,
where $G$ is negative, is then given at the frequency $\nu_M$ where $G/\nu^2$ reaches a \emph{minimum}
\begin{align}
  \frac{G(\nu_M,T)}{\nu_M^2} &= \frac{G(\nu,T)}{\nu^2}\bigg|_{\mathrm{min}}\,.
    \label{nuMdef}
\end{align}
Thus $\nu_M$ is the frequency of \emph{maximal antiscreening}. A fit
of the data to a power law of the form \ref{nuFitExp} gives
\begin{align}
    \frac{\nu_M(T)}{\text{GHz}} &= 3.45 \left( \frac{T}{\text{K}} \right)^{1.08} + 12.90 \,.
\label{nuMFit}
\end{align}
Figure \ref{fig:Ypoints} shows the calculated points overlaid with the fitted curve.

\subsection{Longitudinal mode}
\label{sec:longmodes}

\begin{figure}[p]
  \centering
  \subfloat[The dimensionless longitudinal screening function $F/T^2$.
  As seen in the plot, $F(X)$ is double valued, so 
  the searching algorithm used for the transverse modes had to be modified to look
  for the roots of \eqref{NumAlgoF} in $X$, with fixed values of $F/T^2$ and $T$.
  This root turns out to be unique.
  The curves shown correspond to the temperatures
  $T=5\mathrm K$ (red), $T=8\mathrm K$ (blue) and $T=10\mathrm K$ (green).
  ]{\label{fig:F_at_5K_8K_10K}\includegraphics[width=140mm]{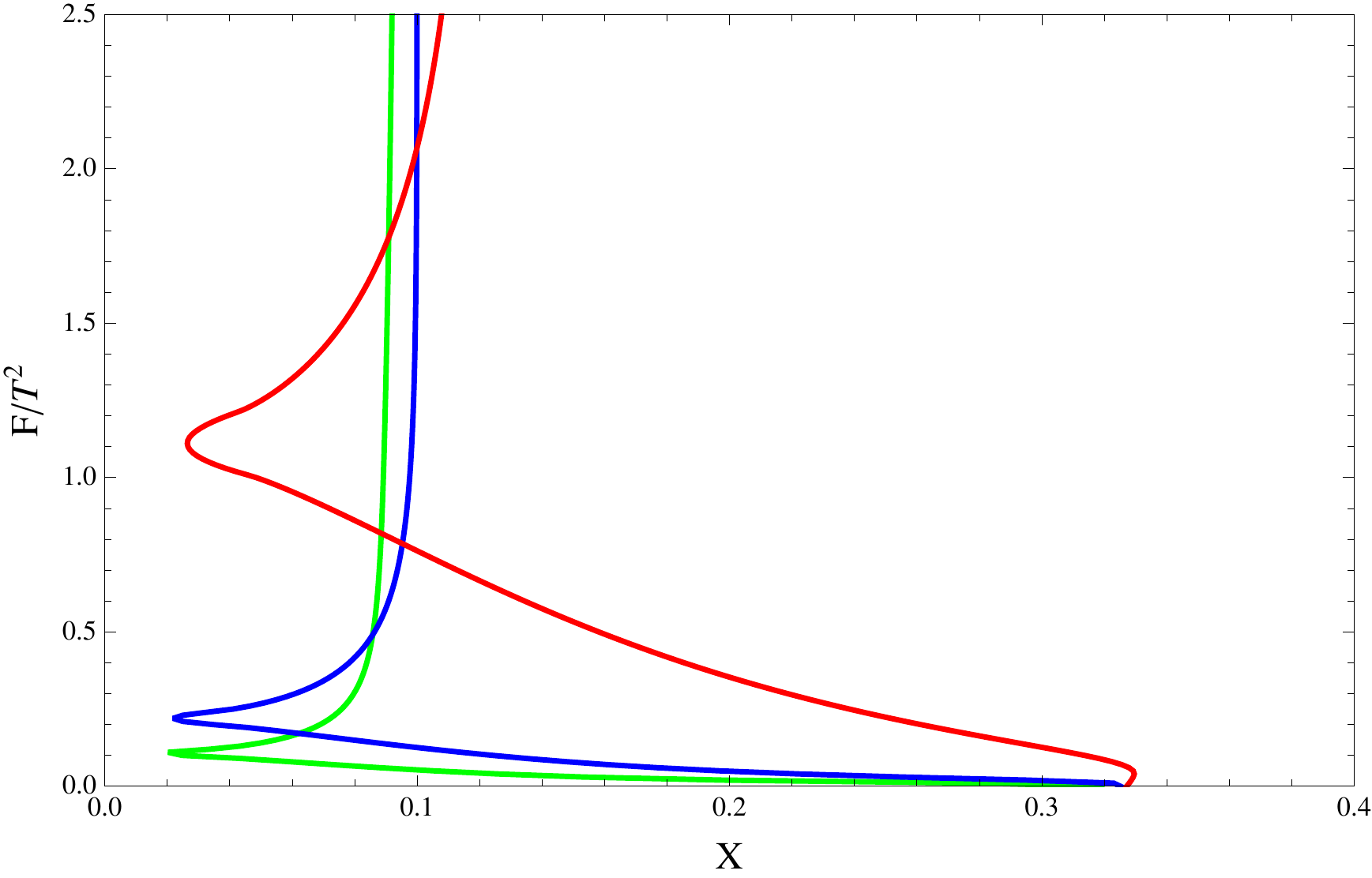}} \\
  \subfloat[The longitudinal dispersion law $Y_L(X)=\sqrt{X^2+F/T^2}$ calculated from $F$.
  Note the two branches, consequence of the double-valuedness of $F$.
  The plotted temperatures are $T=5\mathrm K$ (red), $T=8\mathrm K$ (blue) and $T=10\mathrm K$ (green).
  The U(1) dispersion law $Y=X$ is shown as a black line.
  ]{\label{fig:dispF_at_5K_8K_10K}\includegraphics[width=140mm]{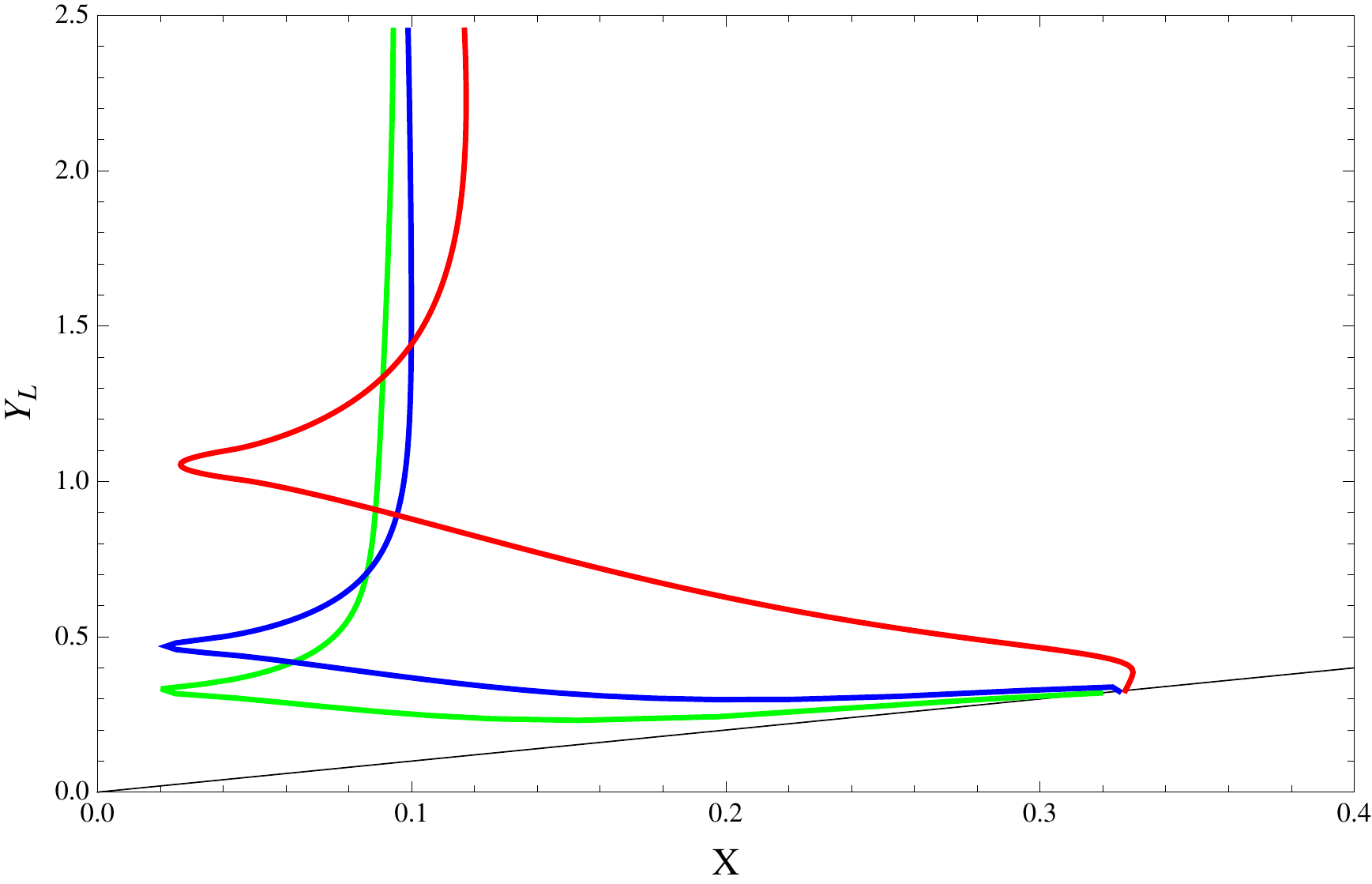}}
  \caption{
  Longitudinal screening mass $F/T^2$ (a) and dispersion relations (b).
  }\label{fig:Fplots}
\end{figure}

Consider now the propagation of longitudinal modes, characterized by the
poles of the term $\propto u_\mu u_\nu$ in \eqref{defFullD}.
On shell, $p^2=F$, and \eqref{FofPi} simplifies to
\begin{align}
  X^2 \, &\equiv \, \frac{\mathbf p^2}{T^2} \, = \, \frac{\Pi_{00}}{T^2}\,.
\label{P00Fonshell}
\end{align}
To calculate this expression numerically, we first assume
that diagram B of figure \ref{1LdiagramB} is the only contribution to the polarization tensor.
Afterwards, we check that diagram A in figure \ref{fig:feynman},
vanishes on-shell $F=p^2$.
The expression for $\Pi^{00}\equiv \Pi^{00}_B$
can be read from equation \eqref{1LdiagramB}
\begin{align}
  X^2 \, &= \, \frac{\Pi^{00}_B}{T^2} \notag \\
         &= \, \int \mathrm d \xi\,\int \mathrm d\rho\,\,  2e^2\lambda^{-3}
                \left(3+\frac{\rho^2+\xi^2}{4e^2}\right)\,\rho\,
                        \frac{n_B\left(2\pi \lambda^{-3/2}\sqrt{\rho^2+\xi^2+4e^2}\right)}{\sqrt{\rho^2+\xi^2+4e^2}}
  \label{defF2} \\
                        &\equiv \, \int \mathrm d \xi\,\int \mathrm d\rho\,\, h_F \left( \xi,\rho, \lambda \right)
                        \,.
  \label{defF}
\end{align}
The integrations are subject, as in the transverse case, to the following constraint
\begin{align}
  s_{\pm}\left(  \xi,\rho,\lambda, X, f \right) \, \le \, 1 \,,
  \label{supportDimlessCondF}
\end{align}                                                                             
with the \emph{support function} $s_{\pm}$ defined as
\begin{align}
  s_{\pm}\left(  \xi,\rho,\lambda, X, f \right) &\equiv
  \left| \frac{f \lambda^{3}}{(2\pi)^2}
  + \frac{\lambda^{3/2}}{\pi}\left(\pm \sqrt{X^2+f} \sqrt{\rho^2+\xi^2+4e^2} - \xi X \right)
        + 4e^2\right| \,,
  \label{supportDimlessF}
\end{align}
with $f \equiv F/T^2$.
As we did in section \ref{sec:transmodes} for the transverse case, we
express the inequality \eqref{supportDimlessCond} by using
the Heaviside step function $\theta(x)$,
and define a function $H_F\left( \lambda, X,f \right)$
as the integral of $h_F\left( \xi,\rho, \lambda \right)$ in \eqref{defF}
summed over the support determined by \eqref{supportDimlessF}
\begin{align}
  H_F\left( \lambda, X,f \right) \,  \equiv \, \sum_{\sigma=+,-} \int_{-\infty}^{\infty}
        \mathrm d \xi\,\int_{0}^{\infty} \mathrm d\rho\,\,
        \theta\left(1-s_\sigma\left(  \xi,\rho,\lambda, X, f \right)\right)\,
        h_F\left( \xi,\rho, \lambda \right)\,.
  \label{defNumIntHF}
\end{align}
Then the consistency condition for the integral equation \eqref{defF2}
is just the expression
\begin{align}
  X^2 - \, H_F\left( \lambda, X, f\right) = 0 \,.
  \label{NumAlgoF}
\end{align}
Using the same method as in section \ref{sec:transmodes},
which is to search for the root of equation \ref{NumAlgoF} in $f$,
given fixed values of $X$ and $\lambda$,
results in numerical difficulties, due to the spectral function
$F(X)$ being \emph{double valued}.
The longitudinal screening function $F$, in contrast to $G$,
\emph{separates in different branches}.
Therefore, the searching algorithm has to be modified to look
for the roots of \eqref{NumAlgoF} in $X$, with fixed values of $f$ and $T$.
This root turns out to be unique.
The search results are displayed in figure \ref{fig:F_at_5K_8K_10K},
where we have plotted $F(X)/T^2$ for different temperatures. Note the existence of
two branches of $F$.
Furthermore, $X$ is confined to a finite domain between $X=0$ and $X=0.4$,
as shown in the plot.
This strange behaviour is reflected in the longitudinal dispersion law
of figure \ref{fig:dispF_at_5K_8K_10K}.

Using similar numerical methods as those described above,
we have checked that diagram A, given by expression \eqref{1LdiagramA}, vanishes on shell $F=p^2$.

In dimensionless quantities the longitudinal dispersion law reads
\begin{align}
  Y_L \, \equiv \, \frac{\omega_L}{T} \,= \, \sqrt{ X^2 + \frac{F(X,T)}{T^2} }\,.
\label{dispLawF}
\end{align}
This relation is plotted in figure \ref{fig:dispF_at_5K_8K_10K} for three different temperatures.
Notice the existence of various branches, consequence of the double-valuedness of $F(X)$.
The lower branch shows a dispersion law with negative slope, 
reaching its maximum available spatial momentum at the point where $Y(X)=X$.
Further, the upper branch of $Y_L$ becomes superluminal at very low $X$,
or large wavelength $X^{-1}$, and diverges to infinity very quickly.
This suggests that the longitudinal modes are unphysical.

Consider also the magnetic-electric dual interpretation
for the thermal SU(2) Yang-Mills theory described in \cite{Hofmann2005},
where the propagating electric field in the SU(2) theory
becomes a magnetic field in the thermal photon plasma.
Recall that longitudinal oscillations describe \emph{electric charge density}
waves \cite{Kapusta}.
In the dual interpretation, the long wavelength branching modes
in figure \ref{fig:dispF_at_5K_8K_10K} become \emph{magnetic} density waves,
and cannot transfer energy in any interaction with electric charges.
We gather that no thermalization can occur for these modes.

From the above arguments we can only conclude that the longitudinal modes
are, if not unphysical, at least unable to propagate nor transport energy.
Thus, we may ignore the longitudinal modes in our further discussion.
The only excitations of the SU(2) thermal photon gas we will consider 
are the transverse modes of the tree-level massless gauge field.

\section{Resummation of one-loop radiative corrections}
\label{sec:resumm}

Apart from the effect on photon propagation,
the transverse screening function $G$ may be used to calculate
thermodynamical properties of the SU(2) gas.
Trough the procedure of \emph{resummation},
corrections to the energy density and pressure induced by interaction
of SU(2) photons with the thermal ground state can be extracted.
We will find such correc-tions to be linear in the temperature.
The thermodynamical consistency of this resummation approach is also discussed in this section.

\subsection{Prerequisites}

Recall from section \ref{sec:thermalGS} that the ground state
of the thermal SU(2) Yang-Mills gas
is composed of a scalar field $\phi$ \eqref{PhiDefL} and a pure-gauge configuration \eqref{DefAPureG}.
From  the effective action \eqref{SU2EffectLag}
the energy-momentum tensor $\theta^{\srm{gs}}_{\mu\nu}$
of this configuration may be calculated,
\begin{align}
  \theta^{\srm{gs}}_{\mu\nu} \, &= \, -4 \pi \Lambda T g_{\mu\nu} \,.
  \label{EffEMT}
\end{align}
Take the SU(2) gas to be a \emph{perfect fluid} in thermal equilibrium,
$\theta_{\mu\nu}^{\srm{gs}}=(P^{\srm{gs}}+\rho^{\srm{gs}}) u_\mu u_\nu + P^{\srm{gs}}g_{\mu\nu}$,
with the restframe specified by the 4-velocity  vector $u_\mu$.
Then we read from \eqref{EffEMT} that $\rho^{\srm{gs}} = - P^{\srm{gs}}=4 \pi \Lambda T$.
As given in \eqref{GSenergy}, the (classical) ground state configuration has a
nontrivial energy density which grows linearly with temperature.
For later comparison with higher-order corrections, we write this result in the form
\begin{align}
  \frac{\rho^{\srm{gs}}}{T^4} \, &= \, -\frac{P^{\srm{gs}}}{T^4} \, = \, 2(2\pi)^4\,\lambda^{-3}\sim 3117.09\,\lambda^{-3}\,,
\label{rho_p_gs}
\end{align}
where as before the dimensionless temperature is defined as $\lambda=\frac{2 \pi T}{\Lambda}$.

Note also that the ground state pressure $P^{\srm{gs}}$ in \eqref{rho_p_gs} is negative.
This may be interpreted microscopically as magnetic monopole-antimonopole
\cite{Kraan19982,Kraan1998,LeeLu,Nahm1983-5}
\emph{attractive} interactions due to the dominance
in thermal fluctuations of small-holonomy caloron configurations \cite{Diakonov2004}.

Equation \eqref{rho_p_gs} gives the tree-level contribution
to the gas thermodynamics. What about higher order corrections?
As shown in \cite{GiacosaHofmann2007}, linear contributions
$\Delta P^{\mathrm{gs,1-loop}}$ and $\Delta\rho^{\srm{gs,1-loop}}$
to $P^{\srm{gs}}$ and $\rho^{\srm{gs}}$, respectively, 
arise from the free fluctuations of tree-level massive modes
at high temperature. This correction is given by
\begin{align}
\frac{\Delta\rho^{\srm{gs,1-loop}}}{T^4} \, &= \, \frac{\Delta P^{\srm{gs,1-loop}}}{T^4} \,=\,
        -\frac14\,a^2 \,= \, -2(2\pi)^4\,\lambda^{-3} \,\sim \, -3117.09^4\,\lambda^{-3}\,,
\label{rho_p_gs_1l}
\end{align}
where $a\equiv\frac{2e|\phi|}{T}=\frac{8\sqrt{2}\pi^2}{\lambda^{3/2}}$ is the
temperature scaled mass \cite{GiacosaHofmann2007}.

Note that, since both $P^{\srm{gs}}$ and $\Delta P^{\srm{gs,1-loop}}$
are linear in temperature, a Legendre transformation to the
energy density $\rho=T\frac{dP}{dT}-P$ must vanish.
But $\rho^{\srm{gs}}+\Delta\rho^{\srm{gs,1-loop}}=0$
also vanishes, therefore both corrections are \emph{thermodynamically consistent}.

In the following, we show how a correction to the base energy and pressure
of the ground state can arise by assuming selfconsistent propagation
at the 1-loop level of tree-level massless modes. This correction will
depend linearly on temperature, again signaling ground state effects
on photon propagation, and vice versa.

\subsection{Resummation}

\begin{figure}[t]
  \centering
  \includegraphics[width=140mm]{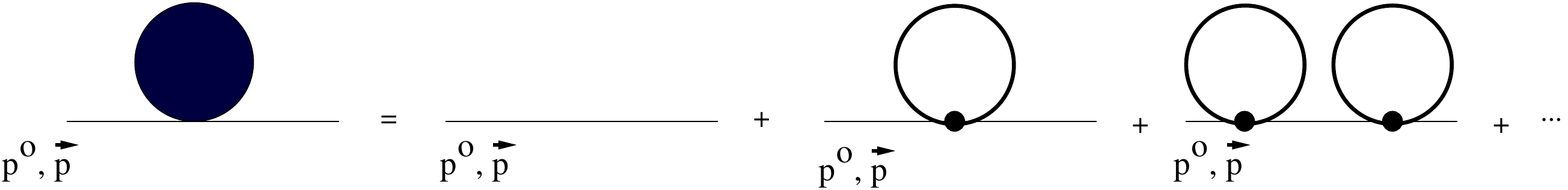}
  \caption{The Dyson series for the dressed propagator
  of the massless mode, containing only tadpole diagrams, since it is
  the only nonvanishing diagramm on the induced mass shell,
  see section \ref{sec:polarization}.
  A thin line refers to the propagation of the tree-level massless,
  a thick line to the propagation of tree-level massive modes.
  Taken from \cite{CF2010}.
  }
  \label{fig:dysomResumm}
\end{figure}

A \emph{resummed} perturbation expansion is given by replacing the bare propagator
with the dressed one. In figure \ref{fig:dysomResumm} the diagrammatic expansion
of the dressed propagator up to one-loop is shown.
Since diagram A in figure \ref{fig:feynman}, as given by \ref{1LdiagramA},
vanishes in the effective theory (see section \ref{sec:polarization}),
we only need to consider the one-loop tadpole (diagram B in figure \ref{fig:feynman}).
We can then calculate resummed thermodynamical quantities by
the usual method of expanding $\ln Z$ in \emph{bubble diagrams},
but replacing the bare propagator with its resummed (dressed) version.

But we already calculated the dressed one-loop propagator
(determined by the one-loop polarization tensor).
As we reviewed in \ref{sec:polarizationIntro}, the dressed propagator
in figure \ref{fig:dysomResumm}, as a result of photon interactions
with the ground state, displays a modified dispersion law for
the propagating, transverse massless modes.
From section \ref{sec:transmodes} we recall the dimensionless
dispersion relation
\begin{align}
  Y \, &= \, \sqrt{X^2 + \frac{G(X,T)}{T^2}} \,,
  \tag{\ref{dispLawY}}
\end{align}
and the definitions $Y\equiv\frac{\omega}{T}$, $X\equiv\frac{|\mathbf p|}{T}$.
Thus, we can insert the modified dispersion relation \eqref{dispLawY}
in the thermodynamical integrals for \emph{energy density} and \emph{pressure}
\begin{align}
  \rho \, &= \, \frac{1}{\pi^2} \int  \mathrm{d} p \, \mathbf{p}^2 \, \frac{\omega}{e^{\beta\omega}-1} \,,
  \label{defE} \\
   P \, &= \, -  \frac{1}{\pi^2} \int  \mathrm{d} p \, \mathbf{p}^2 \,\ln(1-e^{-\beta\omega})\,.
   \label{defP} 
\end{align}
derived from the partition function for the noninteracting (zero chemical potential)
massless boson with two degrees of freedom \cite{Kapusta}
\begin{align}
  \ln{Z} \, &=\, - 2 V \int \frac{\mathrm{d}^3 \mathbf p}{(2 \pi)^3} \left[ \frac{1}{2} \beta\omega + \ln{1-e^{-\beta\omega}} \right]\,.
  \label{defZ}
\end{align}
Then, we get the \emph{1-loop resummed} energy density and pressure
\begin{align}
  \frac{\rho^{\srm{1-loop,res}}}{T^4} \, &= \, \frac{1}{\pi^2}\int_0^\infty dX\, X^2 \frac{\sqrt{X^2+G/T^2}}{e^{\sqrt{X^2+G/T^2}}-1} \,,
  \\
  \frac{P^{\srm{1-loop,res}}}{T^4} \, &= \, -\frac{1}{\pi^2}\int_0^\infty dX\,X^2\ln \,(1-e^{-\sqrt{X^2+G/T^2}}) \,,
\end{align}
where we have rescaled by $T^4$ to make both expressions dimensionless.
Define the dimensionless corrections $\Delta\bar{P}$ and $\Delta\bar{\rho}$ to the contributions of free, massless particles as 
\begin{align}
  \Delta\bar{\rho} \, &\equiv \, \frac{\Delta \rho^{\srm{1-loop,res}}}{T^4} \, = \,
        \frac{\rho^{\srm{1-loop,res}}}{T^4}-\frac{\pi^2}{15}\,,
        \label{correction_rho}\\
  \Delta\bar{P} \, &\equiv \, \frac{\Delta P^{\srm{1-loop,res}}}{T^4} \, = \,
        \frac{P^{\srm{1-loop,res}}}{T^4}-\frac{\pi^2}{45} \,.
        \label{correction_P}
\end{align}

\begin{figure}[p]
  \centering
  \subfloat[The resummed one-loop dimensionless correction to the energy density,
  as defined in equation \eqref{correction_rho}. Fitting of the data
  to a power law of the form \eqref{defPLrho} results in the expression
  $\Delta\bar{\rho} = 3.9577 \cdot \lambda^{-3.02436}$.
  ]{\label{fig:dRho}\includegraphics[width=130mm]{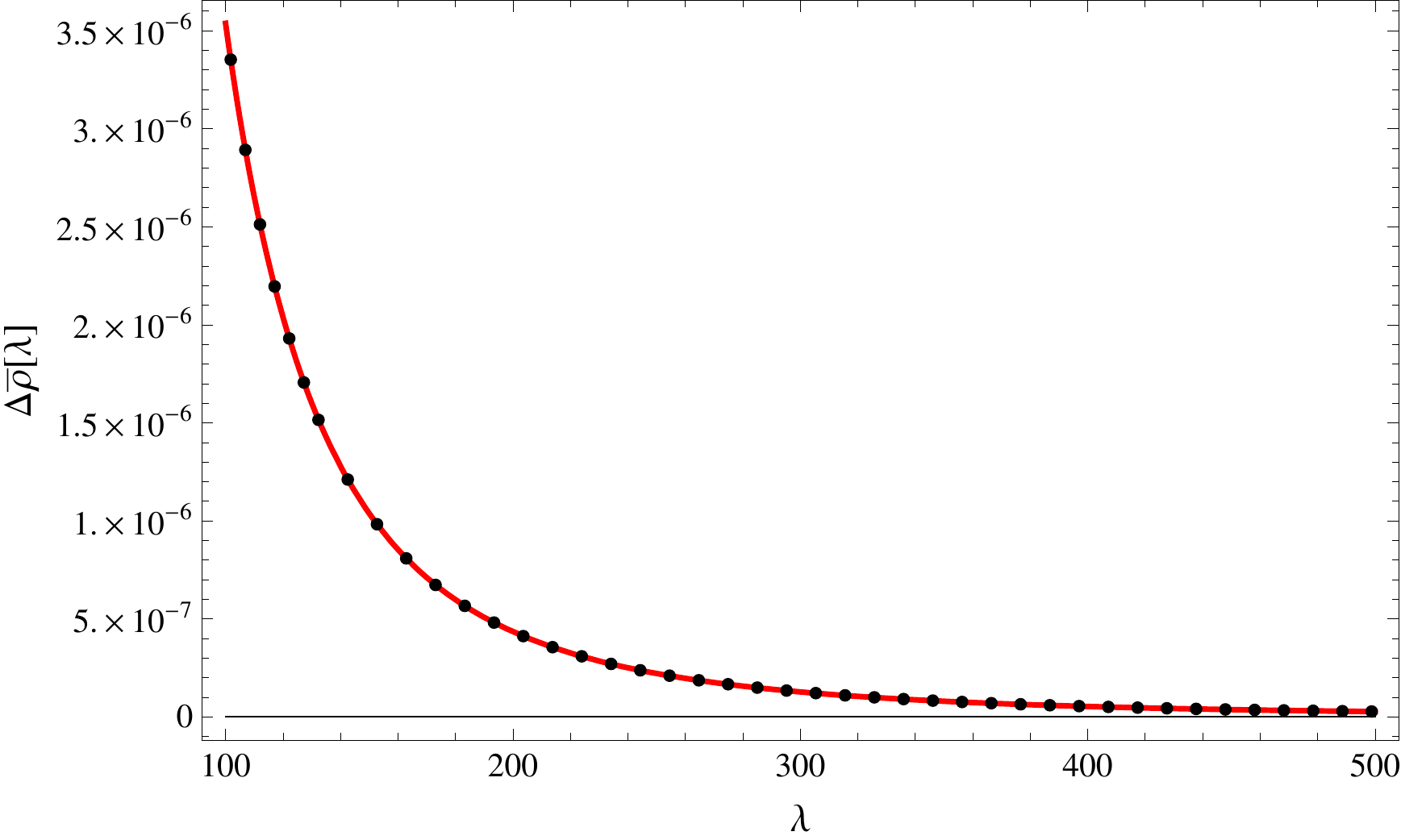}} \\
  \subfloat[The resummed one-loop dimensionless correction to the pressure,
  as defined in equation \eqref{correction_P}. Fitting of the data
  to a power law of the form \eqref{defPLP} results in the expression
  $\Delta\bar{\rho} = 8.4963\cdot \lambda^{-3.00904}$.
  ]{\label{fig:dPre}\includegraphics[width=130mm]{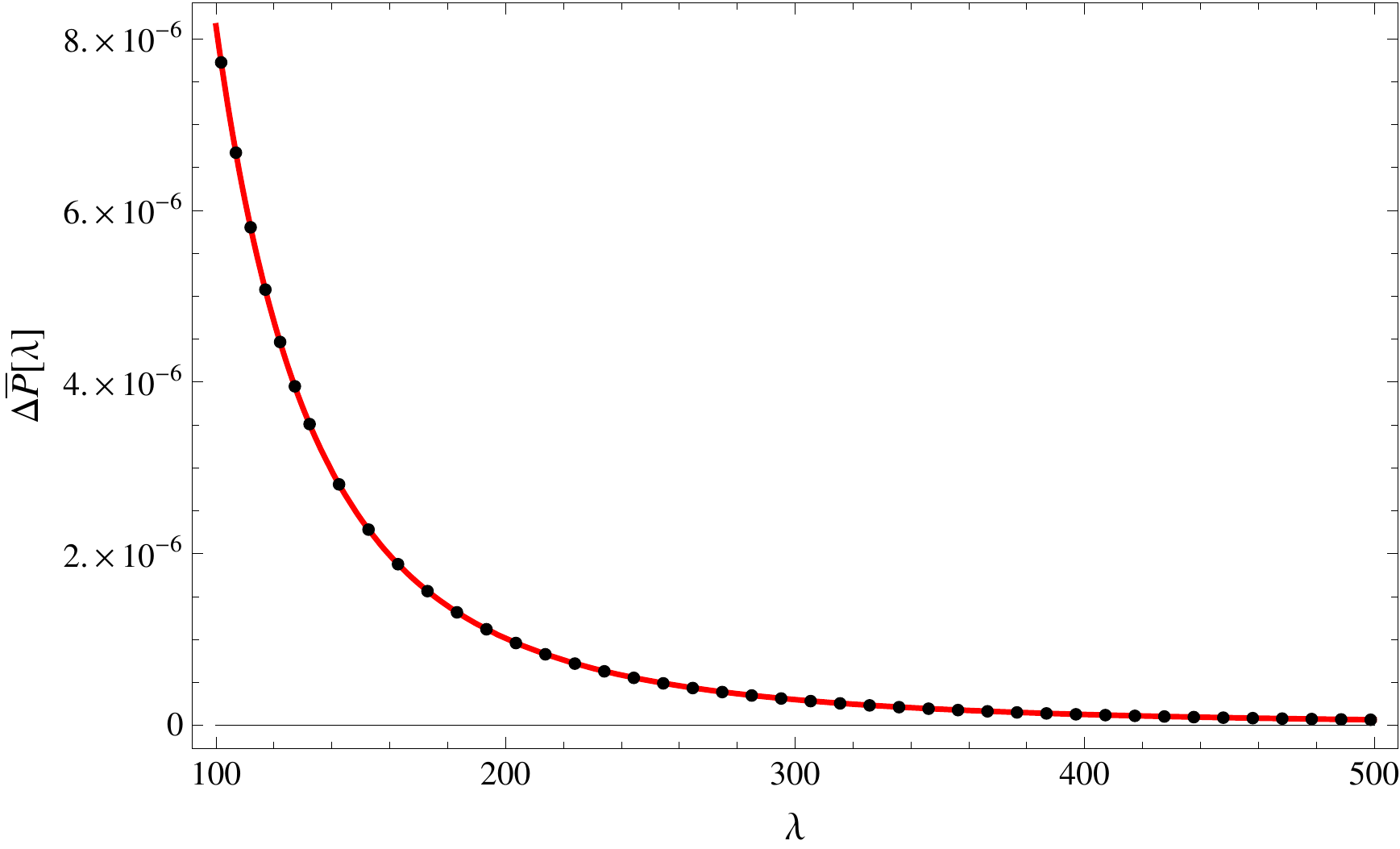}}
  \caption{The calculated high-temperature $\lambda$ dependence of (a) $\Delta\bar{\rho}$ and (b)
  $\Delta\bar{P}$. Dots correspond to computed values, the line is a numerical fit to the power functions defined
  in equations \eqref{defPLrho} and \eqref{defPLP}.
  }\label{Fig-3}
\end{figure}

Corrections \eqref{correction_rho} and \eqref{correction_P} can be calculated,
by first calculating $G$ as shown in section \ref{sec:transmodes}, as a function of temperature.
Figure \ref{Fig-3} depicts both $\Delta\bar{P}$ and $\Delta\bar{\rho}$ 
as functions of the dimensionless temperature $\lambda$ in the high-temperature region
$100\le\lambda\le 500$ (recall that $\lambda_c=13.87$).
The calculated data points can be fitted to a curve assuming power law dependency
\begin{align}
  \Delta\bar{\rho} \, &= \, c_\rho\lambda^{\gamma}\,,
  \label{defPLrho} \\
  \Delta\bar{P} \, &= \, c_P\lambda^{\delta}\,.
  \label{defPLP}
\end{align}
The best fit is given for the parameter values
\begin{align}
  c_\rho \, &= \,3.9577\,,\ \ \ \  \gamma\, = \,-3.02436\,,
  \label{fitvaluesrho} \\
  c_P \, &= \,8.4963\,,\ \ \ \ \delta \, = \, -3.00904\,.
  \label{fitvaluesP}
\end{align}
First note that the resummed corrections are linear, similar to \eqref{rho_p_gs},
but three orders of magnitude smaller.
Equations \eqref{fitvaluesrho}, \eqref{fitvaluesP}
can thus be interpreted as corrections to the tree-level ground state estimates.
The linear dependency is then not surprising,
since selfconsistent propagation is affected by the
quasiparticle mass of tree-level heavy modes in the tadpoles of figure \ref{fig:dysomResumm},
which in turn reflect the a priori estimate for the energy density of the thermal ground state.

In calculating the contribution of higher irreducible 
loop orders to the polarization tensor we observe a hierarchic decrease 
and expect a termination at a finite order.
For the effective theory of SU(2) Yang-Mills thermodynamics described in this chapter,
it can be shown that the contribution of higher irreducible 
loop orders to the polarization tensor
quickly decreases at higher loop orders \cite{Hofmann2006,KavianiHofmann2007}.
Further contributions of these loop orders to the one-loop resummation calculated here
should therefore be small enough to be safely ignored.

\subsection{Thermodynamical consistency}

We have seen that the a priori pressure and energy estimates
for the ground state plus corrections,
which incorporate fluctuations of the tree-level massive modes,
obey the usual Legendre relations of the form $\rho=T\frac{dP}{dT}-P$.
What about the resummed corrections defined in \eqref{correction_rho}, \eqref{correction_P}?

\begin{figure}[t]
  \centering
  \includegraphics[width=120mm]{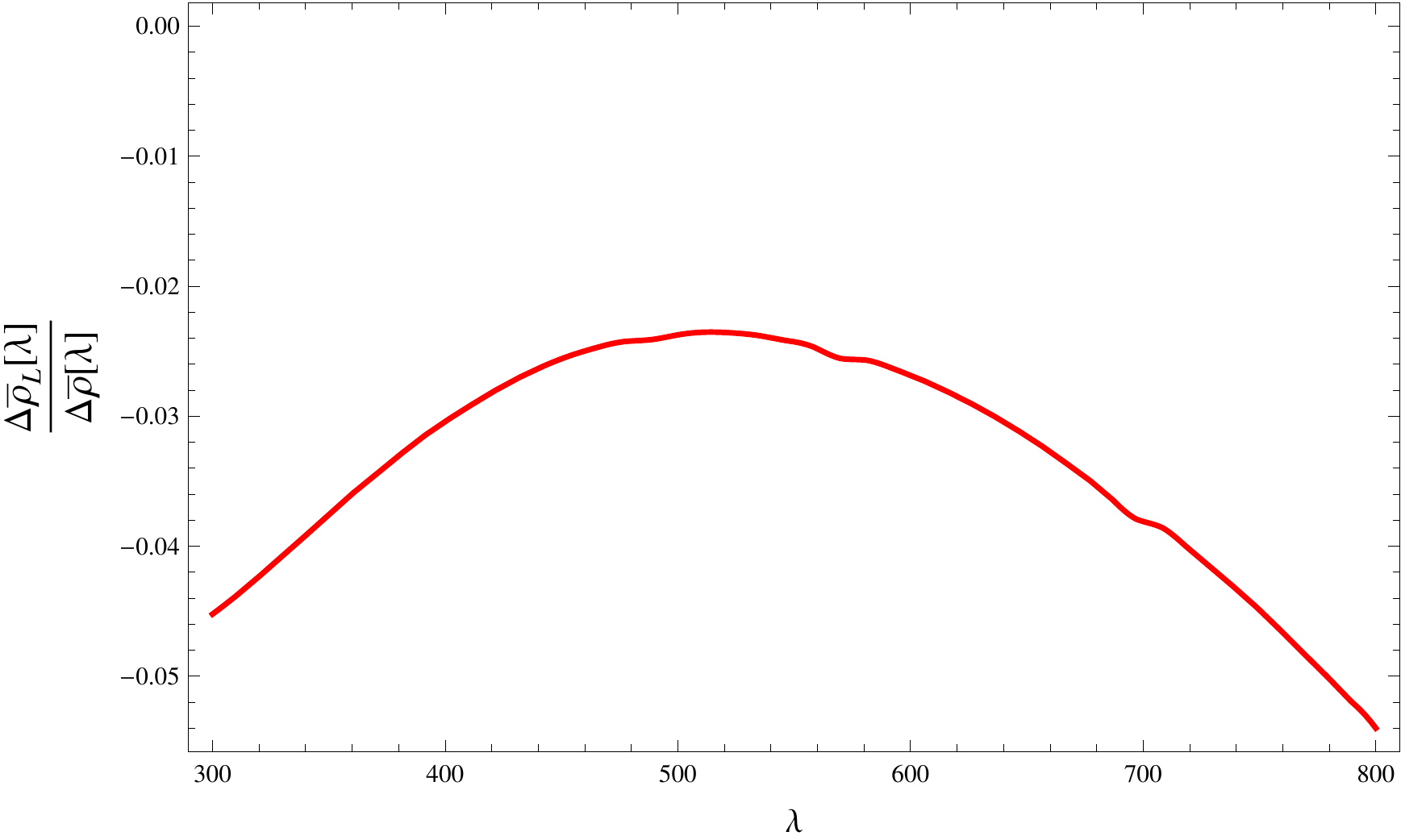}
  \caption{The $\lambda$ dependence of the ratio $\frac{\Delta\bar{\rho}_L}{\Delta\bar{\rho}}$, where
  $\Delta\rho_L^{\mathrm{1-loop,res}}\equiv T^4\,\Delta\bar{\rho}_L$ is defined in equation\,(\ref{rhoL}),
  and $\lambda$ is the dimensionless temperature. 
  The plot shows how the corrections to
  pressure and energy density
  due to one-loop resummation  as calculated in the text
  are not thermodynamically selfconsistent,
  since radiative corrections shifting  the dispersion law of tree-level massive modes
  were not considered in the calculation.}
  \label{Fig-4}
\end{figure}

Directly defining $\Delta\rho_L^{\srm{1-loop,res}}$
as the Legendre transformation of $\Delta P^{\srm{1-loop,res}}$,
\begin{align}
  \Delta\rho_L^{\srm{1-loop,res}} \, &\equiv \,T\frac{d\Delta P^{\srm{1-loop,res}}}{dT}-\Delta P^{\srm{1-loop,res}}\,,
  \label{rhoL}
\end{align}
we can compare both the energy density $\Delta\rho_L^{\srm{1-loop,res}}$
resulting from the Legendre relation \eqref{rhoL} and the resummed
energy density $\Delta\rho^{\srm{1-loop,res}}$ by calculating the ratio of both functions
$\Delta\rho_L^{\srm{1-loop,res}} / \Delta\rho^{\srm{1-loop,res}} = \Delta\bar\rho_L/ \Delta\bar\rho$.
Such a comparison is plotted in figure \ref{Fig-4}.
It is clear that the ratio is far from unity, and both densities show opposite signs.

How should we understand this apparent inconsistency?
Remember that by using the bare (free) propagators
of the TLM vector modes in our calculations,
we ignored one-loop corrections to the propagation of these modes.
Without taking into account radiative corrections for the massive modes,
such inconsistencies were to be expected.
One may stay consistent only up to a given loop order,
and then every resummed propagator must be considered.

Nevertheles, the linear correction given by expressions \eqref{defPLrho} and \eqref{defPLP} with fitted parameters
\eqref{fitvaluesrho} and \eqref{fitvaluesP} can still be interpreted as the
contribution to the thermal gas energy and pressure by photon propagation among \emph{stable}
monopole-antimonopole pairs, which in turn influence the ground state physics.


\chapter{Inhomogeneous Thermodynamics}
\label{chap:inhtd}

In the nonperturbative approach to the thermal SU(2) Yang-Mills gas
discussed in chapter \ref{chap:YM}, the assumption of homogeneous and isotropic thermodynamics
was of central importance.
In particular, it led to the inference that the modulus
$|\phi|$ of the adjoint scalar field $\phi^a$ should be constant
(see section \ref{sec:NonPapp}), and that the ground state
field configuration $a^{\srm{gs}}_\mu$ must be \emph{pure gauge}
(see equation \eqref{DefAPureG}).
The thermal ground state composed by the inert scalar field $\phi$
and the pure gauge solution $a^{\srm{gs}}_\mu$ then serve as a
homogeneous and isotropic background to the dynamics of
topologically trivial fluctuations.

Thus, the question arises as to whether this assumption can be relaxed.
Namely, if it is possible to expand the nonperturbative treatment
of chapter \ref{chap:YM} to consider the case of
inhomogeneous thermodynamics, and whether the physical description
of a thermal ground state made up by a scalar field together
with a background gauge field configuration remains valid.
The second question that arises is, under what circumstances is such
a description \emph{not} valid?
Is it always the case that a ground state $(\phi,a^{\srm{gs}}_\mu)$
exists over which trivial fluctuations propagate?

Both problems are not only of theoretical interest,
but important for experimental considerations as well.
If we hope to realize an SU(2) thermal photon gas in the laboratory,
we need to specify under what experimental circumstances
the theoretical formalism reviewed in chapter \ref{chap:YM} is valid.
Further, and in the same vein, we need to know if there are any possible
experimental conditions where the ground state description of the SU(2) gas breaks down.

This chapter is an attempt to answer both questions
and gain insight into the possibility of experimentally realizing
the thermal SU(2) ground state.

First we tackle the problem of generalizing the approach of chapter \eqref{chap:YM}
to inhomogeneous thermodynamics.
A particular solution to the SU(2) equations of motion
with scalar field has been found,
which in the context of nonperturbative Yang-Mills thermodynamics
and under the adiabatic approximation allows for temperature profiles
without disturbing the ground state.
The derivation of this solution is given in appendix \ref{app:ymeom}. In this chapter,
we use a practical ansatz for the spatial $a^3$ vectors, then calculate with the help of
\eqref{sys1_2c} and \eqref{EofMPhiT} the rest of the gauge and scalar fields.
With a solution at hand, we investigate what kind of thermal gradients are produced,
and under which boundary conditions.

For the second problem, we invoke a recent resonant cavity experiment at low
temperatures by Tada et. al. \cite{Tada2006} as an example of
how the thermal SU(2) ground state may be deformed to be indistinguishable
from the dynamics of a U(1) gas.

\section{The SU(2) thermomagnetic effect}

In this section, we present a solution to the SU(2) Yang-Mills equations of motion
that does not assume space homogeneity. Instead, the temperature profile
is allowed to vary along a coordinate axis.
As we will see, under certain assumptions, this temperature profile
implies the existence of non-trivial gauge field configurations, giving
rise to a \emph{magnetic field}, which may be interpreted as
emerging from the temperature gradient.
This has some parallels with the \emph{thermoelectric effect} in solids.
Indeed, the SU(2) thermomagnetic effect can be understood as a rearrangement of
topological magnetic charges in the ground state because of thermal diffusion.

\subsection{Basic approach}
\label{sec:ITBasApp}

We will abandon the requirement of constant temperature for the thermal system,
and instead assume a space-dependent temperature distribution in one dimension,
$T(\vec x) = T(z)$.
Such a temperature distribution will in turn
affect the thermal ground state through the modulus of the scalar
field $|\Phi|(\vec x) = |\Phi|(z) = \sqrt{\frac{\Lambda^3}{2 \pi T(z)}}$.
Here we use the notation of appendix \ref{app:ymeom} and label the
inhomogeneous scalar field as capital $\Phi$, to differentiate
from the homogeneous inert field $\phi$ of chapter \ref{chap:YM}.
We are interested in the effects of a non-zero temperature gradient
on the gauge field configurations which solve the Yang-Mills equation of motion.

In the SU(2) gas, position-dependence of the modulus of $\Phi$
implies the position-dependence of the mass for the two massive modes.
We expect that the pure-gauge ground state configurations
of the form \eqref{DefAPureG} will no longer be solutions of the
equations of motion with space-dependent ground state,
but will be replaced by a non-homogeneous configuration
reacting to the deformation of $\Phi$.
The non-homogeneities in the gauge field lead to the
emergence of non-scalar (isotropy breaking)
quantities such as a magnetic field $\vec \nabla \times \vec{a^3}$.

The deformation of $\Phi$ will also influence radiative corrections
for the propagation of the transverse massless mode.
Since radiative corrections are small in the effective
theory compared to the tree-level (bare) ensemble, we may ignore such corrections.
Therefore, we only search for classical solutions to the Yang-Mills equations of motion
\eqref{EffEoM} with a space-dependent scalar field $|\Phi|(z)$ taken as a source term.
This is carried out in appendix \ref{app:ymeom}.

\subsection{Adiabatic approximation}
\label{sec:adiabapprox}

An important consideration we must bear in mind is
that the field energy density of such inhomogeneous field configurations described in section \ref{sec:ITBasApp}
\emph{must never be greater} than the energy density of the SU(2) ground state
over the entire domain of the fields.
Such an \emph{energy hierarchy} is condition \emph{sine qua non}
to ensure the existence of the coarse-grained scalar field.
Otherwise, the ensemble dynamics could no longer be described
by topologically trivial fluctuations over a static background,
and the effective theory described in chapter \ref{chap:YM} would not be applicable.
We therefore demand that the energy scale of the scalar field $\Phi$
be larger than any other scale of the system.
This is the \emph{adiabatic approximation}.

In our particular solution to the equations of motion
with $z$-dependent scalar field $\Phi$,
a nontrivial magnetic field $\mathbf B$ with energy density $\rho_{\mathbf B}= \frac{1}{2} |\mathbf B|^2$
emerges. The adiabatic approximation then demands
\begin{align}
  \rho_{\mathbf B}(z) < \rho^{\srm{gs}}(z) \,.
  \label{defAdApp} 
\end{align}
Condition \eqref{defAdApp} guarantees that the effective theory summarized in chapter \ref{chap:YM}
remains a good description of the SU(2) ground state.

\subsection{Restrictions on possible solutions}

\label{sec:ansatzreq}
To use the results of appendix \ref{app:ymeom}, we first need a good ansatz
for the vector $\vec{a}^3$.
To have a physical interpretation
the solutions derived from the ansatz should be such that
\begin{enumerate*}
      \item \label{list:sol1} The temperature field $T(z)$ has a parametrization
        such that a baseline temperature $T_{\mathrm{min}} \equiv \lim_{ z\to \infty} T(z)$
        can always be specified, with $T_{\mathrm{min}}$
        greater than the critical temperature $T_c$ of the deconfining phase,
        $T_{\mathrm{min}}> T_c \sim 2.73\,\mathrm{K}$.
      \item\label{list:sol2} The conditions needed for the validity of the adiabatic approximation,
        described in section \ref{sec:adiabapprox}, particularly equation \eqref{defAdApp}, must hold.
\end{enumerate*}

Note that condition \ref{list:sol1} rules out using simple power laws such as $a^3_i \sim z^\gamma$ as ansatz,
since after inserting in equations \eqref{sys1_2c} and \eqref{EofMPhiT},
the temperature field always shows the limit $\lim_{ z\to \infty} T(z) \sim z^{-2\gamma}$
which is either $0$ or infinity, depending on the sign of $\gamma$.
One could try numerically searching for solutions $a^3(z)$ of the equation
\begin{align}
  \frac{1}{2} (a^3)^2  - \frac{1}{4 e^2} \sqrt{- \frac{ a^3}{\partial^2_z a^3}} \partial^2_z \sqrt{- \frac{\partial^2_z a^3}{ a^3}}
  \, = \, \frac{\Lambda^3}{2 \pi} T^{-1}( z)\,.
  \label{defSolT}
\end{align}
for a specific distribution $T(z)$. Equation \eqref{defSolT} is arrived at by inserting \eqref{sys1_2c} in
\eqref{profT} and taking $a^3_1=a^3_2 \equiv a^3$.

But this was too much of an overhead for our purposes. 
In the next section, we will offer a simple ansatz which fulfills
both requirements \ref{list:sol1} and \ref{list:sol2} using just two parameters.
We then discuss how it describes deformations from homogeneous thermodynamics.

\subsection{A particular ansatz}
\label{sec:partAnsatz}

In this section, we will introduce a simple rational function 
as an ansatz to derive inhomogeneous field and temperature configurations.
We will check that the requirements outlined in section \ref{sec:ansatzreq} be fulfilled,
then discuss the derived field solutions, in particular the resulting temperature distribution.
This exercise will help us build a basic intuition for the possible ways
that an homogeneous temperature distribution may be deformed without disturbing
the ground state physics of the SU(2) gas.

In the following, unless otherwise stated, we work in natural units $\hbar=k_B=c=1$,
so energy and temperature have dimension $\mathrm{eV}$,
length dimension $(\mathrm{eV})^{-1}$, etc.
We use dimensionless coordinate variables, which are made dimensionless by rescaling 
with the factor $\frac{\Lambda}{2\pi}$ (the factor $2\pi$ is for numerical convenience only).
Recall that $\Lambda$ is the SU(2) Yang-Mills scale and has dimension of energy, $[\Lambda]=\mathrm{eV}$,
so lengths are rescaled by multiplication with $\frac{\Lambda}{2\pi}$.
Coordinate variables are then rescaled as $\hat z \equiv \frac{\Lambda}{2\pi} z$.

Define now the $z$-dependent function
\begin{align}
  g\left( \hat z;g_0,g_1 \right) \, &=\, g_0 \, \left( \frac{\Lambda}{2\pi} \right)^2  \, \frac{\hat z^2 + g_1}{\hat z^2 + 1 } \,,
  \label{eqn_defg}
\end{align}
with dimensionless parameters $g_0$ and $g_1$.
By definition, the function \ref{eqn_defg} has dimension of energy squared, $[g]=\mathrm{(eV)^2}$.

From appendix \ref{app:ymeom}, we learn that from a reasonable ansatz for
$\vec a^3(\vec x)=\vec a^3(z)$ we can calculate a complete solution to the equations
of motion \eqref{EofMa}.
Using \eqref{eqn_defg} as an ansatz,
we write
\begin{align}
  a_1^3( \hat z) \,=\, a_2^3( \hat z) \, \equiv \, a^3( \hat z) = \sqrt{\frac{1}{2} g( \hat z)} \,.
  \label{a3Ansatz}
\end{align}
Note that the energy dimensions on the right and left hand side of \eqref{a3Ansatz} coincide.
Definition \eqref{a3Ansatz} requires the parameter $g_0$ to be positive.
We also take $g_1>0$ so as to tentatively allow all $\hat z > 0$.
From \eqref{sys1_2c}, \eqref{EofMPhiT} and \eqref{PhiMod},
we can calculate
\begin{align}
  a_0( \hat z)  \, &= \,\frac{\Lambda}{2 \pi} \sqrt{- \frac{\partial^2_{\hat z} a^3}{2 e^2 a^3}}\,,
  \label{a0Ansatz} \\
  |\Phi|^2( \hat z)        \, &= \, \frac{1}{2} (a^3)^2
                - \left(\frac{\Lambda}{2 \pi}\right)^2 \frac{\partial^2_{\hat z} a_0}{4 e^2 a_0}\,,
  \label{phiAnsatz} \\
  T( \hat z) \, &= \, \frac{\Lambda^3}{2 \pi |\Phi|^2}\,.
  \label{TAnsatz}
\end{align}
The magnetic field for this field configuration can be calculated from \eqref{a3Ansatz}
\begin{align}
  |\mathbf B|\left( \hat z \right) \, &= \,|\mathbf{rot}\, \vec{a^3}| \, = \, \sqrt{2} \frac{\Lambda}{2 \pi} \, \partial_{\hat z}  a^3\,,
  \label{BfieldAnsatz} \\
  \rho_{\mathbf B}\left( \hat z \right) \, &= \, \frac{1}{2} |\mathbf B|^2\,.
  \label{rhoBAnsatz}
\end{align}
Finally, recall the energy density of the thermal ground state,
which is given by equation \eqref{GSenergy}
\begin{align}
  \rho^{\srm{gs}}\left( \hat z \right)\, &= \, 4 \pi \Lambda^3 T\left( \hat z \right) \,.
  \tag{\ref{GSenergy}}
\end{align}

Notice that $\lim_{\hat z\to \infty} |\Phi|^2 = \lim_{\hat z\to \infty} \frac{1}{2} (a^3)^2  = (\frac{\Lambda}{4\pi})^2 g_0$,
since from \eqref{a0Ansatz} we read
$\lim_{\hat z\to \infty} a_0 \sim \frac{\Lambda}{2\pi} {\hat z}^{-1}$
and $\lim_{\hat z\to \infty} \frac{\partial^2_{\hat z} a_0}{a_0} = 0$.
Then from equation \eqref{TAnsatz} it is seen that
\begin{align}
  \lim_{\hat z\to \infty} T(\hat z) \, &= \, 8\pi \Lambda g_0^{-1} \,.
  \label{defTmin1}
\end{align}
So the baseline temperature at infinity can be fixed by adjusting $g_0$.
This fulfills our first condition. In order to have $T_{\mathrm{min}}>T_c$ we need
\begin{align}
  T_{\mathrm{min}} \,= \, 8\pi \Lambda g_0^{-1} \, &> \, T_c \, = \, \frac{\Lambda}{2 \pi}  \lambda_c
  \notag \\
  \implies g_0  \, &< \,\frac{16 \pi^2}{\lambda_c} \,.
  \label{defg01}
\end{align}
For $\lambda_c=13.87$ \cite{Hofmann2009}, this means $g_0<11.39$.
We also demand that the temperature profile decrease asymptotically to $T_c$,
but this means that $a^3$ must be \emph{monotonically increasing} for large $\hat z$.
From \eqref{a3Ansatz} and the definition of $g$ \eqref{eqn_defg}
we see that this can be achieved by having $g_1<1$.

The second condition, that the magnetic field energy density
always be smaller than that of the ground state, is difficult to check
analytically for all $\hat z$. But we can at least make sure that it holds asymptotically.
Recall that in the limit of large $\hat z$, $|\Phi|^2$ behaves as $\frac{1}{2} (a^3)^2$.
Then the ground state energy density at large $\hat z$ is given as
\begin{align}
  \lim_{\hat z\to \infty} \rho^{\srm{gs}}\left( \hat z \right) \, &= \, 32\pi^2 \Lambda^4 g_0^{-1} \,.
  \label{defLimTInf2}
\end{align}
The magnetic field energy density at large $\hat z$ can also be calculated
\begin{align}
  \lim_{\hat z\to \infty}  \rho_{\mathbf B}\left( \hat z \right)
        \, &= \, \left( \frac{\Lambda}{2 \pi} \right)^2 \lim_{\hat z\to \infty} (\partial_{\hat z} a^3)^2
         \, = \, g_0 \left( \frac{\Lambda}{2\pi} \right)^4 \,.
  \label{defLimBInf}
\end{align}
Condition \eqref{defAdApp} then reads
\begin{align}
  g_0  \, &< \, (\sqrt{2}\pi)\, 16 \pi^2 \,.
  \label{defg02}
\end{align}
But $\sqrt{2}\pi>\lambda_c^{-1}$, so the restriction $T_{\mathrm{min}}>T_c$ already
ensures the energy hierarchy required by the adiabatic condition \emph{in the large $\hat z$ limit}.
For small $\hat z$, we will need to test \eqref{defg02} explicitly.

Summarizing, we see that for $0<g_1<16 \pi^2 \lambda_c^{-1}$ and $0<g_1<1$,
the ansatz \eqref{defg02} inserted in \eqref{a3Ansatz}
results in field configurations that
do indeed fulfill the conditions in section \ref{sec:ansatzreq},
at least asymptotically.
We test these configurations in the next section.

\subsection{Some examples}

\begin{figure}[p]
  \centering
  \subfloat[The field $a^3(\hat z)$ plotted in units
  of the SU(2) Yang-Mills scale $\frac{\Lambda}{2 \pi}$,
  calculated from \eqref{a3Ansatz}
  ]{\label{fig:inh01}\includegraphics[width=70mm]{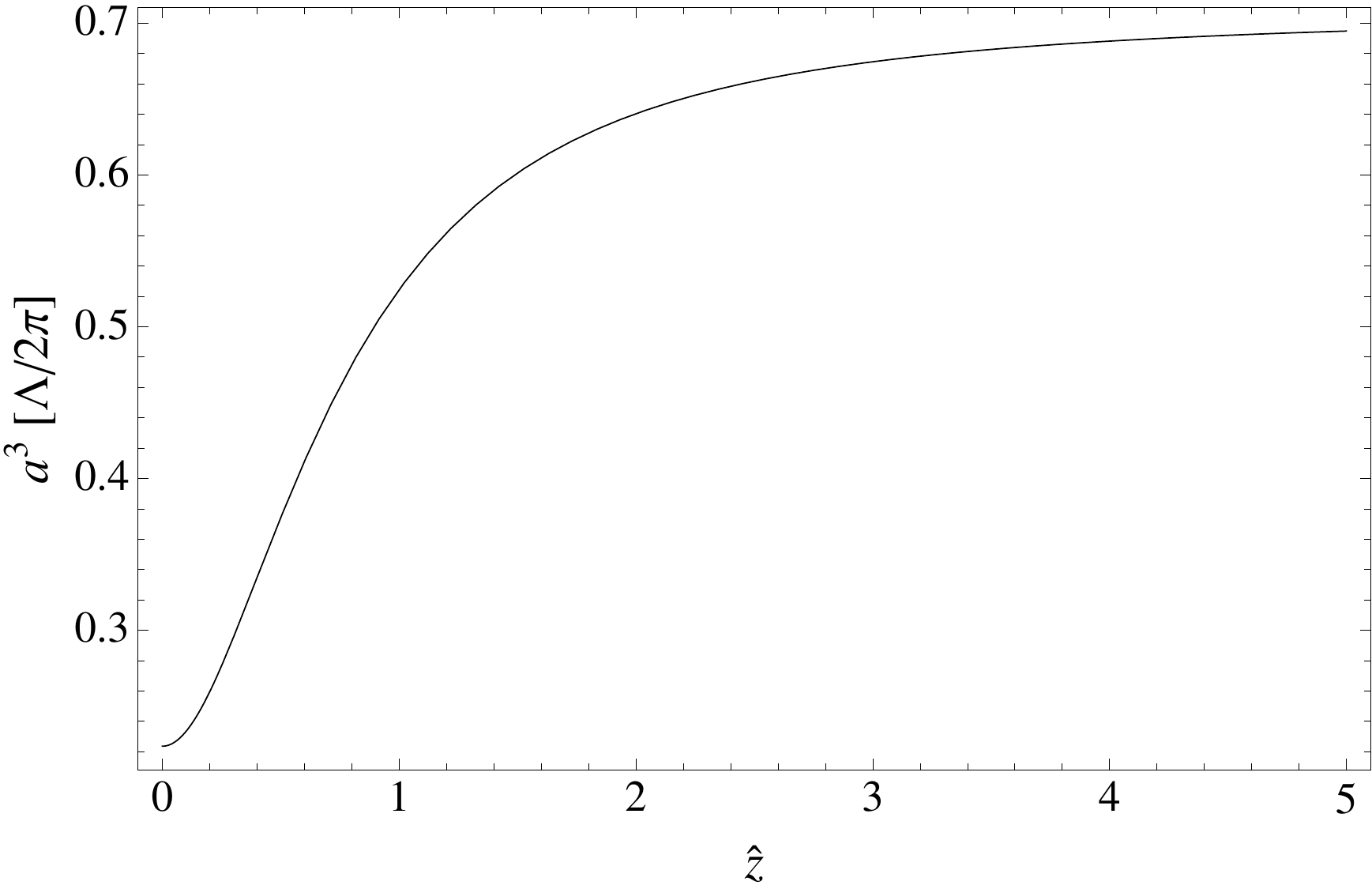}}
  \hspace{5mm}
  \subfloat[The field $a_0(\hat z)$ plotted in units of
  the SU(2) Yang-Mills scale $\frac{\Lambda}{2 \pi}$, calculated from \eqref{a0Ansatz}.
  The real part of $a_0$ is plotted in black, the imaginary part in blue.
  ]{\label{fig:inh02}\includegraphics[width=70mm]{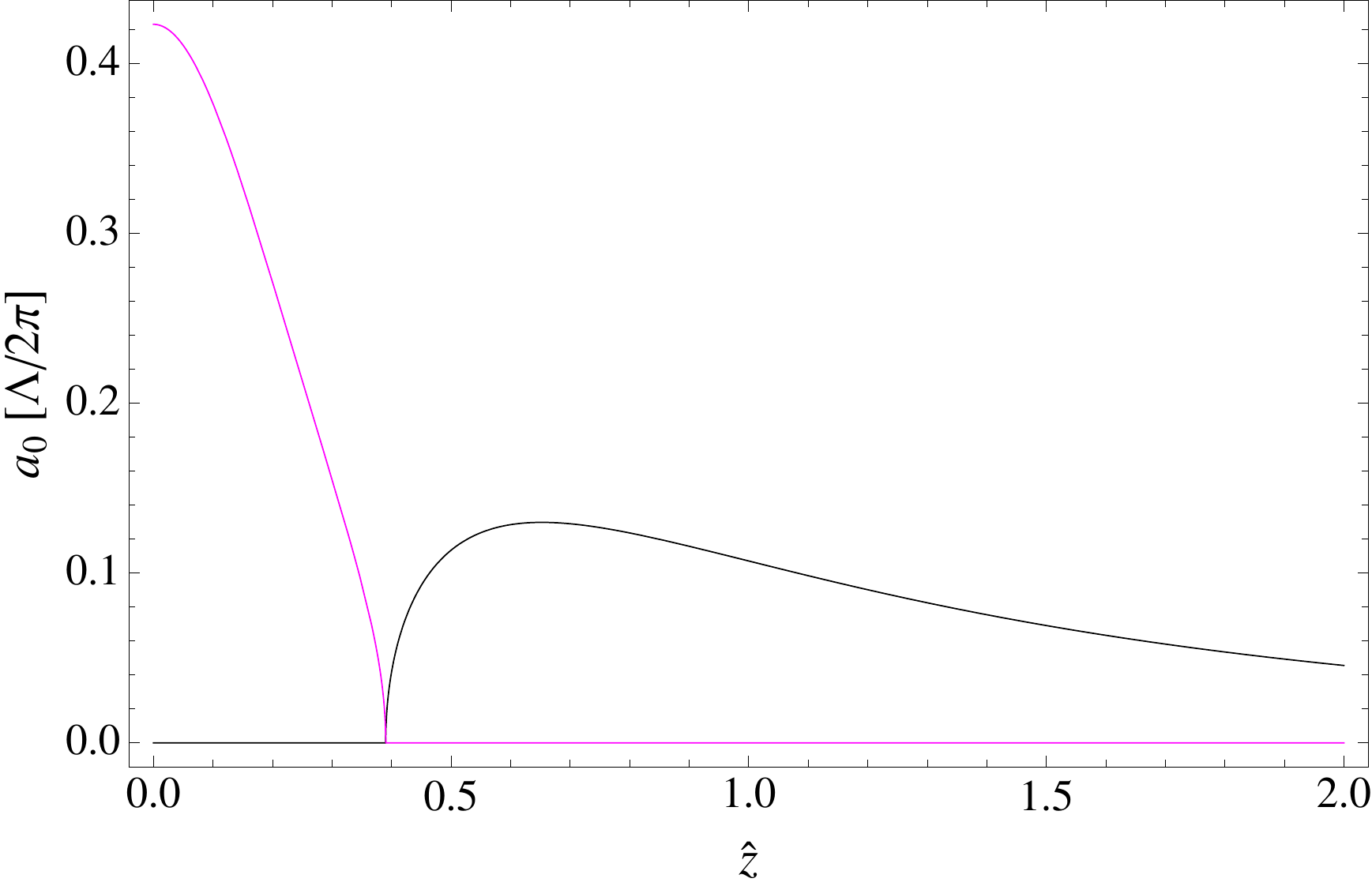}} \\
  \subfloat[The temperature field, plotted in units of the SU(2) Yang-Mills scale $\frac{\Lambda}{2 \pi}$,
  calculated from \eqref{TAnsatz}.
  The maxima of $T(\hat z)$ in the domain where $\mathrm{Im}\;  a_0(\hat z) = 0$
  is shown by the blue line and lies at $\hat z_{max} = 0.836$.
  ]{\label{fig:inh03}\includegraphics[width=120mm]{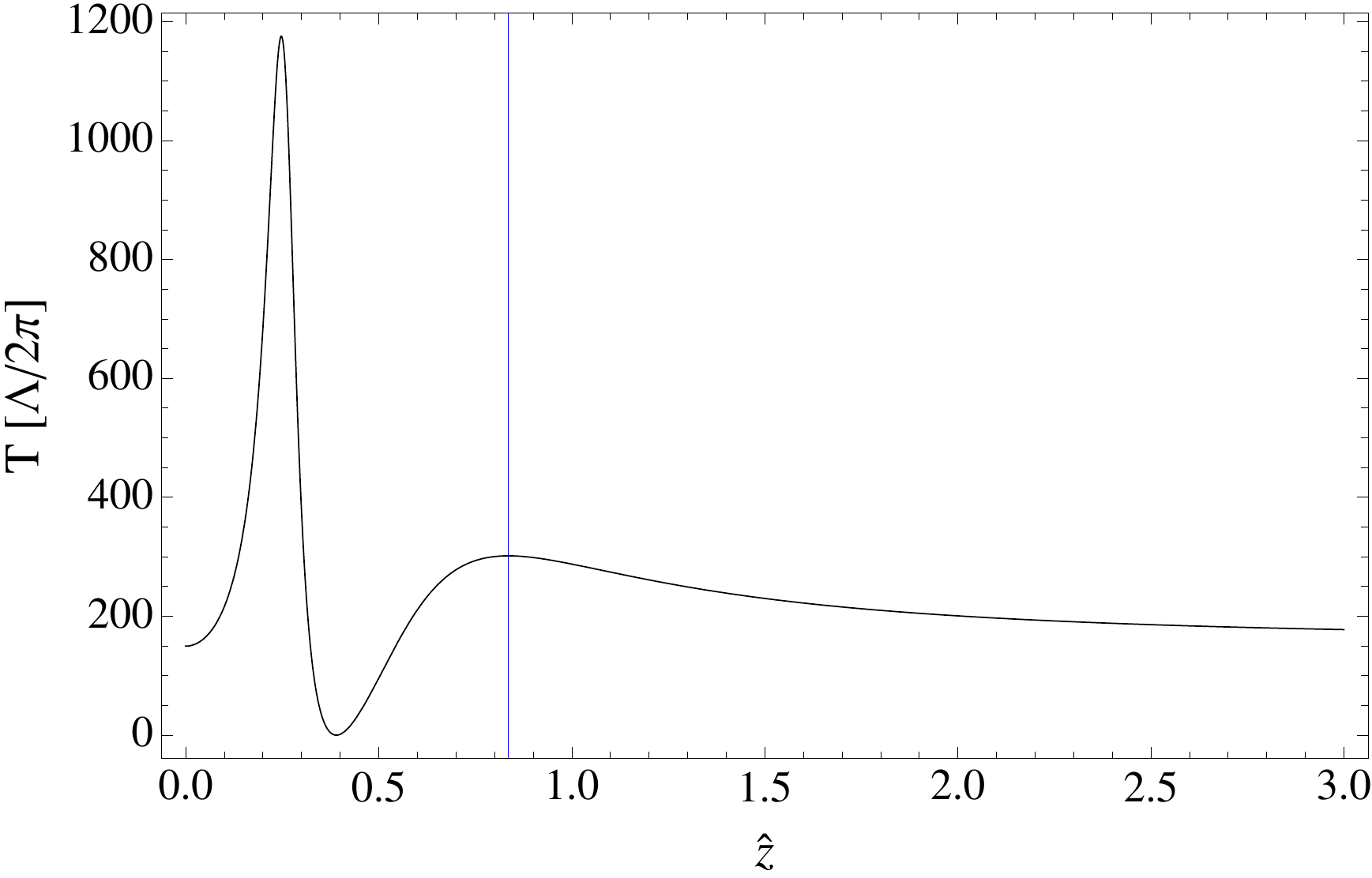}}
  \caption{Various fields calculated from ansatz \eqref{eqn_defg} with parameters $g_0=1$, $g_1=0.1$. $\hat z$ is dimensionless.}
  \label{fig:inh1}
\end{figure}

Now we try our solution using ansatz parameters $g_0=1$, $g_1=0.1$.
The resulting fields are plotted in figure \ref{fig:inh1}, in units of
the SU(2) Yang-Mills scale $\frac{\Lambda}{2 \pi}$ and in dimensionless coordinates.

Note how $a^3(\hat z)$ (figure \ref{fig:inh01}) tends to a constant value
$\frac{1}{\sqrt{2}} \frac{\Lambda}{2 \pi}$
at infinity.
Plotting $a_0(\hat z)$ (figure \ref{fig:inh02}), we encounter our first problem.
$a_0(\hat z)$ becomes imaginary at $\hat z_0 = 0.39$,
which is also the point where the real part of $a_0(\hat z)$ becomes zero.
The temperature field (figure \ref{fig:inh03}) shows non-monotonic behaviour
in the region where $\mathrm{Re}\,a_0(\hat z)=0$.
It assumes a minima at $T=0$,
which contradicts the requirement that $T(\hat z) > T_{\mathrm{min}}$.
This is not surprising.
In section \ref{sec:partAnsatz} we showed
that the function $g$, defined in \eqref{eqn_defg} and used as ansatz for
the field $a^3$ fulfills the requirements outlined in section \ref{sec:ansatzreq}
\emph{asymptotically}.
We need to restrict the domain of our solutions
from $\hat z \in [0,\infty]$ to $\hat z \in [\hat z_{\mathrm{min}},\infty]$
with $z_{\mathrm{min}}$ large enough that the problems outlined above disappear.

Figure \ref{fig:inh03} shows two maxima for the temperature field $T(\hat z)$.
After the second maxima, $T(\hat z)$ appears to decrease asymptotically to
a constant value, which is the behaviour we are looking for.
It seems therefore appropriate to define $z_{\mathrm{min}}$ to be the
value of $\hat z$ at which the second maxima of $T(\hat z)$ occurs,
since this would give us a monotonically decreasing temperature distribution.
In this case then, $\hat z_{max}=0.836$. Note also that  $\hat z_{max}>\hat z_0$.
The domain $[\hat z_{\mathrm{min}},\infty]$ should therefore be free
of imaginary fields.

For simplicity, we define a coordinate translation
$\hat z \rightarrow \hat z - \hat z_{max}$,
and consider only $\hat z>0$. This is the region we are interested in from now on.

\begin{figure}[p]
  \centering
  \subfloat[The field $a_0(\hat z)$ plotted in units of the SU(2) Yang-Mills scale $\frac{\Lambda}{2 \pi}$,
  after translation $\hat z \rightarrow \hat z - \hat z_{max}$.
  ]{\label{fig:inh04}\includegraphics[width=70mm]{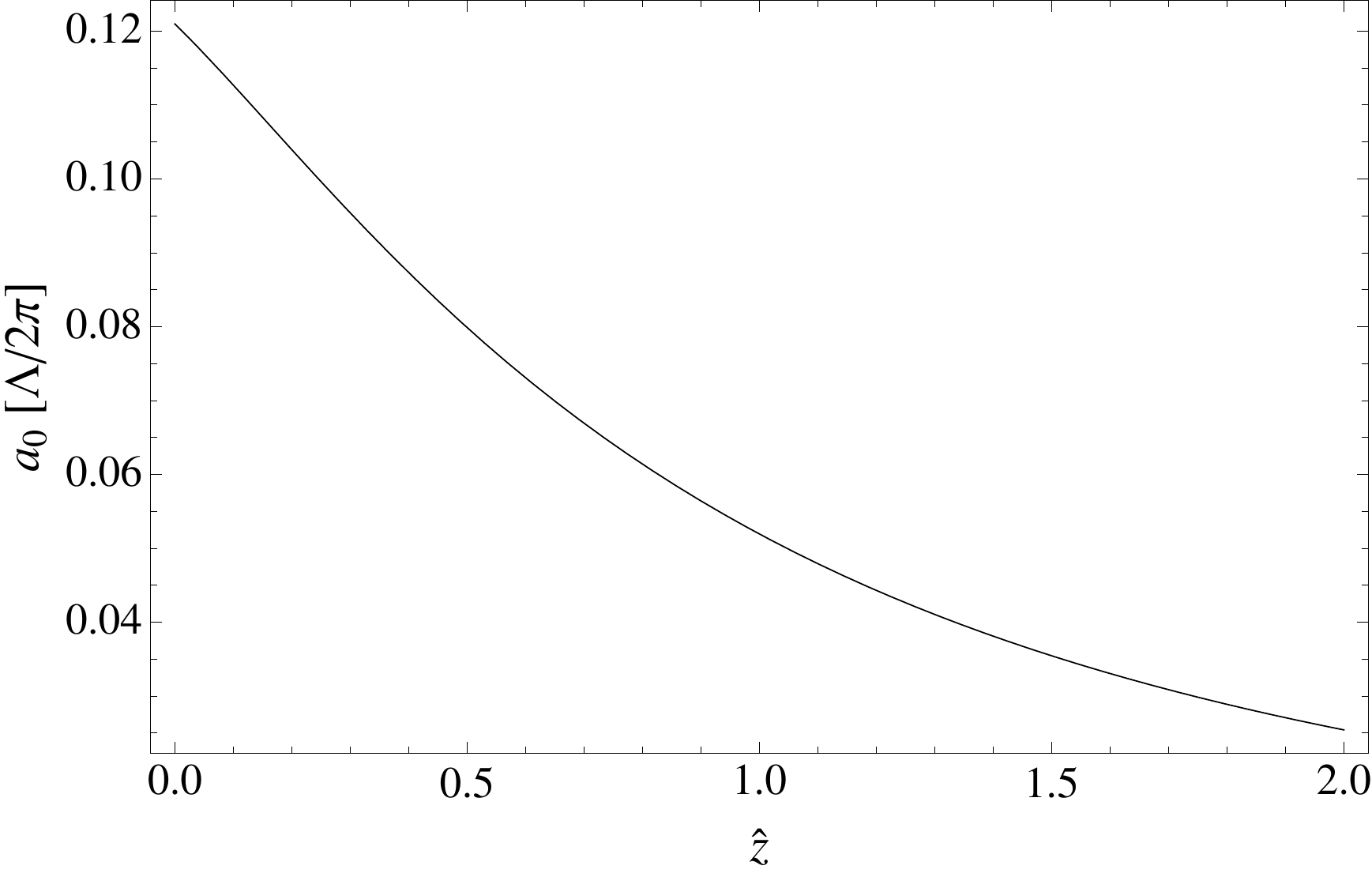}}
  \hspace{5mm}
  \subfloat[The temperature field $T(\hat z)$ plotted in units of the SU(2) Yang-Mills scale $\frac{\Lambda}{2 \pi}$,
  after translation  $\hat z \rightarrow \hat z - \hat z_{max}$ so that it has a maxima $\hat z =0$.
  ]{\label{fig:inh05}\includegraphics[width=70mm]{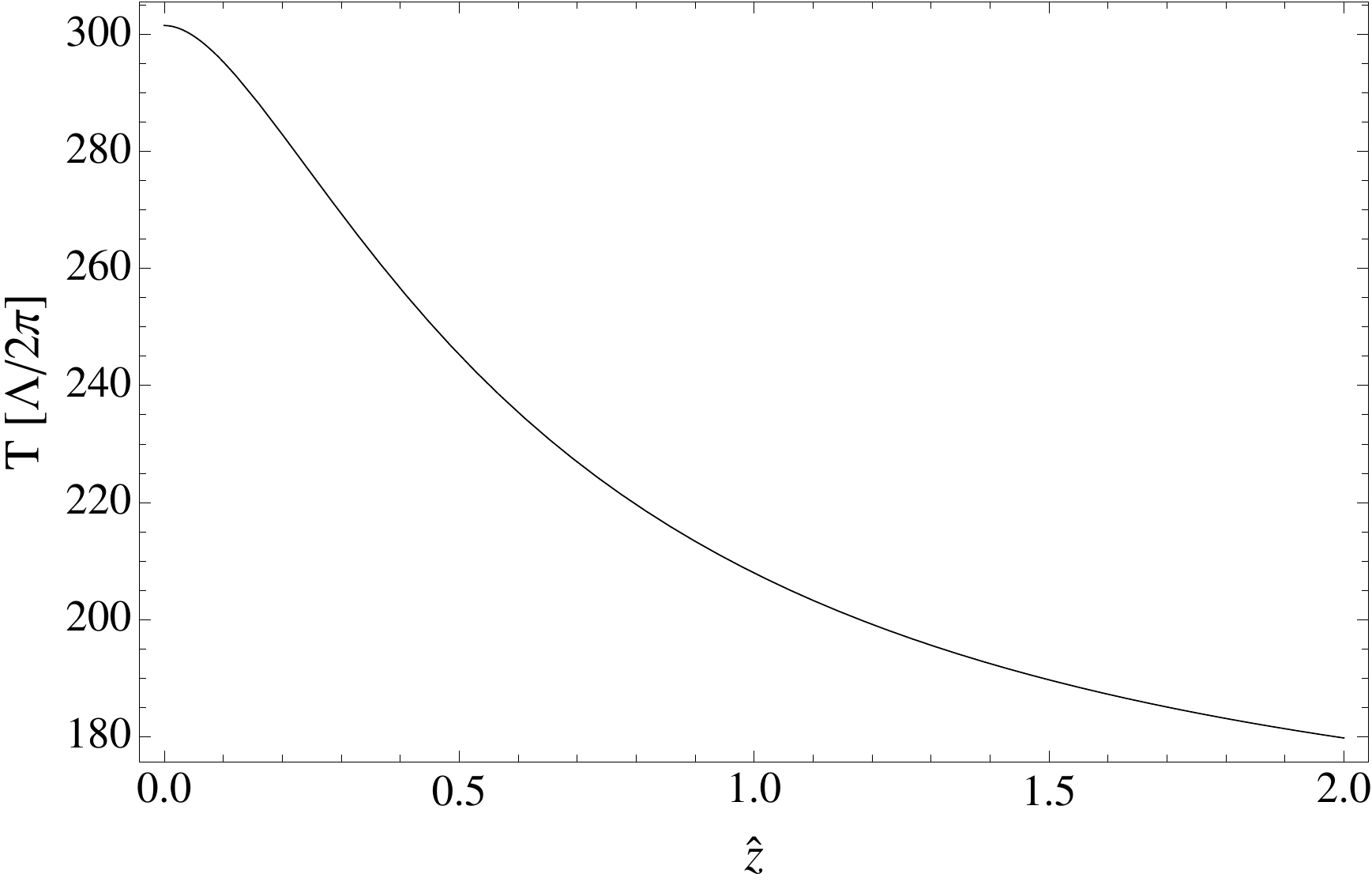}} \\
  \subfloat[The temperature field $T(\hat z)$ from figure \ref{fig:inh05} plotted in SI units.
  It starts from a maximum value of $T_{\mathrm{max}}=59.26 \mathrm{K}$
  and asymptotically decreases to a value $T_{\mathrm{min}}=31.04 \mathrm{K}$.
  This gradient is a result of the ansatz in \eqref{eqn_defg} with parameters $g_0=1$ and $g_1=0.1$.
  ]{\label{fig:inh05b}\includegraphics[width=120mm]{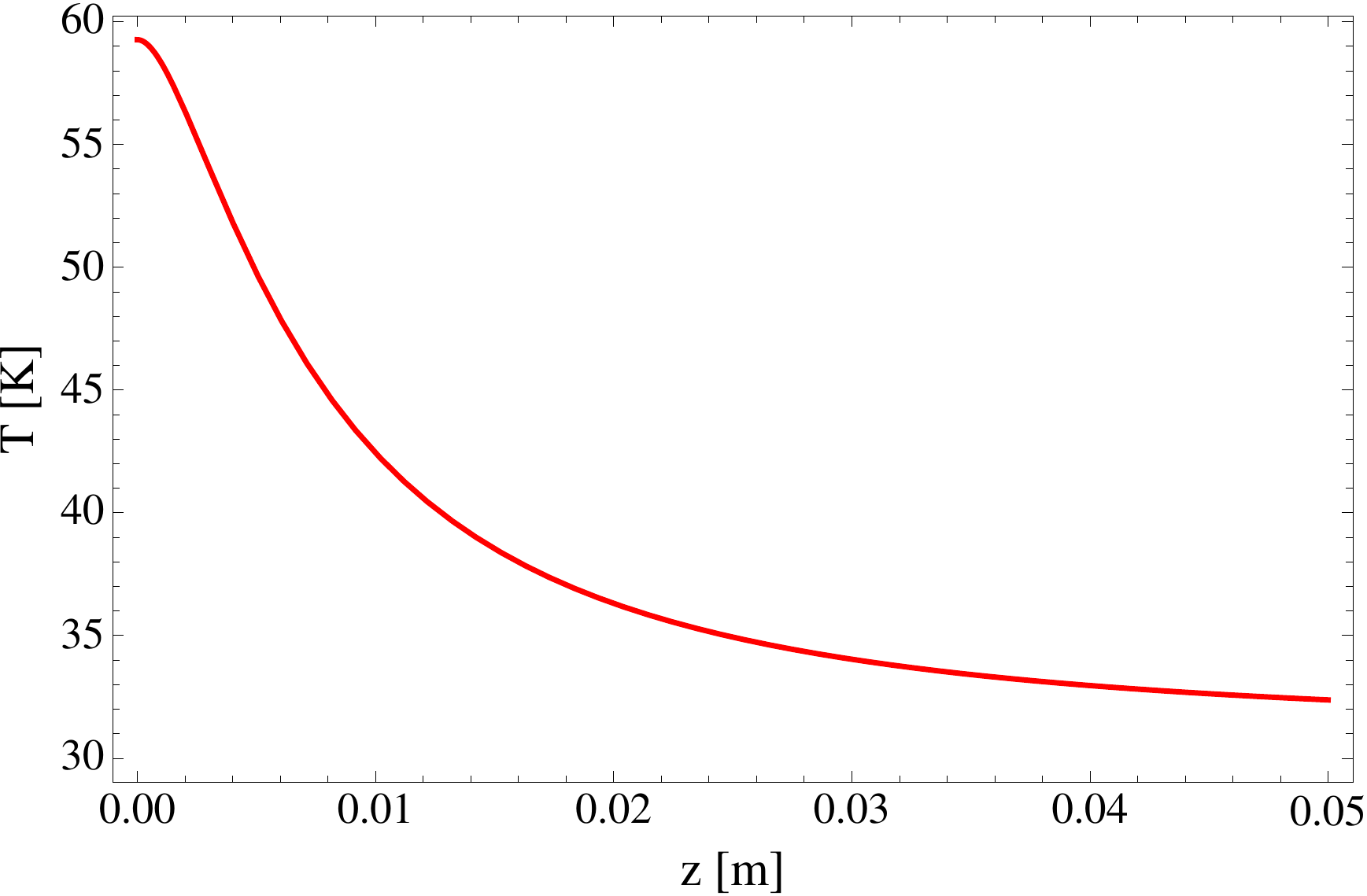}}
  \caption{The fields (a) $a_0(\hat z)$ and (b) $T(\hat z)$ are translated in $\hat z$ so that the maximum
  of the temperature now lies at $\hat z=0$.
  The ansatz parameters are the same as in figure \ref{fig:inh1}, $g_0=1$, $g_1=0.1$. $\hat z$ is dimensionless.
  In (c) we have plotted the temperature field in SI units.
  It decreases from a maxima $T_{\mathrm{max}}=59.26 \mathrm{K}$ to a baseline value $T_{\mathrm{min}}=31.04 \mathrm{K}$.
  }
  \label{fig:inh2}
\end{figure}

In figure \ref{fig:inh2} we have plotted the translated fields $a_0$ and $T$.
We interpret figure \ref{fig:inh05} physically as a temperature gradient.
In figure \ref{fig:inh05b} we have plotted this gradient in SI units.
We see that for parameters $g_0=1$ and $g_1=0.1$, the resulting
temperature distribution decreases from a maxima $T_{\mathrm{max}}=59.26 \mathrm{K}$
to a baseline value $T_{\mathrm{min}}=31.04 \mathrm{K}$. The length scale
of this gradient is of the order of centimeters.

\begin{figure}[p]
  \centering
  \subfloat[The modulus of the magnetic field $B(\hat z)$ plotted in units of the SU(2) Yang-Mills scale $\frac{\Lambda}{2 \pi}$,
  calculated from \eqref{BfieldAnsatz}. The coordinate $z$ is plotted in meters.
  ]{\label{fig:inh06}\includegraphics[width=70mm]{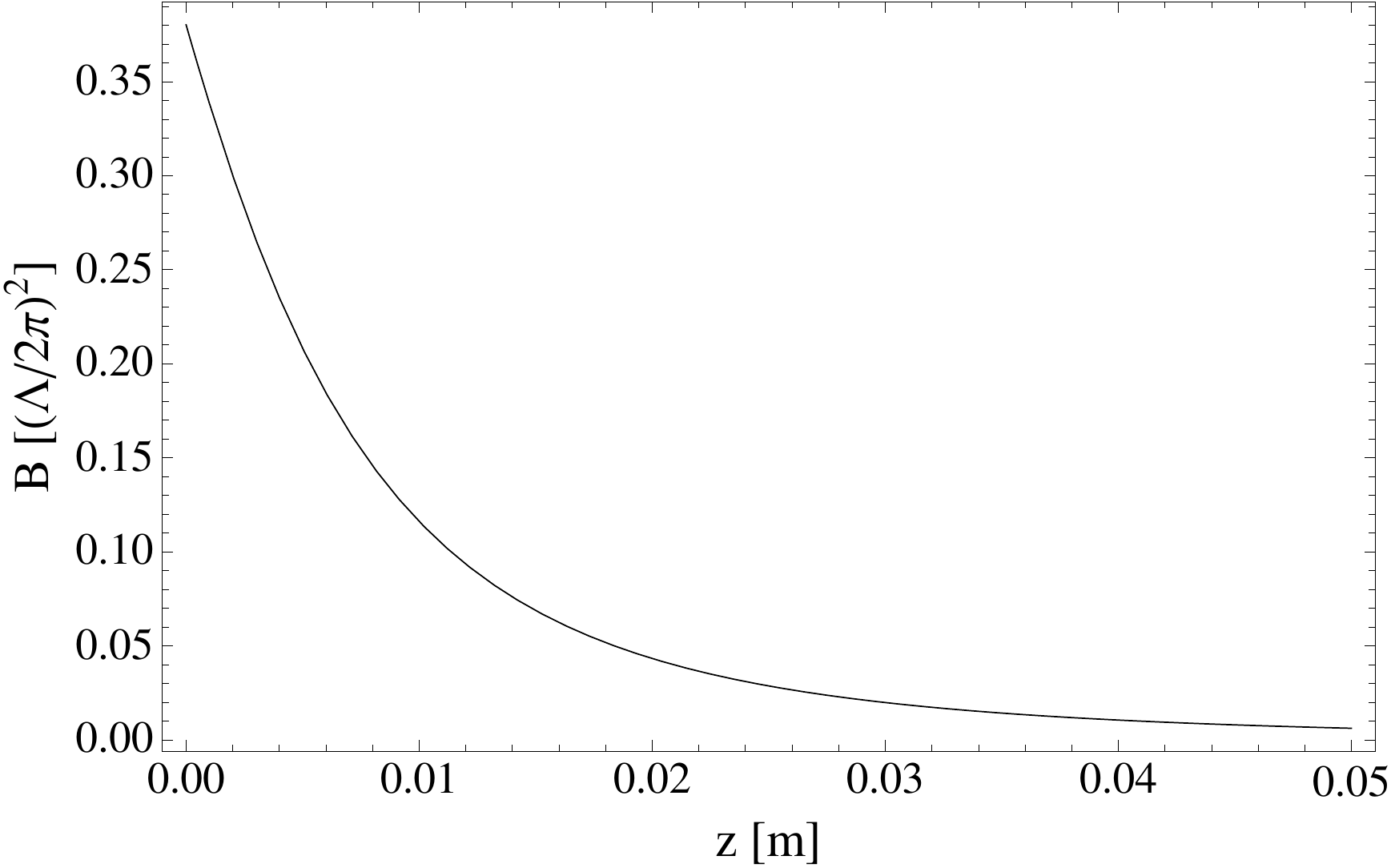}}                
  \hspace{5mm}
  \subfloat[The magnetic energy density $\rho_B(\hat z)$ plotted in units of the SU(2) Yang-Mills scale $\frac{\Lambda}{2 \pi}$,
  calculated from \eqref{rhoBAnsatz}. The coordinate $z$ is plotted in meters.
  ]{\label{fig:inh07}\includegraphics[width=70mm]{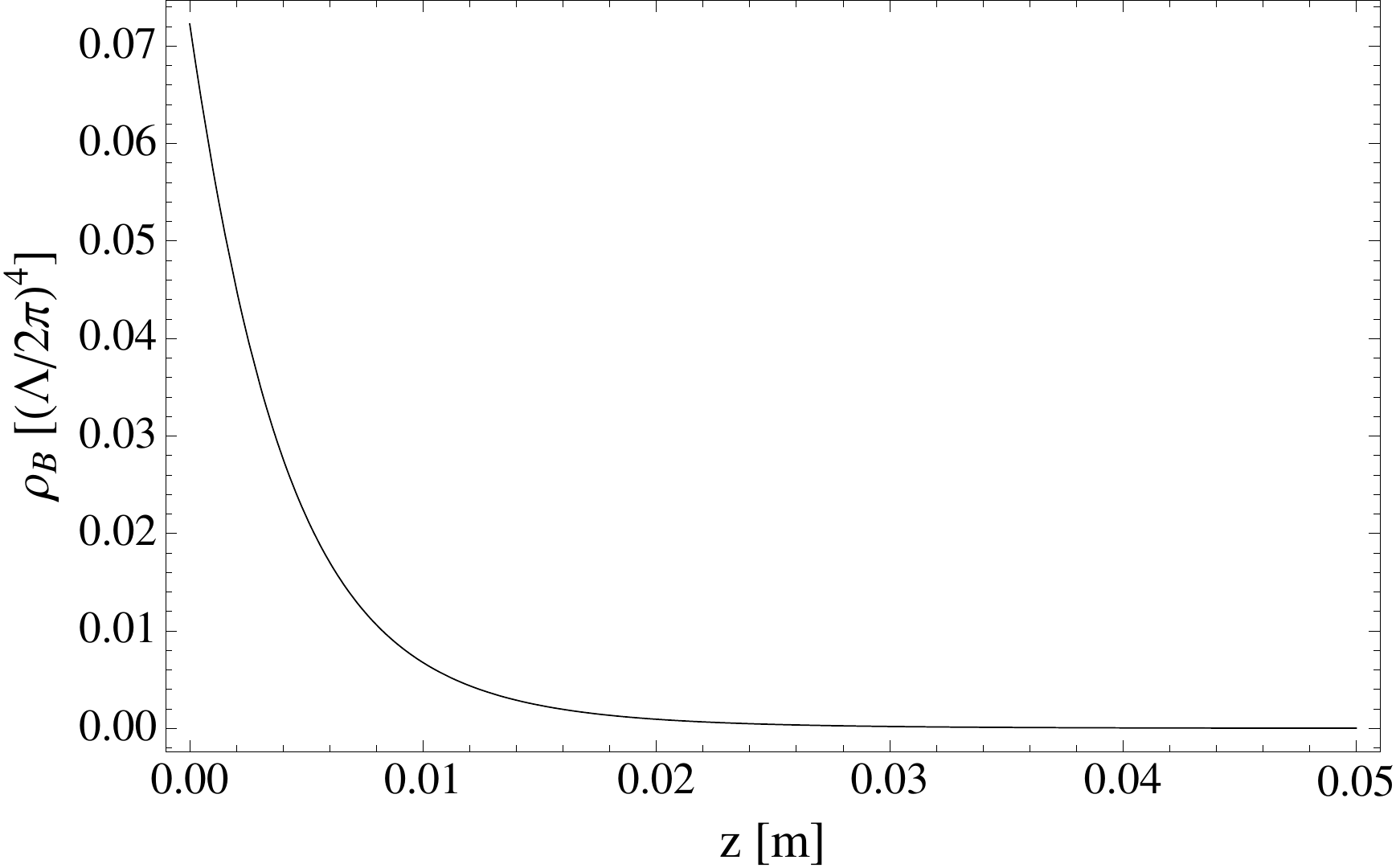}} \\
  \subfloat[The magnetic field energy density $\rho_B(\hat z)$
  (in red) calculated from the generated magnetic field $B(\hat z)$,
  and the energy density of the thermal ground state $\rho^{\scriptstyle{gs}}(T(\hat z))$
  (in black), plotted in units of the SU(2) Yang-Mills scale $\frac{\Lambda}{2 \pi}$.
  $\rho^{\scriptstyle{gs}}$ lies roughly 8 orders of magnitude above $\rho_B$.
  ]{\label{fig:inh08}\includegraphics[width=120mm]{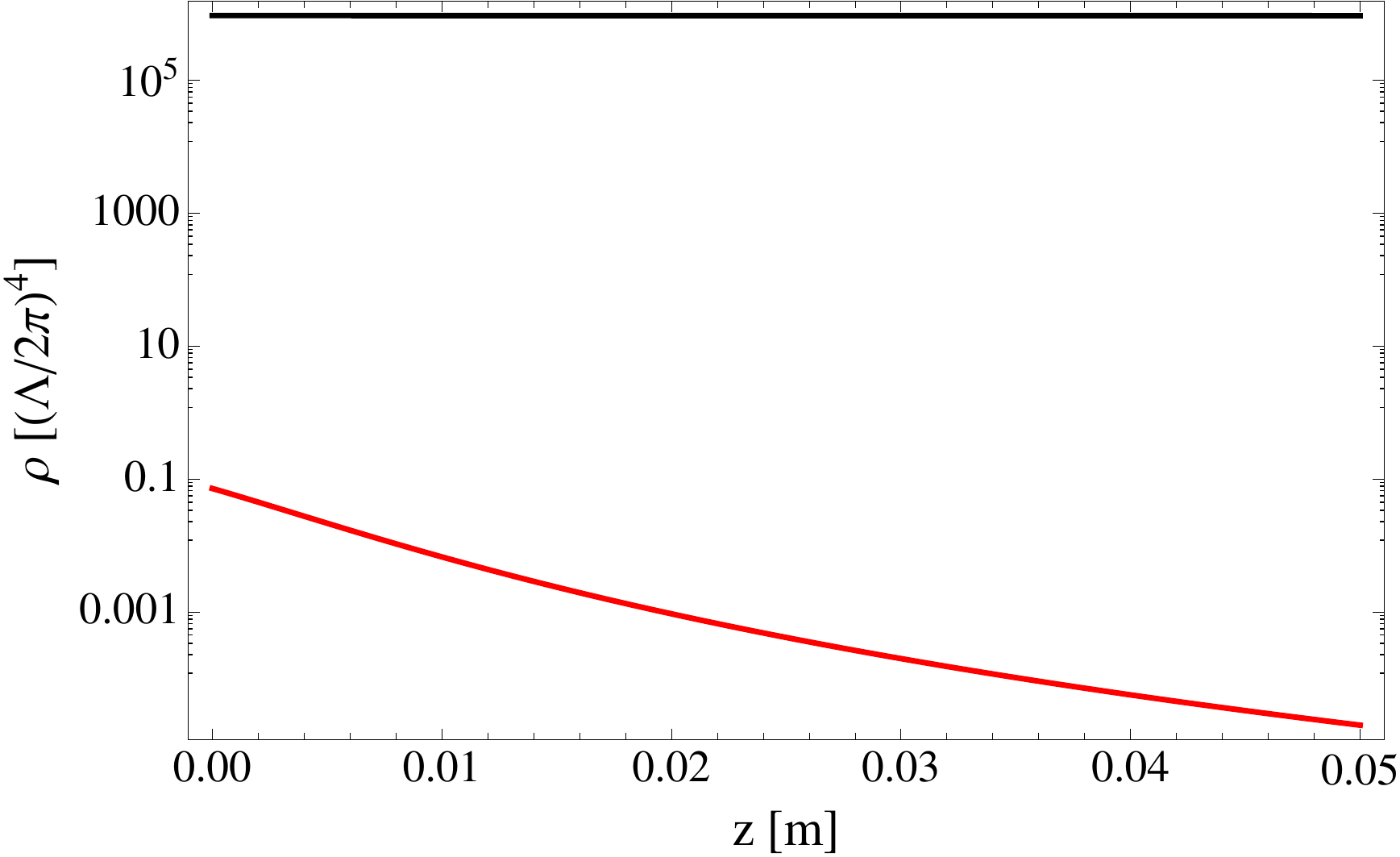}}
  \caption{The magnetic field $B(\hat z)$, its energy density $\rho_B(\hat z)$, and
  a comparison with the ground state energy density $\rho^{\scriptstyle{gs}}(\hat z)$.
  The ansatz parameters are the same as in figure \ref{fig:inh1},
  $g_0=1$, $g_1=0.1$. The coordinate $z$ is plotted in meters.}
  \label{fig:inh3}
\end{figure}

This temperature gradient is a scalar perturbation of the equations of motion
for the SU(2) fields, and \emph{causes} the field configurations \eqref{a3Ansatz}, \eqref{a0Ansatz},
which are static nonzero solutions of the perturbed equations of motion.
Then we can interpret the resulting magnetic field $B(\hat z)$, plotted in figure \ref{fig:inh06},
as a \emph{thermomagnetic field} generated by perturbing the coarse-grained SU(2) Yang-Mills
thermal ground state with a temperature gradient of the form \eqref{TAnsatz}.
The interpretation that the thermal gradient causes the magnetic field and not the
other way around is forced upon us by the requirement that the ground state
energy scale, and therefore the thermal gradient, is the dominant energy scale of the system.
Note from figure \ref{fig:inh06} that it goes to zero very quickly.

The thermomagnetic field in figure \ref{fig:inh06} contributes to the energy density
of the (perturbed) vacuum with $\rho_B$ as given in equation \eqref{rhoBAnsatz}.
This energy density is plotted in figure \ref{fig:inh07}
in units of the SU(2) Yang-Mills scale $\frac{\Lambda}{2 \pi}$.
Comparing $\rho_B$ with the energy density of the ground state (figure \ref{fig:inh08}),
we see $\rho_B$ lies about 8 orders of magnitude \emph{below}  $\rho^{\scriptstyle{gs}}$
over the entire domain.
The condition \eqref{defAdApp} imposed by the adiabatic approximation is satisfied.
Solutions calculated from $g_0\sim 1$ and $g_1\sim 0.1$ are therefore
acceptable deformations of the thermodynamically homogeneous SU(2) Yang-Mills ground state.

\begin{figure}[p]
  \centering
  \subfloat[The temperature field $T(z)$ plotted in SI units.
  It starts from a maximum value of $T_{\mathrm{max}}=31.67 \mathrm{K}$
  and asymptotically decreases to a value $T_{\mathrm{min}}=31.04 \mathrm{K}$.
  This gradient is a result of the ansatz in \eqref{eqn_defg} with parameters $g_0=1$ and $g_1=0.999$.
  ]{\label{fig:inh09}\includegraphics[width=120mm]{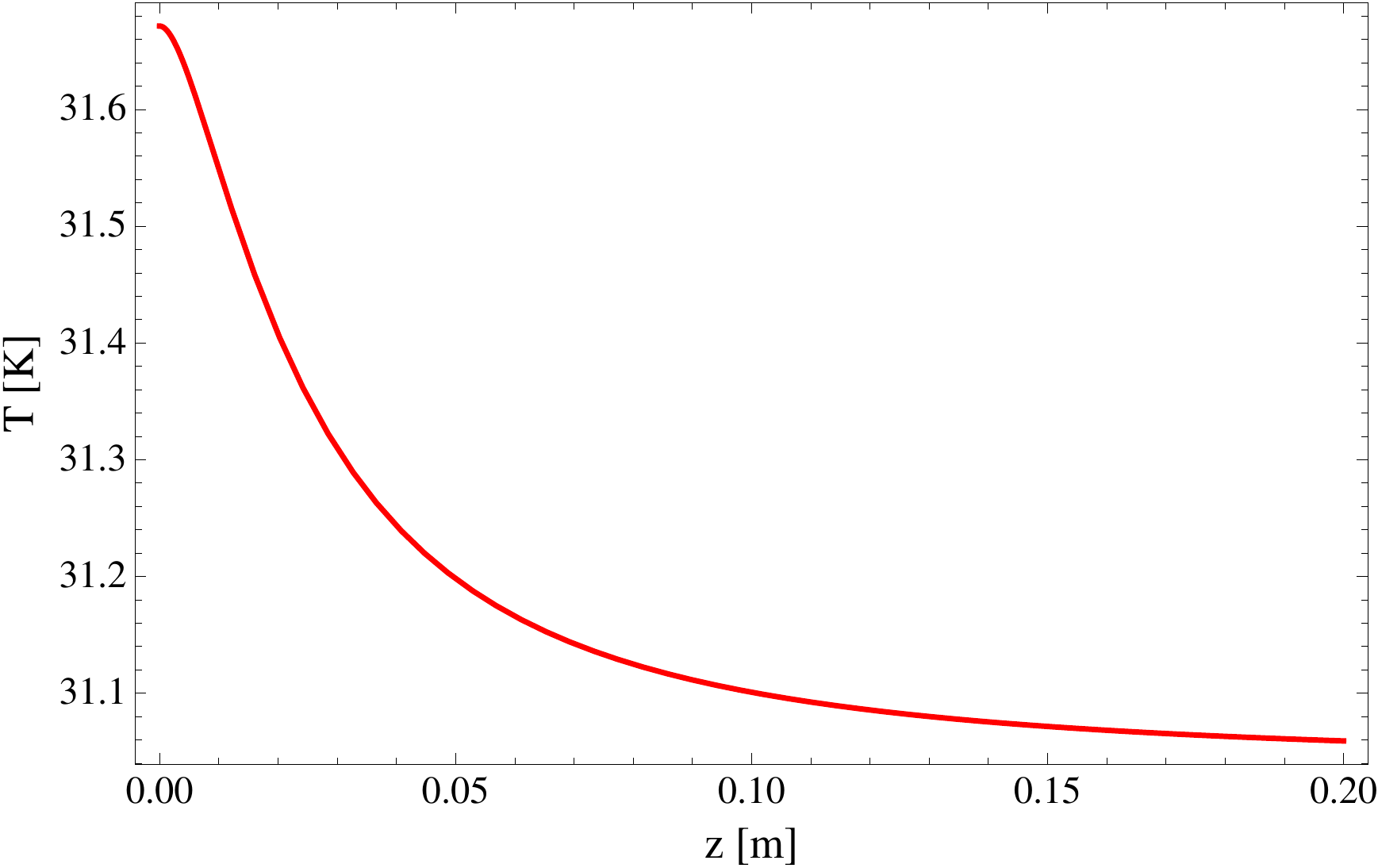}} \\
  \subfloat[The magnetic field energy density $\rho_B( z)$
  (in red) calculated from the generated magnetic field $B(\hat z)$,
  and the energy density of the thermal ground state $\rho^{\scriptstyle{gs}}(T( z))$
  plotted in natural units.
  ]{\label{fig:inh10}\includegraphics[width=120mm]{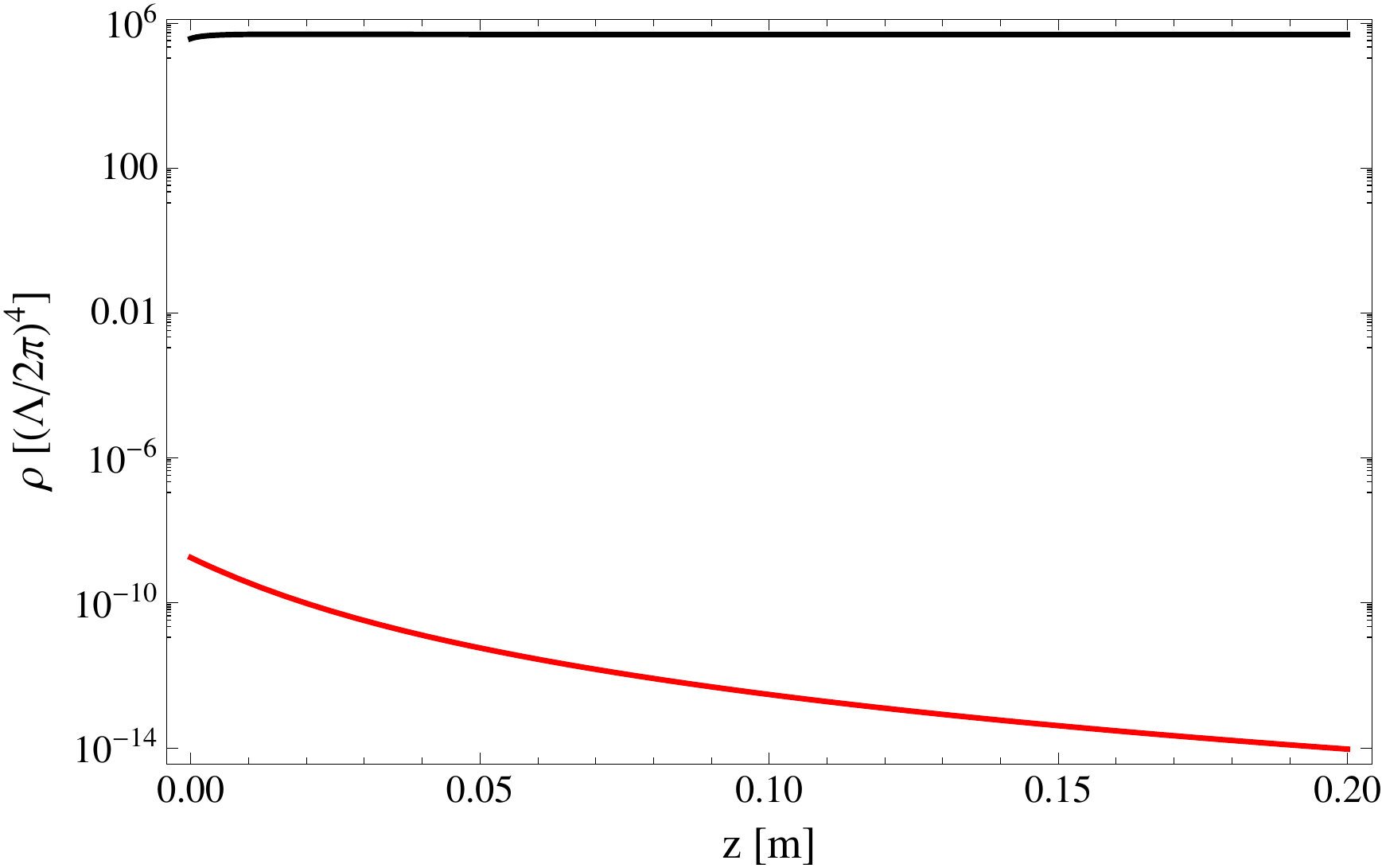}}
  \caption{(a) Temperature field and (b) energy densities calculated
  from ansatz \eqref{eqn_defg} with parameters $g_0=1$, $g_1=0.999$.
  The coordinate $z$ is plotted in meters.}
  \label{fig:inh4}
\end{figure}

To show the effect of parameter $g_1$,
we increase $g_1$ while leaving $g_0$ constant,
and calculate another set of solutions using $g_0=1$ and $g_1=0.999$.
We have plotted the resulting energy density and temperature profile
in figure \ref{fig:inh4}.
Note that the baseline temperature $T_{\mathrm{min}}=31.04\,\mathrm{K}$
remains unchanged, since it only depends on $g_0$ as was shown before.
The new maximum, however, lies at $T_{\mathrm{max}}=31.67 \mathrm{K}$.
Increasing $g_1$ has resulted in a smaller value $T_{\mathrm{max}}$.
This parameter is therefore responsible for setting the maximal temperature of the profile.
It can also be checked numerically that in the limit $g_1 \to 1$, 
$\partial_z T(z) \to 0$ and the temperature is constant over the entire domain,
with $T(z) = T_{\mathrm{max}}=T_{\mathrm{min}}=8\pi \Lambda g_0^{-1}$.
Ansatz \eqref{eqn_defg} shows a continuous limit (numerically)
to the homogeneous case.

Finally, we remark that in the context of nonperturbative SU(2) thermodynamics,
a \emph{dual interpretation} is necessary \cite{Hofmann2005},
in which the magnetic fields in the effective theory are understood as \emph{electric fields} in nature.
The thermomagnetic effect described here is therefore a \emph{thermoelectric effect} in nature.

\section{Decoupling from the ground state}
\label{sec:restoringu1}

We have seen how it is possible to disturb the homogeneity of the thermal ground state
in such a way that it still remains an effective description of the thermal SU(2) gas.
The deciding criterion is the energy scale hierarchy.
As long as the ground state energy density is the dominant scale,
deformations from the totally homogeneous case are still consistent
with the existence of a coarse-grained ground state.
In our particular solution, such distortions were also accompanied by non-trivial
gauge field configurations responsible for the emergence of a magnetic field.
In order to remain consistent with the adiabatic approximation,
the field energy of such configurations must always be considerable lower
than that of the thermal ground state.

In our experimental endeavours, for reasons discussed in chapter \ref{chap:radiometry},
it will not only prove useful to know under what conditions the
temperature or field energy deformations conserve the ground state, but also
under what conditions they cause a decoupling from it, effectively creating U(1) conditions.
In this section, we investigate the possibilities of creating an effective U(1) system.

\subsection{Electromagnetic field decoupling}

As explained in the preceeding section, the energy scale of the ground state energy must
be the largest scale of the system  if the effective theory is to remain valid.
But what happens when this hierarchy is destroyed,
and the SU(2) ground state energy density becomes small
relative to another energy scale?

An example of such an experimental situation is given by
the single-photon counting experiment described in \cite{Tada2006}.
$^{85}$Rb atoms in the 111$_{s_{1/2}}$ Rydberg state flying trough a resonant cavity
are excited upwards to the 111$_{p_{3/2}}$ state by absorption of thermal photons of frequency 2527\,MHz,
and then are counted by a method of selective ionization.
Thus the experiment is sensible to the mean photon number $\bar{n}(T)$
inside the cavity at a given temperature and cavity resonant frequency.
The measurement was performed for temperatures $T$ ranging from 67\,mK up to 1\,K.
During the measurements, a static electric stray field of magnitude
$|\vec{E}|\sim 25\,\frac{\mathrm{mV}}{\mathrm{cm}}$ was present in the cavity.
No deviation of $\bar{n}(T)$ from the U(1) expected Bose-Einstein distribution was observed.
Yet under the SU(2) hypothesis, absolutely no photon propagation should be possible.

From what we have learnt in this chapter,
we can guess that this could have to do with
the energy hierarchy inside the cavity.
The energy density $\rho_E$ of the external electric field in SI units is given by
\begin{equation}
\label{enEfeld}
\rho_E=\frac{\epsilon_0}{2}\,\vec{E}^2\,,
\end{equation}
where $\epsilon_0=8.854 187 817\,10^{-12}\,\mathrm{J/(V^2m)}$.
The energy density of the SU(2) thermal ground state $\rho^{\srm{gs}}$ in SI units is
\begin{equation}
  \rho^{\srm{gs}}=4\pi\,\Lambda^3 \, \frac{k_B}{(\hbar c)^3} T\,,
  \label{rho_gsSI}
\end{equation}
where $\Lambda=2\pi\,\frac{k_B T_c}{\lambda_c}$, $T_c=2.725\,\mathrm K$ and $\lambda_c=13.87$, see section \ref{sec:su2cmb}.
To maintain energy scale hierarchy, the electric field energy density should be
smaller than the energy density of the ground state, $\rho_E < \rho^{\srm{gs}}$.
For an static electric field magnitude of $|\vec{E}|\sim 25\,\mathrm{mV/cm}$,
this would imply a ground state temperature of the order of at least $10^3 \mathrm K$.
That is, the effective radiation temperature implied by the stray
electric field $|\vec{E}|$ is at least three orders of magnitude
above the thermalized radiation at 1\,K (and below) inside the cavity.
With such a large energy disparity,
a decoupling of the ground state physics from
the propagation properties of photons must have occured.
Ground state induced effects on the photon disappear and 
the dispersion law becomes indistinguishable
from that of the conventional U(1) theory. 

As we will see in chapter \ref{chap:radiometry},
this decoupling effect can be used to compare and test for the presence
of an SU(2) thermal ground state.

\section{Summary}

In this section we have investigated the different possibilities of
deforming the thermal ground state.
We have learnt that energy scale hierarchy
determines the validity of the effective theory
and thus the presence of SU(2) effects.

To realize SU(2) black cavities experimentally, we should therefore
be wary of stray electric or magnetic fields with energy densities
much larger than that of the effective ground state
such as those present in the cavity experiments of Tada et. al. \cite{Tada2006}.
For example, at $T=10\,\mathrm K$, the energy density of the ground state
becomes comparable to that of an electric field
of magnitude $|\vec{E}|\sim 2.47\,\mathrm{mV/cm}$.
To protect scale dominance of the thermal ground state,
the cavity should be shielded from stray electromagnetic fields
\emph{much smaller} than $2.47\,\mathrm{mV/cm}$,
of the order magnitude of at most $|\vec{E}|\sim 10^{-3}\,\mathrm{mV/cm}$.

Similarly, to mirror the conditions of the Tada et. al. experiment
and produce a black U(1) emitter at $T=10\,\mathrm K$,
the cavity should be transcended with electric fields of magnitude of at least $|\vec{E}|\sim 10^{3}\,\mathrm{mV/cm}$.

This knowledge will be useful later when we design an experiment
to compare between both regimes.


\chapter{Radiometry and Bolometry of SU(2) Photons}
\label{chap:radiometry}

In this chapter we first review some basic definitions regarding energy transfer,
with the goal of arriving at an expression for the \emph{spectral black-body radiance}
without assuming a U(1) dispersion law.
Next we examine the basic concepts of antenna theory and
re-derive a formula for the \emph{antenna temperature}.
Again, this is done without making any dispersion law explicit.
Both sections are included in this work since, apart from reviewing basic concepts,
we need to check that the usual formulae given in the literature also apply
in the case of a non-trivial dispersion law.
This will, indeed, turn out to be the case
(up to the appearance of a \emph{screening region}, as described in
chapter \ref{chap:YM}),
but it is not \emph{a priori} clear and needs to be shown explicitly.
The third section describes briefly what a radiometer system is, and, most importantly,
the possible sources of measurement error and the maximum possible \emph{radiometer sensitivity}.
Finally, sections \ref{sec:Radiometry} and \ref{sec:Bolometry}
discuss the possibility of detecting the SU(2) thermal
ground state by radiometric means and
possible experimental realizations,
relying on concepts introduced throughout the chapter.

Unless stated otherwise, we always use natural units, $\hbar=c=k_B=1$,
until section \ref{sec:Radiometry} in which we describe calculated predictions of
experimental results using SI units exclusively.

\section{Thermal radiation and energy propagation}
\label{sec:RadDefs}

Following \cite{grum1979optical,wilson2009tools},
in this section we review basic concepts on radiation propagation.
The results derived in this chapter will be important when characterising
\emph{thermal energy transfer}. Specifically the expression for \emph{black-body radiance}
is obtained without making use of any explicit dispersion law.
The SI units of important radiometric quantities introduced in this section
will be written beside their equation number.

Recall that the fundamental self-interacting SU(2) gauge modes become
weakly interacting thermal quasiparticles in the effective theory  \cite{RHLeip2007,HofmannLudescher2010}.
This allows the local \emph{spectral decomposition}
of any quantity $X$ characterising some aspect of energy transfer by photon propagation.
The spectral components $X_k$ or $X_\omega$ of $X$ can always be defined by only considering
contributions from monochromatic radiation of mode $\mathbf k$ or frequency $\omega$. 
Then the total quantity $X$ on a system is given by the integration over its available modes
\begin{align}
  X = \int \mathrm{d} k \; X_k = \int \mathrm{d} \omega \; X_\omega\,.
  \label{defSpectral}
\end{align}
By a change of variable we can relate both $k$-mode and $\omega$-mode spectral decompositions
\begin{align}
  X_{\omega} = \left| \partial_{\mathbf k} \omega_{\mathbf k} \right|^{-1} X_k\,.
  \label{defChangeVar}
\end{align}

\subsection{Radiant power}

In the following we will deal with
the characterisation and measurement of the
electromagnetic field in vacuum.
The labels \emph{energy} or \emph{radiant energy} will always refer to
the energy stored in the propagating photon field as reviewed
in Appendix \ref{app:gauge}.
\emph{Radiant power} $\Phi$ is then defined as the change in radiant energy $Q$ per unit time
\begin{flalign}
  \quad \Phi \equiv \frac{dQ}{dt}\,. & & \mathrm{[W]}
  \label{radDefPower}
\end{flalign}
We will also often refer to "energy rays" or "energy beams".
The ray concept, borrowed from geometrical optics, provides an intuitive characterisation
for the propagation of energy from source to detector.
This intuitive description is made precise by identifying "ray" with the \emph{Poynting vector}
of the propagating electromagnetic field. When expanding in single Fourier modes, the Poynting Vector
is parallel to the \emph{wave vector} $\mathbf k$ of the plane wave,
see section \ref{sec:PlaneWaveExpansion}.

\subsection{Radiance}

\begin{figure}[t]
  \centering
  \includegraphics[width=40mm]{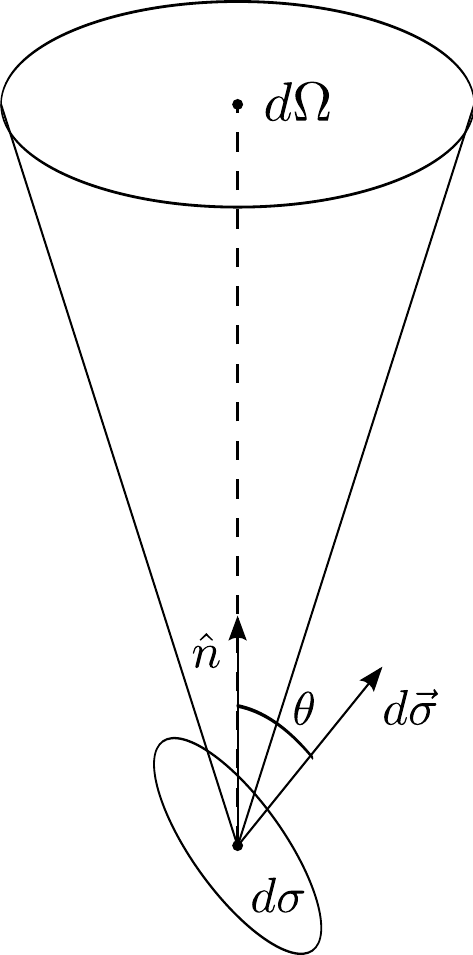}
  \caption{The flux through an area element $d\sigma$
  propagating along the subtended angle $d\Omega$, see text.}
  \label{fig:Radiance2}
\end{figure}

Consider now an elementary beam of incoherent radiant power $\Phi$ in a
homogeneous and isotropic
medium passing through a surface element $d\vec \sigma$,
which propagates further
within the solid angle $d\Omega$ with axis vector $\hat n$,
see Fig. \ref{fig:Radiance2}.
Then the \emph{radiance}
(or \emph{brightness}) at a point in $d \sigma$,
defined as the radiant power per unit solid angle
(in the direction of the ray) and per unit projected area
(perpendicular to the ray), can be written as
\begin{flalign}
  \quad L \equiv \frac{d^2 \Phi}{  d\Omega \left( d\vec \sigma \cdot \hat n \right)}
  = \frac{d^2 \Phi }{  d\Omega \left( d \sigma \cos{\theta} \right) }\,. & & \mathrm{[W\,m^{-2}\,sr^{-1}]}
  \label{radDefL}
\end{flalign}
Consider now two area elements $d\vec \sigma_1, d\vec \sigma_2$, separated by a distance $R$,
through which an incoherent beam of radiant energy propagates along a ray vector $\hat n$.
Let $\theta_1, \theta_2$ be the angles between $\hat n$ and each of the surface's normals.
Then the solid angle subtended by each area element at the opposite end is
\begin{align}
d\Omega_i = \frac{d\vec \sigma_i \cdot \hat n}{R^2} = \frac{d\sigma_i \cos{\theta_i} }{R^2}\,,
\qquad \qquad i=1,2\,\,. \notag
\end{align}
It follows that $d\Omega_1 d\sigma_2 \cos{\theta_2} = d\Omega_2 d\sigma_1 \cos{\theta_1}$.

Since it was assumed that the beam passing through $d\vec \sigma_1$ and $d\vec \sigma_2$ is contained
entirely within each area element,
we conclude from energy conservation that the power $d \Phi$ passing through 
both areas is the same.
Then $d^2 \Phi_1 = d^2 \Phi_2$ and one obtains for the 
radiance $L_{ i}$ at each element
\begin{align}
  L_{ 1} &= \frac{d^2 \Phi }{  d\Omega_2 \left( d \sigma_1 \cos{\theta_1} \right) }
  = \frac{d^2 \Phi }{  d\Omega_1 \left( d \sigma_2 \cos{\theta_2} \right) } 
  = L_{ 2} \notag \,.
\end{align}
$L$ is thus conserved along a ray within an homogeneous and isotropic medium.
This is why radiance is a useful concept for describing energy transfer.

\subsection{Transfer of radiant energy}

Let a radiating source be characterised by its radiance $L$. Then the \emph{power flux}
$S$ emitted by the source of radiance $L$ and passing through an area element $d\sigma$ is given
by an integration over the solid angle $\Omega_s$ subtended by the source
\begin{align}
  S = \int_{\Omega_s} \mathrm{d} \Omega \, \cos{\theta} \, L \left( \theta, \phi \right) \,,
  \label{radS}
\end{align}
where $\theta$ is the angle between the normal $d\vec \sigma$ to the area element
and the axis $\hat n$ of $\Omega_s$, compare with Fig. \ref{fig:Radiance2}.
Accordingly, the \emph{total power} is found by integrating $S_\sigma$
over a total area A of a detecting device
\begin{align}
  \Phi = \int_{A} \mathrm{d} \sigma \, S_\sigma \,.
  \label{radPhi}
\end{align}
Note that the linear superposition of power is only possible by assuming no
interference effects, that is, by assuming \emph{incoherent radiation}. Since we will be
dealing with \emph{thermalized radiation},
this assumption will be met ideally.

\subsection{Radiance formula for thermalized black-body emitters}
\label{sec:radFormula}
We will now derive the known Planck radiance
formula for black-body emitters in thermal equilibrium with radiation \cite{Planck1901}.
Special attention will be paid to the assumptions under which the derivation holds, and
no special dispersion law will be singled out.
We believe it necessary to do this calculation
in detail since most presentations in the literature assume U(1)
dispersion relations.
The dispersion law for the U(1) photon is just
the linear function $\omega_{\mathbf k} = c |\mathbf k |$,
but in the SU(2) case, $\omega_{\mathbf k} = \sqrt{\mathbf k ^2 + G}$
and it is not clear that the classical U(1) result for the black-body
radiance remains valid, since the functional dependence of the energy density for each Fourier mode
of its wave vector is nontrivial (see section \ref{sec:polarization}).
In the following, in order to write thermalized radiation as a sum over plane waves, we will
also need the results from Appendix \ref{app:gauge}.

\begin{figure}[t]
  \centering
  \includegraphics[width=60mm]{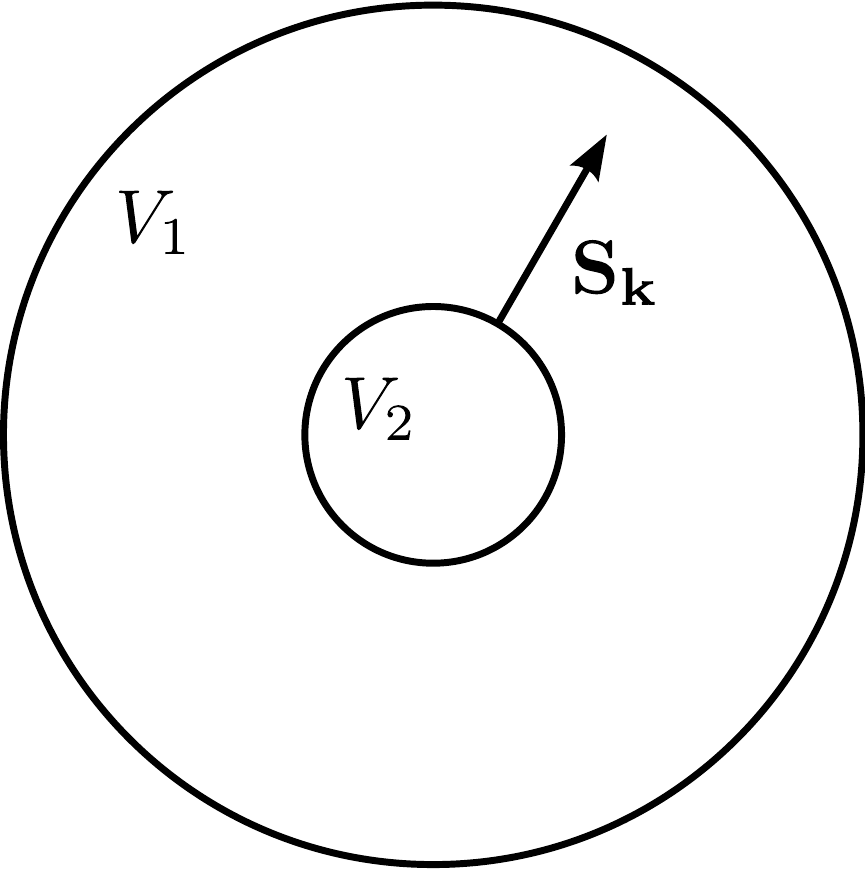}
  \caption{ Spherical volumes $V_1$ and $V_2$}
  \label{fig:RadVol}
\end{figure}

Consider a spherical black-body surface volume element $\partial V_2$ of radius $r_2$ positioned
at the centre of an isothermal spherical black enclosure $V_1$ at the same temperature,
and let the surfaces $\partial V_1, \partial V_2$ be thermally coupled by exchange of radiant energy.
Thermal coupling must hold for every \emph{available} mode of the cavity,
so thermal equilibrium must hold for available mode individually.
Assuming \emph{energy and charge conservation}, the \emph{continuity equation}
\begin{align}
  \partial_t \int_V \mathrm{d^3} \mathbf x \,\rho  + \oint_{\partial V} \mathrm{d} \mathbf \, \sigma \cdot \mathbf S \, =0 \,,
  \tag{\ref{EMTcon}}
\end{align}
must hold, and the radiant power emitted by
$\partial V_2$ in the mode corresponding to the wave vector $\mathbf k$
and collected by $\partial V_1$ must be
\begin{align}
  \Phi_{\mathbf k \; 2 \rightarrow 1} = -\partial_t Q_{\mathbf k\;2}
  = - \partial_t \int_{V_2^{\scriptstyle{c}}} \mathrm{d^3} x \rho_{\mathbf k}
  = \oint_{\partial V_2} \mathrm{d} \mathbf \, \vec \sigma \cdot \mathbf S_{\mathbf k} \,.
  \label{BBRadForm1}
\end{align}
Here the integration volume $V_2^{\scriptstyle{c}}$ is the volume bounded by $\partial V_2$ and infinity.
From the symmetry of the problem and thermodynamical equilibrium, the Poynting vector $\mathbf S$
of the thermalized electromagnetic field at $\partial V_2$ is normal to the surface,
and constant everywhere.
This allows us to write the spectral expression \eqref{BBRadForm1} as
\begin{align}
  \Phi_{\mathbf k \;2 \rightarrow 1} = \left| \mathbf S_{\mathbf k} \right| \oint_{\partial V_2} \mathrm{d} \mathbf \, \sigma 
  = \left| \mathbf S_{\mathbf k} \right| 4 \pi r_2^2 \,,
  \label{BBRadForm}
\end{align}
where $\left| \mathbf S_{\mathbf k} \right|$ is the magnitude of the Poynting vector
for the mode $\mathbf k$ at the boundary $\partial V_2$.

Consider now the spectral power $\Phi_{\mathbf k \; 1 \rightarrow 2}$ emitted by $\partial V_1$ and absorbed at $\partial V_2$.
Using \eqref{radPhi}, we may write it as
\begin{align}
  \Phi_{\mathbf k \; 1 \rightarrow 2} = \int_{ \partial V_2} \mathrm{d} \sigma \, \int_{\Omega_1} \mathrm{d} \Omega \, \cos{\theta_n} \,
  L_{\mathbf k } \left( \theta, \phi \right) \,,
  \label{BBRad2}
\end{align}
where $\theta_n$ is the angle between the normals of each area element at $\partial V_2$ and the
solid angle $\Omega_1=4\Pi$ subtended by $\partial V_1$.
From spherical symmetry, $\theta_n=0$.
Further, it is easy to show \cite{grum1979optical} that for the black-body emitter,
its spectral radiance $\lbk$ must, by definition, be \emph{independent} of the polar angles $\theta, \phi$
\begin{align}
  L_{\mathbf k} \left( \theta, \phi \right) = \lbk \left( \theta, \phi \right) = \lbk \left( 0, 0 \right)
  = \lbk \,.
  \label{bbradiance}
\end{align}
We may then write
\begin{align}
  \Phi_{\mathbf k \; 1 \rightarrow 2} = \int_{ \partial V_2} \mathrm{d} \sigma \, \int_{\Omega_1} \mathrm{d} \Omega \, \lbk
                         = 4 \pi \, \lbk \,\int_{ \partial V_2} \mathrm{d} \sigma
                         = 16 \pi^2 \, \lbk \,r_2^2 \,.
  \label{BBRad3}
\end{align}
Since we have assumed thermalization for each available mode $\mathbf k$ in the cavity,
it follows that $\Phi_{\mathbf k \; 2 \rightarrow 1} = \Phi_{\mathbf k \; 1 \rightarrow 2}$,
and we obtain a general expression for the \emph{spectral black-body radiance}
$\lbk$ in a thermalized cavity
\begin{align}
  \lbk = \frac{1}{4\pi} \left| \mathbf S_{\mathbf k} \right| \,.
  \label{BBRad}
\end{align}
Our conclusion is also valid for any other geometrical configuration. This can be argued as follows.

In general, cavity geometry influences the homogeneity and isotropy of thermal radiation
inside the enclosure.
For a rectangular cavity
of side lengths $L_i$,\,$i=1,2,3$, and volume $V=L_1 L_2 L_3$,
for example,
the correction $\delta D(\nu)$ caused by \emph{geometric discretization effects}
to the ideal \emph{spectral density} $D_0(\nu)=\frac{8\pi V}{c} \left(\frac{\nu}{c}\right)^2$
of electromagnetic modes in a lossless closed cavity of same volume V
is given by the expression (in SI units) \cite{baltes1972problems}
\begin{align}
  \frac{\delta D(\nu)}{D_0(\nu)}\,=\, - \left(\frac{c}{\nu}\right)^2 \, \frac{1}{8\pi V}(L_1+L_2+L_3) \,.
  \label{DiscEff}
\end{align}
But by choosing the geometric dimensions of the cavity to be large enough,
these discretization effects can be made arbitrary small
in the spectral region of interest.
With this in mind, here and in the following we will always presume the cavity to be large enough so that
homogeneity and isotropy may be assumed up to any order.

Take now a black-body cavity with arbitrary geometry, and choose an spherical infinitesimal
volume element $dV_0$ positioned at a point $r_0$ inside the enclosure.
In thermal equilibrium, the radiation field is homogeneous and isotropic,
so that the propagation of radiation from the cavity walls through $dV_0$ may be modelled
by assuming an outer spherical surface $S_0$ centered in $r_0$ emitting
radially at $dV_0$.
By the same argument, since $dV_0$ encloses homogeneous isotropically propagating
thermal radiation, it may be seen as a spherical black-body emitting at $S_0$.
But this results in exactly the same geometrical configuration as before.
We conclude that for a cavity with arbitrary geometry and sufficiently large
dimensions\footnote{What "`sufficiently large"' means depends of course on the cavity geometry.
In this work, we restrict ourselves to rectangular cavities, where corrections are given by \eqref{DiscEff}.}
that discretization effects can be ignored, with walls in \emph{thermal equilibrium with the radiation field}
equation \eqref{BBRad} holds.

We now use the results of Appendix \ref{app:gauge} to express $\lbb$ as an integral over Fourier modes.
Recall the expression for the \emph{energy density} $\rho$ of \emph{thermalized}
U(1) photon radiation at temperature $T \equiv \beta^{-1}$,
first derived by Planck \cite{Planck1901} (here written in natural units $\hbar=c=k_B=1$)
\begin{align}
  \rho\left( \beta \right)= 2\, \int \frac{\mathrm{d^3} \mathbf k}{\left( 2 \pi \right)^3} \,
  \frac{\omega_{\mathbf k}}{e^{\beta \omega_{\mathbf k}} -1 } \,.
  \label{ED_Planck}
\end{align}
To describe the both the thermalized U(1) and SU(2) photon field using the expressions derived in Appendix \ref{app:gauge},
we demand that equation
\begin{align}
  \rho \,&= \, \frac{Q}{V} \, = \, 2\, \frac{1}{V} \int \frac{\mathrm{d^3} \mathbf k}{\, \left( 2 \pi \right)^3}
             \, \left( 2 N_k^2 \, \omega_{\mathbf k} \right) \, \omega_{\mathbf k} \, ,
             \tag{\ref{emEnergy}'}
             \label{emEnergy2}
\end{align}
which gives the energy density of a given U(1) photon configuration in vacuum (defined by $N_k$),
to coincide with \eqref{ED_Planck}.
The factor $2$ in \eqref{emEnergy2} is due to summing over two transverse polarization states.
The energy density of the SU(2) photon gas is given by \eqref{prEnergyT},
which sums only over \emph{transverse} polarization states,
since the longitudinal polarization, which has a very narrow wave vector support,
cannot transport energy (see section \ref{sec:longmodes}).
Equation \eqref{prEnergyT} has the same form of \eqref{emEnergy2}, but with $\omega_{\mathbf k}$
given by the SU(2) transverse dispersion law \eqref{dispRelT} (written here in SI units)
\begin{flalign}
  \qquad \quad
  \left( \hbar \omega_{\mathbf{p}} \right) ^2 = \left( c \, \mathbf p \right)^2 +G \left(\omega, \mathbf{p}; T, \Lambda \right) \,.
  & & \mathrm{[J^2]}
  \label{dispRelTSI}
\end{flalign}
That is, we can use the same spectral decomposition \eqref{emEnergy2}
when describing the total field energy in both theories,
with the only replacement being the non-trivial dispersion law
for the photon of SU(2) Yang-Mills thermodynamics
(see chapter \ref{chap:YM}).
This is important, so we will repeat it here:
\emph{We use the same expression \eqref{emEnergy2} for both the U(1) and the non-abelian SU(2) Yang-Mills photon},
the vast dynamical difference between both is already taken into account by the different dispersion laws.

Comparing both \eqref{ED_Planck} and \eqref{emEnergy2} gives us the correct
\emph{thermal normalization factors} for the spectral decomposition \eqref{emEnergy2}
\begin{align}
  N_k = \sqrt{ \frac{\mathrm V\,  n_B \left( \beta \omega_{ \mathbf k} \right)}{2 \omega_{\mathbf k}} } \,,
  \label{ThermalNorm}
\end{align}
where $n_B (\beta \omega_{\mathbf k}) \equiv \left( e^{\beta \omega_{\mathbf k}} -1 \right)^{-1}$
is the \emph{Bose-Einstein distribution function}.

The \emph{spectral energy density} is then given as
\begin{align}
  \frac{Q}{V} \,=\,
  \int \mathrm{d} k \rho_{\mathbf k} \,&= \,2 \int \frac{\mathrm{d^3} \mathbf k}{\, \left( 2 \pi \right)^3}
  \, n_B(\beta \omega_{\mathbf k}) \, \omega_{\mathbf k} \, 
               = \, \frac{1}{\pi^2} \int \mathrm{d} k \; \mathbf{k}^2  \; n_B(\beta \omega_{\mathbf k}) \, \omega_{\mathbf k} \,,
               \notag \\
  \implies \rho_{\mathbf k} \, &= \, \frac{1}{\pi^2} \; n_B(\beta \omega_{\mathbf k})  \, \mathbf{k}^2 \, \omega_{\mathbf k} \,.
  \label{SpectralEDdef}
\end{align}

The \emph{energy propagation velocity} $\mathbf v_{\mathbf k}$
of a mode $\mathbf k$ can be defined \cite{collin96} so that
\begin{align}
  \rho_{\mathbf k} \, \mathbf v_{\mathbf k} = \mathbf S_{\mathbf k} \,.
  \label{defPropVel}
\end{align}
For each mode, it can be shown that this propagation velocity,
by definition, is just the $k$-gradient of the dispersion law \cite{collin96}
\begin{align}
  \mathbf v_{\mathbf k} = \nabla_{\mathbf k} \omega_{\mathbf k} \,,
  \label{PropVeloc}
\end{align}
\emph{whatever the functional expression for the dispersion law is}.

Inserting \eqref{SpectralEDdef}, \eqref{defPropVel}, and \eqref{PropVeloc} into \eqref{BBRad}, we obtain
the \emph{spectral radiance of a black-body emitter in thermalized radiation}
\begin{align}
  \lbk\, &=\, \frac{1}{4\pi} \left| \rho_{\mathbf k} \; \nabla_{\mathbf k} \omega_{\mathbf k} \right|
        =\, \frac{1}{4 \pi^3} \,\frac{\omega_{\mathbf k} }{e^{\beta \omega_{\mathbf k}} -1 } \, \mathbf k^2 \, 
        \left| \partial_{\mathbf k} \omega_{\mathbf k} \right| \,.
  \label{BBRadK}
\end{align}
Using \eqref{defChangeVar}, we can also write the same expression as a function of \emph{frequency}.
The \emph{radiance per frequency interval of a black-body emitter in thermalized radiation} is then,
written in natural units, given as
\begin{align}
  \lbw\, &=\, \frac{1}{4 \pi^3} \, \frac{\omega_{\mathbf k} }{e^{\beta \omega_{\mathbf k}} -1 } \, \mathbf k^2 \,.
  \qquad \qquad \qquad \qquad \qquad \qquad \qquad \qquad \qquad \qquad
  \mathrm{[eV^{4}]}
  \label{BBL}
\end{align}
To write \eqref{BBL} in SI units, remember that the (angular) frequency and wave vector are rescaled as
$\omega_{\mathbf k} \rightarrow \hbar \omega_{\mathbf k}$, $\mathbf k \rightarrow c \hbar \mathbf k$
and the inverse temperature as $\beta \rightarrow \frac{1}{k_B T}$.
Moreover, radiance has dimension $\mathrm{eV^4}$ in natural units, so we need to divide expression \eqref{BBL}
with $(\hbar c)^2$ to return to SI units. We obtain
\begin{flalign}
  \qquad \quad  \lbv\, &=\, \frac{1}{2 \pi^2} \, \frac{h \nu }{e^{h \nu / k_B T} -1 } \, k^2 \,.
  &  \mathrm{[W\,m^{-2}\,sr^{-1}\,Hz^{-1}]}
  \label{BBL_SI}
\end{flalign}
The additional factor of $2 \pi$  comes from the change of variable $d\omega \rightarrow 2\pi d\nu$.
The formulae \eqref{BBL}, \eqref{BBL_SI} characterise radiation in thermal
equilibrium with a black-body, independently of any functional relation
between $\mathbf k$ and $\omega_{\mathbf k}$.

Later, \eqref{BBL_SI} will be used to calculate the \emph{power output} of any measuring device
observing thermal radiation.

\subsection{Radiance of the SU(2) thermal gas}

In this section we compare radiance functions of both the U(1) and SU(2) theories.
The radiance function $\lbvs{U(1)}(\nu,T)$ for the U(1) photon gas in SI units
can be read off from Eq. \eqref{BBL_SI}.
Inserting the U(1) dispersion law
$\hbar \omega_{\mathbf k} = c \hbar |\mathbf k| = h \nu$
results in the expression
\begin{flalign}
  \qquad \quad  \lbvs{U(1)}(\nu,T) \, &=\, \frac{2 h}{c^2} \, \frac{ \nu^3 }{e^{h \nu / k_B T} -1 } \,.
  &  \mathrm{[W\,m^{-2}\,sr^{-1}\,Hz^{-1}]}
  \label{BBL_U1_SI}
\end{flalign}
To write the radiance function $\lbvs{SU(2)}(\nu,T)$ for the SU(2) case,
we recall the SU(2) dispersion relation \eqref{dispRelT}
$\hbar \omega_{\mathbf{k}} = \sqrt{(c\hbar \mathbf{k})^2 +G \left(\omega, \hbar \mathbf{k}; T, \Lambda \right)}$.
Recall that the SU(2) photon gas shows a \emph{screening region}
so that propagation is forbidden below a cutoff frequency $\nu^*$, see section \ref{sec:polarization},
That is, the only modes allowed in the cavity are those with frequencies $\nu>\nu^*$.
That only those modes can contribute to the spectral radiance is ensured by multiplying with a $\theta$-function.
In the end, we obtain
\begin{flalign}
  \qquad \quad \lbvs{SU(2)}(\nu,T) \,&=\,\frac{2 h}{c^2} \, \frac{ \nu^3 }{e^{h \nu / k_B T} -1 } \, \left( 1 - \frac{c^2 G}{(h \nu)^2} \right)
  \, \theta\left( \nu - \nu^* \right) \notag \\
  &=\, \lbvs{U(1)}\, \left( 1 - \frac{c^2 G}{(h \nu)^2} \right) \, \theta\left( \nu - \nu^* \right)
  \,.
    &  \mathrm{[W\,m^{-2}\,sr^{-1}\,Hz^{-1}]}
  \label{BBL_SU2_SI}
\end{flalign}
Note that outside the screening region, that is, for
$\nu>\nu^*$, the radiance \eqref{BBL_SU2_SI} for the SU(2) gas factorizes
into the factor $(1 - \frac{c^2 G}{(h \nu)^2})$ times the U(1) radiance \eqref{BBL_U1_SI}.
As discussed in \cite{LH2008,SHG2007},
the function $G$ goes to zero power-like for increasing temperature and
exponentially for increasing frequency.
So we expect a return to the U(1) case for high enough temperatures and energies, since
in that limit the factor $(1 - \frac{c^2 G}{(h \nu)^2})\to 1$, $\nu^* \to 0$,
and both radiance functions $\lbvs{U(1)}(\nu,T)$ and $\lbvs{SU(2)}(\nu,T)$ become indistinguishable.

\begin{figure}[p]
  \begin{center}
    \includegraphics[width=150mm]{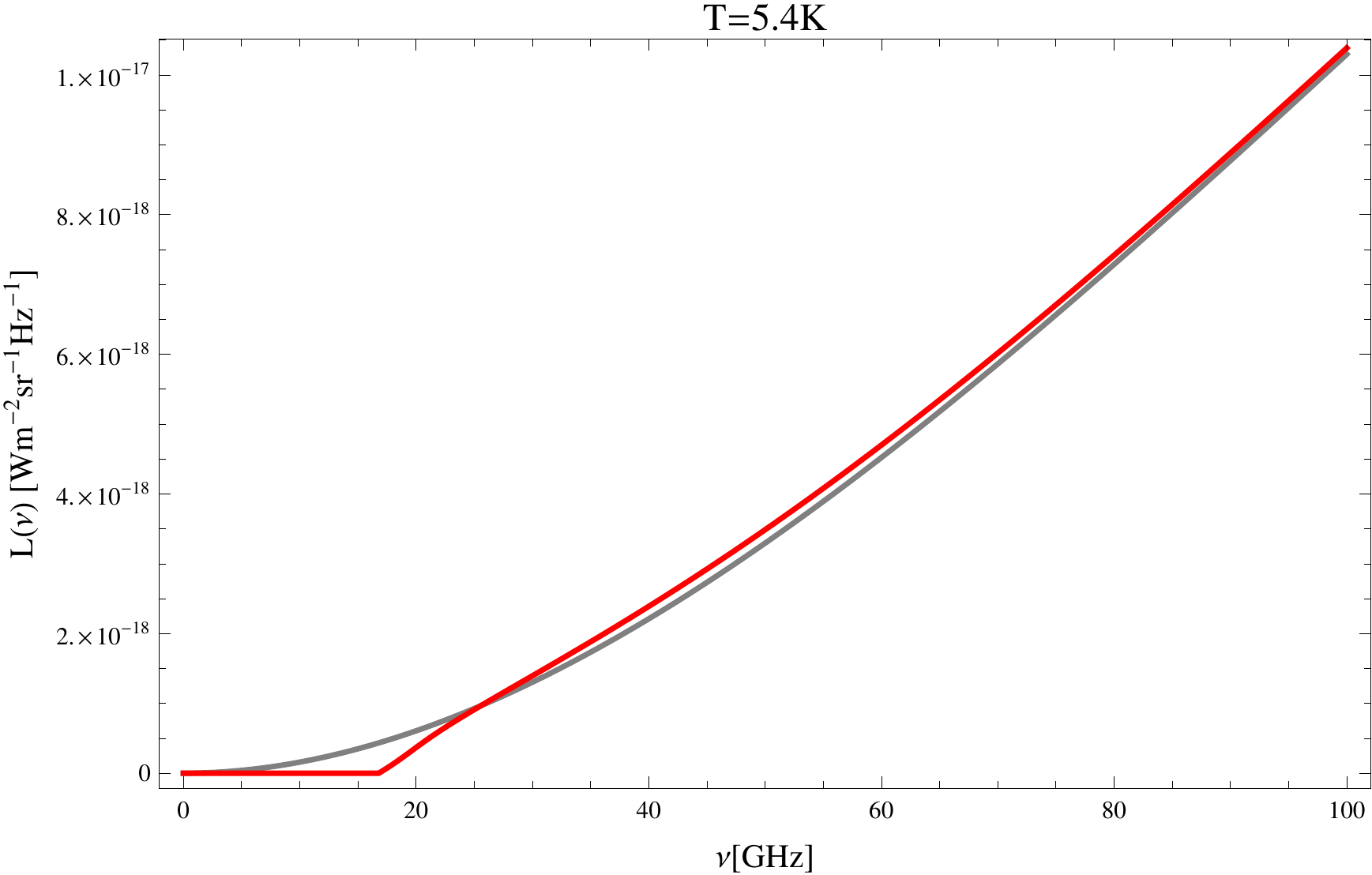}
  \end{center}
  \caption{ Comparison of the spectral radiance of an SU(2) photon gas (red)
  and a U(1) gas (grey) at a temperature $T$=5.4K.
  Note the spectral gap in the low-frequency region.
  }
  \label{fig:SU2rad54K}
\end{figure}

\begin{figure}[p]
  \begin{center}
    \includegraphics[width=140mm]{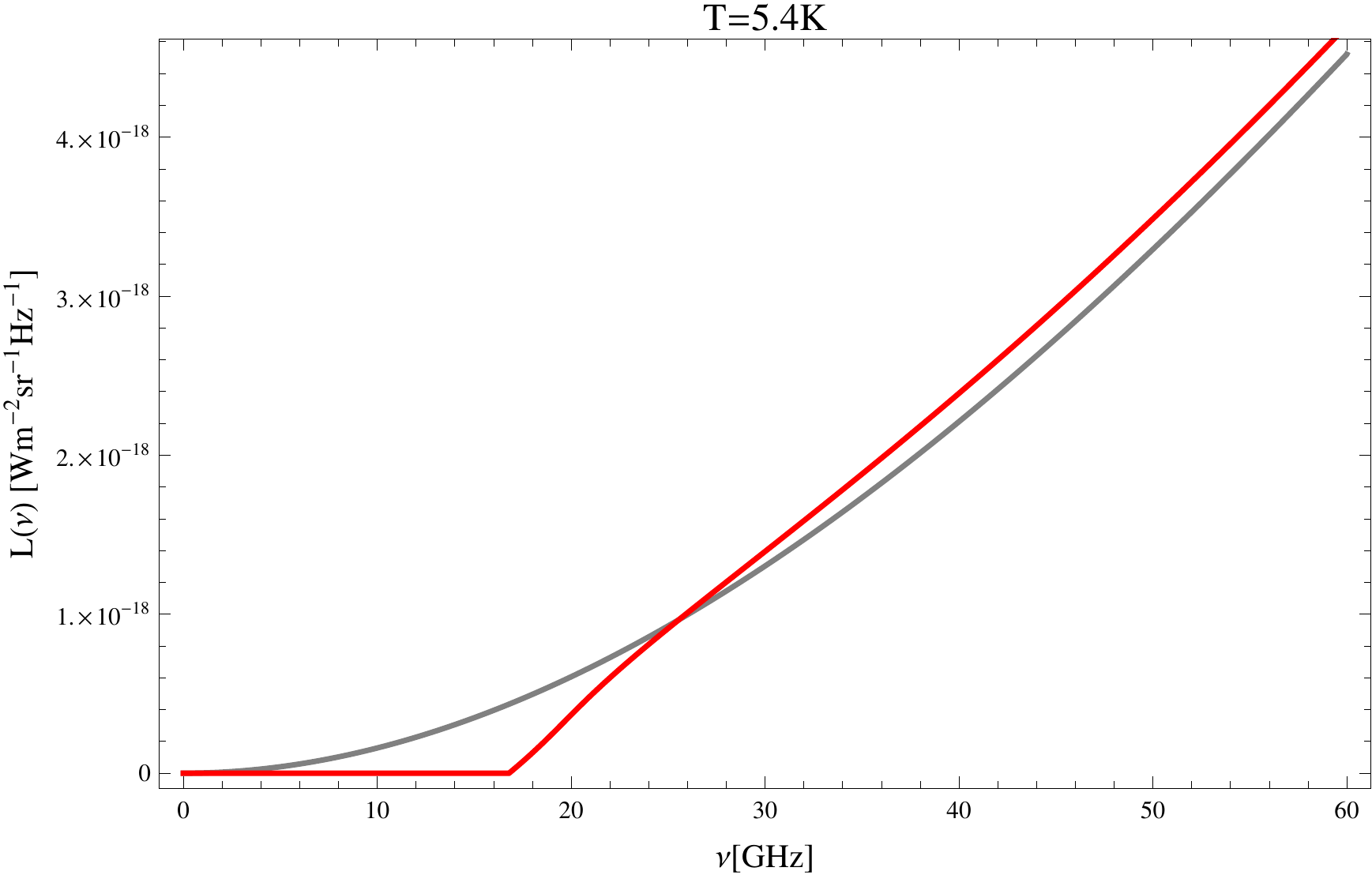}
  \end{center}
  \caption{Zoom-in of figure \ref{fig:SU2rad54K} }
  \label{fig:SU2rad54Kz}
\end{figure}

\begin{figure}[p]
  \begin{center}
    \includegraphics[width=150mm]{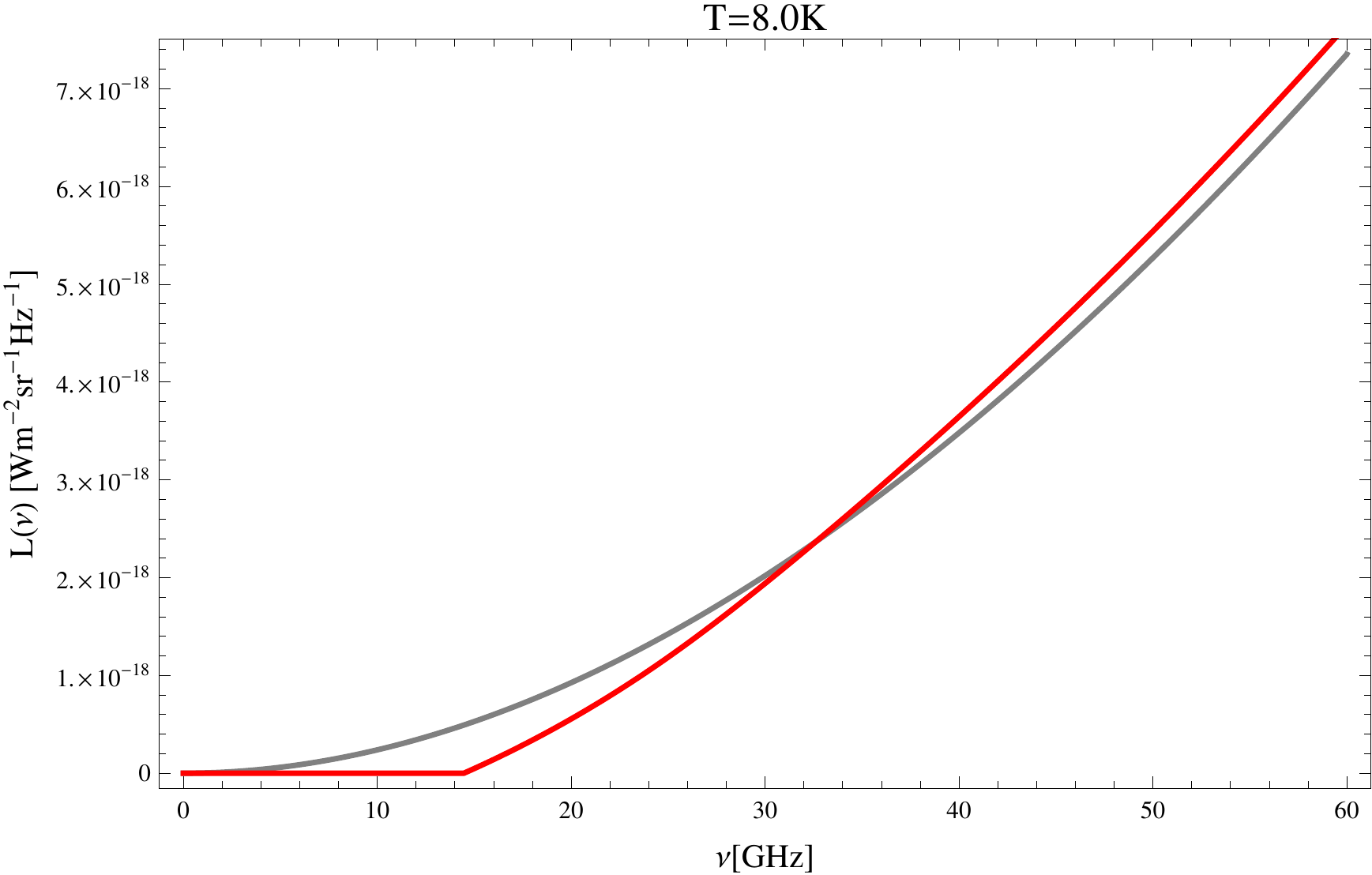}
  \end{center}
  \caption{Comparison of the spectral radiance of an SU(2) photon gas (red)
  and a U(1) gas (grey) at a temperature $T$=8.0K.
  Note the spectral gap in the low frequency region.
  }
  \label{fig:SU2rad8K}
\end{figure}

\begin{figure}[p]
  \begin{center}
    \includegraphics[width=140mm]{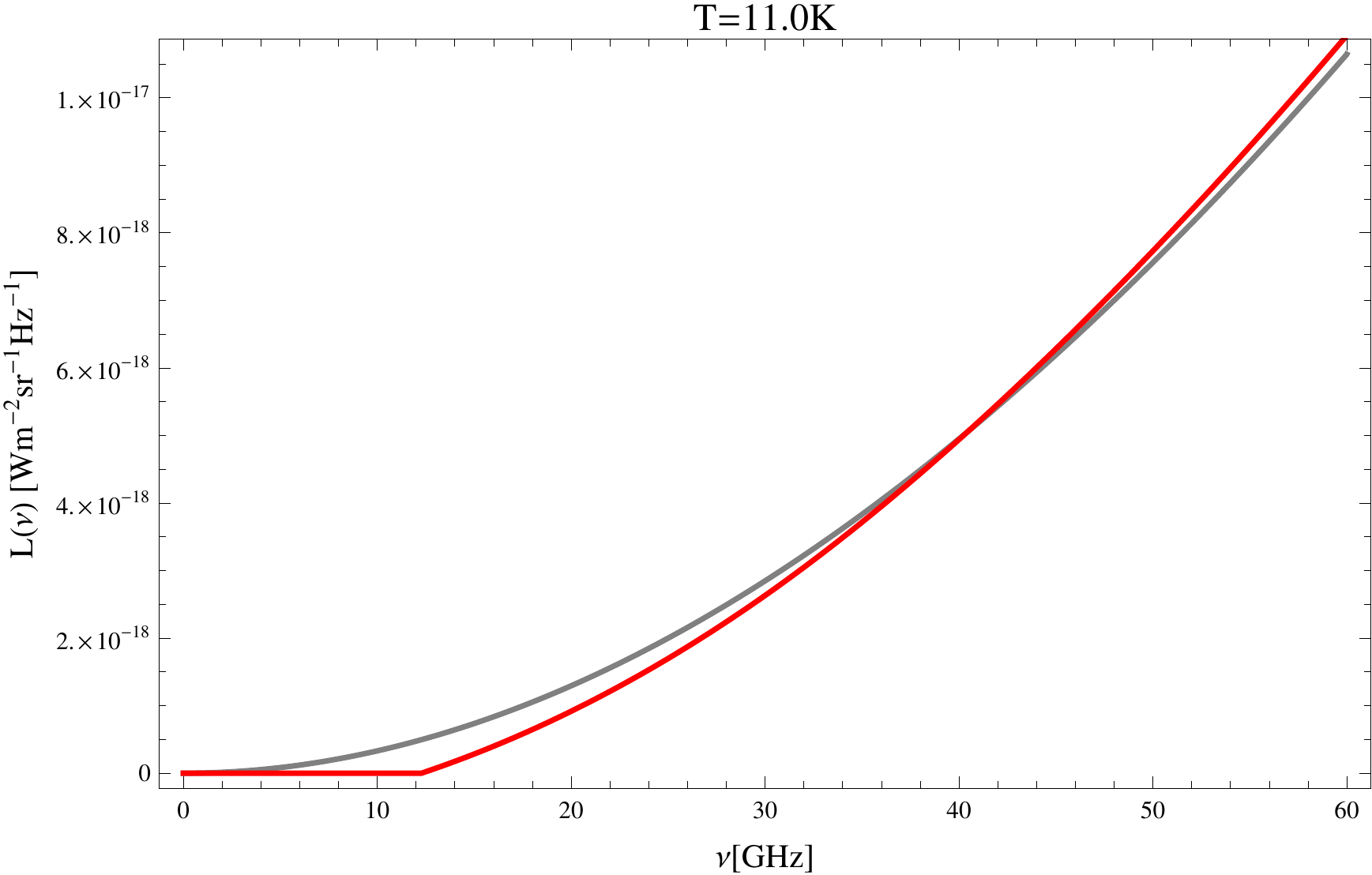}
  \end{center}
  \caption{Comparison of the spectral radiance of an SU(2) photon gas (red)
  and a U(1) gas (grey) at a temperature $T$=11.0K.
  Note the spectral gap in the low frequency region.
  }
  \label{fig:SU2rad11K}
\end{figure}

In the case of low energy and temperature, however, we do expect the SU(2) photon gas
to show considerable deviation from the U(1) case.
This suggests the possibility of \emph{experimental detection} of the SU(2) gas by
radiation measurements at low frequency and temperature. This possibility will be explored further
in section \ref{sec:Radiometry}.

Examples of the calculated radiance for different temperatures are plotted in figures
\ref{fig:SU2rad54K} to \ref{fig:SU2rad11K}.
To plot the SU(2) radiance, the cutoff frequency $\nu^*$ was calculated as shown in section \ref{sec:Gprop}.
Note the screening region at low frequencies where the radiance of the SU(2) gas vanishes.
The \emph{crossover frequency} $\nu_c$, discussed in section \ref{sec:Gprop} is also easily visible in all plots.

\section{Basics of the theory of antennas}
\label{sec:antennas}

In order to test experimentally the most important prediction of the SU(2) photon theory,
and thus ascertain the presence of a thermodynamical ground state composed by topological
field configurations as described in chapter \ref{chap:YM},
we need to be able to detect thermal radiation at low frequencies and temperatures.
In the preceding section we derived an expression for the radiance of
black-body emitters which characterises energy propagation.
Now we turn to the problem of how to measure such an altered propagation.

The experimental science and technology of the measurement of electromagnetic radiation is called
\emph{radiometry} \cite{ulaby1981microwave}.
A typical radiometric set up always consists
of a \emph{radiation source}, an \emph{aperture} or \emph{optical system}, and a \emph{detector}.
The aperture collects the radiation from the source and redirects it to the detector.
The detector converts radiation into a well defined signal that can then be measured
within a certain precision and further processed.

In the microwave region of the electromagnetic spectrum
-- which, as we will later see, is where the predicted differences between U(1) and SU(2) propagation occur --
an \emph{antenna} is often used as the aperture.
An antenna is defined \cite{kraus2002antennas,ulaby1981microwave}
as the region of transition between an electromagnetic wave propagating in free space
and a guided wave propagating in a \emph{transmission line}
\footnote{A \emph{transmission line} is a device allowing for propagation of radio-frequency energy
from one point to another. Antennas \emph{radiate or receive} energy, while transmission lines
\emph{transport} energy. See also section \ref{sec:radiometer}}.

In this section, we review some concepts relating to energy collection through antennas.
The most important result will be an expression for the \emph{antenna temperature} $T_A$,
which quantifies the power output of an antenna in relation to its incident thermal radiation,
derived here without assuming any explicit dispersion law.
The detection of the collected thermal radiation is the subject of section \ref{sec:radiometer}.

Further properties that characterise antennas are geometric in nature, such as the
\emph{antenna radiation pattern} and the \emph{antenna effective area}.
Geometric factors also determine such coefficients as \emph{beam efficiency},
which characterises undesired signal contribution by side-lobe radiation.
These characteristics will play a minor role for us since
we will be dealing with isotropic and homogeneous radiation.
Still, these geometric factors need to be understood before we arrive at a formula for $T_A$.
We briefly review these subjects before arriving at our
main expression for the antenna temperature.

\subsection{Antenna parameters}
\label{sec:AntennaParm}

This section serves as a quick introduction to all antenna parameters
we will require in the following sections to describe and calculate through our experiment.
A thorough review of the subject can be found in \cite{kraus2002antennas}.

Consider a transmitting antenna placed at the origin. For sufficiently large $r$,
we may consider the radiated modes as plane waves.
This zone is called the \emph{far field},
and we always work in this limit unless stated otherwise.

With the antenna positioned at the origin, we expect
the far field electric and magnetic vectors
to be mutually perpendicular and normal to the unit vector $\hat r$.
The Poynting vector $\mathbf S$ must be radial.
In the case of isotropic antenna radiation, the total radiated power given
by integrating over the surface of a spherical volume V
centered at the origin becomes in this limit
\begin{align}
  \Phi = \oint_{\partial V} d \vec \sigma \; \mathbf S = 4 \pi r^2 S_r\,,
  \label{isoRad}
\end{align}
since the radial component $S_r$  is constant over all directions.
In general, antenna radiation patterns are not isotropic, and almost always
distributed around a narrow solid angle called the \emph{main lobe}.
$S_r$ is no longer isotropic and the polar dependency of antenna
radiation is characterised by the \emph{antenna radiation pattern}
\begin{align}
  F\left( r, \theta,\phi \right) \equiv r^2 S_r\left( \theta,\phi \right) \,,
  \label{defARP}
\end{align}
so that the total radiated power in the far field is given by
\begin{align}
   \Phi\, =\, \int \mathrm d \Omega\, F\left( \theta,\phi \right)\,,
  \label{isoRad2}
\end{align}
where we have omitted the $r$-dependence, which in the far-field
reduces to equation \eqref{defARP}, $F( \theta,\phi ) \equiv r^2 S_r$.
For an isotropic antenna, \eqref{isoRad2} reduces, of course, to \eqref{isoRad}.

Imagine now a radiating antenna which we supply with \emph{input power} $P_i$.
From the second law of thermodynamics we may assume that a fraction of $P_i$
is dissipated as heat from the antenna.
Let $P_o$ be the actual \emph{radiated power} from the antenna aperture, and
define the \emph{antenna radiation efficiency} $\eta$ as the ratio $P_o/P_i$.
We can define a polar version of this factor as well, the
\emph{antenna gain function} $G(\theta,\phi)$, as the ratio between
radiated power per solid angle $d\Phi/d\Omega = F\left( \theta,\phi \right)$
and total input power per solid angle $P_i/4\pi$
\begin{align}
  G\left( \theta,\phi \right) \,\equiv\,\frac{4 \pi F\left( \theta,\phi \right)}{P_i}\,.
  \label{defGain}
\end{align}
Defining the \emph{normalized antenna radiation pattern}
\begin{align}
  F_n \left( \theta,\phi \right) = \frac{F\left( \theta,\phi \right)}{F\left( \theta,\phi \right)|_{max}} \,,
  \label{defNRP}
\end{align}
the \emph{pattern solid angle} $\Omega_P$, which characterises
the effective width of the radiation pattern,
\begin{align}
  \Omega_P \,\equiv\, \int_{4 \pi} d\Omega \, F_n \left( \theta,\phi \right)\,,
  \label{defPSA}
\end{align}
the \emph{antenna directivity} $D \left( \theta,\phi \right)$
\begin{align}
  D \left( \theta,\phi \right) \, &\equiv\,
        \frac{F_n \left( \theta,\phi \right)}{\frac{1}{4\pi} \int_{4 \pi} d\Omega \, F_n \left( \theta,\phi \right)} \,
        =\, D_0\, F_n \left( \theta,\phi \right) \,,
  \label{defDirectivity}
\end{align}
and the \emph{maximum directivity} $D_0$
\begin{align}
  D_0 \, &\equiv \, \frac{4\pi}{\Omega_P} \, =\, \frac{4 \pi}{\int_{4 \pi} d\Omega \, F_n \left( \theta,\phi \right)} \,
      =\, \frac{4 \pi S_r \left( \theta,\phi \right)|_{max}}{\int_{4 \pi} d\Omega \, S_r \left( \theta,\phi \right)} \,,
  \label{defD0}
\end{align}
we can write \eqref{defGain} as \cite{ulaby1981microwave}
\begin{align}
  G\left( \theta,\phi \right) \, =\, \eta\, D_0\, F_n\left( \theta,\phi \right) \,
  =\, \eta\, 4 \pi\,\frac{F_n\left( \theta,\phi \right)}{\Omega_P}\,.
  \label{defGain2}
\end{align}
Thus an equation for the radiated power per solid angle emerges
\begin{align}
  \frac{\mathrm d \Phi}{\mathrm d \Omega} \,=\,\frac{\,P_i}{\Omega_P}\, \eta\, \,F_n\left( \theta,\phi \right)\,.
  \label{defRadPo}
\end{align}
Note that \eqref{defRadPo} depends linearly on the input power $P_i$.
Moreover $\eta$, $\Omega_P$ and the polar angle dependency $F_n\left( \theta,\phi \right)$
are all intrinsic characteristics of the radiating antenna. That is, once they
have been calibrated, the power output of an antenna is then fully determined
by the input power $P_i$ feeding the antenna.

So far we have considered \emph{transmitting antennas}, and defined the antenna radiation pattern
and directivity as the characterising angular distribution of \emph{transmitted power}. What about
the \emph{power gain} of an antenna when collecting radiation? The \emph{reciprocity theorem}
\cite{silver1984microwave,wilson2009tools} shows that, under general assumptions (e.g. isotropy of the medium),
the \emph{transmitting} and \emph{receiving} antenna parameters are the same.
The directivity and radiation pattern characterise the \emph{angular cross section} of the antenna,
and it is the same for both transmission and reception of power.
In this sense, problems of radiation emission and radiation detection by an antenna are \emph{interchangeable}.

\subsection{Antenna effective area}
\label{sec:antennaeffarea}

Power transmission and reception by an antenna may be described, in the far field at least,
by borrowing from optics the idea of wave diffraction by a geometric aperture $A$.
This results in the \emph{Fresnel-Kirchhoff diffraction formula} \cite{silver1984microwave},
which reduces the computation of the antenna far field radiation,
given the field at the physical antenna aperture $A_P$,
to a scalar diffraction problem with \emph{effective} geometric antenna aperture $A_E$.
It is found that this effective aperture, or \emph{effective area}, 
depends on the wavelength of the diffracted radiation.
Since we would like to apply the same methods to the radiometry of SU(2) photons,
we must be careful that these results should not depend \emph{explicitly}
on the dispersion law.
This is a priori not clear and needs to be checked.
That the Fresnel-Kirchhoff diffraction formula does not depend explicitly on
a U(1) linear dispersion law and thus holds also for SU(2) photons by implicitly allowing for
functional dependencies such as \eqref{dispRelT} is checked in appendix \ref{app:fresnel}.
Here we use this result to arrive at an expression for the
\emph{antenna effective area} which we need to
relate electromagnetic fields in the far field
to electromagnetic fields in the aperture,
to calculate the \emph{incoming power} at the aperture.
This is done by first calculating the \emph{transmitted power}
through the aperture, and then invoking the reciprocity theorem.
We follow somewhat the presentation in \cite{ulaby1981microwave} but without assuming
any explicit dispersion law.

Consider a transmitting antenna, with its physical aperture area $A_P$,
and a monochromatic field distribution $E_{\mathbf k}$ of constant phase
defined over the surface $A_P$.
This is realised by having the Poynting vector of the monochromatic planar wave be normal to the aperture surface.
We say that the aperture is \emph{illuminated}
by the monochromatic field distribution $E_{\mathbf k}$ of constant phase.
Evaluating the total power $\Phi$ crossing through the aperture gives
\begin{align}
  \Phi &= \int_{A} \mathrm{d} x_a \mathrm{d} y_a \, \hat n \cdot \mathbf S_{\mathbf k} \left(x_a,y_a \right) = 
  \int_{A} \mathrm{d} x_a \mathrm{d} y_a \, S_{\mathbf k} \left(x_a,y_a \right) \notag \\
  &= \frac{1}{2}  \frac{\left|\mathbf k \right|}{ \omega_{\mathbf k}}
  \int_{A_P} \mathrm{d} x_a \mathrm{d} y_a \, \left| E_{\mathbf k}\left(x_a,y_a \right) \right|^2 \,,
  \label{PhiAP}
\end{align}
where $\hat n$ is the normal vector to the aperture, and we have expressed
the modulus of the Poynting vector $S_{\mathbf k}$
by that of the electric field (compare with \eqref{transS}).
The integration variables $x_a$ and $y_a$ parametrize the aperture surface.

Because of energy conservation, \eqref{PhiAP} must be equal to the power radiated in the far field
by the antenna, where radial symmetry holds and the monochromatic field propagates as
an spherical wave after diffraction through the physical aperture
\begin{align}
  \Phi &=  \int_{4 \pi} \mathrm{d} \Omega \, r^2 \, S_r \left(\theta, \phi \right) \,.
  \label{PhiFF}
\end{align}
Hence
\begin{align}
  \int_{4 \pi} \mathrm{d} \Omega \, \, S_r \left(\theta, \phi \right) &=
          \frac{1}{2 r^2} \frac{\left|\mathbf k \right|}{ \omega_{\mathbf k}}
          \int_{A_P} \mathrm{d} x_a \mathrm{d} y_a\, \left| E_{\mathbf k}\left(x_a,y_a \right) \right|^2 \,.
  \label{PhiFFAP}
\end{align}
Take the maximum value $S_0$ of the power density radiated by the antenna.
Assume a symmetric radiation pattern distribution along the z-axis (for example that of a horn antenna),
then the maximum can be calculated by using equations \eqref{Sradial}, \eqref{hIntegral}
and setting $\theta=\phi=0$ \cite{ulaby1981microwave}
\begin{align}
    S_0 &= S_{\mathbf k}\left( 0, 0 \right) = \frac{1}{8 \pi^2} \frac{\left| \mathbf k \right|^2}{r^2}
                        \frac{\left|\mathbf k \right|}{ \omega_{\mathbf k}}
                        \left| h_{\mathbf k} \left( 0, 0 \right)  \right|^2
          = \frac{1}{8 \pi^2} \frac{\left| \mathbf k \right|^2}{r^2}
          \frac{\left|\mathbf k \right|}{ \omega_{\mathbf k}}
                \left| \int_{A_P} \mathrm{d} x_a \mathrm{d} y_a \, E_{\mathbf k}\left(x_a,y_a \right) \right|^2 \,.
  \label{radS0}
\end{align}
Inserting both \eqref{PhiFFAP} and \eqref{radS0} in \eqref{defD0} gives
\begin{align}
  D_0 = \frac{1}{ \pi} \left| \mathbf k \right|^2
  \frac{\left| \int_{A_P} \mathrm{d} x_a \mathrm{d} y_a \, E_{\mathbf k}\left(x_a,y_a \right) \right|^2}{\int_{A_P} \mathrm{d} x_a \mathrm{d} y_a
                                   \, \left| E_{\mathbf k}\left(x_a,y_a \right) \right|^2} \,.
  \label{AD0}
\end{align}
This is the \emph{maximum directivity} $D_0$ for a monochromatic
field $E_{\mathbf k}$ diffracting through the physical antenna aperture $A_P$.
For \emph{constant illumination}, $ E_{\mathbf k} = \mathrm{const} $, the above equation reduces to
\begin{align}
  D_0 = \frac{1}{ \pi} \left| \mathbf k \right|^2 A_P \,.
  \label{ApD0}
\end{align}
Otherwise, from Schwartz's Inequality for complex valued functions follows
\begin{align}
  \left| \int_{A_P} \mathrm{d} x_a \mathrm{d} y_a \, E_{\mathbf k}\left(x_a,y_a \right) \right|^2
  \leq A_P \int_{A_P} \mathrm{d} x_a \mathrm{d} y_a \, \left| E_{\mathbf k} \left(x_a,y_a \right)  \right|^2
  \notag
\end{align}
This motivates the definition of an \emph{effective antenna aperture} $A_E$ \cite{ulaby1981microwave}
\begin{align}
  A_E &\equiv \frac{\left| \int_{A_P} \mathrm{d} x_a \mathrm{d} y_a \, E_{\mathbf k}\left(x_a,y_a \right) \right|^2}{\int_{A_P} \mathrm{d} x_a \mathrm{d} y_a
                                       \, \left| E_{\mathbf k}\left(x_a,y_a \right) \right|^2} \,,
  \label{defAE}
\end{align}
so that in general $A_E \leq A_P$ and
\begin{align}
  D_0 &= \frac{1}{ \pi} \left| \mathbf k \right|^2 A_E \label{EfD0} \,.
\end{align}
Each monochromatic mode $\mathbf k$ interacts with the physical antenna as if its aperture
had the \emph{effective area} $A_E = A_{E,\,\mathbf k}$
and \emph{ directivity}  $D = D_{0,\,\mathbf k} F_{n,\,\mathbf k}\left( \theta,\phi \right)$,
see Eqs. \eqref{defAE} and \eqref{EfD0}.
In the effective Fresnel-Kirchhoff theory,
the antenna \emph{pattern solid angle} also
becomes a function of wave number $\mathbf k$
\begin{align}
  \Omega_{P,\mathbf k} = \frac{4 \pi^2}{A_{\mathbf k}} \, \frac{1}{\left| \mathbf k \right|^2} \,.
  \label{ApOmegaP}
\end{align}
Recall from the reciprocity theorem that this also holds for an antenna \emph{receiving} power.

Now let an antenna sit at the origin while collecting
monochromatic power of wave number $\mathbf k$ from a source emitting with radiance $L_{\mathbf k}$
(with respect to the origin).
What is the power output $P_A$ of the antenna? From \eqref{radS} and \eqref{radPhi} we
can write the expression for spectral power as
\begin{align}
  \Phi_{\mathbf k} &= \int_{A} \mathrm{d} \sigma
  \int_{\Omega_s} \mathrm{d} \Omega \, \cos{\theta} \, L_{\mathbf k} \left( \theta, \phi \right) \,.
  \notag
\end{align}
Before applying this formula to our antenna problem,
we must take the area integrated over as the effective antenna area $A=A_{E,\,\mathbf k}$.
Also, we must replace  $\cos{\theta}$ with $\eta F_{n,\,\mathbf k}$,
since the portion of energy passing through the effective area is no longer given by projecting
the incoming ray to the surface normal, but by the normalized radiation pattern
multiplied with the radiation efficiency factor.
It is $\eta F_{n,\,\mathbf k}$ that projects energy collection, see \eqref{defRadPo}.
Finally we arrive at a formula for the monochromatic power $P_{\mathbf k}$ collected by the antenna
\begin{align}
  P_{\mathbf k} \,&=\, A_{\mathbf k}\,
        \int_{\Omega_s} \mathrm{d} \Omega \, \eta \, F_{n,\,\mathbf k}\left( \theta,\phi \right)\, L_{\mathbf k} \left( \theta, \phi \right) \,,
  \label{radPhik}
\end{align}
with $\Omega_s$ the solid angle subtended by the source.
Then the \emph{total power output} $P_A$ received by an antenna
of radiation efficiency $\eta$ and radiation pattern $F_n \left( \theta,\phi \right)$
from a source with radiance $L_{\mathbf k}\left( \theta,\phi \right)$ subtending the solid angle $\Omega_s$
results from integrating over all available modes
\begin{align}
  P_A\,&=\,\eta \, \int \mathrm{d} |\mathbf{k}| \,P_{\mathbf k} \,
        =\,\eta\,\int \mathrm{d} k A_{\mathbf k}\,
          \int_{\Omega_s} \mathrm{d} \Omega \, \, F_{n,\,\mathbf k}\left( \theta,\phi \right)\, L_{\mathbf k} \left( \theta, \phi \right)\,.
  \label{totPowerOut}
\end{align}

\subsection{Antenna temperature}
\label{sec:ATemp}
The concept of \emph{antenna temperature} was first introduced by R. Dicke
in a famous 1946 article on microwave radiometry \cite{Dicke1946}, where
the power output $P_A$ of an antenna enclosed by a perfect black-body cavity
was related to the temperature $T$ of the radiation inside the enclosure,
by equating $P_A$ with the thermal (Nyquist) noise of a matched resistor.
Here we re-derive the relation between radiation temperature and
antenna power output, which will be needed to characterise our experiment.

To calculate the total power output $P_A$ of an antenna \emph{completely immersed} in thermal
radiation inside a black cavity, we integrate \eqref{totPowerOut} over the full solid angle
$\Omega_s=4\pi$ and each available mode $\mathbf k$.
The thermal radiance is given by $\lbk$.
Since thermal radiation is unpolarized by definition, and antennas are
\emph{polarized} receivers \cite{kraus2002antennas,ulaby1981microwave}
we must divide expression \eqref{totPowerOut} by two, since the antenna can only pick one
of the two available \emph{transverse polarizations}. This gives
\begin{align}
  P_A \, &=\, \frac{1}{2}\, \eta\, \int \mathrm{d} k  A_{\mathbf k}
  \int_{4 \pi} \mathrm{d} \Omega \, F_{n,\,\mathbf k} \left( \theta,\phi \right) \lbk
  \notag \\
       &= \frac{1}{2}\, \eta\,\int \mathrm{d} k \, \lbk \, A_{\mathbf k} \,
       \int_{4 \pi} \mathrm{d} \Omega \, F_{n,\,\mathbf k}\left( \theta,\phi \right) \,, \notag
\end{align}
which by \eqref{defPSA} and \eqref{ApOmegaP} simplifies as
\begin{align}
  P_A \,&=\, \frac{1}{2}\eta \int \mathrm{d} k \, \lbk \, A_{\mathbf k} \, \Omega_{P,\mathbf k} \,
               =\, \frac{1}{2}\eta \int \mathrm{d} k \, \lbk\, \frac{4 \pi^2}{|\mathbf k|^2} \,.\notag
\end{align}

Changing variables to cancel the squared wavelength factor
$\lambda^2 \equiv \frac{(2\pi)^2}{|\mathbf k|^2}$
we obtain finally
\begin{align}
  P_A \, &=\, \frac{1}{2} \eta \int \mathrm{d} \omega_{\mathbf k} \, \lbw\, \frac{4 \pi^2}{|\mathbf k|^2} \notag \\
       \, &=\, \frac{1}{2\pi} \eta \,\int \mathrm{d} \omega_{\mathbf k} \,
                \frac{\omega_{\mathbf k} }{e^{\beta \omega_{\mathbf k}} -1 } \,.
  \label{antPhi}
\end{align}
Note that since equation \eqref{antPhi} is expressed only in terms of the frequency,
it is \emph{independent of any specific dispersion law for photons in the thermal gas}.
\emph{The antenna power output $P_A$ is only sensible to the available propagation modes
of the thermal gas}, which constrain the frequency spectrum available for integration in \eqref{antPhi}.

To write \eqref{antPhi} in SI units recall again that converting from natural units one has
$\omega_{\mathbf k} \rightarrow h\nu$, $\beta \rightarrow \frac{1}{k_B T}$.
The ratio $\eta$ is unit-less.
To have appropriate dimensions, expression \eqref{antPhi} must be divided by $\hbar$.
Then the \emph{total power} $P_A$ collected by an antenna of radiation efficiency $\eta$
fully immersed in thermal radiation is written in SI units as
\begin{align}
  P_A \, &=\,\eta \int \mathrm{d}\nu \, \frac{h\nu}{e^{h\nu/k_B T} -1 } \,.
  \label{antPowerSI1}
\end{align}

In the \emph{Rayleigh-Jeans regime} $h\nu \ll k_B T$, and we can approximate
the exponential function in the denominator of \eqref{antPhi} as
$e^{h\nu / k_B T} = 1 + \frac{h\nu}{k_B T}+ \dots$. In this regime,
\begin{align}
  P_A \, &=\,k_B\,\eta T\, \int \mathrm{d}\nu \,.
  \label{antPowerSI-RJ}
\end{align}
Let the antenna be connected to a receiver of bandwidth $\Delta\nu \equiv \nu_2 - \nu_1$,
with $\nu_2 > \nu_1$ and both frequencies inside the Rayleigh-Jeans zone.
If $\Delta\nu$ lies \emph{ entirely within the available cavity frequencies},
the power delivered from the antenna then results from integrating \eqref{antPowerSI-RJ}
over the whole bandwidth $\Delta\nu$
\begin{align}
  P_A\, &= \, k_B\, \eta T \, \Delta \nu \,.
  \label{antPowerSI}
\end{align}
This simple but important equation confirms what we discussed before:
The \emph{output power} at the terminals of an antenna \emph{fully immersed in thermal radiation}
is \emph{solely determined by the radiation cavity temperature T and accessible bandwidth $\Delta\nu$}.
Let the radiation temperature in the cavity be $T=T_R$.
We can then define the  \emph{temperature} $T_A$ of
an antenna placed inside the cavity,
collecting radiation over a finite bandwidth $\Delta\nu$, as
\begin{align}
  T_A \,=\, \eta T_R\,=\,\frac{P_A}{k_B\Delta\nu}\,.
  \label{defAntennaT}
\end{align}
If $\Delta\nu$ does not lie within the region of allowed frequencies,
there is no power output possible, $P_A=0$ and the antenna temperature is zero.

\subsubsection{The thermal U(1) gas}

For the U(1) thermal photon gas, propagation over the entire spectrum $\nu>0$ is possible
(up to geometrical discretization effects which we may ignore, as discussed in section \ref{sec:radFormula}).
Then \eqref{antPowerSI} for the U(1) case is just
\begin{align}
  P_A^{\scriptscriptstyle{U(1)}} \,&=\, k_B\, \eta T \, \Delta \nu \,.
  \label{PU1}
\end{align}

\subsubsection{The thermal SU(2) gas}

As we saw in section \ref{sec:polarization}, the frequency spectrum available to the
SU(2) photon is bounded from below by a \emph{cutoff frequency} $\nu^*(T)$.
Below $\nu^*$, \emph{no propagation is possible}.
Let the detectable antenna bandwidth be $\Delta\nu=\nu_2-\nu_1>0$.
An antenna completely bathed in thermal SU(2) radiation will provide the power output
\begin{align}
  P_A^{\scriptscriptstyle{SU(2)}} \,&=\,
  \begin{cases}
    k_B\, \eta T \, \Delta \nu  & \text{if } \nu_1 \geq \nu^* \\
    k_B\, \eta T \, \left( \nu_2-\nu^* \right)  & \text{if } \nu_1 < \nu^* < \nu_2 \\
    0                                         & \text{otherwise} \,.
    \end{cases}
  \label{PSU2}
\end{align}
If the bandwidth $\Delta\nu$ lies entirely below the cutoff frequency $\nu^*$,
the antenna power output of a thermal SU(2) photon gas is
zero\footnote{Up to antenna and receiver noise, of course. This is discussed in section \ref{sec:radiometer}.}.
This is in contrast to the U(1) case \eqref{PU1} which always produces a non-zero antenna temperature
proportional to the radiation temperature in the cavity and bandwidth.
The SU(2) gas dynamics manifests themselves as an \emph{antenna temperature gap} below $\nu^*$
This is the \emph{key experimental prediction} from SU(2) thermodynamics,
as developed in chapter \ref{chap:YM}, that is accessible by radiometric means.
To measure the antenna temperature some type of radiometer must be connected to the
output of the antenna. Such experimental configurations are described in the next section.

\section{Radiometer systems}
\label{sec:radiometer}

Before tackling the problem of experimental detection of the
antenna temperature gap described in section \ref{sec:ATemp},
we need to familiarise ourselves with basic radiometric techniques.
The purpose of this section is to give a basic review of the subject,
a deeper treatment can be found in the literature
\cite{ulaby1981microwave,skou2006microwave}.
First we introduce basic concepts, then we describe in brief the
two radiometer configurations one could employ to measure SU(2) effects.
In this section we always take the radiometer bandwidth $\Delta\nu$
to lie outside the screening region of the SU(2) gas and within the Rayleigh-Jeans zone.
This is just to simplify our introductory review. Discussions about screening effects
and their radiometric detection are left to section \ref{sec:Radiometry}.

\subsection{Basic definitions}

A radiometer system is a radiation measurement device
composed of an antenna coupled to a receiver.
The receiver selects a bandwidth $\Delta\nu$ for detection,
and produces a signal characterizing the collected radiation.

In the ideal case, \emph{power radiometers} produce a voltage signal which is directly proportional to
the power output of the connected antenna,
and by way of \eqref{antPowerSI} also proportional to the antenna temperature
\begin{align}
  V_o \,=\, c\,G\,P_A \,= \,c G\, \Delta\nu \, k_B \, T_A \,.
  \label{RadPow0}
\end{align}
Here $G$ is the \emph{amplification factor} of the radiometer,
and $c$ a calibration factor.
A correct calibration is of absolute
necessity if $T_A$ in \eqref{RadPow0} is to be interpreted as a temperature at all.
Different calibration methods are discussed in \cite{skou2006microwave,ulaby1981microwave}.
In the following, we assume that the radiometer
has been correctly calibrated, and absorb the factor $c$ into $G$.

In general, random noise (mostly thermal in nature) intrinsic to the radiometer
apparatus will distort the output voltage \eqref{RadPow0},
so that it is not a strictly linear function of
the antenna output power $P_A$.
It can be shown however \cite{skou2006microwave,ulaby1981microwave}
that radiometer noise is well modelled by considering it as
an additive contribution to the measured signal (the antenna power output).
We thus define the \emph{radiometer noise power\ $P_N$}
as the \emph{power that would be measured by the radiometer
viewing a source at zero absolute temperature}.

We can also write $P_N$ in terms of a temperature,
the \emph{radiometer noise temperature} $T_N \equiv P_N/( G\, k_B \,\Delta\nu)$.
Thus, radiometer noise may be interpreted as an additional temperature $T_N$
collected along with the antenna temperature $T_A$ by an ideal noiseless radiometer.
The noise power (noise temperature) is an intrinsic characteristic of the radiometer.
Allowing for thermal noise, the total radiometer output power then reads
\begin{align}
  V_o \,&= \, G (P_A + P_N) \,= \, G \,\Delta\nu\, k_B \,(  T_A + T_N) \,,
  \notag \\
  &\equiv  G\,\Delta\nu\, k_B \, T_o \,,
  \label{defTRad}
\end{align}
where we have defined the \emph{radiometer output temperature} $T_o = V_o / (G\,\Delta\nu\, k_B)$.
The radiometer output temperature characterizes the \emph{average} power output of a radiometer.
Thermal noise contributes to the stochastic fluctuations in the power output.
These are in turn characterized by the \emph{radiometer sensitivity}
$\Delta T_o$, which gives the minimum detectable change in the radiometer temperature.
In the case of thermal radiation, both radiometer temperature $T_o$
and sensitivity $\Delta T_o$ fully characterize the radiometer
output\footnote{The signal from a radiometer looking at thermal radiation
is normal-distributed \cite{ulaby1981microwave}. See also section \ref{sec:totpowrad}.}.
In the following, we describe two useful radiometer configurations,
and give the corresponding formulas for radiometer temperature and sensitivity.

\subsection{Total power radiometer}
\label{sec:totpowrad}

\begin{figure}[t]
  \begin{center}
    \includegraphics[width=140mm]{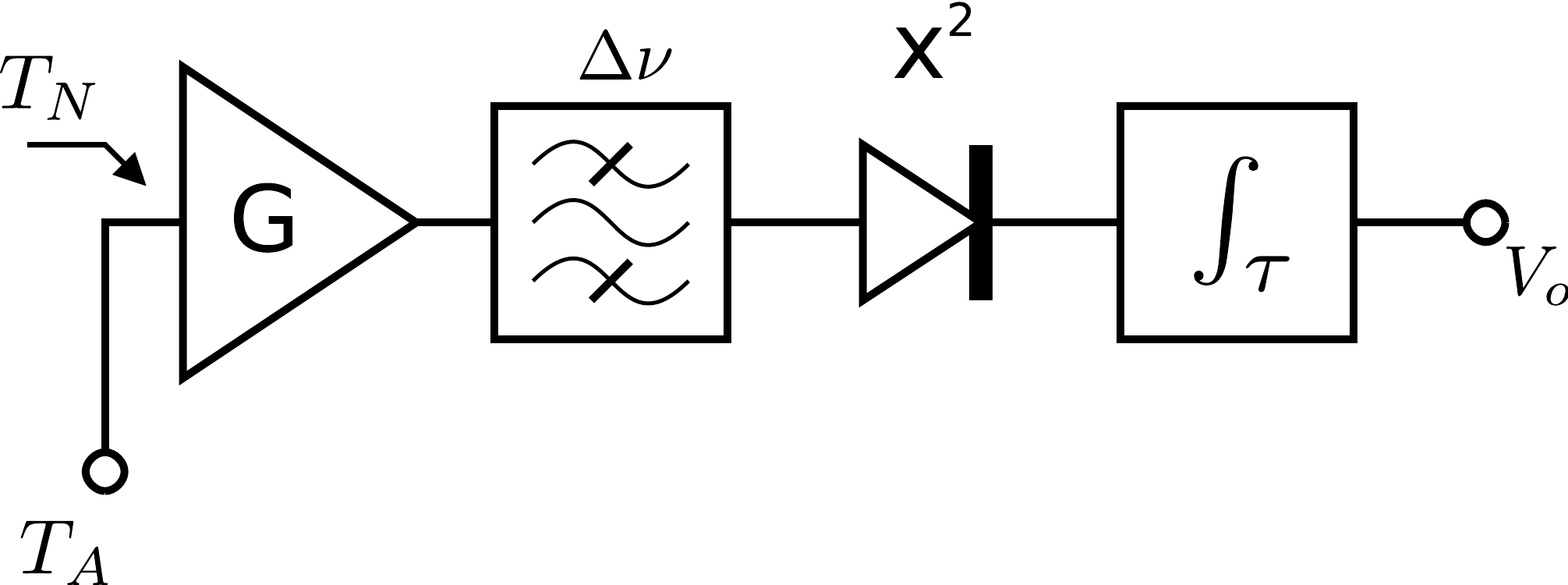}
  \end{center}
  \caption{Diagram of a basic Total Power Radiometer: The input signal to the radiometer is characterized
  by the antenna temperature $T_A$, which is determined by the radiation
  collected in the antenna, and the intrinsic radiometer noise temperature $T_N$,
  which fully characterizes thermal noise in the radiometer.
  The signal is amplified with gain $G$, then filtered to select a narrow bandwidth $\Delta\nu$.
  The (microwave) signal power is detected by a square-law diode.
  Finally, the signal is integrated over a period $\tau$ to smooth out random fluctuations.
  }
  \label{fig:totalrad}
\end{figure}

The \emph{total power radiometer} is the most basic type of radiometer
designed to measure antenna temperature. Its design is sketched in figure \ref{fig:totalrad}.
An antenna of total temperature $T_A$ is coupled via a transmission line  to a receiver consisting of four basic stages.
The first is \emph{signal amplification}, which is carried out by an amplifier of \emph{gain} G.
This is followed by \emph{bandwidth selection} by a filter of narrow bandwidth $\Delta\nu$.
In the\emph{detection} stage, the input signal is converted to a measurable voltage.
For a total power radiometer this is achieved by a semiconductor diode %
working as a \emph{square law detector}: The diode output voltage is proportional to
the input signal power $P_A + P_N$.
In the fourth and final stage, the voltage signal coming out of the detector
is \emph{smoothed} by \emph{integration} over a time period $\tau$ to reduce random fluctuations.
This integrator effectively acts as a low pass filter,
suppressing frequencies higher than $1/\tau$,
where $\tau$ is the \emph{radiometer integration period}.
Note that the integrator causes the radiometer output to be \emph{Gauss-distributed},
independently of the stochastic distribution of the signal.
This can be understood as follows: A single integration over a period $\tau$
of a signal with arbitrary stochastic distribution
is equivalent to taking the average of many single measurements
of said signal. But from the central limit theorem, this average
will tend to be \emph{noramlly distributed}.
The radiometer output after integrations is thus
fully characterized by the average output signal and its variance,
independently of the stochastic distribution of the input signal.

The noise performance of the radiometer system (antenna plus receiver)
can be characterized the radiometer noise temperature $T_N$.
A useful model for $T_N$ is given below.

The total power radiometer as depicted in figure \ref{fig:totalrad} is
therefore fully characterized by specifying a few parameters.
While the stages of bandwidth selection and time integration are self-explanatory,
we believe the subjects of signal amplification and radiometer noise temperature
merit some further discussion.

\subsubsection{Signal amplification}

Signal amplification in modern radiometers is realized by semiconductor devices.
In the microwave region ($\nu\sim\mathrm{1-100GHz}$), High Electron Mobility Transistors (HEMT)
are the ideal solid state amplifier because of high signal gain and low noise.
HEMTs are a form of Field Effect Transistors (FETs) so constructed that
its charge carriers are confined to a very narrow 2-dimensional layer,
forming a \emph{two-dimensional electron gas} (2-DEG).
Transport properties of this 2-DEG are superior to that of a usual FET
since, due to the absence of donor impurities, scattering is greatly reduced,
thus higher electron mobility and low intrinsic noise is achieved \cite{ali1991hemts,silver1984microwave}.
This advantages comes at a cost, however.
HEMTs at cryogenic temperatures suffer from relatively
large fluctuations in the amplification factor (see e.g. \cite{Gallego2004}).
Let the amplification factor variate in time as $G(t)\,=\,G+\delta G(t)$.
Then the standard deviation $\Delta G \equiv \sqrt{ \langle \delta G(t)^2 \rangle }$,
called \emph{gain variation}, is the \emph{dominant} factor influencing
radiometer sensitivity $\Delta T$.
A clever technique conceived by
Dicke \cite{Dicke1946} avoids this difficulty by alternately  measuring
the signal and some other known reference at a higher frequency than the characteristic
gain variation frequency of $G(t)$, called the \emph{knee frequency} $f_{\mathrm{knee}}$.

The defining parameters of an amplifier are thus its
gain $G$, its mean fluctuation or gain variation $\Delta G$ and its
\emph{characteristic frequency} $f_{\mathrm{knee}}$.
The intrinsic amplifier noise contribution to total signal noise
will be included in the \emph{receiver temperature noise}
$T_{\mathrm{rec}}$ to be defined below.

\subsubsection{Noise and attenuation}

Thermal noise due to random motion of electrons
in a conductor is the basic source of random noise in a radiometer,
but not the only one. Other types of noise include \emph{quantum noise},
\emph{shot noise} and \emph{flicker noise} \cite{ulaby1981microwave}.
As the name suggests, quantum noise is a consequence of
quantum-mechanical effects in conducting devices and manifests
itself as discrete random fluctuations in the transported signal.
In the Rayleigh-Jeans limit such quantum effects can be ignored.
Shot noise arises from current flow fluctuations in conducting devices
due to the discrete nature of charge carriers.
Flicker noise is caused by surface irregularities in semiconductors.
All this fluctuations contribute to the noise output power,
which can be  characterised
by a noise temperature $T_N$, as explained above.

A perfect knowledge of each radiometer component would make it possible
to calculate the random noise contribution of each
waveguide, switch, resistor, etc. to the total radiometer noise.
Instead, we use a very simple model that nevertheless is good enough
for most purposes \cite{ulaby1981microwave}, where noise is characterized
by a few measurable parameters.

Let an antenna with \emph{radiation efficiency} $\eta$ (see Sec. \ref{sec:AntennaParm})
be coupled through a transmission line of \emph{loss factor} $L$
to a receiver with characteristic \emph{receiver temperature} $T_{\mathrm{rec}}$.
The antenna and transmission line are assumed to be in
thermodynamical equilibrium, at a \emph{physical temperature} $T_p$.
The \emph{radiometer noise temperature} $T_N$ is then given by \cite{ulaby1981microwave}
\begin{align}
  T_N\,&=\, \left( L - \eta \right) T_{p,A} + L T_{\mathrm{rec}} \,.
  \label{TNDef}
\end{align}
The antenna radiation efficiency $\eta$ and loss factor $L$ of the transmission line
can always be measured.
The temperature $T_{p,A}$ is the \emph{physical temperature}
of the antenna and transmission line coupling at the time of measurement.
This is different from the \emph{antenna temperature} $T_A$
which characterizes the \emph{power} collected by the physical antenna aperture.
The receiver temperature $T_{\mathrm{rec}}$ may be defined
as the temperature that is measured by the receiver when disconnected from the antenna.
It characterizes all noise sources internal to
the receiver.
It is strongly dependent on the \emph{physical temperature} $T_{p,\mathrm{rec}}$
of the receiver at the time of measurement.
The value of $T_{\mathrm{rec}}$ and its dependenc on $T_{p,\mathrm{rec}}$
is specific for every receiver and must always be measured
before any experiment.
Here, however, we can only estimate $T_{\mathrm{rec}}$ based on the values
given in the literature for radiometers designed to measure radiation
in the range of temperatures and frequencies we are interested in.
We use a simple ad-hoc model
\begin{align}
  T_{\mathrm{rec}}(T_{p,\mathrm{rec}}) \, &= \, a_{\mathrm{rec}} T_{p,\mathrm{rec}} + c_{\mathrm{rec}} \,.
  \label{defTrec}
\end{align}
where the receiver noise temperature $T_{\mathrm{rec}}$
is a linear function of its physical temperature $T_{p,\mathrm{rec}}$.
Parameters $a_{\mathrm{rec}}$ and $c_{\mathrm{rec}}$ are chosen
so that $T_{\mathrm{rec}}(T_{p,\mathrm{rec}})$ roughly matches the same order of magnitude
as given in the literature for a receiver working at temperature $T_{p,\mathrm{rec}}$.
This will be explained in more detail in section \ref{sec:Radiometry}.

\subsubsection{Total power radiometer output temperature and sensitivity}

As discussed before, the total power radiometer outputs a voltage signal $V_o$
which may be interpreted as a radiometer output temperature $T_o$,
defined in equation \eqref{defTRad}.
Here and in the following we assume that the radiometer has
been properly calibrated so that we may always calculate $T_o$ from $V_o$.

The measured \emph{radiometer output temperature} $T_o$ of the total power radiometer is then
\begin{align}
  T_o \,&= \, T_A + T_N 
  \notag \\
                   \,&= \,\eta T_A + \left( L - \eta \right) T_p + L T_{\mathrm{rec}}
                   \, = \, \eta T_A + \left( L - \eta \right) T_p +
                   L\left(a_{\mathrm{rec}} T_{p,\mathrm{rec}} + c_{\mathrm{rec}} \right) \,.
  \label{Ttotal}
\end{align}

The radiometer sensitivity is determined by the average fluctuation of the output temperature.
We anticipate that this average fluctuation is affected by the high gain variation of HEMT-type amplifiers.
On the other hand, we expect the integrator to function as a low-pass filter and suppress
fluctuation frequencies much higher than the inverse of the integration period.
Both conjectures turn out to be true. Namely, it can be shown that
the \emph{sensitivity} of the total power radiometer with
bandwidth $\Delta\nu$, integration period $\tau$, gain $G$ and gain variation $\Delta G$
is given by the expression  \cite{ulaby1981microwave}
\begin{align}
  \Delta T_o \, = \, T_o \,\sqrt{\frac{1}{\Delta\nu\,\tau} + \left( \frac{\Delta G}{G} \right) ^2} \,,
  \label{dTtotal}
\end{align}
with $T_o$ given as in \eqref{Ttotal}.

Note that both the integration period and gain variation appear as
independent terms under the square root.
This means that large uncertainties in the
output temperature due to high gain variation $\Delta G$ cannot
be compensated just by choosing a higher integration time.
It is therefore very important to minimise gain variations as much as possible.
Both expressions \eqref{Ttotal} and \eqref{dTtotal} will later
be used to simulate experimental results.

\subsection{Dicke radiometer}

\begin{figure}[t]
  \begin{center}
    \includegraphics[width=140mm]{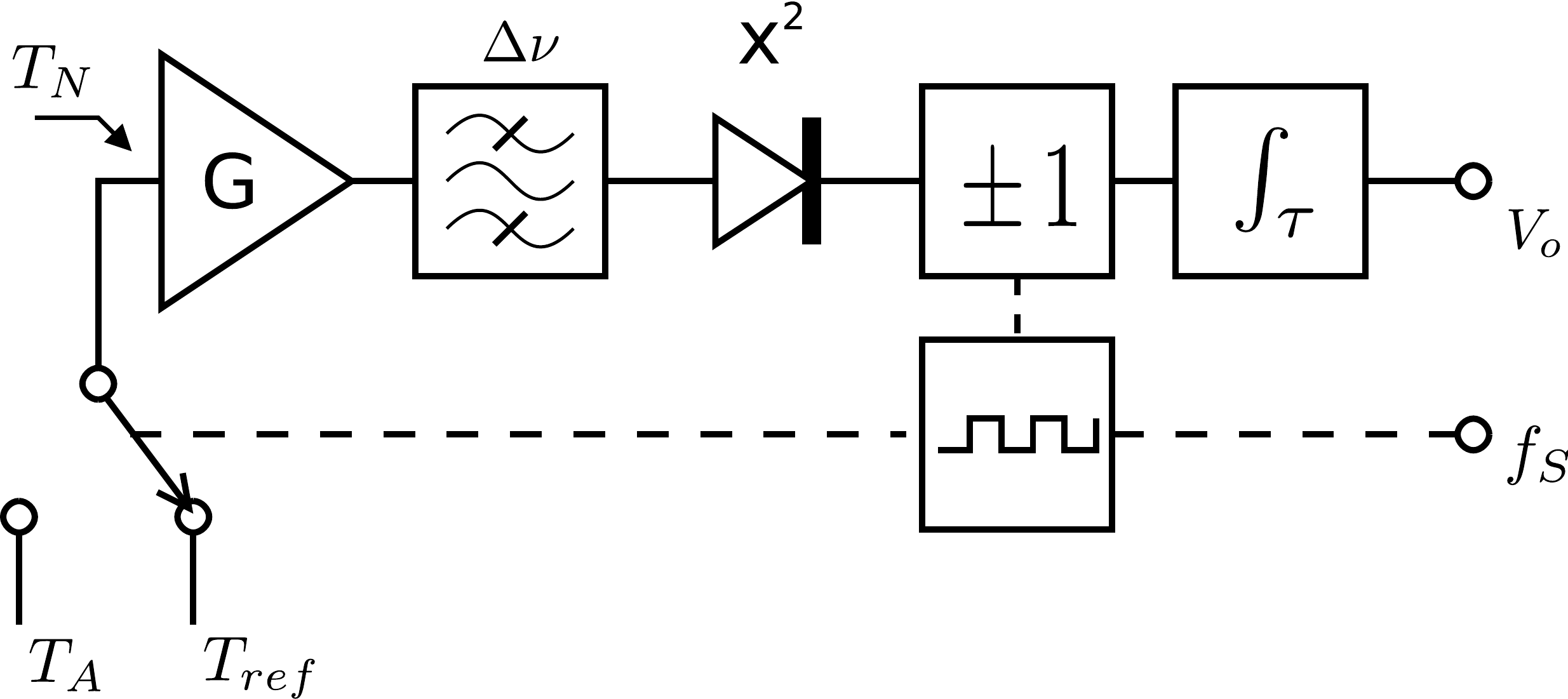}
  \end{center}
  \caption{Diagram of a basic Dicke Radiometer: The receiver input is rapidly
  switched between the antenna signal characterized by $T_A$ and a reference signal
  with temperature $T_{\mathrm{ref}}$. The switching occurs at a frequency $f_S$
  chosen so that it is greater than the characteristic fluctuation frequency of the
  amplifier, $f_S>f_{\mathrm{knee}}$.
  Then for each switching period the amplifier gain may be taken to be constant,
  and a \emph{signal difference} $T_A-T_{\mathrm{ref}}$ can consistently be defined.
  See text for details.
  }
  \label{fig:dickerad}
\end{figure}

A basic Dicke Radiometer configuration is shown in figure \ref{fig:dickerad}.
It is based on a clever switching technique introduced in 1946 by Dicke \cite{Dicke1946}
that suppresses noise power output produced due to amplifier instabilities in total power radiometers.
Dicke's method effectively prevents amplifier gain fluctuations from
degrading the sensitivity of the radiometer \cite{Dicke1946}.
This is realized by \emph{switching} the receiver input
back and forth between the antenna signal $T_A$
and a \emph{reference signal} $T_{\mathrm{ref}}$,
at a \emph{switching frequency} $f_S$ greater than the characteristic amplifier frequency $f_{\mathrm{knee}}$.
After amplification and detection,
the output of the square-law diode is multiplied before integration by $+1$ or $-1$,
depending on the position of the Dicke switch.
The signal integrator then adds (or subtracts) the power coming out of the
square-law diode into the total output signal $V_o$, which will be given by the difference
between antenna and reference signals..
An expression for $V_o$ can be derived as follows:
Let the switching frequency $f_S$ be large enough so 
that the amplifier gain $G(t)= G + \Delta G(t)$ may be regarded as constant over one switch period $1/f_S$,
further let the integration period $\tau$ be much larger than  $1/f_S$.
Then for each switch period,
the \emph{power difference} $G(t) (P_A - P_{\mathrm{ref}})\equiv G(t) \,\Delta\nu\, k_B \,( T_A - T_{\mathrm{ref}})$
at an arbitrary time $t$  may be consistently defined,
since within each period the amplifier gain can be assumed constant, $G(t) \approx G(t + 1/(2f_S))$.

The power difference is integrated over a period $\tau$,
and the average output voltage after integration is given directly by \cite{skou2006microwave}
\begin{align}
  V_o \,&= \, G (P_A + P_N) - G (P_{\mathrm{ref}} + P_N) \,= \, G \,\Delta\nu\, k_B \,( T_A - T_{\mathrm{ref}})
  =  G\,\Delta\nu\, k_B \, T_o \,.
  \label{defTRadD1}
\end{align}
The noise temperature $T_N$ cancels in expression \eqref{defTRadD1},
and the radiometer output temperature $T_o$ is just the difference between antenna and reference temperature.
The average gain $G$ now multiplies a temperature \emph{difference}.
By choosing $T_{\mathrm{ref}}$ sufficiently close to $T_N$, we can thus
suppress the effects of gain fluctuations in the measured signal.
Ideally, $T_A=T_{\mathrm{ref}}$ so that the output signal becomes precisely zero
and the dependence on $G$ vanishes. 

Note however that in Eq. \eqref{defTRadD1} we have assumed  the noise power output of the radiometer
to be just an additive contribution $P_N$ to the total power output $P_o$.
It was further assumed that the generated noise power is constant
over both switch measurements.
Nonlinear effects in the radiometer manifest themselves
as a slight shift from the ideal case \eqref{defTRadD1}.
This we discuss below.

\subsubsection{Dicke radiometer output temperature and sensitivity}

Here we give the output temperature and sensitivity formulas for a properly calibrated Dicke radiometer.
We use \eqref{TNDef} and \eqref{defTrec} to model radiometer noise.
A refinement of Eq. \eqref{defTRadD1} must be made to account for nonlinear effects \cite{Jarosik2000}.
The \emph{radiometer output temperature} $T_o$ of a Dicke radiometer is then given by the expression
\begin{align}
  T_o \,= \, T_A - T_{\mathrm{ref}} + T_{\mathrm{off}} \,.
  \label{Tdicke}
\end{align}
The switching noise temperature $T_{\mathrm{off}}$ results from non-idealities
in radiometer components, even when comparing between identical signal temperatures
\cite{Jarosik2000}.
It cannot be included within the receiver noise temperature $T_N$, since $T_N$
cancels in our simple linear noise model. As given in \cite{Arcade2}, it is of the order of $10^{-1}$\,K.

The minimum detectable temperature difference of a Dicke radiometer is given
by a modification of \eqref{dTtotal} to take into account the diminishing influence of
gain fluctuations
\begin{align}
  \Delta T_o \, =  \,\sqrt{\frac{2}{B\tau}\left[ \left( T_A+T_N \right)^2 + \left( T_{\mathrm{ref}} + T_N \right)^2 \right]
  + \left( \frac{\Delta G}{G} \right) ^2 \left( T_A-T_{\mathrm{ref}} \right)^2} \,,
  \label{dTdicke}
\end{align}
where $T_N$ is defined as in equation \eqref{TNDef}.
As was expected, the effect of $\Delta G$ can be made negligible,
by choosing $T_A-T_{\mathrm{ref}}$ to be as close to zero as possible.
This is the great advantage of a Dicke radiometer setup.

\section{Radiometric detection of the SU(2) ground state}
\label{sec:Radiometry}

We now arrive at the main section of this chapter, where we describe possible
ways to detect the presence of a thermal ground state.

\subsection{Detection principle}

There are two ways in which the problem of experimentally testing the SU(2) photon hypothesis can be approached.
\begin{enumerate*}
    \item \label{list:dp1} Measure the total output power of the photon gas directly
      at the \emph{radiation temperature} $T_R$ and frequency $\nu$. From \eqref{PSU2}, a very sharp decline
      of output power below the cutoff frequency $\nu^*(T_R)$ would corroborate the predicted
      \emph{spectral gap}, and the SU(2) hypothesis.
    \item \label{list:dp2} Measure the output power difference between a prepared U(1)
      and an SU(2) photon gas cavity, both at \emph{radiation temperature} $T_R$.
      From \eqref{PU1} and \eqref{PSU2}, this difference should
      be zero above the cutoff frequency $\nu^*(T_R)$ and non-zero below.
\end{enumerate*}
We saw in chapter \ref{chap:inhtd} that the thermal ground state
can be deformed by temperature gradients and still remain
a good a priori description of the SU(2) gas in the adiabatic approximation.
However, the thermal radiation formulas derived in this chapter always
assume a homogeneous temperature distribution.
To ensure the validity of these formulas,
temperature gradients within the cavity should be carefully controlled
to be as \emph{small as possible}.

From section \ref{sec:restoringu1} we also know that it is possible to deform
the SU(2) ground state with static electric fields
so that the thermodynamical assumptions from which it was derived are no longer valid.
The emitted radiation then becomes physically indistinguishable from that of the U(1) gas.
When probing for SU(2) thermodynamics one should be very careful
with the experimental conditions so as to avoid this.
The cavity containing the SU(2) thermal gas should be shielded
from any stray electromagnetic radiation.
For the low temperature region of experimental interest,
this shielding should guarantee an electromagnetic field energy density
of at most  $|\vec{E}|=10^{-3}\,\mathrm{mV/cm}$.

On the other hand, the destruction of the SU(2)
ground state gives us the possibility of generating U(1)
behaviour to use as a reference.
This will be exploited in the design of the experiment.
The U(1) gas is realized by applying a sufficiently large static electric field.
This is explained in the next section.

\subsection{Experimental setup}

\begin{figure}[t]
  \begin{center}
    \includegraphics[width=150mm]{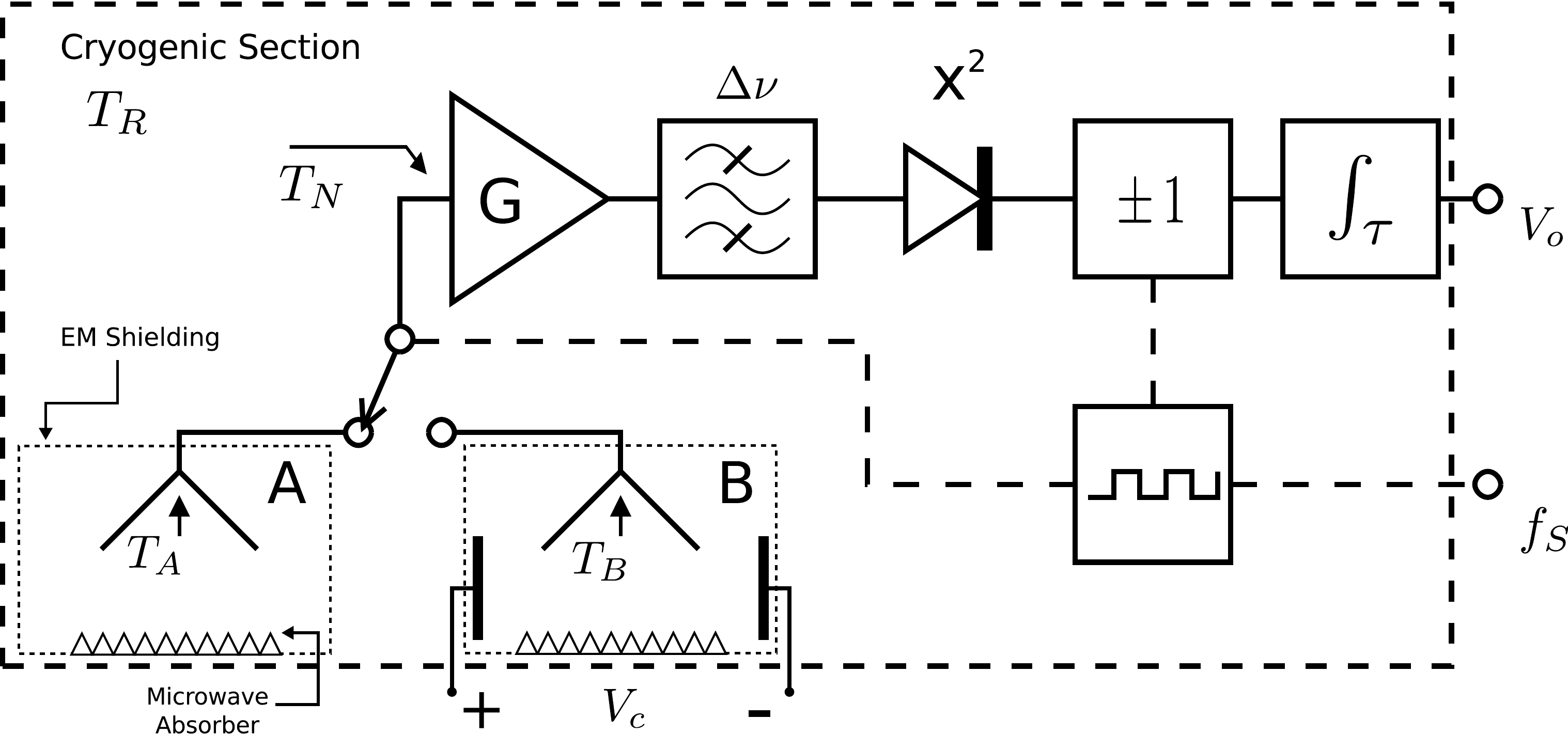}
  \end{center}
  \caption{Basic Radiometer Setup for the detection of thermal SU(2) effects.
  Inside a cryogenic chamber at temperature $T_R$ we place two identical cavities A and B,
  of side dimensions $L \sim 1\mathrm m$. The interior cavity walls of both cavities
  are fully coated with a thick enough layer of microwave absorber with $\varepsilon \geq 0.99$ emissivity.
  (This is shown in the diagram only for one wall). Cavity B is further equipped with capacitor
  plates opposite to each other. They connect to a voltage source $V_c$ to generate a
  static electric field of adjustable energy density inside the cavity.
  Each cavity is also shielded from stray electromagnetic fields.
  Identical horn antennas are placed inside each cavity, both of which connect to the radiometer
  through a Dicke switch. The radiometer is also placed inside the cryogenic chamber, so that
  the whole system: cavities, antennas, couplers and radiometer is at the same thermodynamic temperature $T_R$.
  }
  \label{fig:radiometersu2}
\end{figure}
The basic setup of our experiment is sketched in figure \ref{fig:radiometersu2}.
Take a cryogenic chamber
capable of operating in a specified temperature range $T_{min}\dots T_{max}$.
This range will be determined later when the radiometer bandwidth is chosen.
The dimensions of the cryogenic chamber should be large enough to accommodate two
identical rectangular cavities.
Let each cavity be of side length $L_i\,,\,i=1,2,3$, and take each $L_i$ to be
large enough so that modifications due to cavity geometry
to the Planck expression for black-body spectral energy density \eqref{ED_Planck}
may be safely ignored.
These corrections are given in \eqref{DiscEff}.
The spectral region of experimental interest is $5\, \mathrm{GHz} < \nu < 15\, \mathrm{GHz}$ (see below).
In this frequency range, taking each length to be of the order of $L_i \sim 1\mathrm m$
ensures that discretization corrections \eqref{DiscEff} are at most of the order $10^{-4}$
\cite{baltes1972problems} and may safely be ignored.

Label each cavity A and B, as shown in figure \ref{fig:radiometersu2}.
The walls of A and B are covered with a thick enough layer of \emph{microwave absorber}
of the largest possible \emph{emissivity} $\varepsilon \geq 0.9999$,
so that radiation within each cavity is completely thermalized.
As discussed in chapter \ref{chap:inhtd}, in order to avoid ground state
decoupling effects, A and B must further be shielded from stray electromagnetic radiation.
We will be working with temperatures of at least 8\,K.
The ground state energy density at this temperature is equivalent
to the electric field density of an electric field of magnitude $|\vec{E}|=2.216\,\mathrm{mV/cm}$
(see Eqs. \eqref{enEfeld} and \eqref{rho_gsSI}).
We therefore require the cavity to be shielded from stray electromagnetic radiation
down to $|\vec{E}|=10^{-3}\,\mathrm{mV/cm}$, to be certain that no decoupling effects occur at all.

Cavity B is further equipped
with two capacitor plates
positioned at opposite sides within the cavity,
connected to an adjustable voltage source $V_c$,
so that a constant  electric field
of  energy density $\rho_c$
can be generated inside B.

A Dicke Radiometer is also brought inside the cryogenic chamber.
The radiometer parameters are discussed in the next sections.
Dicke switching occurs between the terminals of two identical \emph{horn antennas},
labelled A and B. Each antenna is placed inside their corresponding cavity,
as shown in the figure.

In order to specify the necessary experimental parameters, we must first choose
the spectral region and working temperature the radiometer is going to probe.
This is done in the next section.

\subsection{Working frequency and temperatures}

Here we determine the optimal frequency bandwidth for the receiver,
and the corresponding radiation temperature range.

In order to test for screening effects, we need  to measure the propagation
of thermal radiation inside and outside the 
photon screening region (see section \ref{sec:Gprop}). This region is determined
by a frequency $\nu^*=\nu^*(T)$, above and below of which photon propagation is
allowed and forbidden, respectively.

We first define the \emph{middle frequency} $\nu_m$ and \emph{bandwidth} $\Delta\nu$
which will be investigated in this experiment.
Following \cite{Arcade2}, we choose the 10\,GHz Low channel which is used there
with $\nu_m\,=\,9.72\, \mathrm{GHz}$ and $\Delta\nu\,=\,860\,\mathrm{MHz}$.
Now use the results of section \ref{sec:Gprop} to plot $\nu^*$ as
a function of $T$ and select an adequate temperature region.

\begin{figure}[t]
  \begin{center}
    \includegraphics[width=140mm]{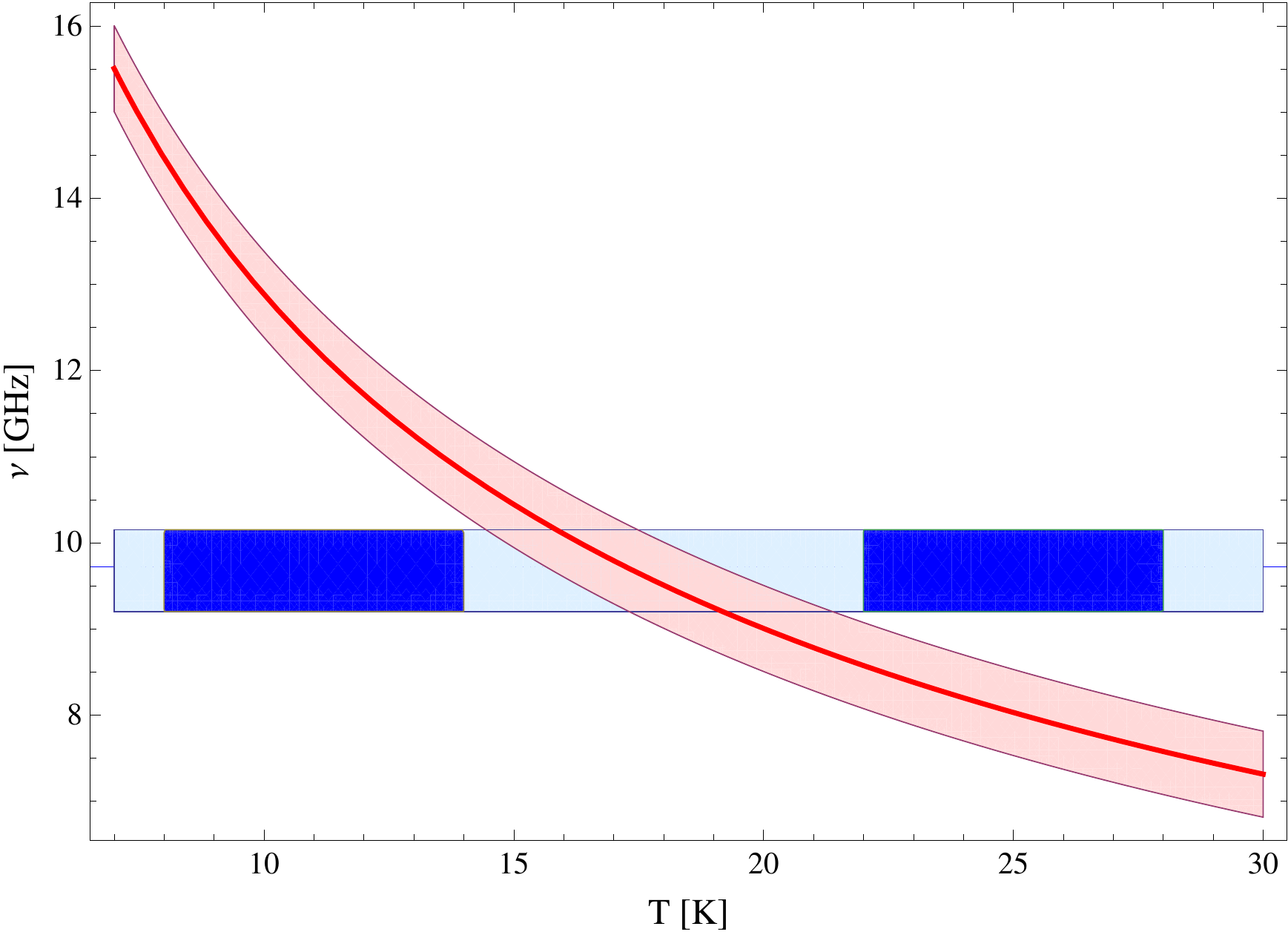}
  \end{center}
  \caption{The red line plots the cutoff frequency $\nu^*$, calculated as in section \ref{sec:Gprop}.
  The region 500\,MHz above and below $\nu^*$ is plotted in light red.
  The 860\,MHz bandwidth available to the radiometer is shown as light blue, centred around the
  frequency $\nu_m\,=\,9.72\, \mathrm{GHz}$. The region lying within available bandwidth and
  working temperature is shown in blue. The temperature regions were so chosen that the measured
  bandwidth lie at least   500\,MHz above or below the temperature-dependent cutoff frequency $\nu^*$
  }
  \label{fig:arbeitspunkt}
\end{figure}

Since we would like to avoid boundary effects such as having the frequency $\nu^*$ lie within the
bandwidth region of the above 10\,GHz channel,
we avoid temperatures for which the bandwidth boundary would lie within 500\,MHz from the cutoff.
The temperature ranges $8\,\mathrm K \, \leq\, T_R \,\leq\, 14\,\mathrm K$ and
$22\,\mathrm K \, \leq\, T_R \,\leq\, 28\,\mathrm K$ match this condition.
In figure \ref{fig:arbeitspunkt} we have plotted the cutoff frequency $\nu^*$ and the
region 500\,MHz above and below it. The working bandwidth of  860\,MHz around
the middle frequency $\nu_m\,=\,9.72\, \mathrm{GHz}$ is visible, along
with the chosen temperature ranges.
We therefore need a cryogenic chamber capable of working between
temperatures $T_{min}=8\,\mathrm K$ and $T_{max}=28\,\mathrm K$.

Note also that with our choice of frequencies and temperatures,
we still lie within the Rayleigh-Jeans region, since for the lowest
temperature, $(h \nu_m)/(k_B T_{min}) = 0.0583 \ll 1$.

The rest of the experimental parameters are taken to be
of the order of magnitude as those in the radiometer configuration used in \cite{Arcade2}.
This is discussed in the next section.

\subsection{Radiometer parameters}

In this section, we will simulate the results of radiometer measurements
predicted from the $\sucmb$ hypothesis described in section \ref{sec:su2cmb}.
The most important radiometric consequence of this hypothesis
is the existence of a frequency cutoff $\nu^*$ below which photon propagation is impossible.
Using the formulas for radiometer temperature and standard deviation, \eqref{Ttotal}, \eqref{Tdicke},
\eqref{dTtotal}, and \eqref{dTdicke}, we can generate expected measurement plots,
so as to have an intuitive idea of the order of magnitude of both the expected signal average and stochastic variation.
But before using the radiometer formulas given above, we must specify certain \emph{characteristic parameters}
of the radiometer.

The parameters used to generate simulation plots are given in table \ref{table:radparms}.
These are taken to be of the same order of magnitude as those described in the
literature \cite{artalradiometers,Jarosik2000,jarosik2003design,pospieszalski2005extremely,Arcade2,singal2005design,smithstathemt96,weinreb1998noise}
for radiometers used in experimental situations closely matching our own.
That is, low temperatures and frequencies in the microwave region.
We have divided them in three sets.
Set I is the parameter set which corresponds to an ideal radiometer with above-average sensitive.
The parameter set II corresponds to a radiometer with average performance.
Set III results in a very noisy radiometer.
This parameter set is further divided in IIIa and IIIb,
the only difference being the gain variation.
The choice of parameters is explained in detail below.

\begin{table}[t]
  \begin{center}
    \setlength{\extrarowheight}{2mm}
    \begin{tabular}{lccccc}
    \toprule
    Radiometer Parameters     &   & I          & II           & IIIa      & IIIb \\
    \cmidrule{1-1}
    Antenna efficiency          &  $\eta$                   & 0.99       &   0.99       &    0.99       &   0.99 \\
    Middle frequency            &  $\nu_m$                  & 9.72   GHz & 9.72   GHz   & 9.72   GHz    & 9.72   GHz\\
    Bandwidth                   & $\Delta\nu$               & 0.86  GHz  & 0.86  GHz    & 0.86  GHz     & 0.86  GHz\\
    Radiometer noise     & $T_{\mathrm{rec}}$        & 0.1$T$+10K & 0.2$T$ + 15K & $T$ + 15K     & $T$ + 15K\\
    Radiometer offset    & $T_{\mathrm{off}}$        & 200  mK            & 500mK        & 1K            & 1K       \\
    Gain variation       & $\left(\Delta G/G\right)$ &-30\,dB=$10^{-3}$   & -20\,dB=$10^{-2}$ & -10\,dB=$10^{-1}$ & -20\,dB=$10^{-2}$   \\
    Loss factor          & $L$                       & 0.5\,dB =$10^{0.05}$ & 1\,dB=$10^{0.1}$   & 5\,dB=$10^{0.5}$ & 5\,dB=$10^{0.5}$     \\
    Integration time     & $\tau$                    & 0.1s       & 1s           & 10s           & 10s      \\
    Switching frequency  & $f_S$                     & 75\,Hz      & 75\,Hz        & 75\,Hz       & 75\,Hz   \\
    \bottomrule
   \end{tabular}
   \caption{Radiometer parameters, see text.  Note that $N_{dB} \equiv 10\,\log{(N)}$.
   \label{table:radparms}}
\end{center}
\end{table}

\emph{Corrugated horn antennas} are a popular choice for microwave applications because
of high beam symmetry, low side-lobes, low crosspolarization and low reflection
over a wide broadband. Horn antennas may be designed to satisfy requirements
such as working frequency, antenna gain and directivity \cite{hornantennas}.
This in turn specifies the expected antenna radiation efficiency, which
we take to be $0.99$.
This sole parameter characterises the antenna response in our simplified example.
Of course, a full characterisation including
radiation beam pattern is needed to precisely describe an actual experiment, see for example \cite{singal2005design}.
Antenna efficiency is taken to be the same in all cases.

\emph{Middle frequency and bandwidth} of the radiometer were chosen following \cite{Arcade2}.
Radiometric equipment in this frequency region is commercially readily available, further, the temperature
region it imposes is such that radiometer and amplifier noise remains low enough
for our purposes.
Higher frequencies show too narrow temperature ranges to test for screening effects,
compare with figure \ref{fig:arbeitspunkt},
and we would like to measure screening effects over a relatively wide range of temperatures.
On the other hand, a lower frequency would imply
increasingly large temperatures ranges to test for non-screening
(note the decreasing slope of $\nu^*(T)$ in figure \ref{fig:arbeitspunkt}).
Noise and random measurement error increase with temperature, so we would like to
avoid relatively large temperatures. We find that $\nu_m=$9.72\,GHz is a nice compromise.
The temperature ranges for testing above and below $\nu^*$ are then 
given by $8\,\mathrm K \, \leq\, T_R \,\leq\, 14\,\mathrm K$ and
$22\,\mathrm K \, \leq\, T_R \,\leq\, 28\,\mathrm K$.

\emph{Receiver noise} is determined by the thermodynamical temperature of the physical radiometer system.
The HEMT amplifier noise is also dependent on the working frequency band \cite{weinreb1998noise}.
Since we plan to access the SU(2) ground state by probing different temperatures $T_i$,
receiver noise will in general change for each temperature. In the actual experiment this
will become  a source of systematic error, since for each run $T_{\mathrm{rec}}$ will vary,
however slightly, as a function of the working temperature.
This variation should be calibrated for each $T_i$,
so that for every measurement run we may extract the \emph{output power temperature} $T_o$,
whatever the radiometer working temperature. Further, the receiver noise temperature
$T_{\mathrm{rec}}$ crucially limits the experimental precision of the radiometer.
Here, we assume the simple linear model given in \eqref{defTrec}
\begin{align}
  T_{\mathrm{rec}}(T_{p,\mathrm{rec}}) \, &= \, a_{\mathrm{rec}} T_{p,\mathrm{rec}} + c_{\mathrm{rec}}\,,
  \label{defTrec2}
\end{align}
where $T_{p,\mathrm{rec}}$ is the physical receiver temperature.
The coefficients $a_{\mathrm{rec}}$ and $c_{\mathrm{rec}}$
are chosen such that in the temperature range we work with,  $T_{min}=8\,\mathrm K$ and $T_{max}=28\,\mathrm K$,
they are of the same order of magnitude as the receiver noise found in the literature, e.g.
\cite{artalradiometers,jarosik2003design,pospieszalski2005extremely,Arcade2,smithstathemt96,weinreb1998noise},
which we have estimated into best, average and worst performance (sets I, II and III).

\emph{Radiometer offset} is the baseline noise in Dicke switching measurements \cite{Jarosik2000}.
We have taken it to be of the order of $10^{-1}$K \cite{Arcade2}.

The \emph{power gain variation} $\Delta G / G$ is the most important determinant of
radiometer precision, as can be seen from the equations \eqref{dTtotal} and \eqref{dTdicke}.
(And as we will later confirm in the simulations).
Commercially available HEMT amplifiers come with
gain variation of the order of $10^{-3}$ \cite{Berkshire}.
We take this value for the best case scenario I.
We assume a poorer mean fluctuation of the order of $10^{-2}$
for the average case II and "better" worst case IIIb.
For the worst case scenario IIIa we take fluctuations of the order $10^{-1}$

\emph{Loss factor} gives the signal power loss between antenna and receiver.
Here we have taken to be of the order of 1\,dB$=10^{0.1}$ for the average scenario, based on the \cite{Arcade2} setup.
The values for the I and III cases were taken to be some orders of magnitude above an below 1\,dB.

The \emph{integration time} can be calculated so as to minimise radiometer uncertainty
\cite{wilson2009tools}. Here we take it to be $\tau\,=\,1\,\mathrm s$ in the average case
and vary the value accordingly for best and worst case.
Finally, the \emph{switching frequency} $f_S$ when working in Dicke mode should be chosen
such as to be greater than the characteristic fluctuation frequency of the  amplifier, $f_S>f_{\mathrm{knee}}$.
Otherwise the Dicke switching formulas \eqref{Tdicke} and \eqref{dTdicke} do not apply.
Here we use the value given in \cite{Arcade2}, $f_S=75$\,Hz.

\subsection{Experimental signature}

Having defined the most important experimental parameters in table \ref{table:radparms},
we may now use those values to estimate the radiometer output in the experiment.
As already described, two experimental procedures are possible, total power mode
and Dicke switching mode.

We simulate an experimental run of 600 measurements.
The integration time of the radiometer is given in table \ref{table:radparms} for each set of parameters,
so each measurement run has a duration of
1 minute, 10 minutes and 100 minutes for each case I, II and III.
Output temperatures for each single measurement were simulated by using a
Gauss-distributed random generator with mean $T_o$ and standard deviation $\Delta T_o$,
both of which were calculated using the appropriate formulae \eqref{Ttotal}, \eqref{dTtotal}
for a total power run and \eqref{Tdicke}, \eqref{dTdicke} for a Dicke switch run.
The radiometer noise temperature $T_N$ given in \eqref{TNDef} was used
in both cases to calculate the noise contribution to the output signal.

The temperature of the cryogenic chamber \emph{defines} the radiation temperature $T_R$,
since the black-boy cavities then have thermodynamical temperature $T_R$.
The physical temperature of the receiver and of the antenna and transmission-line coupling,
were also taken to be $T_R$, since the radiometer is placed inside the cryogenic chamber.
That is, $T_p$ and $T_{p,\mathrm{rec}}$ are taken equal to the radiation temperature
$T_p=T_{p,\mathrm{rec}}=T_R$.

Recall that the antenna temperature $T_A$ is
by definition the temperature of the collected radiation.
Inside the black-body cavity, radiation is at thermal equilibrium
with the cavity walls and $T_A=T_R$.
This means that the antenna efficiency $\eta$ \emph{cancels} in the expression
\eqref{Ttotal}, since in our case, $T_A=T_p$.

The results of the simulations are discussed below.

\subsubsection{Total power mode}

In total power mode, we set the cryogenic chamber to a temperature $T_R$ then let the radiometer
observe the output A. For each $T_R$ we run 600 measurements.
The results are plotted below.
We first describe the experimental procedure, then comment on each case.

Start the first measurement run at $T_R=28$K, then decrease the temperature to $T_R=26$K
and measure again, and so on until $T_R=22$K.
If the radiometer parameters are such that
the relative signal error is small enough for a distinct linear
relationship between $T_o$ and $T_R$ to be identified,
we may fit the four pairs $(T_R,T_o)$ to a line
\begin{align}
  T_o \, &= \, A_{ac}T_R + C_{ac}\,.
  \label{defLinTotAC}
\end{align}
The subscript $ac$ stands for "`\emph{above cutoff}"'.
We call this measurement series the \emph{above-cutoff} series.

Decrease now the temperature further, to  $T_R=14$K, and start the
measurement series again going down to $T_R=8$K.
As above, if the measurement errors are small enough, we can fit the four measured data points to a line
\begin{align}
  T_o \, &= \, A_{bc}T_R + C_{bc}\,,
  \label{defLinTotBC}
\end{align}
where $bc$ stands for "\emph{below cutoff}". Similarly, we call this measurement series the \emph{below-cutoff} series.

As an example, take the average radiometer (II) first.
It has integration period $\tau=$1\,s, so each measurement run
takes 10 minutes. The results are plotted in figure \ref{fig:radioDataTP2} for all temperatures $T_R$.
The red lines represent simulated experimental runs assuming the SU(2) hypothesis,
the grey lines would be the result of U(1) dynamics.

The standard error $\Delta T_o$ is also visible in figure \ref{fig:radioDataTP2} as
the width of the measurement.
For parameter set II, it is of the order of $10^{-2}$, the dominant contribution coming from the 
average gain fluctuation.
This is small enough to fit Eqs. \eqref{defLinTotAC}, \eqref{defLinTotBC}.

Now consider the \emph{above-cutoff} series.
In this temperature range, the cutoff temperature lies below the radiometer frequency bandwidth,
compare with figure \ref{fig:arbeitspunkt}.
We should therefore measure a radiometer output temperature $T_o$ consistent
with a finite $T_R$ temperature and equation \eqref{Ttotal}, independently
of whether the photon obeys SU(2) or pure U(1) dynamics.
Figure \ref{fig:radioDataTP2} shows how both predictions overlap in this temperature region.

Take now the \emph{below-cutoff} series. Assume SU(2) dynamics.
Then, in this temperature range, the radiometer is now directly looking
at the \emph{screening} region, compare with figure \ref{fig:arbeitspunkt}.
There can be no propagation inside the cavity within the frequency bandwidth of the radiometer.
The effective radiation temperature $T_R$ the radiometer sees is therefore zero.
The expected output temperature inside this region is then given by \eqref{Ttotal} 
with $T_R$ set to zero, $T_R=0$K. From this same formula, we see that the only output
measured is radiometer noise. This is now the key experimental fact.
If we compare the fitted parameters in \eqref{defLinTotBC}, \eqref{defLinTotAC},
for the SU(2) case, the should \emph{differ}. The parameters \emph{above-cutoff}
characterize a finite antenna signal plus noise, while those \emph{below-cutoff}
pick up only the noise signal.
For small standard variation in $T_o$ both sets of parameters should differ beyond
curve fitting errors.

In the case of U(1) dynamics, on the contrary, both linear fits should be similar
enough so as to have each fitted parameter lie within each others error bars.

All of this is visible in figure \ref{fig:radioDataTP2}. The red lines show
a slope in the non-screening region of about $A_{ac}=1.6$,
compared to the slope in the screening domain of $A_{bc}=0.3$.
The gray lines, on the other hand, both show the same slope $A_{bc}=A_{ac}=1.6$
above and below the screening region.

This is therefore the \emph{experimental signature} of SU(2) Yang-Mills thermal ground state
when probed by a total power radiometer.

We now discuss the other sets of radiometer parameters.
The results of the \emph{worst case} radiometer IIIa measuring
in total power mode are given in figure \ref{fig:radioDataTP3a}.
We see that in this case, the signal error is so great that no
definitive linear relationship may be attributed between
measured radiometer temperature $T_o$ and the cavity radiation temperature $T_R$.
It is still of note that an \emph{absolute difference} is visible in the
screening region between the U(1) case (grey) and SU(2) case (red), they
do not overlap. But this depends on the calculation of predicted
U(1) signals, which in turns depends on the full characterization
of the radiometer noise temperature function. This would induce enough error
dependencies that such a signal would lose any experimental significance.

Amusingly, this worst case radiometer may be immediately made useful
by just replacing the $10^{-1}$ gain variation amplifier with a quieter one of gain variation $10^{-2}$.
Since gain fluctuations $\Delta G$ are the dominant error source in the receiver,
this change is enough to make the IIIa radiometer a good enough IIIb.
This is seen in figure \ref{fig:radioDataTP3b}.
The high receiver noise still influences the absolute value of the output temperature
(compare the $T_o \sim 225$K at $T_R=28$K with $T_o \sim 61$K at the same radiation
temperature for the II radiometer), but the different slopes are clearly visible:
$A_{ac}\approx 4.75$ outside the screening region and $A_{bc}\approx 2.35$
inside it. This shows that the most important radiometer parameter in our case is
is the average gain fluctuation $\Delta G$.

Figure \ref{fig:radioDataTP1} shows the results from simulating a \emph{best case} radiometer
measurement in total power mode. The SU(2) signal is clearly visible, 
with $A_{ac}\approx 0.93$ \emph{above-cutoff} and $A_{bc}\approx 0.08$ below it.
It is remarkable that such a strong signal may be recovered from
a very simple radiometric technique. This of course assumes the availability
of a \emph{best case} radiometer as listed in table \ref{table:radparms}.

\subsubsection{Dicke switching mode}

We now discuss the radiometer working in Dicke switching mode.
As explained before, in this mode the radiometer switches with frequency  $f_S$
between measuring the antenna temperature $T_A$ and a reference temperature $T_{\mathrm{ref}}$
In this experiment we will use a \emph{U(1) cavity as reference}.
Such a cavity is prepared as described in section \ref{sec:restoringu1}.

A voltage is applied to the condensator plates inside the shielded cavity B,
so that the ground state is destroyed, and the thermal radiation behaves identically as
in a U(1) gas. The field should be of the order of $|\vec{E}|\sim 1\,\mathrm{V/cm}$
or greater, compare equations \eqref{enEfeld} and \eqref{rho_gsSI}. This ensures
a U(1) radiating cavity.

The antenna B may now be used as a reference emitter for the Dicke switching experiment.
Each experimental run of 600 individual measurements is done, as before, for temperatures
between 8K and 14K (the \emph{below-cutoff} series) and temperatures between 22K and 28K
(the \emph{above-cutoff} series).

Since both cavities find themselves inside the cryogenic chamber,
both radiate at the same temperature $T_R$, \emph{outside the screening region}.

That is, for the  \emph{above-cutoff} series, we expect from formula \eqref{Tdicke}
that the output temperature of the radiometer be just $T_{\mathrm{off}}$, which is of the order of 0.5\,K.
This should be the case wether the $\sucmb$ hypothesis
is true or not, since above the cutoff frequency both U(1) and SU(2) cavities
have the same power output. Since both signal and reference temperature are the same,
we further expect the influence of gain fluctuation errors $\Delta G$ on radiometer sensitivity to vanish
in this temperature region.

It is inside the screening region, that is for the \emph{below-cutoff} measurement series,
where we expect large differences between the U(1) and SU(2) behaviour.
In this temperature range the cavity B still emits U(1) radiation at a temperature $T_{\mathrm{ref}}=T_R$.
If the $\sucmb$ hypothesis is correct, however,
no photon propagation can occur in cavity A.
The effective antenna temperature \emph{vanishes}, $T_A=0$,
and formula  \eqref{Tdicke} now results in $T_o= T_{\mathrm{off}} -T_R$
(in our plots we have redefined $T_o$ to be positive, for convenience).
The influence of gain fluctuation errors $\Delta G$ cannot vanish anymore, on the contrary,
it should return to be the dominant fluctuation source.
If the $\sucmb$ hypothesis is false, no change from the
\emph{above-cutoff} series should be noticeable, since both cavities
would still have non vanishing output power.

\begin{figure}[p]
  \centering
  \subfloat[Output temperature $T_o$ measured by a total power radiometer.
  The graphic was generated by using a Gauss-distributed random generator
  of mean $T_o$ and standard deviation $\Delta T_o$,
  which were calculated using the corresponding formulas \eqref{Ttotal} and \eqref{dTtotal}
  ]{\label{fig:radioDataTP2}\includegraphics[width=137mm]{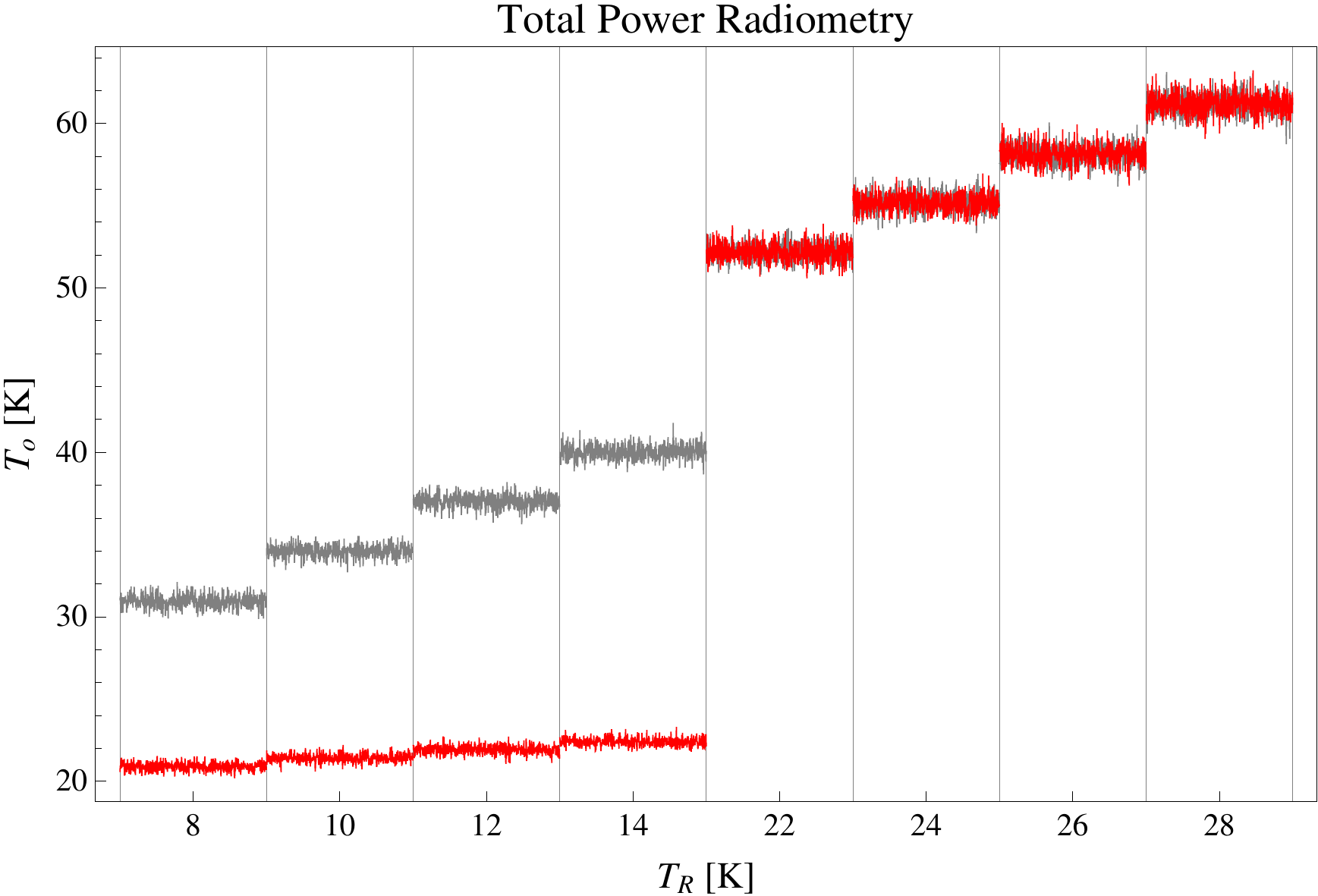}} \\
  \subfloat[Output temperature $T_o$ measured by a Dicke switching radiometer.
  The graphic was generated by using a Gauss-distributed random generator
  of mean $T_o$ and standard deviation $\Delta T_o$,
  which were calculated using the corresponding formulas \eqref{Tdicke} and \eqref{dTdicke}
  ]{\label{fig:radioDataDS2}\includegraphics[width=137mm]{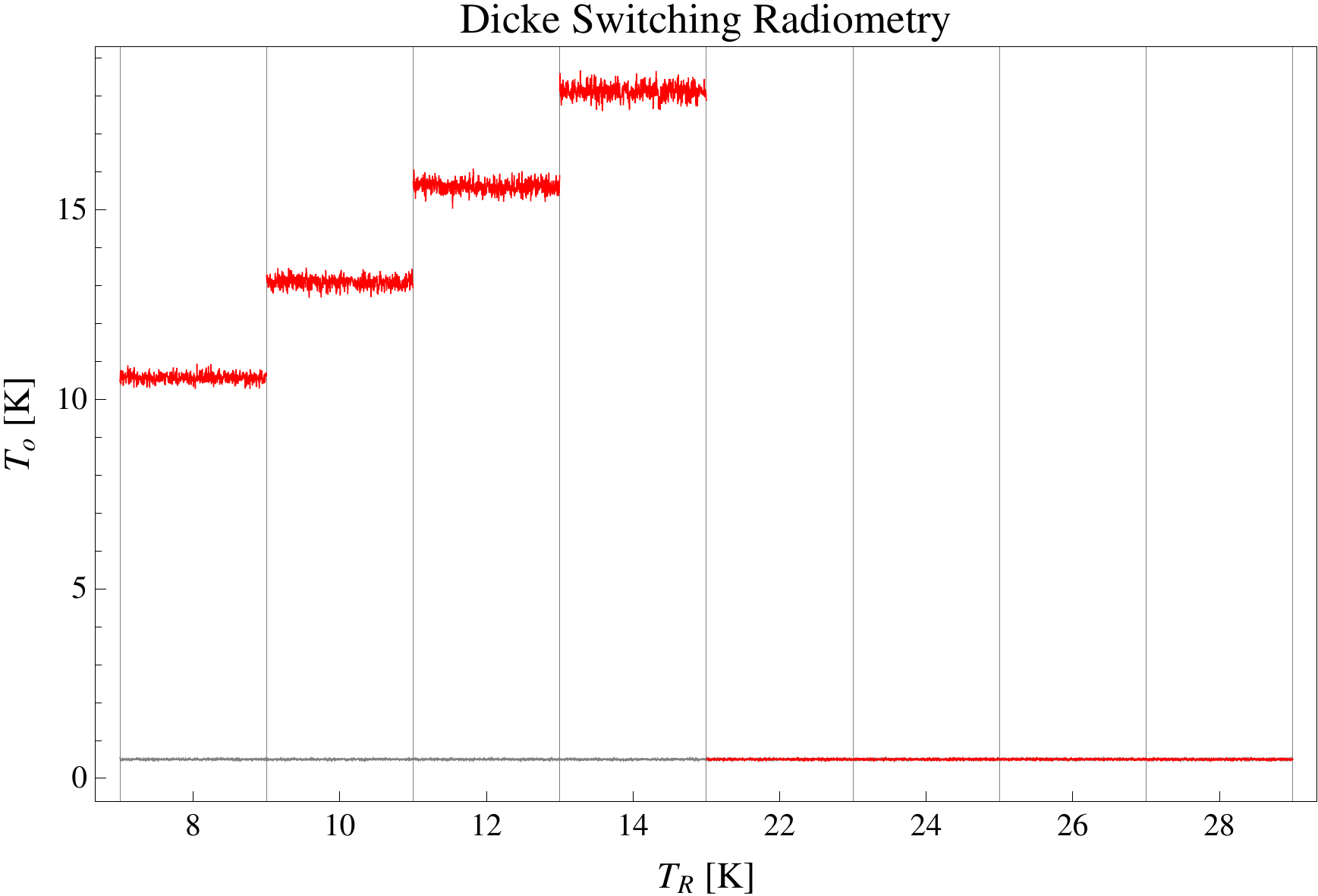}}
  \caption{Simulated results for the set of radiometer parameters II given in table \ref{table:radparms}
  }\label{fig:radioData2}
\end{figure}

As an example, take radiometer II with average parameters. The output of a Dicke switching
measurement run is given in figure \ref{fig:radioDataDS2}.
Outside the screening region, the generated values for $T_o$ hover above the
radiometer offset noise of  $T_{\mathrm{off}}=0.5$K, as expected.
Inside the screening region, however, there is a big jump in the signal in the case
of SU(2) dynamics. This is the \emph{experimental signature} for the Dicke switching experiment.
If photon dynamics are indeed described by an effective thermal theory with nontrivial ground state,
then the signature event would be a nontrivial signal, of the order of 10K \emph{below-cutoff}.

\begin{figure}[p]
  \centering
  \subfloat[Output temperature $T_o$ measured by a total power radiometer.
  The graphic was generated by using a Gauss-distributed random generator
  of mean $T_o$ and standard deviation $\Delta T_o$,
  which were calculated using the corresponding formulas \eqref{Ttotal} and \eqref{dTtotal}
  ]{\label{fig:radioDataTP3a}\includegraphics[width=137mm]{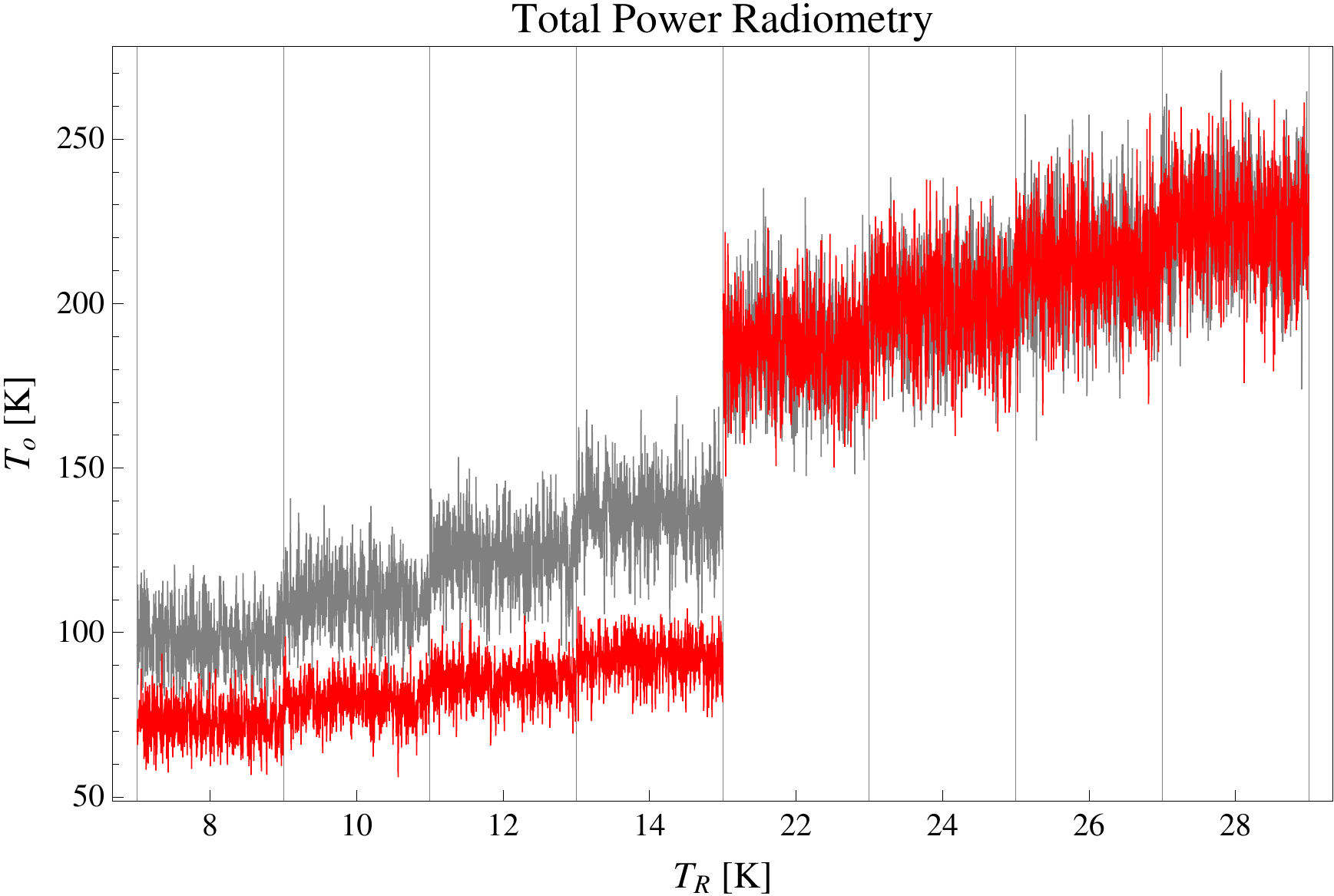}} \\
  \subfloat[Output temperature $T_o$ measured by a Dicke switching radiometer.
  The graphic was generated by using a Gauss-distributed random generator
  of mean $T_o$ and standard deviation $\Delta T_o$,
  which were calculated using the corresponding formulas \eqref{Tdicke} and \eqref{dTdicke}
  ]{\label{fig:radioDataDS3a}\includegraphics[width=137mm]{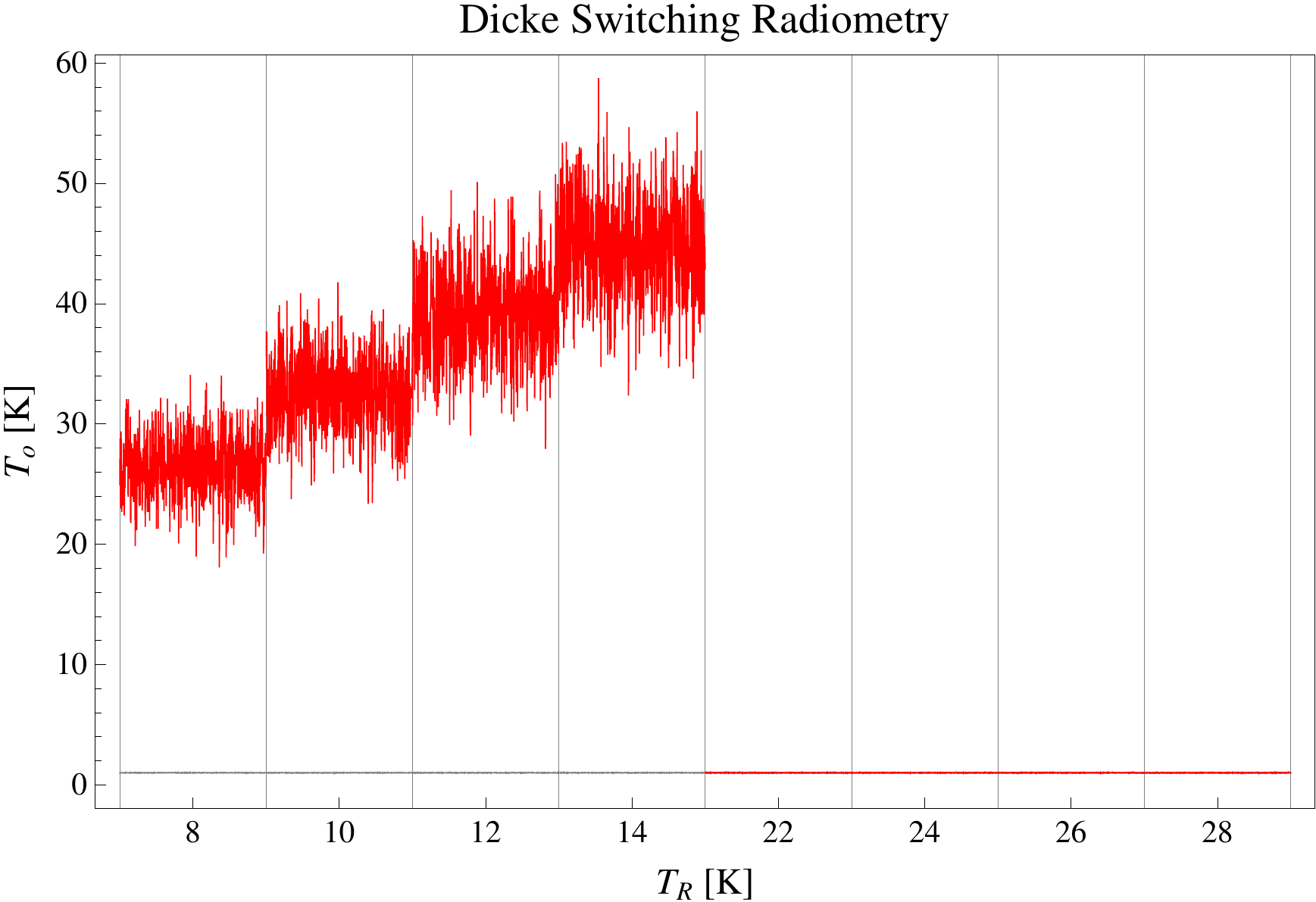}}
  \caption{Simulated results for the set of radiometer parameters IIIa given in table \ref{table:radparms}
  }\label{fig:radioData3a}
\end{figure}

For the average radiometer II such a signal is easily detectable. Figure \ref{fig:radioDataDS3a}
shows the generated temperature output for the radiometer with parameter set III.
And here is where the Dicke switching technique really shines. As long as the switching frequency
has been chosen correctly (that is, smaller than the characteristic gain fluctuation frequency of the amplifier)
the SU(2) signal is also clearly visible!
The measured signal shows a large error inside the screening
region but that is not a problem, since we can automatically detect the presence of SU(2) dynamics
just by having a non trivial signal at all!
Compare the difference in orders of magnitude from $T_o  \sim 40$K
below the cutoff, to 1K above it.
Such a huge disparity in the Dicke signal output would immediately signal
the presence of the SU(2) ground state.
Since this signal is visible even for the worst of parameters, it is the ideal
configuration for a radiometric experiment probing for the existence in nature of SU(2) physics.
On the other hand, if the radiometer output below 14K  stayed close to zero,
the $\sucmb$ hypothesis would be automatically falsified.

\begin{figure}[p]
  \centering
  \subfloat[Output temperature $T_o$ measured by a total power radiometer.
  The graphic was generated by using a Gauss-distributed random generator
  of mean $T_o$ and standard deviation $\Delta T_o$,
  which were calculated using the corresponding formulas \eqref{Ttotal} and \eqref{dTtotal}
  ]{\label{fig:radioDataTP3b}\includegraphics[width=137mm]{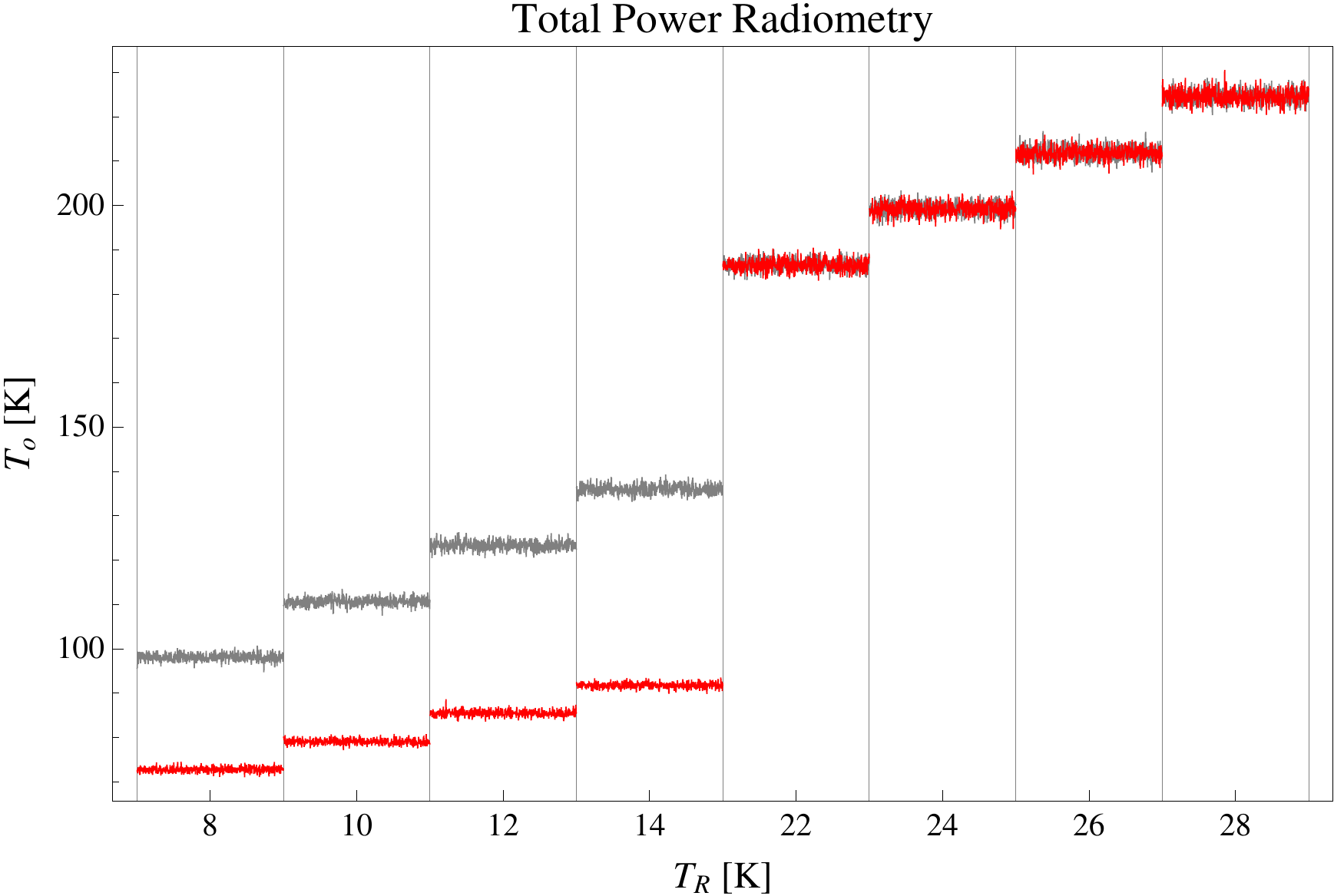}} \\
  \subfloat[Output temperature $T_o$ measured by a Dicke switching radiometer.
  The graphic was generated by using a Gauss-distributed random generator
  of mean $T_o$ and standard deviation $\Delta T_o$,
  which were calculated using the corresponding formulas \eqref{Tdicke} and \eqref{dTdicke}
  ]{\label{fig:radioDataDS3b}\includegraphics[width=137mm]{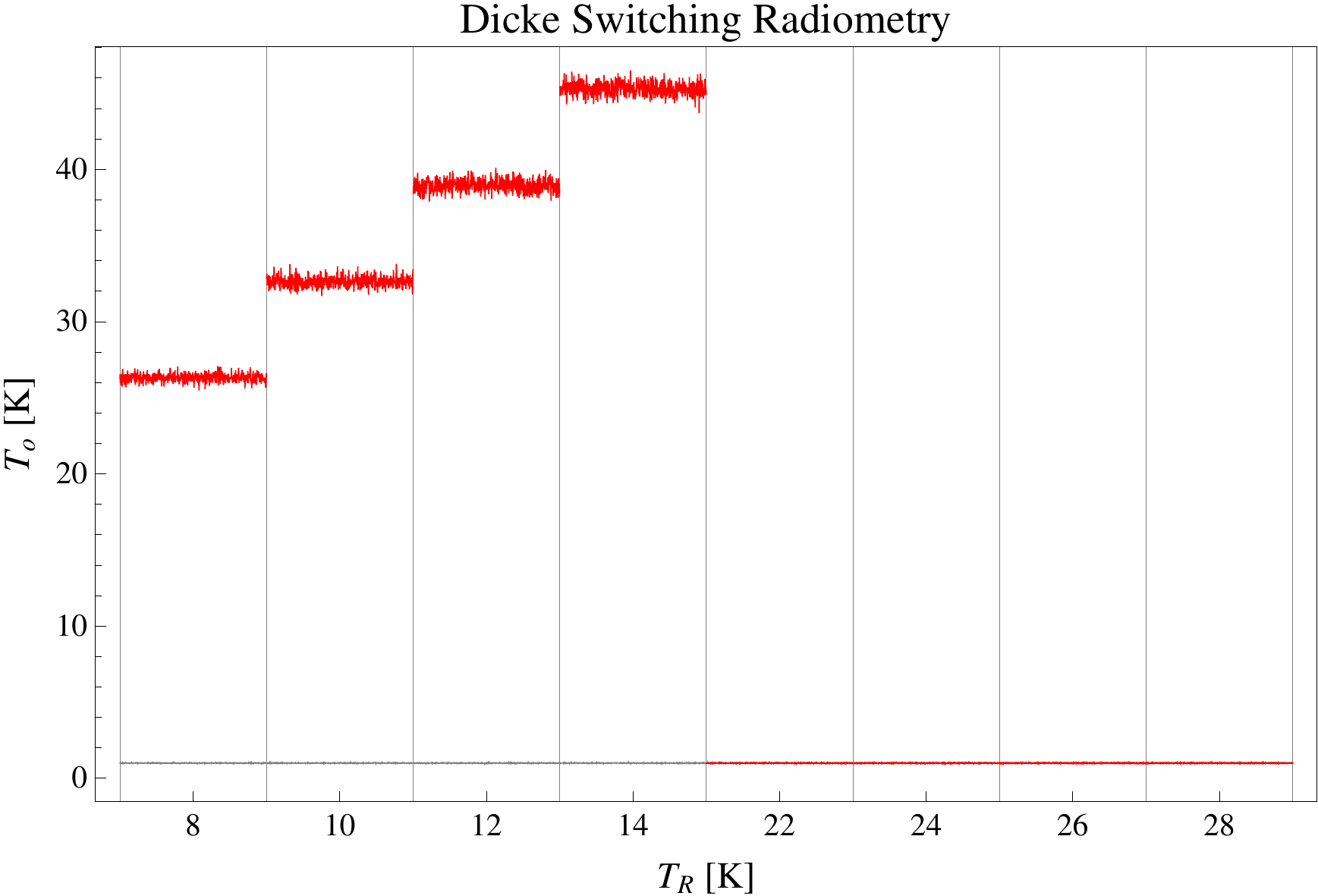}}
  \caption{Simulated results for the set of radiometer parameters IIIb given in table \ref{table:radparms}
  }\label{fig:radioData3b}
\end{figure}

\begin{figure}[p]
  \centering
  \subfloat[Output temperature $T_o$ measured by a total power radiometer.
  The graphic was generated by using a Gauss-distributed random generator
  of mean $T_o$ and standard deviation $\Delta T_o$,
  which were calculated using the corresponding formulas \eqref{Ttotal} and \eqref{dTtotal}
  ]{\label{fig:radioDataTP1}\includegraphics[width=137mm]{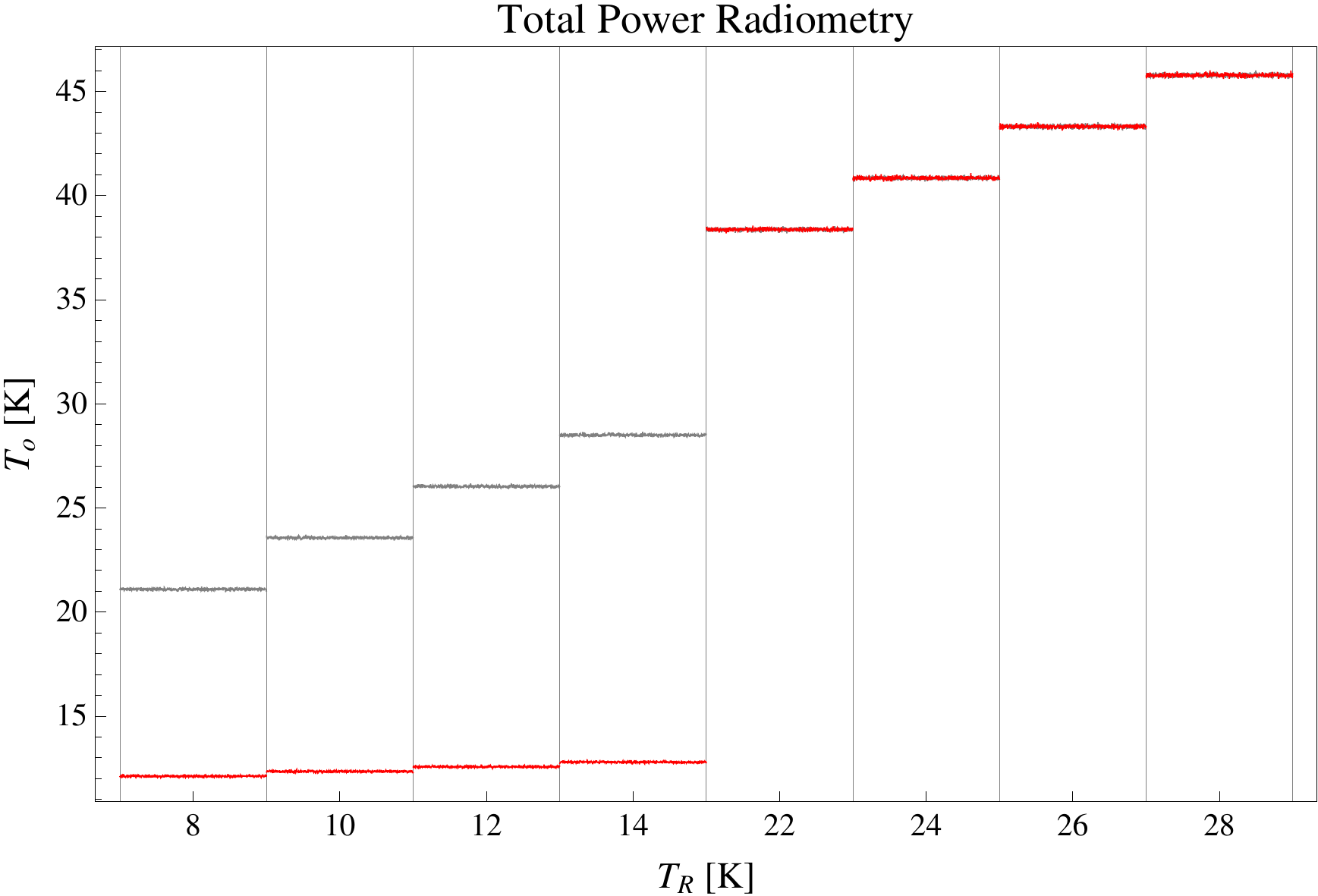}} \\
  \subfloat[Output temperature $T_o$ measured by a Dicke switching radiometer.
  The graphic was generated by using a Gauss-distributed random generator
  of mean $T_o$ and standard deviation $\Delta T_o$,
  which were calculated using the corresponding formulas \eqref{Tdicke} and \eqref{dTdicke}
  ]{\label{fig:radioDataDS1}\includegraphics[width=137mm]{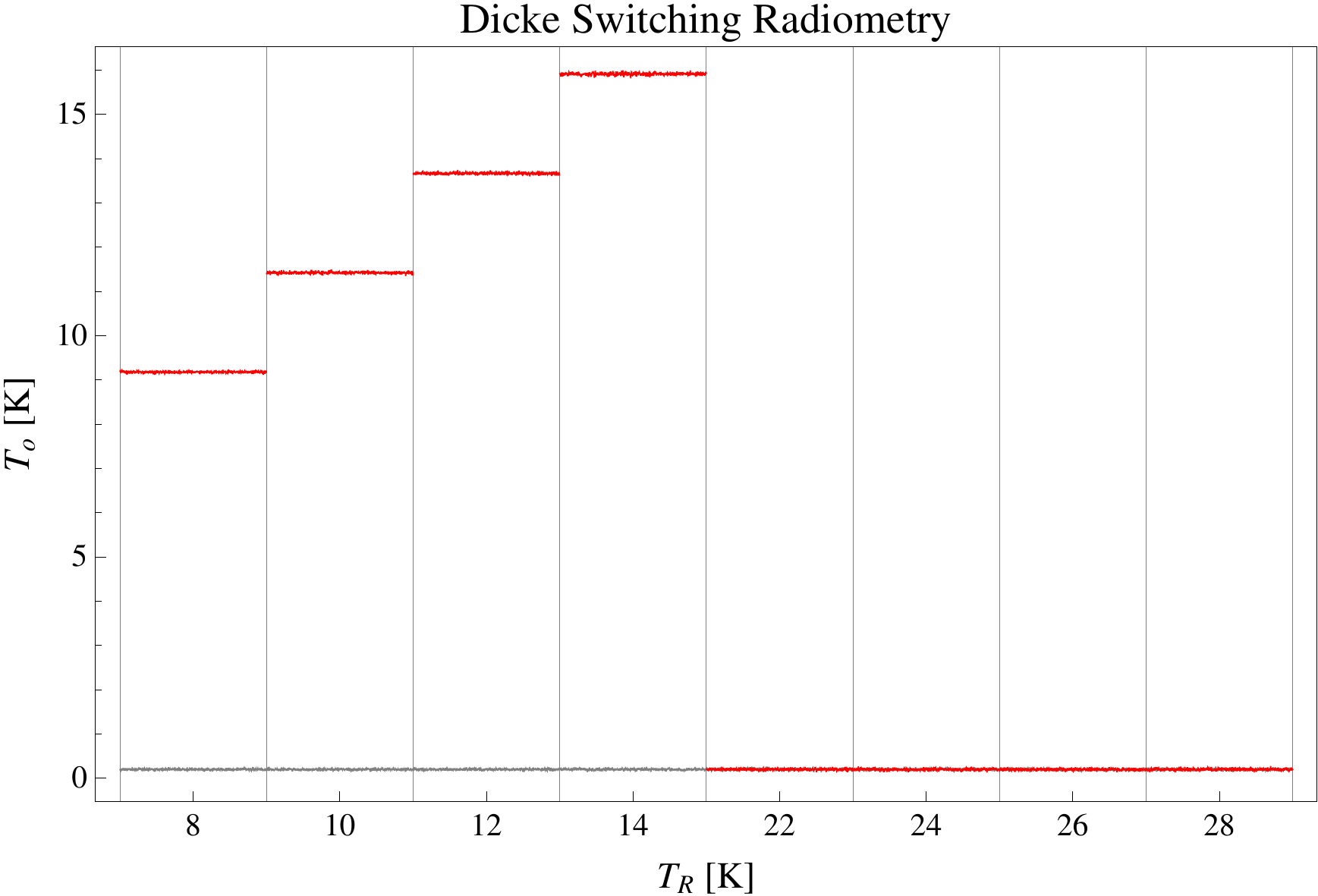}}
  \caption{Simulated results for the set of radiometer parameters I given in table \ref{table:radparms}
  }\label{fig:radioData1}
\end{figure}

For comparison, diagrams \ref{fig:radioDataDS3b} and \ref{fig:radioDataDS1}
show the simulated outputs for a radiometer with parameters from the set IIIb and I respectively.

\section{Bolometry of SU(2) photons}
\label{sec:Bolometry}

Aside from the radiometric experiment described in section \ref{sec:Radiometry},
a type of bolometric experiment can be conceived.
Unlike radiometric experiments where a narrow measurement frequency band is used,
bolometric experiments are by design generally \emph{broadband} experiments,
probing a wide region in the frequency spectrum.

Here we review a type of bolometric experiment first discussed in \cite{falquez2010modification}
that detects the presence of SU(2) ground state dynamics by \emph{thermal cooling}.
The down side of this experimental setup is the need for special filters,
which should ideally absorb all radiation above (or below) the cutoff frequency $\nu^*$.
The experimental setup is as follows.

Let the apertures of an isolated low-temperature U(1) black-body at temperature $T_1$ and
that of an SU(2) black-body of identical characteristics at temperature $T_2$ face each other,
and exchange radiant energy. The U(1) black-body is prepared as before, by means of a
static electric field. At low temperatures, of the order of 5K-10K,
the magnitude of the electric field should be of the order of $|\vec{E}|\sim 1\,\mathrm{V/cm}$
to guarantee a decoupling from the SU(2) ground state (compare with section \ref{sec:restoringu1}).
Couple the SU(2) black-body to a large heat reservoir,
so as to keep its thermodynamical wall temperature $T_2$ constant,
and insert a \emph{low-pass filter} within the common aperture
so that photons above the cutoff frequency of the SU(2) cavity $\nu^*(T_2)$ are completely absorbed.
The filter works bi-directionally, that is, no matter from which direction the photons come from,
they are absorbed if they propagate above $\nu^*(T_2)$.

Our bolometric experiment is then so configured that,
if the $\sucmb$ hypothesis is correct,
power is transmitted \emph{unidirectionally} from the U(1) cavity to the SU(2) cavity.
Because of the spectral gap below $\nu^*$, the SU(2) cavity cannot radiate in this region of the spectrum.
Energy is transfered continuously from the U(1) cavity to the SU(2) cavity only, and thermal equilibrium can never be reached.
This effect can then be detected as the \emph{lowering} of the U(1) cavity temperature $T_1$.

Consider now the case $T_1=T_2\equiv T$.
Replace the low-pass filter from before
with a \emph{high-pass filter} which absorbs all radiation
below the cutoff frequency $\nu^*(T)$.
Let a monochromatic radiant beam of frequency $\nu>\nu^*(T)$
propagate from the SU(2) cavity into the U(1) region
trough the aperture, along an axis $\hat n$ with solid angle $d\Omega_{\srm{U(1)}}$
defined on a sphere centered at a point within the common aperture (recall Fig. \ref{fig:Radiance2}).
Similarly, have another radiant beam of same frequency $\nu$
propagate from the U(1) cavity into the SU(2) region,
along an axis $- \hat n$ with solid angle $d\Omega_{\srm{SU(2)}}$.

First note that by Snell's law \cite{krall1986principles}, we can relate both solid angles as
\begin{align}
\label{snell's law}
\frac{d\Omega_{\srm{SU(2)}}}{d\Omega_{\srm{U(1)}}}=\frac{v^2_{ph}}{c^2}\,,
\end{align}
where the phase velocity $v_{ph}$ of energy propagation inside the SU(2) black-body is given by
$v_{ph}\equiv\frac{\omega}{k}=c\left(\sqrt{1-\frac{c^2 G}{(h\nu)^2}}\right)^{-1}$.
But this results in the identity
\begin{align}
\label{matchL}
\lbvs{SU(2)}(\nu,T)\,d\Omega_{\srm{SU(2)}}=\lbvs{U(1)}(\nu,T)\,d\Omega_{\srm{U(1)}} \,.
\end{align}
From Eqs. \eqref{radS} and \eqref{radPhi} we recognize the left and right hand side of expression \eqref{matchL}
as the total power emitted respectively by the SU(2) and U(1) cavities.
That is, for temperature $T$ and frequency $\nu>\nu^*(T)$,
the total power transfer from the U(1) cavity into the SU(2) cavity
is equal to the  total power transfer from the SU(2) cavity into the U(1) cavity.
Above the cutoff frequency $\nu^*(T)$, the system is in equilibrium, and no temperature change takes place.


\chapter{Summary and Outlook}
\label{chap:summary}

The main subject of this thesis was the investigation of possible
experimental methods to detect the presence of a nontrivial effective thermal ground state,
under the hypothesis that the abelian U(1) gauge group
responsible for photon dynamics in the Standard Model
is not fundamental, but emerges from a pure SU(2) Yang-Mills theory
as the unbroken gauge symmetry at finite temperature
in the effective theory of the deconfining phase after coarse-graining.
In this thesis, we refer to this identification as the $\sucmb$ hypothesis,
following the literature.
The CMB in the subscript refers to the fact that the single parameter of
the nonperturbative approach to SU(2) thermodynamics
can be fixed by the observation that the temperature of the CMB radiation gas
filling the universe should be very close to the critical temperature
for the deconfining/preconfining transition of the thermal SU(2) gas in its deconfining phase.

Under the $\sucmb$ hypothesis, thermalized photons are understood
as free propagating thermal quasiparticles
whose dispersion law in the effective theory
is determined by interactions with massive vector modes.

In this thesis, we have calculated these dispersion laws and discussed the results,
in the context of possible experimental detection.
It was found that the \emph{transverse polarization modes} propagate via a modified dispersion relation.
The modified dispersion relation reflects the fact that
a temperature- and momentum-dependent dynamical photon mass $\sqrt{G(\mathbf p, T)}$
emerges in the effective theory.
Further, this dispersion law shows a temperature dependent minimal
propagation frequency, which we call the \emph{cutoff frequency} $\nu^*(T)$.
It is shown that below this frequency, no propagation
of transverse polarized photons is possible.
This defines a spectral \emph{screening region} for frequencies below $\nu^*(T)$.
We have calculated numerically a functional relationship for $\nu^*(T)$ 
by fitting to a power law.
A novel result of this thesis is the calculation of the dispersion law
for the \emph{longitudinal mode}.
The dispersion relation shows that longitudinal mode
propagation is possible only for very short momentum (or long wavelength)
and therefore cannot be probed by neither radiometric nor bolometric means.

The prediction of a temperature-dependent screening region
has important experimental consequences, which were investigated in this thesis.
It was shown that an antenna immersed in an SU(2) photon gas of temperature $T$
collecting radiation within a bandwidth $\Delta \nu\equiv \nu_2 - \nu_1 > 0$
about a middle frequency $\nu_m \equiv \nu_1 + \frac{\Delta \nu}{2}$
would measure a power output $P_o$ identical to that of the U(1) gas,
as long as the entire bandwidth $\Delta \nu$ lies above the cutoff, $\nu_1>\nu^*(T)$.
By decreasing the temperature to $T'<T$, however, an experimental situation can be achieved in which
the entire bandwidth $\Delta \nu$ lies \emph{below} the cutoff, $\nu_2<\nu^*(T')$.
In this case, photon propagation is forbidden and the antenna would deliver zero power output.
This fundamental difference between U(1) and SU(2) behaviour
becomes the \emph{experimental signature} for the existence of SU(2) photon dynamics.
Namely, since this region can be probed by radiometrical or bolometrical methods,
the $\sucmb$ hypothesis becomes experimentally testable, and thus falsifiable.
This is the main result of our thesis.

Two experimental techniques proposed in this thesis
for probing the power output of a thermal photon gas at low temperatures and frequencies
are total power and Dicke switching radiometry.
It is shown that both can be made sensitive enough to detect SU(2) ground-state effects
on photon propagation.
Dicke switching, however, is preferable because by design it suppresses gain variations of the amplifier.

A bolometric technique is also discussed in which the observable effect is cooling
by power emission. This technique depends on the availability of a low-pass
filter which should bidirectionally absorb radiation above the cutoff frequency $\nu^*(T)$.

Beside our main experimental result, we offer a novel theoretical investigation
on the possibility of studying thermally inhomogeneous ground-state dynamics.
This subject is by itself interesting (thermomagnetic effect), but also
necessary to elucidate questions of experimental parameter tolerance.
We have found a solution to the pure SU(2) Yang-Mills equations of motion
in Minkowski signature with an adjoint scalar field,
which in the context of nonperturbative Yang-Mills thermodynamics and under the
adiabatic approximation allows for the generalization of the thermal ground state
to space-dependent temperature profiles.
It was shown that such temperature gradients are capable of generating magnetic fields.
This is the SU(2) thermomagnetic effect.

Finally, an investigation on the consistency of 1-loop resummation of
the energy density and pressure of the SU(2) gas was made. We found
that thermodynamical consistency cannot be guaranteed if radiative corrections
to the vector modes in the effective theory are ignored.

It is remarkable that the experimental detection of SU(2) effects
and thus confirmation of the $\sucmb$ hypothesis is achievable
by a simple tabletop experiment consisting of  a cryogenic chamber,
black-body cavities and a well calibrated radiometer.
The experimental realization of the black-body experiment described in this thesis
is a worthwhile endeavour because of its considerable implications were the $\sucmb$ hypothesis be confirmed.

We introduced this work by referring to the low-frequency power-like deviation
of the CMB spectrum from perfect black-body physics, which can be explained
by assuming the  $\sucmb$ hypothesis.
Another important consequence of SU(2) dynamics can be read in \cite{JHEP2007}.
There it is shown that the realization in nature of the $\sucmb$ hypothesis
would invalidate the Higgs mechanism for electroweak symmetry breaking in the Standard Model.
$\sucmb$ yields six additional degrees of freedom to the total number of
cosmologically active relativistic degrees of freedom $g_*$
at Big-Bang nucleosynthesis. This in turn would imply a value for the Fermi coupling
at zero temperature $G_F$ of about 12\% larger than predicted by electroweak SM physics.

It is our hope that this thesis aids further experimental investigations regarding the $\sucmb$ hypothesis.

\cleardoublepage


\appendix
\appendixpage
\addappheadtotoc


\chapter{SU(2) Equations of Motion with Scalar Field}
\label{app:ymeom}

\section{Introdution}
Here we derive a special solution to the SU(2) Yang-Mills Equations of Motion
with a scalar field $\Phi$ taken as perturation.
The derivation is subject to some simplifying assumptions.
Our only purpose here is to present the mathematical details,
the physical interpretation of the result is discussed in chapter \ref{chap:inhtd}.

We work with Minkowski metric, defined here as $g^{\mu\nu}=\mathrm{diag}(-1,1,1,1)$
to streamline calculations

\section{A particular solution}

We now embark in our search for particular solutions of the
SU(2) Yang-Mills equations of motion \eqref{EffEoM} 
with scalar field. These are given by
\begin{align}
  D_\mu G^{\mu\nu} &= ie \left[ \Phi , D^{\nu} \Phi \right]\,.
  \label{EofMa}
\end{align}
We write the scalar field in capital greek $\Phi$ so as to differentiate
it from the inert scalar $\phi$ described in chapter \ref{chap:YM},
since  we will allow the scalar field to variate in space.
Here we work in the physical unitary gauge \eqref{GSgauge},
$\Phi^a=2\,\delta^{a3}|\Phi|$

Recall now that both $\Phi$ and the trivial topology gauge field $a_\mu=a_\mu^{a}t^a$
live in the adjoint representation, so the covariant derivative
acts on them as $D_\mu=\partial_\mu \cdot - ie[a_\mu,\cdot]$.
With this in mind we expand \eqref{EofMa}
\begin{align}
  \partial_\mu \left( \partial_\mu a_\nu - \partial_\nu a_\mu \right) \notag \\
  - 2ie \partial_\mu \left[ a_\mu , a_\nu  \right] + ie \left[ \partial_\mu a_\mu , a_\nu \right] -ie \left[ \partial_\nu a_\mu, a_\mu \right] 
  - e^2 \left[ a_\mu, \left[ a_\mu, a_\nu \right] \right] \notag \\
  - ie \left[ \Phi, \partial_\nu \Phi \right] + e^2 \left[ \Phi , \left[ \Phi, a_\nu \right] \right] &= 0 \,.
  \label{eqn_expanded}
\end{align}

\subsection{The Yang-Mills equations for static fields in Minkowski space-time}

We are going to look for \emph{stationary solutions}, that is, $\partial_0 \equiv 0$.
The time-like  $\nu=0$ and space-like $\nu=k$ components of \eqref{eqn_expanded} then become
\begin{align}
  \partial^2_j a_0 
  -2 ie \partial_j \left[ a_j , a_0 \right] + ie \left[ \partial_j a_j , a_0 \right] \notag \\
  -e^2 \left[ a_\mu, \left[ a_\mu, a_0 \right] \right] + e^2 \left[ \Phi, \left[ \Phi, a_0 \right] \right] &=  0 \,,
  \label{eqn_0} \\
  \notag \\
  \partial^2_j a_k -\partial_k \partial_j a_j 
  -2 ie \partial_j \left[ a_j , a_k \right] + ie \left[ \partial_j a_j , a_k \right] \notag \\
  -ie \left[ \partial_k a_\mu , a_\mu \right] -e^2 \left[ a_\mu, \left[ a_\mu, a_k \right] \right] +
        e^2 \left[ \Phi, \left[ \Phi, a_k \right] \right] &=  0 \,.
  \label{eqn_j}
\end{align}
To simplify our notation, we define the complex fields
\begin{align}
  a^\pm_j \equiv \frac{1}{\sqrt{2}} \left( a^1_j \pm i a^2_j \right)\,.
  \label{defap}
\end{align}
Now write equations \eqref{eqn_0} and \eqref{eqn_j} explicitely in color indices
\begin{flalign}
  \left[ \partial^2 + 2ie a_j^3 \partial_j +ie \partial_j a_j^3 -e^2 a_\mu^3 a_\mu^3 + 4 e^2 |\Phi|^2 \right] a_0^+
  &= \, ie \left[ a_0^3 \partial_j + 2 \partial_j a_0^3  \right] a_j^+ \notag \\
  & \quad + e^2 \left[ a_0^+ a_\mu^- - a_0^- a_\mu^+ - a_0^3 a_\mu^3 \right] a_\mu^+ \,,
  \label{eqn0_12} \\
  \notag \\
  \left[ \partial^2 - 2e^2 a_\mu^+ a_\mu^- \right] a_0^3 + e^2 \left[ a_0^+ a_\mu^- + a_0^- a_\mu^+ \right] a_\mu^3 
  &= -ie \left[ a_0^- \partial_j + 2 \partial_j a_0^- \right] a_j^+ \notag \\
  & \quad + c.c. \,,
  \label{eqn0_3} \\
  \notag \\
  \left[ \partial^2 + 2ie a_j^3 \partial_j +ie \partial_j a_j^3 -e^2 a_\mu^3 a_\mu^3 + 4 e^2 |\Phi|^2 \right]a_k^+ \notag \\
  - ie \left[ a_k^3 \partial_j + 2 \partial_j a_k^3  \right] a_j^+ 
  + ie \left[ a_\mu^3 \partial_k - \partial_k a_\mu^3  \right] a_\mu^+ -\partial_k \partial_j a_j^+ 
  &= \, e^2 \left[ a_k^+ a_\mu^- - a_k^- a_\mu^+ - a_k^3 a_\mu^3 \right] a_\mu^+ \,,
  \label{eqnk_12} \\
  \notag \\
  \left[ \partial^2 - 2e^2 a_\mu^+ a_\mu^- \right] a_k^3 +  e^2 \left[ a_k^+ a_\mu^- + a_k^- a_\mu^+  \right] a_\mu^3
        - \partial_k \partial_j a_j^3
  &= -ie \left[ a_k^- \partial_j + 2 \partial_j a_k^- \right] a_j^+ -ie a_\mu^- \partial_k a_\mu^+ \notag \\
  &\quad + c.c. \,.
  \label{eqnk_3}
\end{flalign}
The symbol $c.c.$ denotes the complex conjugate of the expression on the
right hand side of equations \ref{eqn0_3} and \eqref{eqnk_3}.

\subsection{Simple configurations with U(1) magnetic field}

In order to solve the above equation system, we make further simplifying
assumptions. Namely, that the time-like-component of the tree-level massless gauge field (photon),
as well as the space-like components of the complex tree-level heavy fields \eqref{defap}, both vanish.
\begin{align}
  a_0^3 = a_k^\pm = 0 \,.
  \label{eqn_annahme}
\end{align}
Since we have already assumed stationarity, this restriction limits the set of possible
solutions to pure magnetic field configurations.
Under our assumptions, the electric field is always zero
$\mathbf E = -\mathbf \nabla A^0 - \partial_t \mathbf A = 0$,
and only magnetic configurations $\mathbf B = \mathbf \nabla \times\mathbf A$
are possible.
From \eqref{eqn_annahme} it follows that the expression $a_\mu^\pm a_\mu^3 = 0$ vanishes,
equations \eqref{eqn0_3} and \eqref{eqnk_12} are thus trivially satisfied.
Inserting \eqref{eqn_annahme} in equation \eqref{eqn0_12} we obtain (recall our metric convention)
\begin{align}
  \left[ \partial^2 + 2ie a_j^3 \partial_j +ie \partial_j a_j^3 -e^2 a_j^3 a_j^3 + 4 e^2 |\Phi|^2 \right] a_0^+ = 0 \,.
  \label{eqn_1}
\end{align}
Recall the identity $f \partial^2 f = \frac{1}{2}\partial^2 f^2 - ( \partial f )^2$,
which we use to write the above expression as
\begin{align}
  \left[ \frac{1}{2} \partial^2 + e^2 \left( 4 |\Phi|^2 - a_j^3 a_j^3 \right) \right] \left( a_0^+ \right)^2 =
     \left( \partial_j a_0^+ \right)^2 -ie \partial_j\left[ \left( a_0^+ \right)^2 a_j^3\right] \,.
     \label{eqn_2b}
\end{align}
Reordering terms, \eqref{eqn_2b} becomes
\begin{align}
     e^2 \left( 4 |\Phi|^2 - a_j^3 a_j^3 \right) 
                                - \left( \frac{\partial_j a_0^+}{a_0^+} \right)^2
                                + \frac{1}{2} \partial^2 {a_0^+}^2 / {a_0^+}^2
                                +ie \partial_j \left[ { a_0^+}^2 a_j^3\right] / {a_0^+}^2 = 0 \,.
  \label{eqn_2}
\end{align}
Expanding $a_0^+$ in a positive real magnitude $a_0(x)$ and a real phase $\varphi(x)$
\begin{align}
  a_0^+ \equiv a_0 e^{i\varphi} \,,
  \label{eqn_a0phase}
\end{align}
we can simplify the gradient, Laplacian and divergence terms in \eqref{eqn_2}
\begin{align} 
  \partial_j \left[ { a_0^+}^2 a_j^3\right] / {a_0^+}^2 &= \partial_j a_j^3 + 2 a_j^3 \frac{\partial_j a_0^+}{a_0^+} \,,
  \notag \\
  \frac{\partial_j a_0^+}{a_0^+} &= \frac{\partial_j a_0}{a_0} + i \partial_j \varphi \,,
  \notag \\
  \partial^2 {a_0^+}^2 / {a_0^+}^2 &= 2 \left( \frac{\partial_j a_0}{a_0} \right)^2 + 2 \frac{\partial^2 a_0}{a_0} -4 \left( \partial_j \varphi \right)^2
  +2i\left( \partial^2 \varphi + 4 \partial_j \varphi \frac{\partial_j a_0}{a_0} \right) \,.
  \notag
\end{align}
Separate \eqref{eqn_2} into real and imaginary components
\begin{align} 
  e^2 \left( 4 |\Phi|^2 - a_j^3 a_j^3 \right) - \left( \partial_j \varphi \right)^2 - 2 e a_j^3 \partial_j \varphi + \frac{\partial^2 a_0}{a_0} \notag \\
  +i \left[ \partial^2 \varphi + 2 \partial_j \varphi \frac{\partial_j a_0}{a_0} +e\partial_j a_j^3 
        + 2 e a_j^3 \frac{\partial_j a_0}{a_0}  \right] &= 0 \,.
  \label{eqn_3}
\end{align}
Taking Re[\eqref{eqn_3}]=0 and Im[\eqref{eqn_3}]=0 we obtain two real equations
\begin{align}
  e^2 \lvert\vec{a}^3\rvert^2 - \frac{\partial^2 a_0}{a_0} + \partial_j \varphi \left[ \partial_j \varphi + 2 e  a_j^3 \right]
   &=4 e^2 |\Phi|^2 \,,
  \label{sys1_1} \\
  e \partial_j a_j^3 + 2e a_j^3 \frac{\partial_j a_0}{a_0} + \partial^2 \varphi + 2 \partial_j \varphi \frac{\partial_j a_0}{a_0} 
  &= 0 \,.
  \label{sys1_2}
\end{align}
Inserting \eqref{eqn_annahme} in \eqref{eqnk_3} gives us
\begin{align}
  \left[ \partial^2 + 2e^2 a_0^+ a_0^- \right] a_k^3  - \partial_k \partial_j a_j^3
        =& ie \left[ a_0^- \partial_k a_0^+ - a_0^+ \partial_k a_0^- \right] \,.
\end{align}
Writing $a_0^+$ in components as in \eqref{eqn_a0phase}, we finally get
\begin{align}
  \left[ \partial^2 + 2e^2 \left( a_0 \right)^2 \right] a_k^3  - \partial_k \partial_j a_j^3
        =& -2e \left( a_0 \right)^2  \partial_k \varphi \,.
        \label{eq:reord}
\end{align}
Reordering \eqref{eq:reord}
\begin{align}
  2e \left( a_0 \right)^2  \left[  e a_k^3 + \partial_k \varphi \right]  = \partial_k \partial_j a_j^3 - \partial^2 a_k^3 \,.
  \label{sys1_3}
\end{align}
Now, from \eqref{sys1_2} one may easily see that
\begin{align}
  \partial_j \left[ \left( a_0 \right)^2\left( e a_j^3 + \partial_j \varphi \right) \right] = 0 \,.
  \notag
\end{align}
That is, there is some vector field $\vec{V}(x)$ such that
\begin{align}
  \left( a_0 \right)^2\left( e a_k^3 + \partial_k \varphi \right) = \mathbf{rot}_k \vec{V} \,,
  \tag{\ref{sys1_2}'} 
    \label{sys1_2b}
\end{align}
where we have defined $\mathbf{rot}_k \vec{V} \equiv (\mathbf \nabla \times \vec V)_k$.

Moreover, from \eqref{sys1_3} we infer
\begin{align}
  2 e \;\mathbf{rot}_k \vec{V}  = \partial_k \partial_j a_j^3 - \partial^2a_k^3 \,.
  \tag{\ref{sys1_3}'}
  \label{sys1_3b}
\end{align}
So $\vec{V}(x)$ must be of the form
\begin{align}
   2e \vec{V}_k &= \mathbf{rot}_k \, \vec{a}^3 \, 
   \overset{!}{=} \, \mathbf B_k \,.
   \label{sys1_3c2} 
\end{align}
Equation \eqref{sys1_3c2} can be used to rewrite \eqref{sys1_1} as
\begin{align}
  4 e^2 |\Phi|^2 &=  e^2 \lvert\vec{a}^3\rvert^2 - \frac{\partial^2 a_0}{a_0}
                    + \partial_j \varphi \frac{\mathbf{rot}_j \mathbf B}{e \left( a_0 \right)^2}
                    - \left( \partial_j \varphi \right)^2 \,.
  \tag{\ref{sys1_1}'}
  \label{sys1_1b}
\end{align}

\subsection {One-dimensional temperature distribution}

We now specialize to the case of spatial dependency
in just one dimension, and assume that the scalar source $\Phi$
and all $\Phi$-coupled physical fields to be functions
of coordinate $z\equiv x_3$ only, so that $|\Phi(\vec x)|=|\Phi(z)|$, and so on.
Our double primed equations will hold for this special case only
(together with stationarity and unitary gauge fixing \eqref{GSgauge}, remember).

Under this assumptions, we can write equation \eqref{sys1_3c2} in the form
\begin{align}
 a_k^3 &= a_k^3(z) \,,
 \notag \\
 \implies 2e \mathbf{rot}_k \vec{V} \, &= \,\mathbf{rot}_k \mathbf B  \, = \, \left( \delta_{k3} - 1 \right) \partial^2_z a_k^3 \,.
 \label{sys1_3c}
\end{align}
The magnetic field $\mathbf B$ is perpendicular to the $z$-axis.
Further, together with equation \eqref{sys1_2b}, equation \eqref{sys1_3c}
directly implies
\begin{align}
  e a_3^3(x) \,=\, - \partial_z \varphi(x) \,.
  \label{eqa3phi}
\end{align}
It follows that $\varphi(x)$ functions as a U(1) gauge parameter,
and can be set to zero (together with $a_3^3$).
The identity \eqref{eqa3phi} allows us to use equation \eqref{sys1_2b} to solve for $a_0$
\begin{align}
  \left( a_0 \right)^2 = - \frac{\partial^2_z a^3_1}{2 e^2 a^3_1} \, \overset{!}{=} \, - \frac{\partial^2_z a^3_2}{2 e^2 a^3_2}\,,
  \tag{\ref{sys1_2}''} 
    \label{sys1_2c}
\end{align}
since under our assumptions $\partial_1 \varphi = \partial_2 \varphi = 0$.
Equation \eqref{sys1_2c} implies the identity $a^3_2 \partial^2_z a^3_1 = a^3_1 \partial^2_z a^3_2$,
and can be used to determine $a_0$ from the fields $a^3_{1,2}$.
These fields also define the magnetic field $\mathbf B$,
$B_1 = -\partial_z a^3_2$, $B_2 = \partial_z a^3_1$.

The expression \eqref{sys1_1b} for $|\Phi|^2$ also simplifies in this special case.
From \eqref{sys1_3c} we read $\mathbf{rot}_3 \mathbf B \equiv 0$,
which together with $\partial_1 \varphi = \partial_2 \varphi = 0$ and equation \eqref{eqa3phi},
reduces \eqref{sys1_1b}  to
\begin{align}
  |\Phi|^2  \,&=\, \frac{1}{4} \sum_{i=1,2} a^3_i a^3_i - \frac{\partial^2_z a_0}{4 e^2 a_0} \,.
  \label{EofMPhiT}
\end{align}

The temperature profile can be calculated from \eqref{PhiMod}.
Inserting \eqref{EofMPhiT} we obtain ($T\equiv \beta^{-1}$)
\begin{align}
  T(z) \, &= \, \frac{\Lambda^3}{2 \pi |\Phi|^2} \, = \, \frac{\Lambda^3}{2 \pi} \,
  \left( \frac{1}{4} \sum_{i=1,2} a^3_i a^3_i - \frac{\partial^2_z a_0}{4 e^2 a_0} \right)^{-1} \,.
  \label{profT}
\end{align}
This means that if we find a reasonable Ansatz for $a^3(z)$,
we automatically obtain the fields $a_0$, $\mathbf B$, $|\Phi|$
and the profile for the temperature $T(z)$.
Such solutions and their physical interpretations are discussed in chapter \eqref{chap:inhtd}.

%
%


\chapter{Review of Classical Gauge Fields}
\label{app:gauge}

The basic result of the nonperturbative approach to SU(2) Yang-Mills thermodynamics in the deconfining phase
reviewed in chapter \ref{chap:YM} is the emergence of an inert scalar field $\phi$, which together
with a pure gauge configuration $a^{\srm{gs}}_\mu$ (see equation \eqref{DefAPureG}) defines an effective
thermal ground state acting as a background for the dynamics of weakly interacting thermal quasiparticles.
The scalar field $\phi$ is also responsible for breaking the original SU(2) gauge symmetry down to
its abelian subgroup U(1). Two of the original off-Cartan gauge modes acquire a temperature dependent mass,
while a third one is left massless at tree-level.
Under the $\sucmb$ hypothesis, this tree-level massless (TLM) gauge mode is identified with the SM photon.
(see section \ref{sec:su2cmb}).
Radiative corrections, however, induce an effective \emph{screening mass}
$m = \sqrt{G}$ for the transverse polarizations of the TLM mode,
which modifies their dispersion law (see section \ref{sec:transmodes}).
Further, a possible longitudinal polarization is rendered nonpropagating (see section \ref{sec:longmodes}).

Massive particles of spin 1 are described by the \emph{Proca equations}.
But the standard radiometric results used in this work where derived for
massless spin 1 fields, described by the \emph{Maxwell equations}.
We must therefore explicitly check how these results are altered
for spin 1 (vector boson) particles of nonzero dynamical mass.

In this Appendix, we review the theory of classical spin-1 (vector boson) fields,
following the presentation in \cite{greiner1996field}.
First we introduce the Lagrangians for the massless and the massive case, i.e., the Maxwell and Proca fields.
The respective equations of motion are derived, along with expressions for the energy and momentum density for each field.
In section \ref{sec:PlaneWaveExpansion}, polarization vectors for the massless and the massive case
are introduced, we then expand the energy and momentum density in Fourier modes.
Section \ref{sec:contEq} recalls the continuity equation.
Finally, since propagating excitations of the SU(2) TLM mode may only be transverse,
transversality must therefore be imposed on these Proca solutions.
This is done in section \ref{sec:transproca}.

The results of this appendix are mostly needed to rederive 
Planck's black-body formula (see section \ref{sec:radFormula})
and the Fresnel-Kirchhoff diffraction integral used in calculating
the effective antenna area (see appendix \ref{app:fresnel} and section \ref{sec:antennaeffarea}).
It will be seen that the standard results remain unchanged,
save for the modified dispersion relation
\begin{align}
  \left( \hbar \omega_{\mathbf{p}} \right) ^2 \, &= \, \left( c\, \mathbf p \right)^2 +G \left(\omega, \mathbf{p}; T, \Lambda \right) \,.
  \label{dispRelTSI2}
\end{align}

\section{Lagrangian formulation}
\label{sec:LagForm}
\subsection{The Maxwell field}

As we saw in chapter \ref{chap:GaugeSym}, the dynamics of the massless spin-1 field $A^\mu$
subject to an external current density $j^\mu = \left( \rho, \mathbf j \right)$
are described by the Lagrangian density
\begin{align}
  \mathcal L_{\tss{M}} \, \equiv\, -\frac{1}{4} F^{\mu\nu} F_{\mu\nu} - e j^\mu A_\mu\,.
  \label{emLag}
\end{align}
with the \emph{gauge field} $A_\mu=(A_0,\mathbf A)$,
the \emph{field strength}  $F_{\mu\nu} = \partial_\mu A_\nu - \partial_\nu A_\mu$
and the \emph{coupling} $e$ is the charge of the external field
inducing the current density $j^\mu$.
The physical electric and magnetic fields are defined as in \eqref{defEfromF} and \eqref{defBfromF}.

By the principle of least action
\begin{align}
  \delta S \, = \, \delta \int d^4x\, \mathcal L \, \overset{!}{=} \, 0 \quad \implies \quad
  \frac{\delta \mathcal L}{\delta A_\nu} -  \partial_\mu \frac{\delta \mathcal L}{\delta \partial_\mu A_\nu} \, = \, 0 \,,
 \notag
\end{align}
we obtain the Maxwell field equations of motion
\begin{align}
  \partial_\mu F^{\mu\nu} \, &= \, j^\nu \,,
 \label{MaxwellF1}
\end{align}
which for the massless gauge field $A^\mu$ read
\begin{align}
  \Box A^\nu + \partial^\nu \partial_\mu A^\mu = j^\nu \,.
  \label{emFieldEq}
\end{align}
Taking the divergence of \eqref{MaxwellF1} leads to the \emph{continuity equation}
\begin{align}
  \partial_\nu j^\nu = 0 \,,
  \label{MaxxwellCE}
\end{align}
since $F^{\mu\nu}$ is antisymmetric. Thus, in the massless spin-1 field theory,
current conservation is \emph{automatically guaranteed}.
This is a direct consequence of the conservation of U(1) gauge symmetry.

Writing \eqref{emFieldEq} in terms of the electromagnetic fields  \eqref{defEfromF} and \eqref{defBfromF},
we recover the familiar \emph{inhomogeneous} Maxwell equations, which we write here in SI units
\begin{subequations}
\begin{align}
  \mathbf \nabla \cdot \mathbf E  &= \frac{ \rho } { \varepsilon_0 } \,, \\
  \mathbf \nabla \times \mathbf B - \varepsilon_0 \mu_0 \, \partial_t \mathbf E &= \mu_0 \, \mathbf j \,.
\end{align}
  \label{MaxwellEq1}
\end{subequations}
Recall also from chapter \ref{chap:GaugeSym} that the \emph{Bianchi identity}
satisfied by the field strength tensor \cite{atiyah1979,frankel2004geometry,ryder1996quantum}
\begin{align}
  \partial_\mu \tilde{F}^{\mu\nu} \, &= \, 0 \,,
  \label{MaxwellDualDiv}
\end{align}
results in the \emph{homogeneous} Maxwell equations, here given in SI units
\begin{subequations}
  \begin{align}
    \mathbf \nabla \cdot \mathbf B  &= 0 \,,
    \\
    \mathbf \nabla \times \mathbf E + \partial_t \mathbf B &= 0 \,.
  \end{align}
  \label{MaxwellEq2}
\end{subequations}
In the following, we will be mostly interested in \emph{vacuum} configurations, that is, $j^\mu \equiv 0$.

Note first that the covariant formulation in \eqref{emLag} introduces redundant degrees of freedom
that allow for \emph{local gauge transformations} without affecting the Lagrangian
density
\begin{align}
  A_\mu \rightarrow A'_\mu &= A_\mu + \partial_\mu \Lambda(x) \,,
  \label{emGaugeTraf}  \\
  \mathcal L \rightarrow \mathcal L' &= \mathcal L \,,
  \label{emGaugeTrafL}
\end{align}
for a Lorentz scalar function $\Lambda(x)$.
In order to interpret the gauge field $A^\mu$ physically,
we must \emph{fix the degrees of freedom} so as to eliminate spurious polarization modes.
This is called \emph{gauge fixing}, and can be done in various ways.
A manifestly covariant gauge fixing is given by the \emph{Lorentz gauge}
\begin{align}
  \partial_\mu A^\mu \, &= \, 0\,,
  \label{emLorentzGauge}
\end{align}
In this gauge, the field equations for the Maxwell Field \eqref{emFieldEq} become
\begin{align}
  \Box A^\mu \, &= \, j^\mu \,.
  \label{emFieldEqL}
\end{align}
The Lorentz condition \eqref{emLorentzGauge} however still does not fix the gauge completely.
Gauge equivalent field configurations may still be found that are
related to each other by gauge transformation functions $\Lambda(x)$
which solve the Laplace equation, $\Box \Lambda(x) = 0$.
The Lorentz constraint $\partial_\mu A^\mu = 0$ cannot fix $A^\mu$ uniquely,
as the Maxwell field possesses just two independent degrees of freedom, not four.

\subsubsection{Energy and momentum density}

The familiar expression for the \emph{energy-momentum tensor}
\begin{align}
  \mathcal \theta^{\mu\nu} = \frac{\delta \mathcal L}{\delta \partial_\mu A_\sigma} \delta^\nu A_\sigma -g^{\mu\nu} \mathcal L \,,
  \label{EMtensor}
\end{align}
turns out to be not symmetric (also not gauge invariant) when
the Lagrangian \ref{emLag} is inserted directly in \eqref{EMtensor}.
To construct a physical energy-momentum tensor $T^{\mu \nu}$ we add a surface term to \eqref{EMtensor}
\begin{align}
  T^{\mu\nu} &= \theta^{\mu\nu} + \partial_\sigma \left( F^{\mu\sigma} A^\nu \right) \notag \\
  & = \frac{1}{4} g^{\mu \nu} F^{\alpha\beta} F_{\alpha \beta} + F^\mu{}_\sigma F^{\sigma \nu}
  + g^{\mu\nu} j_\sigma A^\sigma - j^\mu A\nu \,.
  \label{emEMtensor}
\end{align}
Writing \eqref{emEMtensor} in terms of the physical fields \eqref{defEfromF} and \eqref{defBfromF},
and taking $j^\mu=0$, we recover the familiar expression for the electromagnetic \emph{energy density} in vacuum
\begin{align}
    \rho \equiv T^{00}
    &= \frac{1}{4} \left( 2 F^{0i}F_{0i} +F^{ij}F_{ij}\right) + F^0{}_\sigma F^{\sigma 0} \notag\\
    &=\frac{1}{2}\mathbf E^2 + \frac{1}{4} \left(\delta^j_m\delta^k_n-\delta^k_m\delta^j_n\right)F_{jk}F^{mn} \notag\\
    &=\frac{1}{2}\mathbf E^2 + \left(-\frac{1}{2}\varepsilon^{ijk}F_{jk}\right) \left(-\frac{1}{2}\varepsilon_{imn}F^{mn}\right) \notag\\
    &=\frac{1}{2}\left(\mathbf E^2+\mathbf B^2\right) \,,
    \label{emED}
\end{align}
with $\varepsilon^{ijk}\varepsilon_{imn} = \delta^j_m\delta^k_n-\delta^k_m\delta^j_n$.
The \emph{Poynting vector} of the Maxwell field in vacuum,
that is, the vacuum \emph{momentum density},
can also be easily calculated
\begin{align}
  S^k &\equiv T^{0k}
  =  F^{0 \sigma} F_\sigma{}^k = E^i F_i{}^k = E^i \frac{1}{2} \left( \delta^m_i\delta^k_n-\delta^n_i\delta^k_m \right) F_m{}^n
  \notag\\
  &= \varepsilon_{ijk} E^i \left(-\frac{1}{2}\varepsilon^{jmn}F_{mn}\right) \,,
  \notag \\
  \mathbf S &= \mathbf E \times \mathbf B \,.
  \label{emMD}
\end{align}
In SI units the fields are rescaled as
$\mathbf E \rightarrow \frac{\mathbf E}{c}$,
$\mathbf S \rightarrow \frac{\mathbf S}{c}$,
$T^{\mu\nu} \rightarrow \frac{1}{\mu_0} T^{\mu\nu}$ and
\begin{align}
  \rho &=\frac{1}{2} \left( \varepsilon_0 \mathbf E^2+ \frac{1}{\mu_0} \mathbf B^2 \right) \,,
  \label{emEDSI} \\
  \mathbf S &= \frac{1}{\mu_0} \mathbf E \times \mathbf B \,.
  \label{emMDSI}
\end{align}
This basic results can now be compared with the same expressions for
the Proca field, which we review next.

\subsection{The Proca field}

The massive spin-1 field, or \emph{Proca} field, is described by the Lagrangian density
\eqref{emLag} plus a mass term quadratic in the gauge field
\begin{align}
  \mathcal L = - \frac{1}{4} F_{\mu\nu} F^{\mu\nu} -j_\mu A^\mu + \frac{1}{2} m^2 A_\mu A^\mu \,.
  \label{prProcaLag}
\end{align}
The electric and magnetic fields are again defined as in \eqref{defEfromF} and \eqref{defBfromF}.

By the principle of least action we obtain the Proca field equations of motion
\begin{align}
  \partial_\mu F^{\mu\nu} + m^2 A^\nu \, &= \, j^\nu \,.
  \label{prEoM}
\end{align}
In terms of the field $A^\mu$ this reads
\begin{align}
  \left( \Box + m^2 \right) A^\nu = \partial^\nu \partial_\mu A^\mu + j^\nu \,.
  \label{prFieldEq2}
\end{align}
The \emph{inhomogeneous} Proca equations result from writing \eqref{prEoM}
in terms of the electric and magnetic fields, as in the Maxwell case.
In SI units, they are given by
\begin{subequations}
\begin{align}
  \mathbf \nabla \cdot \mathbf E  &= \frac{ \rho } { \varepsilon_0 } - \left(\frac{m c}{\hbar}\right)^2  A^0 \,,
  \label{ProcaEq11} \\
  \mathbf \nabla \times \mathbf B - \varepsilon_0 \mu_0 \, \partial_t \mathbf E &= \mu_0 \mathbf j -  \left(\frac{m c}{\hbar}\right)^2 \mathbf A \,.
  \label{ProcaEq12}
\end{align}
  \label{ProcaEq1}
\end{subequations}
These are the inhomogeneous Maxwell equations \eqref{MaxwellF1} modified by a mass term $\left(\frac{m c}{\hbar}\right)^2$.
Note the appearance of $\hbar$ along with $c$.
As in he massless case, the Bianchi identity \eqref{MaxwellDualDiv} results in
the \emph{homogeneous} Proca equations
\begin{subequations}
  \begin{align}
    \mathbf \nabla \cdot \mathbf B  &= 0 \,,
    \label{ProcaEq21} \\
    \mathbf \nabla \times \mathbf E + \partial_t \mathbf B &= 0 \,.
    \label{ProcaEq22}
  \end{align}
  \label{ProcaEq2}
\end{subequations}
which remain unchanged from the massless case

The addition of a mass term to the spin-1 field Lagrangian \eqref{emLag} breaks the U(1) gauge invariance.
As consequence of this, the current $ j^\nu$ is no longer inevitably conserved.
Taking the four-divergence from \eqref{prEoM}, we get
\begin{align}
m^2 \partial_\nu A^\nu \, &= \,  \partial_\nu j^\nu \,.
  \label{ProcaCurrCons}
\end{align}
This expression does not automatically vanish, in contrast to the massless case,
and one needs to explicitly postulate $\partial_\nu j^\nu \overset{!}{=} 0$
if current conservation is to be maintained.
By demanding current conservation, the massive spin-1 field automatically satisfies the \emph{Lorentz condition}
\begin{align}
  \partial_\nu A^\nu = 0 \,.
  \label{prLorentz}
\end{align}
There is no gauge freedom left as in the Maxwell case, and the Proca field equations become
\begin{align}
  \left( \Box + m^2 \right) A^\mu = j^\mu \,.
  \label{prFieldEq}
\end{align}

\subsubsection{Energy and momentum density}

The energy momentum tensor is given by \eqref{emEMtensor} with an added mass term
\begin{align}
  T^{\mu\nu} &= \frac{1}{4} g^{\mu \nu} F^{\alpha\beta} F_{\alpha \beta} + F^\mu{}_\sigma F^{\sigma \nu}
  + g^{\mu\nu} j_\sigma A^\sigma - j^\mu A^\nu \notag \\
  & \qquad + m^2 \left( A^\mu A^\nu - \frac{1}{2} g^{\mu\nu} A_\sigma A^\sigma \right) \,.
  \label{prEMtensor}
\end{align}
Setting $j^\mu = 0$ we obtain the energy density in vacuum
\begin{align}
  \rho \equiv T^{00} &= \frac{1}{2}\left(\mathbf E^2 + \mathbf B^2\right) + \frac{1}{2} m^2 \left( A_0^2 + \mathbf A^2 \right) \,.
    \label{prED}
\end{align}
Similarly, the \emph{Poynting vector} of the Proca field in vacuum 
can be calculated to be
\begin{align}
  S^i  \, & \equiv \,T^{0i} \,,
  \notag \\
  \mathbf S \, &= \, \mathbf E \times \mathbf B + m^2 A_0 \mathbf A \,.
  \label{prMD}
\end{align}
By converting to SI units, the fields are rescaled as
$\mathbf S \rightarrow \frac{\mathbf S}{c}$,
$\mathbf E \rightarrow \frac{\mathbf E}{c}$, $A^0 \rightarrow \frac{A^0}{c}$,
$T^{\mu\nu} \rightarrow \frac{1}{\mu_0} T^{\mu\nu}$.
The energy and momentum density of the Proca field are given in SI units by
\begin{align}
  \rho &=\frac{1}{2} \left( \varepsilon_0 \mathbf E^2+ \frac{1}{\mu_0} \mathbf B^2 \right)
  +\frac{1}{2} \left(\frac{m c}{\hbar}\right)^2 \left(\varepsilon_0 A_0^2 + \frac{1}{\mu_0} \mathbf A^2 \right)  \,,
  \label{prEDSI} \\
  \mathbf S &= \frac{1}{\mu_0} \left( \mathbf E \times \mathbf B + \left(\frac{m c }{\hbar}\right)^2 A_0\, \mathbf A \right) \,.
  \label{prMDSI}
\end{align}

\section{Plane wave expansion and dispersion relations}
\label{sec:PlaneWaveExpansion}

\subsection{Polarization vectors}

To ease calculations, we will expand the vector field $A^\mu$ in its Fourier modes,
\begin{align}
  A _{k, \lambda}{}^\mu(x) = N_k \, \epsilon_{k, \lambda}{}^\mu \, e^{-i k x} \,,
  \label{peFE}
\end{align}
with $\epsilon_{k, \lambda}{}^\mu$ a set of four 4-dimensional \emph{polarization vectors}
labeled by $\lambda$.
Demand that the polarization vectors $\epsilon_{k, \lambda}{}^\mu$ build a 4-dimensional orthonormal system
\begin{align}
  \epsilon_{k, \lambda}{}^\mu \epsilon_{k, \lambda' \mu} = g_{\lambda \lambda'} \,.
  \label{emPolVec1}
\end{align}
Following \cite{greiner1996field}, we may define a set of polarization vectors obeying \eqref{emPolVec1}
for the Maxwell and Proca fields.

\subsubsection{Polarization vectors for the massless field}

In the Lorentz frame where the plane wave has 3-momentum $\mathbf k$,
we may pick two space-like \emph{transverse polarization vectors}
\begin{subequations}
\begin{align}
  \epsilon_{k, 1} = ( 0, \vec{\epsilon}_{k, 1}) \,,  \\
  \epsilon_{k, 2} = ( 0, \vec{\epsilon}_{k, 2}) \,,
\end{align}
  \label{emDefPol12}
\end{subequations}
such that for $i,j \in \{1,2\}$ it holds
\begin{subequations}
\begin{align}
  \mathbf k \cdot \vec{\epsilon}_{k, i} &= 0 \,,
  \\
  \vec{\epsilon}_{k, i} \cdot \vec{\epsilon}_{k, j} &= \delta_{ij} \,.
\end{align}
\label{emPol12}
\end{subequations}
For the \emph{longitudinal polarization} $\epsilon_{k, 3}$ a covariant expression can be written
by first defining the time-like unit vector $n$ with $n^2 = 1$ and $n=(1,\vec 0)$ in the local frame.
Then
\begin{align}
  \epsilon_{k, 3} &= \frac{k - n \left( k \cdot n \right)}{\sqrt{\left( k \cdot n \right)^2 - k^2}} \,.
  \label{emPol3}
\end{align}
In the local frame therefore $\epsilon_{k, 3} = (0,\frac{\mathbf k}{|\mathbf k|} )$.
The remaining \emph{time-like polarization} degree can be taken to be $n$,
\begin{align}
  \epsilon_{k, 0} &= n \,.
  \label{emPol0}
\end{align}
One can easily check that this set of four polarization vectors satisfy the normalization condition \eqref{emPolVec1}.

\subsubsection{Polarization vectors for the massive field}

Again, working in the Lorentz frame where the plane wave has 3-momentum $\mathbf k$,
we define \emph{transverse polarization} vectors as in the massless case \eqref{emDefPol12}, \eqref{emPol12}.
The third vector may be chosen so that its spatial components are parallel to $\mathbf k$.
Its zero-component can be further specified by the condition that $\epsilon_{k, 3}$ is orthogonal
to the momentum four-vector $k_\mu \epsilon_{k, 3}{ }^\mu = 0$. This completely defines the
\emph{longitudinal polarization vector}
\begin{align}
  \epsilon_{k, 3} = \left( \frac{|\mathbf k|}{m},  \frac{\mathbf k}{|\mathbf k|}  \frac{k^0}{m} \right) \,.
  \label{prPol3}
\end{align}
Note that this definition was impossible in the massless case, where 4-dimensional
transversality would have implied infinite norm.

The fourth \emph{time-like polarization vector} can be defined as
\begin{align}
  \epsilon_{k, 0} = \frac{1}{m} k \,.
  \label{prPol0}
\end{align}
This set of polarization vectors also satisfy condition \eqref{emPolVec1}.

\subsection{Plane wave expansion of the Maxwell field}

In the \emph{Lorentz gauge} $\partial_\mu A^\mu = 0$ the equations of motion \eqref{emFieldEq}
for the Maxwell field $A^\mu$ become \eqref{emFieldEqL}
and the general solution may be given as a sum over polarizations $\lambda$,
which labels the set of four polarization vectors $\epsilon_\lambda{}^\mu$ for the massless case
given in \eqref{emDefPol12}, \eqref{emPol12}, \eqref{emPol0}, \eqref{emPol3}.
\begin{align}
  A^\mu(x) \sim \sum_{\lambda} \sum_k \epsilon_{k, \lambda}{}^\mu e^{-ikx} \Big|_{k^0 = \omega_{\mathbf k}} \,,
    \label{emAsol}
\end{align}
The familiar dispersion relation for the Maxwell field in vacuum can easily be recovered
in the Lorentz gauge $\partial_\mu A\mu =0$ by inserting \eqref{emAsol} into \eqref{emFieldEqL} and taking $j^\mu=0$,
\begin{align}
  \omega_{\mathbf k} \, &= \, | \mathbf k | \,.
  \label{emDispRel}
\end{align}

Consider now the \emph{Coulomb gauge} $\nabla \cdot \mathbf A = 0$. Then \eqref{emFieldEq} becomes
\begin{subequations}
\begin{align}
 \Box \mathbf A + \partial_t \nabla A^0 = 0 \,,
 \\
 \nabla^2 A^0 = 0 \,.
\end{align}
  \label{emFieldEqC}
\end{subequations}
We can take $A^0 = 0$ and expand $\mathbf A$ in its Fourier modes.
Taking the limit
\begin{align}
  \lim_{V\to \infty} \frac{1}{V} \sum_{\mathbf k} \rightarrow \int \frac{\mathrm{d^3} \mathbf k}{\, \left( 2 \pi \right)^3} \,,
  \label{FourierEq}
\end{align}
we can write the sum over all modes $\mathbf k$ as
\begin{align}
  \mathbf A (x) &= \sum_{\lambda} \int \frac{\mathrm{d^3} \mathbf k}{\, \left( 2 \pi \right)^3} \, N_k 
  \left( \vec \epsilon_{k, \lambda} e^{-ikx} + \vec {\epsilon}^*_{k, \lambda} e^{ikx} \right)\Big|_{k^0 = \omega_{ \mathbf k}} \,.
  \label{emFieldFourier}
\end{align}
Since $\nabla \cdot \mathbf A = 0$ we sum only over the transverse (physical) modes $\lambda = 1,2$.
Choose linear polarization so we can take $\epsilon_{k, \lambda}$ real. The fields are then given by
\begin{subequations}
\begin{align}
  \mathbf A (x) &= \sum_{\lambda = 1}^2 \int \frac{\mathrm{d^3} \mathbf k}{\, \left( 2 \pi \right)^3} \, N_k 
  \vec \epsilon_{k, \lambda} \left( e^{-ikx} + e^{ikx} \right)\Big|_{k^0 = \omega_{\mathbf k}} \,,
  \\
  \mathbf E (x) &= \sum_{\lambda = 1 }^2 \int \frac{\mathrm{d^3} \mathbf k}{\, \left( 2 \pi \right)^3} \, N_k 
  \, i \omega_{\mathbf k} \, \vec \epsilon_{k, \lambda} \left( e^{-ikx} - e^{ikx} \right)\Big|_{k^0 = \omega_{\mathbf k}} \,,
  \\
  \mathbf B (x) &= \sum_{\lambda = 1}^2 \int \frac{\mathrm{d^3} \mathbf k}{\, \left( 2 \pi \right)^3} \, N_k 
  \, i \mathbf k \times \vec \epsilon_{k, \lambda} \left( e^{-ikx} - e^{ikx} \right)\Big|_{k^0 = \omega_{\mathbf k}} \,.
\end{align}
    \label{emCoulFields}
\end{subequations}
We may now express the total energy $Q$ of the Maxwell field as the integral of \eqref{emED} over some
finite volume $V$
\begin{align}
  Q \, &= \, \int_V \mathrm{d^3} \mathbf x \, \rho
  \notag \\
         &= \int_V \mathrm{d^3} \mathbf x \, \frac{1}{2} \left( \mathbf E^2 + \mathbf B^2 \right) \,.
\end{align}
Inserting the Fourier expansion of the fields and taking $V \rightarrow \infty$ we get
\begin{multline}
  Q \, = \, \frac{1}{2} \sum_{\lambda = 1}^2 \sum_{\lambda' = 1 }^2 \int_V \mathrm{d^3} \mathbf x
                        \int \frac{\mathrm{d^3} \mathbf k}{\, \left( 2 \pi \right)^3}
                        \int \frac{\mathrm{d^3} \mathbf q}{\, \left( 2 \pi \right)^3} \, N_k N_q
     \left( \omega_{\mathbf k} \omega_{\mathbf q} \vec \epsilon_{k, \lambda} \cdot \vec \epsilon_{q, \lambda'} +
     (\mathbf k \times \vec \epsilon_{k, \lambda}) \cdot ( \mathbf q \times \vec \epsilon_{q, \lambda'} )
   \right)\\
  \left( e^{ix(k-q)} + e^{-ix(k-q)} - e^{-ix(k+q)} - e^{ix(k+q)} \right) \bigg|_{\substack{k^0 = \omega_{\mathbf k}\\ q^0 = \omega_{\mathbf q}}} \,.
  \notag
\end{multline}
The resulting  delta functions are integrated over $\mathrm{d^3} \mathbf q$. Recalling the vector identity
$(\mathbf a \times \mathbf b) \cdot ( \mathbf c \times \mathbf d ) = (\mathbf a \cdot \mathbf c)( \mathbf b \cdot \mathbf d )
-(\mathbf a \cdot \mathbf d)( \mathbf b \cdot \mathbf c )$, the expression simplifies further to
\begin{multline}
  Q \, = \, \frac{1}{2} \sum_{\lambda = 1 }^2 \sum_{\lambda' = 1}^2 \int \frac{\mathrm{d^3} \mathbf k}{\, \left( 2 \pi \right)^3}
     \left\{ 2 N_k^2
     \left( \omega_{\mathbf k}^2 \vec \epsilon_{k, \lambda} \cdot \vec \epsilon_{k, \lambda'} +
     \mathbf k^2 \epsilon_{k, \lambda} \cdot \vec \epsilon_{k, \lambda'}
     - (\mathbf k \cdot \vec \epsilon_{k, \lambda}) \cdot ( \mathbf k \cdot \vec \epsilon_{k, \lambda'} )
     \right) \right.\\
     -\left.  \left(  e^{-2it\omega_{\mathbf k}} + e^{2it\omega_{\mathbf k}} \right)
    \left( \omega_{\mathbf k}^2 \vec \epsilon_{k, \lambda} \cdot \vec \epsilon_{k, \lambda'} -
     \mathbf k^2 \epsilon_{k, \lambda} \cdot \vec \epsilon_{k, \lambda'}
     + (\mathbf k \cdot \vec \epsilon_{k, \lambda}) \cdot ( \mathbf k \cdot \vec \epsilon_{k, \lambda'} )
     \right)
     \right\} \,.
     \notag
\end{multline}
Remembering the dispersion relation \eqref{emDispRel} and the polarization tensor definition \eqref{emPol12}
\begin{align}
  Q \, &= \, \sum_{\lambda = 1 }^2 \sum_{\lambda' = 1 }^2 \int \frac{\mathrm{d^3} \mathbf k}{\, \left( 2 \pi \right)^3}
          \, 2 N_k^2 \omega_{\mathbf k}^2 \, \vec \epsilon_{k, \lambda} \cdot \vec \epsilon_{k, \lambda'} \,
       = \, \sum_{\lambda = 1 }^2 \sum_{\lambda' = 1}^2 \int \frac{\mathrm{d^3} \mathbf k}{\, \left( 2 \pi \right)^3}
          \, 2 N_k^2 \omega_{\mathbf k}^2 \, \delta_{\lambda \lambda'} \notag \\
       &= \, \sum_{\lambda = 1}^2 \int \frac{\mathrm{d^3} \mathbf k}{\, \left( 2 \pi \right)^3}
          \, \left( 2 N_k^2 \, \omega_{\mathbf k} \right) \, \omega_{\mathbf k} \,.
  \label{emEnergy}
\end{align}
A similar calculation for the total momentum vector results in
\begin{align}
  \mathbf P \, &= \, \sum_{\lambda = 1}^2 \int \frac{\mathrm{d^3} \mathbf k}{\, \left( 2 \pi \right)^3}
  \, \left( 2 N_k^2\, \omega_{\mathbf k} \right) \, \mathbf k \,.
  \label{emP}
\end{align}

\subsection{Plane wave expansion of the Proca field}

The vacuum solutions of the equations of motion for the Proca field \eqref{prFieldEq} may also be
written as plane waves
\begin{align}
  A^\mu (x) &= \sum_{\lambda} \int \frac{\mathrm{d^3} \mathbf k}{\, \left( 2 \pi \right)^3} \, N_k 
  \left( \epsilon_{k, \lambda}{}^\mu e^{-ikx} + {\epsilon}^*_{k, \lambda}{}^\mu e^{ikx} \right)\Big|_{k^0 = \omega_{ \mathbf k}} \,.
  \label{prFieldFourier} 
\end{align}
The dispersion relation for the Proca field of mass $m$ in vacuum results from
replacing \eqref{prFieldFourier} into \eqref{prFieldEq} and taking $j^\mu=0$,
\begin{align}
  \omega_{ \mathbf k} \, &= \, \sqrt{\mathbf{k}^2 + m^2} \,.
  \label{prDispRel}
\end{align}

Recalling the consistency condition $\partial_\mu A^\mu =0$ and again choosing
$\epsilon_{k, \lambda}{}^\mu$ real this can be written as
\begin{align}
   A^0 (x) &= \sum_{\lambda=1}^3 \int \frac{\mathrm{d^3} \mathbf k}{\, \left( 2 \pi \right)^3} \, N_k 
   \frac{1}{\omega_{ \mathbf k}} \mathbf k \cdot \vec \epsilon_{k, \lambda} \left( e^{-ikx} + e^{ikx} \right)\Big|_{k^0 = \omega_{ \mathbf k}} \,,
   \\
  \mathbf A (x) &= \sum_{\lambda=1}^3 \int \frac{\mathrm{d^3} \mathbf k}{\, \left( 2 \pi \right)^3} \, N_k 
  \vec \epsilon_{k, \lambda} \left( e^{-ikx} + e^{ikx} \right)\Big|_{k^0 = \omega_{ \mathbf k}} \,.
  \label{prFieldA}
\end{align}
The electric and magnetic fields expand as
\begin{align}
  \mathbf E (x) &= \sum_{\lambda=1}^3 \int \frac{\mathrm{d^3} \mathbf k}{\, \left( 2 \pi \right)^3} \, N_k 
        i\, \omega_{ \mathbf k} \, \left( \vec \epsilon_{k, \lambda} - \frac{\mathbf k \cdot \vec \epsilon_{k, \lambda}}
                {\omega_{ \mathbf k}^2} \mathbf k \right)
        \left( e^{-ikx} - e^{ikx} \right)\Big|_{k^0 = \omega_{ \mathbf k}} \,,
        \\
  \mathbf B (x) &= \sum_{\lambda=1}^3 \int \frac{\mathrm{d^3} \mathbf k}{\, \left( 2 \pi \right)^3} \, N_k 
        i\, \mathbf k \times \vec \epsilon_{k, \lambda} \left( e^{-ikx} - e^{ikx} \right)\Big|_{k^0 = \omega_{ \mathbf k}} \,.
  \label{prFieldEB}
\end{align}
Then the total energy may be calculated as before from \eqref{prED} to give
\begin{align}
  Q \, &= \, \sum_{\lambda=1}^3 \int \frac{\mathrm{d^3} \mathbf k}{\, \left( 2 \pi \right)^3} \, \left( 2 N_k^2 \, \omega_{ \mathbf k} \right)
                \, \omega_{ \mathbf k} \,.
  \label{prEnergy}
\end{align}
This is the same expression as \eqref{emEnergy} but with an added (longitudinal) polarization.

The total momentum also follows from \eqref{prMD} after a little calculation, given by
\begin{align}
  \mathbf P \, &= \, \sum_{\lambda=1}^3 \int \frac{\mathrm{d^3} \mathbf k}{\, \left( 2 \pi \right)^3} \,
  \left( 2 N_k^2 \, \omega_{ \mathbf k} \right)    \, \mathbf k \,,
  \label{prP}
\end{align}
summing over all 3 modes $\lambda$. Note that the total momentum it is always longitudinal,
even though the gauge field and the electromagnetic fields are not always transverse.

\section{Energy density and Flux}
\label{sec:contEq}

Recall that the \emph{energy-momentum tensor} derived from the invariance of the Lagrangian under
space-time translations obeys the conservation equation
\begin{align}
  \partial_\mu T^{\mu\nu} = 0 \,,
  \label{EMTcon1}
\end{align}
which integrated over a volume $V$ in 3-space becomes
\begin{align}
  0 \, &= \, \int_V \mathrm{d^3} \mathbf x \, \partial_\mu T^{\mu\nu}
  \notag \\
    &= \, \int_V \mathrm{d^3} \mathbf x \, \left( \partial_0 T^{00} + \partial_i T^{0i} \right)
    \notag \\
    &= \, \partial_t \int_V \mathrm{d^3} \mathbf x \,\rho  + \int_V \mathrm{d^3} \mathbf x \, \nabla \cdot \mathbf S \,.
    \notag
\end{align}
Applying the known Gauss theorem we recover the familiar \emph{equation of continuity}
\begin{align}
   \partial_t \int_V \mathrm{d^3} \mathbf x \,\rho  + \oint_{\partial V} \mathrm{d} \mathbf \, \sigma \cdot \mathbf S \, &= \, 0\,,
  \label{EMTcon}
\end{align}
which is valid for both Maxwell and Proca fields.

\section{Transversal solutions for the Proca field}
\label{sec:transproca}
It will become necessary for us to ignore the longitudinal modes of the Proca Equation
since those modes are unexcitable in the thermal Yang Mills theory in the deconfining phase
we reviewed in chapter \ref{chap:YM}. From \eqref{prFieldEq} it is clear that we can choose
transverse-only solutions for the sourceless case so long as we respect the condition
$\partial_\mu A^\mu = 0$. Choosing $A^0 = 0$ immediately imposes the Coulomb "gauge"
$\nabla \cdot \mathbf A = 0$, making the gauge field transverse. The electric field
also recovers transversality since $\nabla \cdot \mathbf E = -m^2 A^0 = 0$. The degrees
of freedom are now restricted to just two, and we recover all
the same results from the Maxwell theory in Coulomb gauge but with a modified
dispersion relation $\omega_{\mathbf k} = \sqrt{\mathbf k^2 + m^2}$
\begin{align}
  Q \, &= \, \sum_{\lambda=1}^2 \int \frac{\mathrm{d^3} \mathbf k}{\, \left( 2 \pi \right)^3} \, \left( 2 N_k^2 \, \omega_{ \mathbf k} \right)
                \, \omega_{ \mathbf k} \,,
  \label{prEnergyT} \\
  \mathbf P \, &= \, \sum_{\lambda=1}^2 \int \frac{\mathrm{d^3} \mathbf k}{\, \left( 2 \pi \right)^3} \,
  \left( 2 N_k^2 \, \omega_{ \mathbf k} \right)    \, \mathbf k \,.
  \label{prPT}
\end{align}

Consider now a single mode $\mathbf A = \mathbf A_{\mathbf k} e^{-ikx} |_{k^0 = \omega_{\mathbf k}}$.
The electric and magnetic fields are given by
\begin{align}
  \mathbf E_{\mathbf k} &= i \omega_{\mathbf k} \mathbf A_{\mathbf k} \,,
  \\
  \mathbf B_{\mathbf k} &= i \mathbf k \times \mathbf A_{\mathbf k} \,.
  \label{transProca}
\end{align}
The \emph{time averaged Poynting vector} of this mode is given by
\begin{align}
  \mathbf S_{\mathbf k} &= \frac{1}{2} \mathrm{Re}\,\left( \mathbf E_{\mathbf k} \times \mathbf B_{\mathbf k}^* \right)
  = \frac{1}{2} \omega_{\mathbf k} \left( \mathbf A_{\mathbf k}^2 \mathbf k - 
        \mathbf A_{\mathbf k} \cdot \mathbf k \mathbf A_{\mathbf k} \right) \notag \\
        &= \frac{1}{2} \omega_{\mathbf k} \left| \mathbf A_{\mathbf k} \right|^2 \mathbf k \,,
        \\
        \left| \mathbf S_{\mathbf k} \right| &=
        \frac{1}{2} \frac{\left|\mathbf k \right|}{ \omega_{\mathbf k}}
        \left| \mathbf E_{\mathbf k} \right|^2 \,.
  \label{transS}
\end{align}
We will use these expressions in our discussions of antennas in chapter \ref{chap:radiometry}.


\chapter[The Fresnel-Kirchhoff Diffraction Integral]{The Fresnel-Kirchhoff Diffraction Integral from the Proca Equations}
\label{app:fresnel}

Here, we sketch a derivation of the \emph{Fresnel-Kirchhoff Diffraction Integral}
for the case of massive spin-1 particles.
This integral formula is needed to calculate
the effective antenna area, see section \ref{sec:antennaeffarea}.
We follow the presentation given in \cite{silver1984microwave,ulaby1981microwave},
but accounting for nonzero mass.

\section{Helmholtz equations}

Take the curl of \eqref{ProcaEq22} and eliminate $\mathbf B$ with \eqref{ProcaEq12} to
get an expression containing only $\mathbf E$. Do likewise for $\mathbf B$ with \eqref{ProcaEq12},
we arrive at
\begin{subequations}
  \begin{align}
    \nabla \times \nabla \times \mathbf E + \left( \partial_t^2 + m^2 \right) \mathbf E  
        &= - \partial_t \mathbf j + \nabla \left( \nabla \cdot \mathbf E \right) \,,
        \\
    \nabla \times \nabla \times \mathbf B + \left( \partial_t^2 + m^2 \right) \mathbf B &= \nabla \times \mathbf j \,.
  \end{align}
  \label{HelmholtzEq1}
\end{subequations}
Using the identity $\nabla \times \nabla \times \mathbf V = \nabla \left( \nabla \cdot \mathbf V \right) - \nabla^2 \mathbf V$
and recalling $\nabla \cdot \mathbf B = 0$ we get finally
\begin{subequations}
  \begin{align}
    \left( \nabla^2 -\partial_t^2 - m^2 \right) \mathbf E &= \partial_t \mathbf j \,,
    \\
    \left( \nabla^2 -\partial_t^2 - m^2 \right) \mathbf B &= -\nabla \times \mathbf j \,.
  \end{align}
  \label{HelmholtzEq2}
\end{subequations}
Consider now the Fourier expansion of the components $E_i$, $B_i$.
It is clear that, in the absence of charges, each such Fourier mode $E_{\mathbf k}$, $B_{\mathbf k}$
obeys the \emph{homogeneous Helmholtz Equations}
\begin{subequations}
  \begin{align}
    \left( \nabla^2 + \mathbf k^2 \right) E_{\mathbf k} &= 0 \,,
    \\
    \left( \nabla^2 + \mathbf k^2 \right) B_{\mathbf k} &= 0 \,,
  \end{align}
  \label{HelmholtzEq}
\end{subequations}
with \emph{dispersion relation}
\begin{align}
    \omega_{ \mathbf k} \, &= \, \sqrt{\mathbf{k}^2 + m^2} \,.
    \tag{\ref{prDispRel}}
    \label{FKdisp}
\end{align}
We have omitted the component indices for clarity.

Finally, recall that the \emph{Green's function} $G_{\mathbf k} (x,x_0 )$ to the Helmholtz Equation
\begin{align}
  \left( \nabla^2 + \mathbf k^2 \right) G_{\mathbf k}\left(x,x_0  \right) = - 4 \pi \delta\left( x_0 \right) \,,
  \label{GK1}
\end{align}
is given by the spherical wave $G_{\mathbf k}$
expanding about a point $\mathbf x_0$
\begin{align}
  G_{\mathbf k} \left(x,x_0  \right) \equiv \frac{e^{-i|\mathbf k||x - x_0|}}{|x - x_0|} \,.
  \label{defGK}
\end{align}

\section{Integral theorem of Helmholtz and Kirchhofff}

We now investigate the problem of determining the field Vectors $\mathbf E$, $\mathbf B$ at a
point $x_0$, given the values of the electric and magnetic field vectors over an equiphase surface S.

Recall \emph{Green's theorem}: Let two scalar functions $F$, $G$ be defined over some volume V,
further let them have continuous first and second derivatives in V.
Then $F$, $G$ satisfy the following identity
\begin{align}
  \int_V \mathrm{d}V \left( F \nabla^2 G - G \nabla^2 F \right) +
  \oint_{\partial V} d \vec A \left( F \nabla G - G \nabla F \right) = 0 \,.
  \label{ScalarGreenTheorem}
\end{align}

Let $\mathbf E$, $\mathbf B$ be defined over a volume V which contains $x_0$,
and have $u_{\mathbf k}(x)$ stand for any Fourier mode $E_{\mathbf k}$, $B_{\mathbf k}$.
Inserting $u_{\mathbf k}$ and $G_{\mathbf k}$ as scalar potentials in \eqref{ScalarGreenTheorem},
we get\footnote{The function $G_{\mathbf k}(x,x_0 )$
is not continuous at $x_0$. In this particular case, use of Green's theorem may be made rigorous
by a simple limiting process \cite{stratton2007electromagnetic}}
\begin{multline}
  \int_V \mathrm{d}V \left( G_{\mathbf k}\left(x,x_0  \right) \nabla^2 u_{\mathbf k}(x)
  - u_{\mathbf k}(x) \nabla^2 G_{\mathbf k}\left(x,x_0  \right) \right) \\
  =  - \oint_{\partial V} d \vec A \left( G_{\mathbf k}\left(x,x_0  \right) \nabla u_{\mathbf k}(x)
  - u_{\mathbf k}(x) \nabla G_{\mathbf k}\left(x,x_0  \right) \right) \,.
  \label{IT1}
\end{multline}
Using the identities \eqref{HelmholtzEq}, \eqref{GK1}
\begin{align}
  u_{\mathbf k}(x_0) = - \frac{1}{4 \pi} \oint_{\partial V} d \vec A \left( G_{\mathbf k}\left(x,x_0  \right) \nabla u_{\mathbf k}(x)
    - u_{\mathbf k}(x) \nabla G_{\mathbf k}\left(x,x_0  \right) \right) \,.
  \label{ITHK}
\end{align}
The fields at point $x_0$ have thus been expressed as an integral over a boundary surface
$\partial V$ which must contain $x_0$.

In application to antenna radiation problems, the surface $\partial V$ (called antenna \emph{aperture})
for which we know the field distribution (called \emph{illumination}) is not closed.
To calculate anything useful, formula \eqref{ITHK} needs to be generalized to open surfaces.
We expect linear integrals over the open surface boundary to appear as extra terms.
However, it turns out that under still reasonable conditions we may ignore the extra terms
and integrate directly with \eqref{ITHK} over the open surface \cite{silver1984microwave}.
Namely, in the \emph{Frauenhofer (far-field) approximation} the \emph{diffraction field} $u_{\mathbf k}$
over the antenna aperture A is given by
\begin{align}
  u_{\mathbf k}(x_0) = - \frac{1}{4 \pi} \int_{A} d \vec A \left( G_{\mathbf k}\left(x,x_0  \right) \nabla u_{\mathbf k}(x)
    - u_{\mathbf k}(x) \nabla G_{\mathbf k}\left(x,x_0  \right) \right) \,.
  \label{ITHK2}
\end{align}
The Frauenhofer approximation is valid provided the point $x_0$ is far enough from the aperture,
\begin{align}
  R \, \geq \, \frac{1}{\pi} \, d^2\, |\mathbf k| \,,
  \label{defFrh}
\end{align}
where $R$ is the distance from $x_0$ to the aperture surface $A$, $d$ the maximum linear dimension
of the aperture, and $\mathbf k$ the wave vector of the monochromatic wave illuminating the aperture \cite{ulaby1981microwave}.
Expression \eqref{defFrh} defines the \emph{Frauenhofer region}.
Integral expression \eqref{ITHK2} allows the calculation of $u_{\mathbf k}$ in the Frauenhofer region,
\emph{given the values of $u_{\mathbf k}$ at the aperture}. Thus, we may use \eqref{ITHK2}
to solve antenna aperture problems by replacing $u_{\mathbf k}$ with the corresponding
electromagnetic field component $E_{\mathbf k}$, $B_{\mathbf k}$.

Let us now approximate the field illuminating the aperture by using simple monochromatic wave
\begin{align}
  u_{\mathbf k}(x) = A\left( x \right) e^{i k L(x)} \,,
  \label{ApField}
\end{align}
with amplitude $A(x)$ and phase $L(x)$. Then
\begin{align}
  \frac{\partial u_{\mathbf k} }{\partial n} &\equiv \mathbf {\hat n} \cdot \nabla u_{\mathbf k} 
                                = ik u_{\mathbf k} \mathbf {\hat n} \cdot \nabla L
                        + u_{\mathbf k} \frac{1}{A} \frac{\partial A }{\partial n} \notag \\
                        &\approx ik u_{\mathbf k} \mathbf{\hat n} \cdot \nabla L \,,
  \label{ApField1}
\end{align}
where $\mathbf {\hat n}$ is defined as the normal to the surface element in \eqref{ITHK2},
$d \vec A\equiv dA\,\mathbf {\hat n}$.
It can further be shown that $\mathbf s \equiv \nabla L$
is the normal unit vector to the propagation wavefront \cite{silver1984microwave}. 
It defines the direction of propagation of the wave illuminating the aperture A.

The second expression we need is
\begin{align}
  \frac{\partial G_{\mathbf k} }{\partial n} \,&=\, \mathbf n \cdot \nabla G_{\mathbf k}
  \,=\, \left( \frac{1}{r} - ik \right) \mathbf n \cdot \mathbf{\hat r} \frac{a^{ikr}}{r} \,,
  \label{PhiField}
\end{align}
where $\mathbf r \equiv x - x_0 $. Finally, we define $\cos{\theta_r} = \mathbf n \cdot \mathbf{\hat r}$,
$\cos{\theta_s} = \mathbf n\cdot \mathbf s$.
After joining terms we get
\begin{align}
  u_{\mathbf k}\left( x,y,z \right) = \frac{1}{4 \pi} \int_{A} \mathrm{d} x_a \mathrm{d} y_a
  \left[ \left( \frac{1}{r} -i k \right) \cos{\theta_r} -i k \cos{\theta_s} \right]
  \frac{e^{i k r}}{r} u_{\mathbf k} \left( x_a,y_a \right) \,.
  \label{FKIntegral2}
\end{align}

\section{Fresnel-Kirchhoff diffraction integral}

Since our derivation of \eqref{ITHK2} depended only on the field $u_{\mathbf k}$
obeying the Helmholtz equation, we may replace it with any solution of \eqref{HelmholtzEq},
including the vector components of the electric and magnetic fields. We can then write
the \emph{Fresnel-Kirchhoff Diffraction Integral}
\begin{align}
  E_{\mathbf k}\left( x,y,z \right) = \frac{1}{4 \pi} \int_{A} \mathrm{d} x_a \mathrm{d} y_a
  \left[ \left( \frac{1}{r} -i k \right) \cos{\theta_r} -i k \cos{\theta_s} \right]
  \frac{e^{i k r}}{r} E_{a,\mathbf k} \left( x_a,y_a \right) \,.
  \label{FKIntegral}
\end{align}
Here $ E_{a,\mathbf k} \left( x_a,y_a \right)$ is the electric field at the aperture.
In the Frauenhofer region, this expression can be further simplified to give
\begin{align}
  E_{\mathbf k}\left( r, \theta, \phi \right) = \frac{1}{2\pi i}\frac{e^{i k r}}{r} \left| \mathbf k \right|
  \, h_{\mathbf k} \left( \theta, \phi \right) \,,
  \label{EkIntegral}
\end{align}
with $r, \theta, \phi$ polar coordinates defined
with the aperture at the origin and normal to the z-axis \cite{ulaby1981microwave}.
The polar function $h_{\mathbf k} \left( \theta, \phi \right)$ is defined as surface integral over the
illumination at the aperture A
\begin{align}
  h_{\mathbf k} \left( \theta, \phi \right) = \int_{A} \mathrm{d} x_a \mathrm{d} y_a
  e^{-ik \sin{\theta}\left( x_a \cos{\phi} + y_a \sin{\phi} \right)} E_{a,\mathbf k} \left( x_a,y_a \right) \,.
  \label{hIntegral}
\end{align}
In this region, the propagating modes are simply spherical waves, with the Poynting vector
$\mathbf S$ parallel to $ \mathbf k$ and $\mathbf{ \hat r}$, $\mathbf S \parallel \mathbf k \parallel \mathbf{ \hat r}$.
The radial component is then read from \eqref{transS} as
\begin{align}
  S_{\mathbf k}\left( \theta, \phi \right) = \frac{1}{8 \pi^2} \frac{\left| \mathbf k \right|^2}{r^2}
                        \frac{\left|\mathbf k \right|}{ \omega_{\mathbf k}}
                        \left| h_{\mathbf k} \left( \theta, \phi \right)  \right|^2 \,.
  \label{Sradial}
\end{align}


\chapter{Fundamental Radiometric Quantities}
\label{app:defrad}

\begin{center}
\scalebox{1.1}{
  \setlength{\extrarowheight}{5pt}
  \begin{tabular}{lrclll}
    \toprule
    Quantity & \multicolumn{3}{c}{Symbol / Definition} & Units \\
    \midrule
    Radiant Energy                & Q     &   &                                     & J (Joules)\\
    Radiant Energy Density        & $w$   &$=$& $dQ/dV$                             &  J m$^{-3}$   \\
    Radiant Power (Flux)          & $\Phi$&$=$& $dQ/dt$                             &  W (Watt)   \\
    Radiant Exitance              & $M$   &$=$& $d\Phi/dA$                          &  W m$^{-2}$   \\
    Irradiance                    & $E$   &$=$& $d\Phi/dA$                          &  W m$^{-2}$   \\
    Radiant Intensity             & $I$   &$=$& $d\Phi/d\Omega$                     &  W sr$^{-1}$    \\
    Radiance                      & $L$   &$=$& $d^2\Phi/(d\Omega dA \cos{\theta})$ &    W m$^{-2}$ sr$^{-1}$ \\
    Emmisivity                    & $\varepsilon$&$=$& $M/M_{bb}$                             &  -   \\
    \bottomrule
   \end{tabular}
   }
\end{center}

Taken from \cite{grum1979optical}

\cleardoublepage

\phantomsection
\addcontentsline{toc}{chapter}{Bibliography}

\begin{thebibliography}{10}

\bibitem{ali1991hemts}
F.~Ali and A.~K. Gupta.
\newblock {\em {HEMTs and HBTs: devices, fabrication, and circuits}}.
\newblock Artech House, 1991.

\bibitem{artalradiometers}
E.~Artal, B.~Aja, M.~de~la Fuente, N.~Roddis, D.~Kettle, F.~Winder, L.~Pradell,
  and P.~De~Paco.
\newblock {Radiometers at 30 and 44 GHz for the Planck Mission}.
\newblock {\em Conference Proceedings of Microwave Technology and Techniques
  Workshop}, pp.~41--48.

\bibitem{atiyah1979}
M.~Atiyah.
\newblock {\em {Geometry of Yang-Mills fields}}.
\newblock Lezioni fermiane. Scuola normale superiore, 1979.

\bibitem{baltes1972problems}
H.~Baltes and P.~Stettler.
\newblock {Problems and design of black-body references}.
\newblock {\em Infrared detection techniques for space research}, 1972.

\bibitem{Berkshire}
{Berkshire Technologies}.
\newblock Cooled amplifier products.
\newblock \url{http://www.quinstar.com/berkshiretech/cooled.htm}.
\newblock Last accessed February 2011.

\bibitem{bogomol1976stability}
E.~B. Bogomol'nyi.
\newblock {The stability of classical solutions}.
\newblock {\em Sov. J. Nucl. Phys.} 24:449, 1976.

\bibitem{collin96}
R.~Collin.
\newblock {\em {Field Theory of Guided Waves}}.
\newblock IEEE/OUP Series on Electromagnetic Wave Theory. Oxford University
  Press, USA, 1996.

\bibitem{Diakonov2009}
D.~{Diakonov}.
\newblock {Topology and Confinement}.
\newblock {\em Nuclear Physics B - Proceedings Supplements} 195:5--45, 2009.

\bibitem{Diakonov2004}
D.~{Diakonov}, N.~{Gromov}, V.~{Petrov}, and S.~{Slizovskiy}.
\newblock Quantum weights of dyons and of instantons with nontrivial holonomy.
\newblock {\em Phys. Rev. D} 70(3):036003, 2004.

\bibitem{Dicke1946}
R.~H. {Dicke}.
\newblock {The Measurement of Thermal Radiation at Microwave Frequencies}.
\newblock {\em Review of Scientific Instruments} 17:268--275, 1946.

\bibitem{DolanJackiw}
L.~{Dolan} and R.~{Jackiw}.
\newblock Symmetry behavior at finite temperature.
\newblock {\em Phys. Rev. D} 9(12):3320--3341, 1974.

\bibitem{CF2010}
C.~{Falquez}, R.~{Hofmann}, and T.~{Baumbach}.
\newblock {Improved ground-state estimate by thermal resummation}.
\newblock 2010, \href{http://arxiv.org/abs/1009.1715}{arXiv:1009.1715}.

\bibitem{falquez2010modification}
C.~Falquez, R.~Hofmann, and T.~Baumbach.
\newblock Modification of black-body radiance at low temperatures and
  frequencies.
\newblock {\em Annalen der Physik} 522(12):904--911, 2010,
  \href{http://arxiv.org/abs/1006.3011}{arXiv:1006.3011}.

\bibitem{Arcade2a}
D.~J. {Fixsen}, A.~{Kogut}, S.~{Levin}, M.~{Limon}, P.~{Lubin}, P.~{Mirel},
  M.~{Seiffert}, J.~{Singal}, E.~{Wollack}, T.~{Villela}, and C.~A. {Wuensche}.
\newblock {ARCADE 2 Measurement of the Extra-Galactic Sky Temperature at 3-90
  GHz}.
\newblock 2009, \href{http://arxiv.org/abs/0901.0555}{arXiv:0901.0555}.

\bibitem{frankel2004geometry}
T.~Frankel.
\newblock {\em {The geometry of physics: an introduction}}.
\newblock Cambridge University Press, 2004.

\bibitem{Gallego2004}
J.~D. {Gallego}, I.~{L{\'o}pez-Fern{\'a}ndez}, C.~{Diez}, and A.~{Barcia}.
\newblock {Experimental results of gain fluctuations and noise in microwave
  low-noise cryogenic amplifiers}.
\newblock {\em Society of Photo-Optical Instrumentation Engineers (SPIE)
  Conference Series}, pp.~402--413, Presented at the Society of Photo-Optical
  Instrumentation Engineers (SPIE) Conference 5470, 2004.

\bibitem{PSA2005}
F.~Giacosa and R.~Hofmann.
\newblock {A Planck-scale axion and SU(2) Yang Mills dynamics: present
  acceleration and the fate of the photon}.
\newblock {\em European Physical Journal C} 50:635--646, 2007,
  \href{http://arxiv.org/abs/hep-th/0512184}{arXiv:hep-th/0512184}.

\bibitem{GiacosaHofmann2007}
F.~{Giacosa} and R.~{Hofmann}.
\newblock {Linear growth of the trace anomaly in Yang-Mills thermodynamics}.
\newblock {\em Phys. Rev. D} 76(8):085022, 2007,
  \href{http://arxiv.org/abs/hep-th/0703127}{arXiv:hep-th/0703127}.

\bibitem{gilmore2008}
R.~{Gilmore}.
\newblock {\em {Lie groups, physics, and geometry: an introduction for
  physicists, engineers and chemists}}.
\newblock Cambridge University Press, 2008.

\bibitem{greiner1996field}
W.~Greiner and J.~Reinhardt.
\newblock {\em {Field Quantization}}.
\newblock Springer, 1996.

\bibitem{GrossYaffe1981}
D.~J. Gross, R.~D. Pisarski, and L.~G. Yaffe.
\newblock {QCD and instantons at finite temperature}.
\newblock {\em Rev. Mod. Phys.} 53(1):43--80, 1981.

\bibitem{grum1979optical}
F.~Grum and R.~Becherer.
\newblock {\em {Optical Radiation Measurements}}.
\newblock Optical Radiant Energy. Academic Press, 1979.

\bibitem{HerbstHofmann2004}
U.~{Herbst} and R.~{Hofmann}.
\newblock Asymptotic freedom and compositeness.
\newblock 2004,
  \href{http://arxiv.org/abs/hep-th/0411214}{arXiv:hep-th/0411214}.

\bibitem{hofmann20062}
R.~{Hofmann}.
\newblock {A strongly interacting SU(2) pure gauge theory and the nature of
  light}.
\newblock 2005,
  \href{http://arxiv.org/abs/hep-ph/0508176}{arXiv:hep-ph/0508176}.

\bibitem{Hofmann2005}
R.~Hofmann.
\newblock Nonperturbative approach to {Yang-Mills thermodynamics}.
\newblock {\em International Journal of Modern Physics A} 20:4123--4216, 2005,
  \href{http://arxiv.org/abs/hep-th/0504064}{arXiv:hep-th/0504064}.

\bibitem{Hofmann2006}
R.~{Hofmann}.
\newblock {Loop expansion in Yang-Mills Thermodynamics}.
\newblock 2006,
  \href{http://arxiv.org/abs/hep-th/0609033}{arXiv:hep-th/0609033}.

\bibitem{RHLeip2007}
R.~{Hofmann}.
\newblock {SU(2) Yang-Mills thermodynamics and photon physics}.
\newblock 2007, \href{http://arxiv.org/abs/0710.1169}{arXiv:0710.1169}.

\bibitem{Hofmann2007fd}
R.~Hofmann.
\newblock {Yang-Mills Thermodynamics}.
\newblock {\em Symmetry, Integrability and Geometry: Methods and Applications}
  3:1--38, 2007, \href{http://arxiv.org/abs/0710.0962}{arXiv:0710.0962}.

\bibitem{Hofmann2007}
R.~{Hofmann}.
\newblock {Yang-Mills Thermodynamics}.
\newblock 2007, \href{http://arxiv.org/abs/0710.0962}{arXiv:0710.0962}.

\bibitem{RH2009}
R.~Hofmann.
\newblock {Low-frequency line temperatures of the CMB (Cosmic Microwave
  Background)}.
\newblock {\em Annalen der Physik} 18(9):634--639, 2009,
  \href{http://arxiv.org/abs/0902.2700}{arXiv:0902.2700}.

\bibitem{Hofmann2009}
R.~{Hofmann}.
\newblock {Onset of magnetic monopole-antimonopole condensation}.
\newblock 2009, \href{http://arxiv.org/abs/0908.4027}{arXiv:0908.4027}.

\bibitem{hofmann2011thermodynamics}
R.~Hofmann.
\newblock {\em {The Thermodynamics of Quantum Yang-Mills Theory: Theory and
  Applications}}.
\newblock World Scientific Publishing Company, 2011.

\bibitem{Jarosik2000}
N.~Jarosik.
\newblock {The Use of Cryogenic HEMT Amplifiers in Wide Band Radiometers}.
\newblock {\em {Gallium Arsenide applications symposium. GAAS 2000, 2-6 october
  2000, Paris.}}, 2000.

\bibitem{jarosik2003design}
N.~Jarosik, C.~Bennett, M.~Halpern, G.~Hinshaw, A.~Kogut, M.~Limon, S.~Meyer,
  L.~Page, M.~Pospieszalski, D.~Spergel, et~al.
\newblock {Design, Implementation, and Testing of the Microwave Anisotropy
  Probe Radiometers}.
\newblock {\em The Astrophysical Journal Supplement Series} 145:413, 2003.

\bibitem{Kapusta}
J.~Kapusta and C.~Gale.
\newblock {\em {Finite-temperature Field Theory: Principles and Applications}}.
\newblock Cambridge monographs on mechanics and applied mathematics. Cambridge
  University Press, 2006.

\bibitem{KavianiHofmann2007}
D.~{Kaviani} and R.~{Hofmann}.
\newblock {Irreducible Three-Loop Contributions to the Pressure in Yang-Mills
  Thermodynamics}.
\newblock {\em Modern Physics Letters A} 22:2343--2352, 2007,
  \href{http://arxiv.org/abs/0704.3326}{arXiv:0704.3326}.

\bibitem{Kraan19982}
T.~C. {Kraan} and P.~{van Baal}.
\newblock {Exact T-duality between calorons and Taub-NUT spaces}.
\newblock {\em Physics Letters B} 428(3-4):268--276, 1998.

\bibitem{Kraan1998}
T.~C. {Kraan} and P.~{van Baal}.
\newblock Periodic instantons with non-trivial holonomy.
\newblock {\em Nuclear Physics B} 533(1-3):627--659, 1998.

\bibitem{krall1986principles}
N.~Krall and A.~Trivelpiece.
\newblock {\em {Principles of plasma physics}}.
\newblock International series in pure and applied physics. San Francisco
  Press, 1986.

\bibitem{kraus2002antennas}
J.~Kraus and R.~Marhefka.
\newblock {\em {Antennas for all applications}}.
\newblock McGraw-Hill series in electrical engineering. McGraw-Hill, 2002.

\bibitem{LandsmanWeert}
N.~{Landsman} and C.~G. {van Weert}.
\newblock Real- and imaginary-time field theory at finite temperature and
  density.
\newblock {\em Physics Reports} 145(3-4):141--249, 1987.

\bibitem{LeBellac}
M.~{Le Bellac}.
\newblock {\em {Thermal Field Theory}}.
\newblock Cambridge monographs on mathematical physics. Cambridge University
  Press, 1996.

\bibitem{LeeLu}
K.~Lee and C.~Lu.
\newblock {SU(2) calorons and magnetic monopoles}.
\newblock {\em Phys. Rev. D} 58(2):025011, 1998.

\bibitem{Lenz05}
F.~{Lenz}.
\newblock {Topological Concepts in Gauge Theories}.
\newblock {\em Topology and Geometry in Physics}, p.~7, Lecture Notes in
  Physics, Berlin Springer Verlag 659, 2005,
  \href{http://arxiv.org/abs/hep-th/0403286}{arXiv:hep-th/0403286}.

\bibitem{hornantennas}
L.~Lucci, G.~Pelosi, S.~Selleri, and R.~Nesti.
\newblock Corrugated horn antennas.
\newblock {\em Encyclopedia of RF and Microwave Engineering}. John Wiley \&
  Sons, Inc., 2005.

\bibitem{LH2008}
J.~Ludescher and R.~Hofmann.
\newblock {Thermal photon dispersion law and modified black-body spectra}.
\newblock {\em Annalen der Physik} 18:271--280, 2009,
  \href{http://arxiv.org/abs/0806.0972}{arXiv:0806.0972}.

\bibitem{HofmannLudescher2010}
J.~Ludescher, J.~Keller, F.~Giacosa, and R.~Hofmann.
\newblock {Spatial Wilson loop in continuum, deconfining SU(2) Yang-Mills
  thermodynamics}.
\newblock {\em Annalen der Physik} 19(1-2):102--120, 2010,
  \href{http://arxiv.org/abs/0812.1858}{arXiv:0812.1858}.

\bibitem{COBE1994}
J.~C. {Mather}, E.~S. {Cheng}, D.~A. {Cottingham}, R.~E. {Eplee}, Jr., D.~J.
  {Fixsen}, T.~{Hewagama}, R.~B. {Isaacman}, K.~A. {Jensen}, S.~S. {Meyer},
  P.~D. {Noerdlinger}, S.~M. {Read}, L.~P. {Rosen}, R.~A. {Shafer}, E.~L.
  {Wright}, C.~L. {Bennett}, N.~W. {Boggess}, M.~G. {Hauser}, T.~{Kelsall},
  S.~H. {Moseley}, Jr., R.~F. {Silverberg}, G.~F. {Smoot}, R.~{Weiss}, and
  D.~T. {Wilkinson}.
\newblock {Measurement of the cosmic microwave background spectrum by the COBE
  FIRAS instrument}.
\newblock {\em {Astrophys. J.}} 420:439--444, 1994.

\bibitem{COBE1990}
J.~C. {Mather}, E.~S. {Cheng}, R.~E. {Eplee}, Jr., R.~B. {Isaacman}, S.~S.
  {Meyer}, R.~A. {Shafer}, R.~{Weiss}, E.~L. {Wright}, C.~L. {Bennett}, N.~W.
  {Boggess}, E.~{Dwek}, S.~{Gulkis}, M.~G. {Hauser}, M.~{Janssen},
  T.~{Kelsall}, P.~M. {Lubin}, S.~H. {Moseley}, Jr., T.~L. {Murdock}, R.~F.
  {Silverberg}, G.~F. {Smoot}, and D.~T. {Wilkinson}.
\newblock {A preliminary measurement of the cosmic microwave background
  spectrum by the Cosmic Background Explorer (COBE) satellite}.
\newblock {\em {Astrophys. J. L.}} 354:L37--L40, 1990.

\bibitem{moriyasu1983elementary}
K.~Moriyasu.
\newblock {\em {An elementary primer for gauge theory}}.
\newblock World Scientific, 1983.

\bibitem{Nahm1983-5}
W.~{Nahm}.
\newblock {Self-dual monopoles and calorons}.
\newblock {\em Trieste Group Theor. Method 1983}, p.~189, 1993.

\bibitem{PenziasWilson}
A.~A. {Penzias} and R.~W. {Wilson}.
\newblock {A Measurement of Excess Antenna Temperature at 4080 Mc/s.}
\newblock {\em {Astrophys. J.}} 142:419--421, 1965.

\bibitem{peskin1995introduction}
M.~Peskin and D.~Schroeder.
\newblock {\em {An introduction to Quantum Field Theory}}.
\newblock Addison-Wesley Pub. Co., 1995.

\bibitem{Planck1901}
M.~Planck.
\newblock {\"Uber das Gesetz der Energieverteilung im Normalspectrum}.
\newblock {\em Annalen der Physik} 309(3):553--563, 1901.

\bibitem{Polyakov1977}
A.~M. Polyakov.
\newblock {Quark confinement and topology of gauge theories}.
\newblock {\em Nuclear Physics B} 120(3):429--458, 1977.

\bibitem{pospieszalski2005extremely}
M.~Pospieszalski.
\newblock {Extremely low-noise amplification with cryogenic FETs and HFETs:
  1970-2004}.
\newblock {\em Microwave Magazine, IEEE} 6(3):62--75, 2005.

\bibitem{ryder1996quantum}
L.~{Ryder}.
\newblock {\em {Quantum Field Theory}}.
\newblock Cambridge University Press, 1996.

\bibitem{JHEP2007}
M.~Schwarz, R.~Hofmann, and F.~Giacosa.
\newblock Gap in the black-body spectrum at low temperature.
\newblock {\em Journal of High Energy Physics} 2:91, 2007,
  \href{http://arxiv.org/abs/hep-th/0603174}{arXiv:hep-th/0603174}.

\bibitem{SHG2007}
M.~{Schwarz}, R.~{Hofmann}, and F.~{Giacosa}.
\newblock {Radiative Corrections to the Pressure and the One-Loop Polarization
  Tensor of Massless Modes in SU(2) Yang-Mills Thermodynamics}.
\newblock {\em International Journal of Modern Physics A} 22:1213--1237, 2007,
  \href{http://arxiv.org/abs/hep-th/0603078}{arXiv:hep-th/0603078}.

\bibitem{silver1984microwave}
S.~Silver, editor.
\newblock {\em {Microwave Antenna Theory and Design}}.
\newblock McGraw-Hill, 1949.

\bibitem{Arcade2}
J.~{Singal}, D.~J. {Fixsen}, A.~{Kogut}, S.~{Levin}, M.~{Limon}, P.~{Lubin},
  P.~{Mirel}, M.~{Seiffert}, T.~{Villela}, E.~{Wollack}, and C.~A. {Wuensche}.
\newblock {The ARCADE 2 Instrument}.
\newblock {\em Astrophys.~J.} 730:138, 2011,
  \href{http://arxiv.org/abs/0901.0546}{arXiv:0901.0546}.

\bibitem{singal2005design}
J.~Singal, E.~Wollack, A.~Kogut, M.~Limon, P.~Mirel, P.~Lubin, and M.~Seiffert.
\newblock {Design and performance of sliced-aperture corrugated feed horn
  antennas}.
\newblock {\em Review of Scientific Instruments} 76:124703, 2005.

\bibitem{skou2006microwave}
N.~Skou and D.~Vine.
\newblock {\em {Microwave radiometer systems: design and analysis}}.
\newblock Artech House, 2006.

\bibitem{smithstathemt96}
P.~M. Smith.
\newblock {Status of InP HEMT technology for microwave receiver applications}.
\newblock {\em Microwave Theory and Techniques, IEEE Transactions on}
  44(12):2328--2333, 1996.

\bibitem{stratton2007electromagnetic}
J.~A. Stratton.
\newblock {\em {Electromagnetic theory}}.
\newblock IEEE Press series on electromagnetic wave theory. John Wiley \& Sons,
  Inc., 2007.

\bibitem{Tada2006}
M.~Tada et~al.
\newblock {Single-photon detection of microwave blackbody radiations in a
  low-temperature resonant-cavity with high Rydberg atoms}.
\newblock {\em Physics Letters A} 349(6):488--493, 2006.

\bibitem{ulaby1981microwave}
F.~T. Ulaby, R.~K. Moore, and A.~K. Fung.
\newblock {\em {Microwave Remote Sensing: Microwave remote sensing fundamentals
  and radiometry}}.
\newblock Microwave remote sensing. Addison-Wesley Pub. Co., Advanced Book
  Program/World Science Division, 1981.

\bibitem{weinberg2008cosmology}
S.~Weinberg.
\newblock {\em {Cosmology}}.
\newblock Oxford University Press, 2008.

\bibitem{weinreb1998noise}
S.~Weinreb.
\newblock {Noise temperature estimates for a next generation very large
  microwave array}.
\newblock {\em IEEE MTT-S Symposium Digest}, vol.~2, pp.~673--676, 1998.

\bibitem{wilson2009tools}
T.~Wilson, K.~Rohlfs, and S.~H\"uttemeister.
\newblock {\em {Tools of Radio Astronomy}}.
\newblock Astronomy and astrophysics library. Springer, 2009.

\bibitem{YangMills54}
C.~N. {Yang} and R.~L. {Mills}.
\newblock {Conservation of Isotopic Spin and Isotopic Gauge Invariance}.
\newblock {\em Phys. Rev.} 96(1):191--195, 1954.

\end{thebibliography}

\cleardoublepage

\chapter*{Acknowledgments}

\thispagestyle{empty}

I would like to manifest my deepest gratitude towards Prof.~Dr.~Tilo Baumbach for giving me the opportunity
of working at the Laboratorium f\"ur Applikationen der Synchrotronstrahlung (LAS)
in this very interesting subject
and for providing the financial means to visit the Physikalisch-Technische Bundesanstalt (PTB) in Braunschweig.

Likewise, I am particularly grateful to Priv.~Doz.~Dr.~Ralf Hofmann, who introduced me to the subject of this thesis.
Dr. Hofmann is an exemplary mentor and teacher, without whose invaluable help and encouragement this work would not have been possible.

We would also like to thank Dr.~J\"org Hollandt and Dr.~Gerhard Ulm
of PTB in Braunschweig for an unforgettable tour through the facilities
which once housed the turn-of-the-century black-body radiation measuring experiments
stimulating the discovery of Planck's famous formula,
and for kindly answering many questions concerning black-body experimental techniques.

We are very grateful to Dr.~Dale Fixsen and Dr.~Alan Kogut from the ARCADE collaboration
for their generous advice regarding radiometric measurement techniques.

Many thanks also to Josef Ludescher for providing the basic Mathematica code
to calculate the transverse component of the polarization tensor,
to Dr.~Markus Schwarz for many helpful and instructive discussions during the work of this thesis
and for providing the numerical data used to calculate the temperature-dependence of the effective coupling,
and to Julian Moosmann for proof-reading and correcting an earlier version of this manuscript.

Finally, I would like to thank my family: my father, my mother, my sister and my brother
for their unwavering support during all these years.

\end{document}